\documentclass{aa}
\usepackage[colorlinks=true,allcolors=blue]{hyperref}
\usepackage{xcolor}
\usepackage{graphicx}
\usepackage{txfonts}
\usepackage{colortbl}
\usepackage{multirow}
\usepackage{parskip}

\usepackage{placeins}
\usepackage{afterpage}

\begin{document}

   \title{SFR estimations from $z=0$ to $z=0.9$}

   \subtitle{A comparison of SFR calibrators for star-forming galaxies}

   \author{M. Figueira
          \inst{1,2},
          A. Pollo
          \inst{1,3},
          K. Małek
          \inst{1,4},
          V. Buat
          \inst{4},
          M. Boquien
          \inst{5},
          F. Pistis
          \inst{1},
          L. P. Cassarà
          \inst{6,7},
          D. Vergani,
          \inst{8}
          M. Hamed
          \inst{1}
          \and S. Salim
          \inst{9}
          }

   \institute{National Centre for Nuclear Research, ul. Pasteura 7, 02-093, Warszawa, Poland
   \and Institute of Astronomy, Faculty of Physics, Astronomy and Informatics, Nicolaus Copernicus University, ul. Grudziądzka 5, 87-100 Toruń, Poland
         \and Astronomical Observatory of the Jagiellonian University, Orla 171, 30-244 Kraków, Poland
         \and Aix Marseille Univ, CNRS, CNES, LAM, Marseille, France
         \and Centro de Astronomía (CITEVA), Universidad de Antofagasta, Avenida Angamos 601, Antofagasta, Chile
         \and INAF–IASF Milano, via Bassini 15, 20133, Milano, Italy
         \and Institute for Astronomy, Astrophysics, Space Applications and Remote Sensing, National Observatory of Athens, Penteli,15236 Athens, Greece
         \and INAF – Osservatorio di Astrofisica e Scienza dello Spazio, Via P.Gobetti 93/3, 40129 Bologna, Italy
         \and Department of Astronomy, Indiana University, Bloomington, IN, 47404
             }

   \date{Received 02/07/2021; accepted 04/08/2022}

 
  \abstract
   {The star formation rate (SFR) is a key ingredient for studying the formation and evolution of galaxies. Being able to obtain accurate estimations of the SFR, for a wide range of redshifts, is crucial for building and studying galaxy evolution paths over cosmic time.}
   {Based on a statistical sample of galaxies, the aim of this paper is to constrain a set of SFR calibrators that are able to work in a large redshift range, from $z=0$ to $z=0.9$. Those calibrators will help to homogenize SFR estimations of star-forming galaxies and to remove any possible biases from the study of galaxy evolution.}
   {Using the VIMOS Public Extragalactic Redshift Survey (VIPERS), we estimated a set of SFR based on photometric and spectroscopic data. We used, as estimators, photometric bands from ultraviolet (UV) to mid-infrared (mid-IR), and the spectral lines H$\beta$, [O{\,\sc{ii}}]$\lambda 3727,$ and [O{\,\sc{iii}}]$\lambda 5007$. Assuming a reference SFR obtained from the spectral energy distribution reconstructed with Code Investigating GALaxy Emission (CIGALE), we estimated the reliability of each band as an SFR tracer. We used the \textit{GALEX}-SDSS-WISE Legacy Catalog (GSWLC, $z<0.3$) to trace the dependence of these SFR calibrators with redshift.}
  {The far and near UV (FUV and NUV, respectively), \textit{u}-band and 24-$\mu$m bands, as well as $L_{TIR}$, are found to be good SFR tracers up to $z\sim0.9$ with a strong dependence on the attenuation prescription used for the bluest bands (scatter of SFR of 0.26, 0.14, 0.15, 0.23, and 0.24~dex for VIPERS, and 0.25, 0.24, 0.09, 0.12, and 0.12~dex for GSWLC). The 8-$\mu$m band provides only a rough estimate of the SFR as it depends on metallicity and polycyclic aromatic hydrocarbon (PAH) properties (scatter of 0.23~dex for VIPERS). We estimated the scatter of rest-frame luminosity estimations from CIGALE to be 0.26, 0.14, 0.12, 0.15, and 0.20~dex for FUV, NUV, ugriz, K$_{\mathrm{s}}$, and 8-24~$\mu$m-$L_{\mathrm{TIR}}$. At intermediate redshift, the H$\beta$ line is a reliable SFR tracer (scatter of 0.19~dex) and the [O{\,\sc{ii}}]$\lambda$3727 line gives an equally good estimation when the metallicity from the $R_{23}$ parameter is taken into account (0.17 for VIPERS and 0.20~dex for GSWLC). A calibration based on [O{\,\sc{iii}}] retrieves the SFR only when additional information such as the metallicity or the ionization parameter of galaxies are used (0.26 for VIPERS and 0.20~dex for GSWLC), diminishing its usability as a direct SFR tracer. Based on rest-frame luminosities estimated with CIGALE, we propose our own set of calibrations from FUV, NUV, \textit{u}-band, 8, 24~$\mu$m, $L_{TIR}$, H$\beta$, [O{\,\sc{ii}}], and [O{\,\sc{iii}}].}
   {}

   \keywords{Galaxy: evolution $-$ Galaxy: general $-$ galaxies: photometry $-$ galaxies: star formation $-$ techniques: spectroscopic}

\titlerunning{SFR calibrators from $z=0$ to $z=0.9$}
\authorrunning{Figueira et al.}
\maketitle

%

\section{Introduction}
The formation and evolution of galaxies is a complex process guided by the buildup of stellar mass ($M_{*}$) through the formation of stars. By measuring the star formation rate (SFR), the evolution of galaxies over cosmic time can be quantified through the cosmic star formation history (e.g., \citealt{lil95,mad96,smo09,mad14,dri18}). In other words, the measurement of SFR over a large range of redshift is crucial in order to acquire a broad and accurate understanding of the Universe. Because different surveys are characterized by different sets of observed wavelengths and/or spectroscopic features, it is necessary to be able to estimate the SFR consistently from several bands.\\
Over several decades a variety of methods have been developed to measure the SFR using different bands (e.g., \citealt{sea73,don84,ken98a,ken09,kenn12,dav16,bro17}). The simplest method, and at the same time the most common, assumes that the flux in a certain rest-frame band is related to the flux emitted by high-mass stars, either directly or indirectly. It implies that a simple scaling exists between the luminosity in this band and the SFR, and so far this assumption has been found to work very well. Bands from X-rays to radio, including also spectral lines, have been used and calibrated, especially in the local Universe where a high statistical precision has been reached using the prolific Sloan Digital Sky Survey (SDSS) database.\\

In the local Universe, the H$\alpha$ line ($\lambda$ $\sim656$~nm) is a common SFR tracer (e.g., \citealt{kew04,bri04,sal07,sal16,bro17}) used through the SFR calibrations of \citet{ken83}, \citet{ken94}, and \citet{ken98a}, and is sensitive to timescales of a few Myr ($<20$~Myr, \citealt{cal13,dav16}). This emission line represents the ionized (H{\,\sc{ii}}) regions created through the emission of Lyman continuum photons ($E$ $>$ 13.6~eV) from high-mass stars. In addition to being a relatively direct tracer of high-mass stars, it does not depend on metallicity or star formation history (SFH), and the dust attenuation is relatively low.
However, beyond $z$ $\sim$ 0.5, the H$\alpha$ line is redshifted out of the optical window. While some studies make use of H$\alpha$ above this limit (e.g., \citealt{mcc99,tre02,mai15,vil21}), these samples generally contain a small number of galaxies and have a narrow redshift range. SFR estimations beyond the local Universe require the use of different bands and calibrations, along with a method to estimate the dust attenuation suffered by galaxies.\\ The H$\beta$ line ($\lambda$ $\sim486$~nm), for instance, can be used as an SFR tracer if we assume that it can be related to H$\alpha$ through the case B recombination. It suffers, however, from higher dust attenuation than H$\alpha$ and from stellar absorption. Another spectral line commonly used when H$\alpha$ is not available is the [O{\,\sc{ii}}]$\lambda 3727$ line. This spectral line is less related to the emission of photons from high-mass stars, suffers a high attenuation, and can strongly depend on the ionization parameter and the metallicity. This latter parameter is more difficult to estimate at intermediate redshift and depends on the chosen calibration \citep{kew04}. The [O\,{\sc{iii}}]$\lambda 5007$ line has also been calibrated as an SFR tracer but gives a poor estimation due to its strong dependence on metallicity and the ionization parameter (e.g., \citealt{mou06b,vil21}). Nonetheless, a proper SFR calibration with [O\,{\sc{iii}}]$\lambda 5007$ will become necessary for future observations at high redshift, such as with the \textit{James Webb Space Telescope} (JWST). \\ As spectroscopic observations are more costly than imaging, they are not available for every galaxy. Instead, surveys are generally performed in the continuum and these continuum bands need to be calibrated to give a reliable estimation of the SFR.\\\\
The ultraviolet (UV) continuum represents a direct tracer of the emission of high-mass stars, sensitive to short timescales ($\sim$100~Myr, \citealt{bua15,dav16}). However, the calibration factor linking $L_{UV}$ to the SFR shows a dependence on metallicity, which is expected to be more important at higher metallicity \citep{mad14}. In addition, UV bands suffer from very high dust attenuation, making them particularly sensitive to the prescription used to correct it. The Balmer decrement method, together with an attenuation law (e.g., \citealt{car89,odo94,cal00}), are generally used to correct for dust attenuation through the comparison of the observed H$\alpha$ to H$\beta$ ratio with the dust-free theoretical value \citep{ber36,gro06}. Other methods can be used if those spectroscopic lines are not observed: it is possible to link the attenuation correction to the magnitude in the near-ultraviolet (NUV) and far-ultraviolet (FUV) bands \citep{hao11}, use the UV spectral slope ($\beta_{UV}$, \citealt{meu99}), or simply assume an average correction based on the literature. Thanks to the huge amount of observed galaxies by the UV space telescope \textit{GALEX} \citep{mart05}, several works have calibrated the FUV and NUV bands in the local Universe.\\\\
The \textit{u}-band is less affected by dust attenuation than UV bands but can be contaminated by stellar emission from the old stellar population \citep{hop03,pre09,boq14,dav16}, which becomes more and more important as the redshift decreases. However, unlike the previous bands, \textit{u}-band measurements can easily be obtained from ground-based telescopes, making it a powerful SFR tracer, as it reflects the UV part of galaxies past a certain redshift.\\\\
On the other side of the spectrum, infrared (IR) bands can also be used as SFR tracers. Because the dust is heated by the UV flux from high-mass stars and it reprocesses the emission in the IR, these bands trace, indirectly, high-mass stars, and therefore the SFR, with a higher timescale ($\sim$1~Gyr, \citealt{boi13}) compared to bluer bands. One advantage is that dust attenuation at these wavelengths is not as significant as in the blue part of the spectrum, and therefore the correction can generally be safely ignored.\\\\
At 8~$\mu$m, the flux is dominated by the emission of polycyclic aromatic hydrocarbons (PAHs), which are found around H{\,\sc{ii}} regions. The luminosity at this wavelength roughly correlates with total infrared luminosity ($L_{TIR}$, $8-1000$~$\mu$m) \citep{pop08,nor12} and can be used as an SFR tracer, even though it is known that the metallicity has an impact on the PAH strength, and therefore on the derived calibrations (e.g., \citealt{mad06,dra07,cie14,sch18}). The luminosity at $\sim$20~$\mu$m allows for a robust estimation of $L_{TIR}$ when accounting for the nonlinear behavior (e.g., \citealt{cal10}) at high SFRs. The observed flux is also often used to estimate $L_{TIR}$ based on templates up to $z\sim4$ (e.g., \citealt{cha01,wuy08,mag12,dal14,tal15,boq21}) and is linked to the SFR trough calibrations (e.g., \citealt{ken98a}). We note that the accuracy of $L_{TIR}$ estimation based on templates depends on the redshift and the band used. \citet{elb10} found a scatter of 0.15~dex for $L_{TIR}$ estimation when using the 24-$\mu$m band and \citet{cha01} templates up to $z\sim 1.5$. The overestimation above redshift 1.5 increases up to a factor of seven. Using \citet{mag12} templates, \citet{tal15} found a scatter for $L_{TIR}$ of 0.19~dex for galaxies up to $z\sim3$. The scatter of $L_{TIR}$ and SFR from a single band were estimated by \citet{boq21} for \textit{Spitzer}, Wide-field Infrared Survey Explorer (WISE), JWST, and \textit{Herschel} measurements, and ranges from 0.05 to 0.23~dex for $L_{TIR}$, and 0.13 to 0.24~dex for the SFR.\\
The \textit{Spitzer} satellite enabled a robust calibration of the SFR from the 24-$\mu$m band and the WISE satellite \citep{wri10} increased the number of galaxies observed at 22~$\mu$m\footnote{The effective wavelength of the WISE-4 band is closer to 22.8~$\mu$m (see \citealt{wri10} and \citealt{bro14a,bro14b})} with its all-sky survey, but also at 12~$\mu$m from which the SFR can also be derived (e.g., \citealt{clu14,clu17}). Reliable measurements of $L_{TIR}$ were made possible with the advent of satellites such as the Infrared Space Observatory (ISO) (up to $240$~$\mu$m), \textit{Spitzer} (up to $160$~$\mu$m), AKARI (up to $168$~$\mu$m), and \textit{Herschel} (up to 500~$\mu$m).\\

Because bluer bands represent the emission of high-mass stars and the IR bands represent the dust reprocession, it is possible to use them in combination to obtain the total SFR (e.g., \citealt{bel05,ken09,boq14,cla15}). These composite calibrations are useful as they are able to give an estimation of the SFR without relying on dust attenuation corrections.\\

The fact that H$\alpha$ is more difficult to observe beyond the local Universe represents an obstacle regarding the choice of a reference SFR. Another common method, also used in the local Universe, relies on the reconstruction of the spectral energy distribution (SED) of the galaxies, from which several properties such as the SFR, $M_{*}$, and rest-frame luminosity can be estimated. One of the main advantages of SED fitting over pre-calibrated relations at a given wavelength is that the red shifting of the observed data is naturally accounted for.\\
Such reconstructions are based on stellar population libraries \citep{bru03,mara05}, dust emission templates (e.g., \citealt{cha01,dal02,dra14}), active galactic nucleus (AGN) models \citep{fri06,sta12,sta16}, SFHs (e.g., \citealt{boi03,bua08,cie16,cie17}), attenuation laws (\citealt{cha00,cal00}), initial mass function (IMF, \citealt{sal55,kro02,cha03}), and metallicity. Several codes are available to perform broad-band SED fittings of galaxies, such as LePhare \citep{arn13}, Multi-wavelength Analysis of Galaxy Physical Properties (MAGPHYS, \citealt{dac12}), \textit{hyperz} \citep{bol00,bol10}, or Code Investigating GALaxy Emission (CIGALE, \citealt{bur05,nol09,boq19}). The obvious limitation is that panchromatic data are mandatory in order to accurately reconstruct the SEDs and obtain reliable estimations of the physical properties of galaxies, and a multiwavelength set of data is not always available. However, a lot of effort was made in order to build multiwavelength catalogs such as the \textit{Herschel} extragalactic legacy project (HELP, \citealt{shi21}), a catalog of $\sim$170 million objects over 1270~deg$^2$, based on 51 surveys from optical to the far-infrared (FIR), the \textit{GALEX}-SDSS-WISE legacy catalog (GSWLC, \citealt{sal16,sal18}), containing $\sim$700~000 galaxies at $z<0.3$, the GAMA survey \citep{dri16}, with $\sim$250~000 objects at $z<0.5$, or the Deep Extragalactic VIsible Legacy Survey (DEVILS, \citealt{dav18}), with $\sim$60~000 galaxies at $0.3 < z < 1.0$. The existence of more unique, statistically important multiwavelength catalogs makes possible a new calibration of the true SFR.\\

The goal of this paper is to estimate the SFRs of a sample of star-forming galaxies at intermediate redshifts ($0.5<z<0.9$) using the VIMOS Public Extragalactic Redshift Survey (VIPERS), a spectroscopic statistically significant survey with very rich auxiliary data and for which spectral lines are available. We used indicators known to be accurate in the local Universe and measured their reliability at moderate redshifts. To make a comparison at lower redshifts, we made use of GSWLC ($z<0.3$, \citealt{sal16,sal18}). Based on VIPERS and GSWLC, we derived a new set of calibrations from UV to $L_{TIR}$, and from the spectral lines H$\beta$, [O{\,\sc{ii}}]$\lambda 3727,$ and [O{\,\sc{iii}}]$\lambda 5007$.\\

The paper is organized as follows. In Section~\ref{Sect:sample_presentation}, we describe the data samples used in this study. In Section~\ref{Sect:data_metodology}, we present the criteria applied to select the star-forming samples and the parameters of the CIGALE run used to derive the reference SFR and rest-frame luminosities. The analysis of each SFR tracer is presented in Section~\ref{Sect:Analysis} and the discussion of the results, as well as the derivation of new calibrations, are presented in Section~\ref{Sect:discussion}. We present our conclusions in Section~\ref{Sect:conclusion}. \\Throughout the paper, we used the WMAP7 cosmology \citep{kom11} with $\Omega_M=0.272$, $\Omega_{\Lambda}=0.728$, and $H_{0}=70.4$~km~s$^{-1}$~Mpc$^{-1}$. The SFR was estimated based on the \citet{cha03} IMF. The conversion from the \citet{sal55} and \citet{kro02} IMFs was done by applying a multiplicative factor of 0.64 and 0.94, respectively \citep{mad14}. All magnitudes are given in the AB system \citep{oke74}.

\section{The samples of data}\label{Sect:sample_presentation}

\subsection{The intermediate-redshift sample: VIPERS}

VIPERS is a spectroscopic survey performed with the VIsible Multi-Object Spectrograph (VIMOS, \citealt{lef03}) and is primarily dedicated to the study of large-scale structures of the Universe through redshift space distortions \citep{guz13,gari14,sco18}. Nonetheless, the VIPERS catalog can be equally used in the field of galaxy formation and evolution thanks to its statistically significant spectrophotometric sample of $\sim$10$^{5}$ galaxies. The volume surveyed by VIPERS ($\sim$5$\times 10^{7}$~h$^{3}$~Mpc$^{-3}$) is comparable to 2dFGRS at $z\sim0.1$ \citep{col01,col03}. Additionally, because VIPERS targeted galaxies in the range $0.5<z<1.2$, this survey can be seen as a unique counterpart of SDSS ($z<0.5$, \citealt{yor00,ahn14,bla17}) at intermediate redshifts.\\ 

VIMOS is a 4-channel imaging spectrograph with a 224~arcmin$^{2}$ field of view at the European Southern Observatory's  Very Large Telescope (ESO/VLT). Observations were performed with the multiobjects-spectroscopic mode and the low-resolution red grism, giving a spectral resolution of $R=220$ and spectral coverage from 5500 to 9500~\AA\, in the observed frame.\\ The selection of targets was based on the "T0005" release of the Canada-France-Hawaii Telescope Legacy Survey (CFHTLS, \citealt{gor09}) Wide photometric catalog toward a sub-area of the W1 ($\sim$16~deg$^{2}$) and W4 fields ($\sim$8~deg$^{2}$). To limit the sample to galaxies with $z<1.2$, only sources from the photometric parent catalog having $i_{\mathrm{AB}}<22.5$~mag (after Galactic extinction correction) were selected. To remove galaxies below $z\sim0.5$, a simple but effective color-color selection based on the \textit{g}, \textit{r}, and \textit{i} bands was applied. Stellar decontamination was performed based on the half-light radius and the reconstruction of the spectral energy distribution from CFHTLS \textit{ugriz} bands. In the end, VIPERS galaxies cover a range of $17.5<i_{AB}<22.5$ in magnitude.\\
The efficiency of these selection criteria was tested with the help of two control samples based on previous studies with VIMOS: VVDS-Deep and VVDS-Wide \citep{lef05,gar08} using galaxies observed both in VIPERS and in the control samples. Even though the above selection rules are effective to select intermediate-redshift galaxies from amongst low-redshift galaxies and stars, a perfect sample is not achievable and some contamination was expected. The use of these simple selections, however, allowed the sampling rate to be doubled, reaching $\sim$47\%, compared to a selection based only on magnitude ($\sim$23\% for VVDS-Wide, \citealt{gar08}).\\
Using the EZ code \citep{gar10,gar14}, the observed spectrum was compared to galaxy templates and the contribution of a set of emission lines in order to assign a redshift, which was then independently checked by two members of the VIPERS team. In this work, we used the VIPERS PDR-2\footnote{\url{http://vipers.inaf.it/}} catalog consisting of 91~507 objects (galaxies, AGN, and stars).

\subsection{The low-redshift sample: GSWLC}\label{Subsect:GSWLC_selection}

GSWLC\footnote{\url{https://salims.pages.iu.edu/gswlc/}} \citep{sal16,sal18} is a local galaxy catalog based on the tenth release of SDSS data \citep{ahn14}. Three different catalogs were produced depending on the \textit{GALEX} exposure time (GSWLC-A, -M, and -D for all-sky shallow, medium, and deep exposure time), with a total of 658~911 objects ($\sim$90\% of SDSS DR10 objects) at $0.01<z<0.3$. Photometry was taken from the \textit{GALEX} GR6/7 final release (\citealt{bia14} with additional corrections, see \citealt{sal16}), the 2MASS Extended Source Catalog (XSC, \citealt{jar00}), SDSS DR10, and WISE from the AllWISE Source Catalog and unWISE \citep{lan16}. SDSS and \textit{GALEX} were corrected for Galactic extinction based on \citet{pee13} and \citet{yua13} coefficients. All the details concerning the construction of the catalog can be found in \citet{sal16}.\\

We chose to use GSWLC for consistency because the associated catalog of physical parameters was obtained with the CIGALE code, the same as for VIPERS (Sect.~\ref{Sect:Cigale}), reducing the bias associated with the use of different SED fitting methods. In the second version of the GSWLC catalog, the WISE-4 (or WISE-3 if not available) band from unWISE was used to estimate $L_{TIR}$ from the \citet{cha01} templates, allowing us to better constrain the SED fitting in the IR part of the spectrum. The $L_{TIR}$ was corrected for AGN contamination based on a systematic trend observed for galaxies classified as AGNs in GSWLC-A1, between the SFR estimated from stellar emission and the SFR estimated from WISE-4 (or WISE-3 if not available) observed fluxes \citep{sal18}. By using this new method, referred to as the SED+$L_{TIR}$ fitting where the $L_{TIR}$ acts as an additional constraint, the dust emission is less dependent on the reconstructed stellar emission. \\

We point out here that \citet{cha01} templates were constructed two decades ago and it might be risky to use them in order to estimate $L_{TIR}$. When constructing the catalog, \citet{sal16} checked the consistency of $L_{TIR}$ estimated using the templates of \citet{cha01} with those of \citet{dal02}, finding a marginal difference of 0.01~dex between them. In addition, these estimations were compared with $L_{TIR}$ estimated using \citet{dal02} templates for galaxies belonging to the SDSS Stripe82 \citep{ros02} from which they found a very good agreement, with a mean difference of 0.01~dex and a scatter of 0.07~dex.\\
Estimations of $L_{TIR}$ from 24~$\mu$m using the \citet{cha01} templates are reliable up to $z=1.5$ \citep{elb10} but can be overestimated by a factor of seven above this redshift \citep{nor10} due to the different ISM conditions \citep{mag10}, and above $L_{TIR}\ge 10^{12}$~$L_{\odot}$ \citep{ber13}. Using the \citet{mag12} templates, it is possible to correctly estimate $L_{TIR}$ up to $z\sim3$ (scatter of 0.2~dex, \citealt{tal15}).\\ Because this work focuses on galaxies at $z<1$, there should not be any problem in using the GSWLC, whose SEDs were constrained by $L_{TIR}$ from the \citet{cha01} templates. When estimating $L_{TIR}$ from mid-IR bands in this work, however, we use the new set of templates\footnote{\url{https://salims.pages.iu.edu/bosa/}} from \citet{boq21} (see Appendix.~\ref{Subsect:wise4}). We refer the reader to \citet{boq21} for a more detailed discussion on the comparison with previous templates, and the scatter for $L_{TIR}$ and SFR when estimated using a single band.\\
In this work, we made use of GSWLC-A2.1, an updated version\footnote{This version uses updated SDSS filters while \citet{fuk96} filters were used in the previous ones.} (Salim, private communication) of the current GSWLC-A2 public catalog, with a total of 640~659 objects \citep{sal18}.

\section{Data preparation and methodology}\label{Sect:data_metodology}

\subsection{Selection of sources}

\subsubsection{VIPERS}\label{subsubsect:VIPERS_BPT}

The VIPERS PDR-2 catalog \citep{sco18} contains galaxies, AGNs, and stars, which can be either primary objects (main targets of the survey) or secondary objects (targets falling into main targets slits), each of them having an associated redshift and a redshift flag ($z_{flag}$). The first step of the selection consists in excluding stars from our sample by setting a lower limit on the redshift and broad-band AGNs based on the specific VIPERS $z_{flag}$. In addition, we required a well-defined spectroscopic redshift ($3<z_{flag}<4.5$, confidence level on the redshift estimation at the level of 99\%).
We also chose to include secondary objects in our sample, with the same criteria for the redshift quality, under the condition that these objects fulfill the same constraint used to select the main target galaxies ($17.5<i_{AB}<22.5$~mag).\\

\begin{figure}[t]
\centering
\includegraphics[angle=0,width=0.5\textwidth]{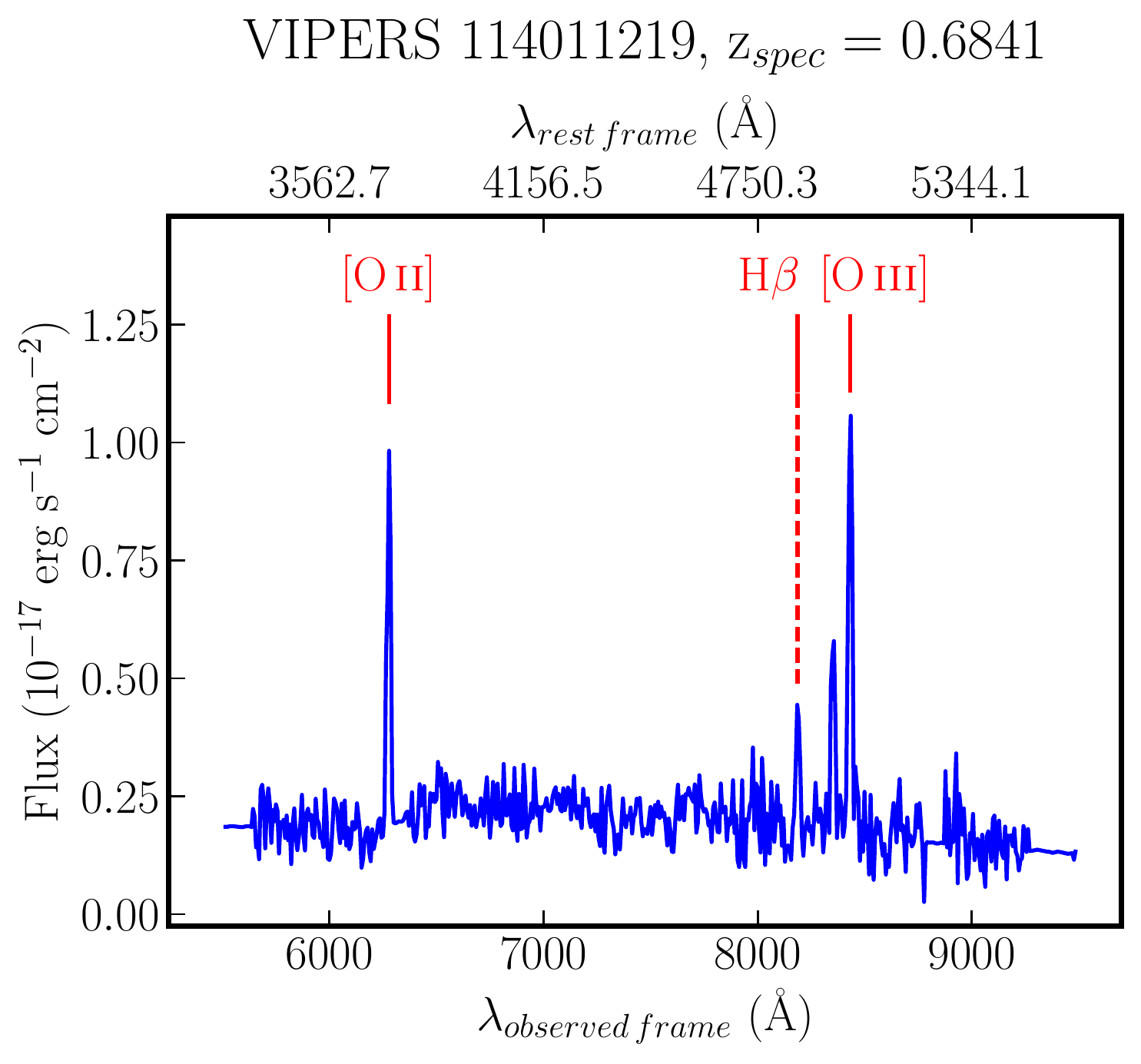}
\caption{Exemplary spectrum of a VIPERS star-forming galaxy based on the BPT diagram, at the median redshift of our sample. Line measurements of all four emission lines ([\ion{O}{ii}]$\lambda$3727, H$\beta,$ and [\ion{O}{iii}]$\lambda\lambda$4959,5007) are considered to be reliable based on the catalog flags (flag 1112, see text).}
\label{Fig:Spectrum_SED}
\end{figure}

The second step of the selection is based on the detection of the lines observed in VIPERS: H$\beta$, [O{\,\sc{ii}}]$\lambda$3727, and [O{\,\sc{iii}}]$\lambda\lambda 4959,5007$ (see Fig.~\ref{Fig:Spectrum_SED} for a typical spectrum of VIPERS\ star-forming galaxies close to the median redshift). In the catalog, the quality of line detection and fitting is characterized by four criteria: the difference between the observed and Gaussian peak should be lower than 7~\AA~(flag $x$), the full width at half maximum should be between 7 and 22~\AA~(flag $y$), the difference between the observed and Gaussian amplitude should be $<30$\% (flag $z$), the signal-to-noise (S/N) ratio of the flux or S/N ratio of the equivalent width (EW) should be higher than seven and three, respectively (flag $t=1$), or 3.5 and eight, respectively (flag $t=2$). The flags are equal to 1 (for $xyz$) if the condition is fulfilled and 0 otherwise.

\noindent In this work, we selected galaxies having flags higher or equal to 1111, considered reliable in VIPERS. Taking $t=1$ instead of $t=2$  increases the size of our sample by 1015 additional galaxies. This selection affects, in particular, the weakest lines [O{\,\sc{iii}}]$\lambda$4959 (699 galaxies with flags 1111) and H$\beta$ (440 galaxies), but much less [O{\,\sc{iii}}]$\lambda$5007 (21 galaxies) and [O{\,\sc{ii}}]$\lambda$3727 (six galaxies).\\ We ended up with a sample of 4795 galaxies with well-defined redshift and lines measurements. The number of primary and secondary galaxies for the W1 and W4 fields after each selection step is shown in Table~\ref{tab:selection}. \\

\begin{table*}[h!]
\caption{Number of VIPERS galaxies after each selection and for both fields.}
\label{tab:selection}      
\centering                          
\begin{tabular}{c|c|c|c|c}
\hline
 \hline
Field & VIPERS Galaxies  & $3.0\le z_{flag}\le 4.5$ & Flag(Line)$\ge$1111  &  \% of total \\
\hline
W1 & 57970 (+287) & 35221 (+152) & 3407 (+11) & 5.8\%\\
W4 & 28112 (+243) & 16301 (+106) & 1371 (+6) & 4.8\%\\
\hline
W1+W4 & 86082 (+530) & 51522 (+258) & 4778 (+17) & 5.5\%\\
\hline
 \hline
\end{tabular}
\tablefoot{The numbers in brackets correspond to the secondary objects observed by VIPERS. The first column describes the field, and the second column shows the number of primary and secondary galaxies on that field. The third column shows the number of galaxies after the cut for the redshift range, while the fourth shows an additional cut for the quality of the line measurements. The last column presents the final percentage of the initial sample used for this work. The initial VIPERS galaxies sample (first column) is defined so that: 1) stars are removed; 2) galaxies have redshift measurements ($1<z_{flag}<29.5$); and 3) the i-band flux of secondary galaxies respects the same constraint as target galaxies ($17.5<i_{\mathrm{AB}}<22.5$). This sample represents all of the potential galaxies that could have ended up in our final sample. See \citet{guz13} and \citet{sco18} for the explanation of the flag system.}
\end{table*}

To select the star-forming galaxies, we used the blue BPT diagram \citep{bal81,lam10} based on the [O{\sc{ii}}]$\lambda3727$, [O{\sc{iii}}]$\lambda5007,$ and H$\beta$ spectral lines. Contrary to the original BPT diagram, based on [N{\sc{ii}}], [O{\sc{iii}}]$\lambda5007$, H$\beta,$ and H$\alpha$, which is insensitive to reddening because these lines are relatively close to each other, the blue BPT diagram is affected by extinction, in particular for the [O{\sc{ii}}]$\lambda3727$/H$\beta$ ratio. We used the equivalent width instead of fluxes of the emission lines (only for the purpose of selecting our sample of star-forming galaxies), which significantly reduced the dependence on reddening \citep{lam10,bong10}. Another effect to take into account is the absorption by the stellar photosphere to the $n=2$ level for Balmer lines \citep{hop03,tre04,dav16}, which reduces the equivalent width of the line by a few \AA. This value depends on the temperature and gravity of the stars in the galaxy and can go up to 15~\AA\, \citep{gro12}. To correct H$\beta$ for Balmer absorption, we assumed a general value of EW$_{c}=2$~\AA~\citep{mil02,got03} (increasing the flux by 20\% in average), our sample being composed of blue galaxies with strong emission lines by construction.\\

\begin{figure*}[t]
\centering
\includegraphics[angle=0,width=0.506\textwidth]{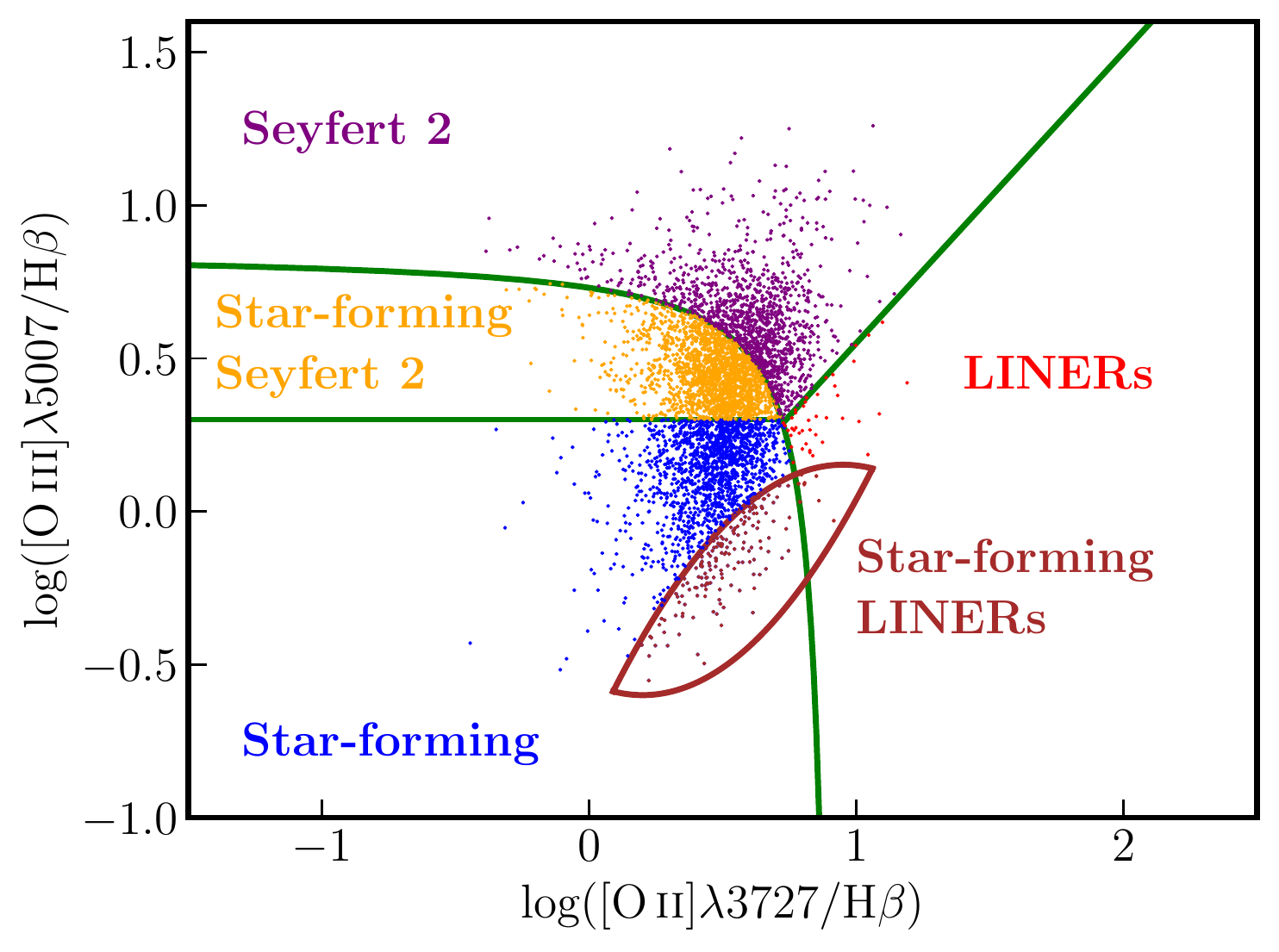}\includegraphics[angle=0,width=0.492\textwidth]{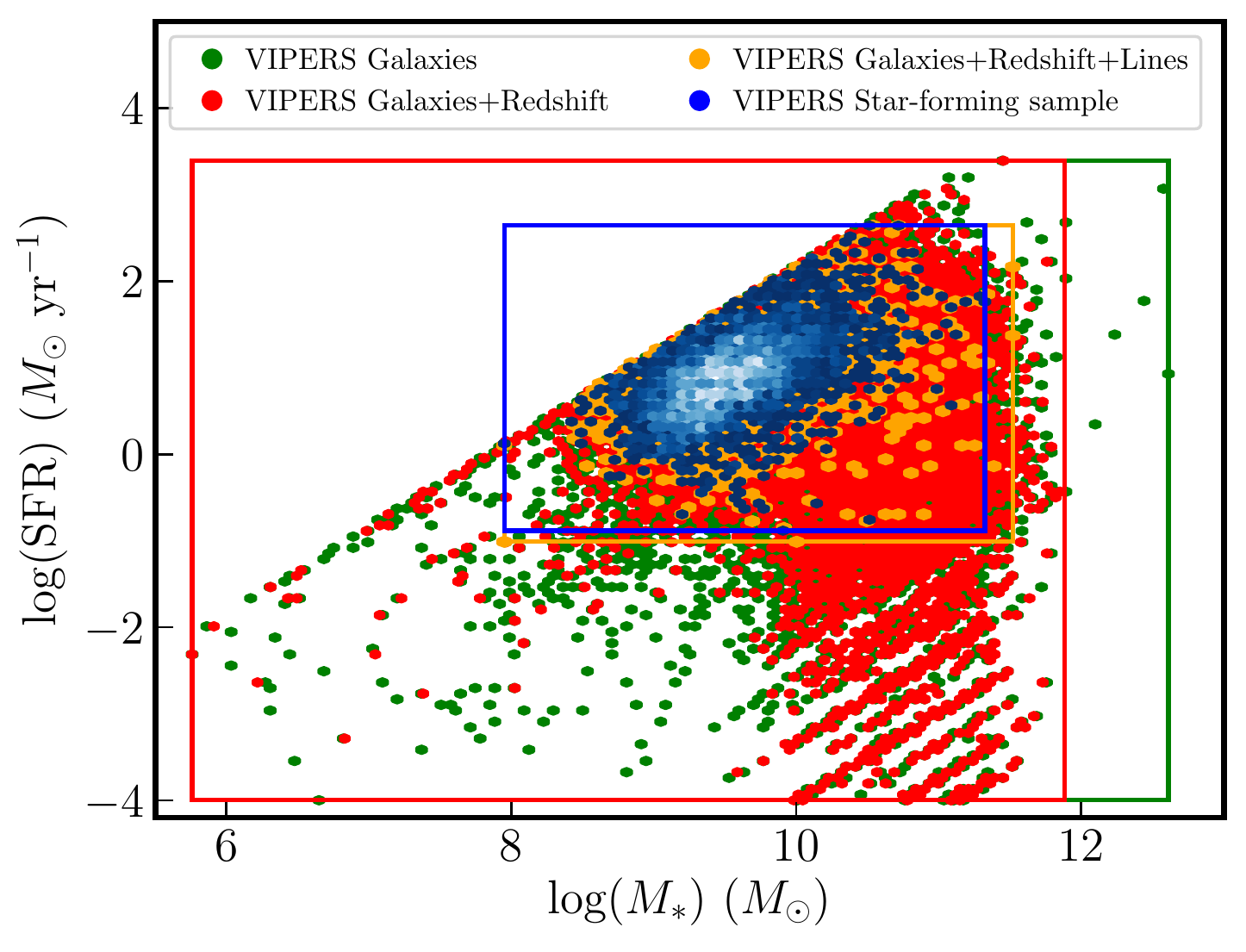}
\caption{Distribution of VIPERS galaxies throughout the selection process. Left panel: BPT diagram for the sample of VIPERS galaxies. Star-forming (blue), Star-forming/Seyfert~2 composite (yellow), and Seyfert~2 (purple) galaxies, and LINERs (red), and star-forming/LINERs composites (brown) are separated by green lines whose equations are given in \citet{lam10}. Right panel: SFR versus $M_{*}$ for VIPERS galaxies after each selection (SFR and $M_{*}$ are estimated using \textit{hyperz} in this plot): all galaxies (green), selection on redshift (red), on lines (orange), and star-forming sample (blue). Visible green, red, and orange points therefore represent galaxies excluded after each selection. Rectangles represent the extent in SFR and $M_{*}$ after each selection.} 
\label{Fig:BPT_VIPERS}
\end{figure*}

Figure~\ref{Fig:BPT_VIPERS} (left) shows that the BPT diagram accounts for two composite regions where star-forming galaxies and Seyfert~2, and star-forming galaxies and low-ionization nuclear emission-line regions (LINERs) are mixed. The star-forming and Seyfert~2 composite region should be dominated by star-forming galaxies (up to 30\% of Seyfert~2 in \citealt{lam10}) so we decided to include this region in our star-forming sample as it increases the number of galaxies by a factor of two. In the same way, we considered the star-forming and LINERs composite region located inside the star-forming region as occupied by star-forming galaxies. The number of galaxies in each category is given in Table~\ref{tab:SFG_selection}.\\

\begin{table}[h]
\caption{Number of VIPERS galaxies in each of the classes defined by the BPT diagram of \citet{lam10}.}            
\label{tab:SFG_selection}      
\centering
\begin{tabular}{c|c}
Class & Number \\
\hline
\hline
All & 4795 \\
\hline
Star-forming & 1633 \\
Star-forming/Seyfert~2 & 1824 \\
Seyfert~2 & 1286 \\
LINERs & 52 \\
Star-forming/LINERs \tablefootmark{$1$} & 176 \\
\hline
\hline
Our work & 3457 \\
\hline
\end{tabular}
\tablefoot{
\tablefoottext{$1$}{The star-forming/LINERs galaxies are classified either as star-forming galaxies or LINERs.}
}
\end{table}

We note that the selection of star-forming galaxies might be different at higher redshifts because of the evolution of the [O{\sc{iii}}]$\lambda5007$/H$\beta$ ratio. An increase in this ratio was observed at $z>1$ (e.g., \citealt{kew13a,hol16}) and confirmed by \citet{ste14} at $z\sim2.3$. The evolution of [O{\sc{iii}}]$\lambda5007$/H$\beta$ between $0.2<z<0.6$ is on the order of 0.2 to 0.3~dex, and is due to an increase in the ionization parameter \citep{kew15}, in agreement with models of ionization parameter and ionization parameter + ISM pressure evolution \citep{cul16}. As a result, some of the star-forming galaxies would be shifted at a higher position in the BPT diagram, leading to a reduction of the star-forming sample and contamination by AGNs. Nevertheless, we are confident as this blue BPT diagram was already used to classify VVDS galaxies up to $z\sim0.8$ \citep{lam09,lam10}.\\

The relations between $M_{*}$ and SFR, the so-called main sequence (MS), in terms of the different selections applied to obtain the VIPERS star-forming galaxies is plotted in Fig.~\ref{Fig:BPT_VIPERS} (right). This is the selection of the lines that significantly affects the distribution of galaxies in the SFR-$M_{*}$ space and biases our sample toward more active galaxies for which log(SFR) $>$ $-$1. The SFR and $M_{*}$ values used in Fig.~\ref{Fig:BPT_VIPERS} (right) are public and were obtained from the SED fitting with \textit{hyperz} \citep{dav13}. Several stripes can be seen, and are particularly visible at low SFR. They are due to the low number of templates used in the SED fitting process, and degeneracies between the SED templates and the physical properties of the galaxies. This leads to a poorer estimation of the physical parameters. For this reason and for consistency with the GSWLC catalog, we reconstructed the SED of galaxies using the CIGALE code (see Sect.~\ref{Sect:Cigale}). Nonetheless, this reconstruction was performed only for the star-forming sample, as it is the main sample studied in this paper.

\subsubsection{GSWLC}

To construct the sample of low-redshift galaxies, we made use of the catalog flags quantifying the quality of the SED, the UV, and MIR data availability, and data the belonging to the SDSS main galaxy sample (MGS). We first selected a subsample of galaxies with good SEDs (no broad-line spectrum, no missing SDSS photometry, and an SED fitting characterized by a reduced $\chi^2$ ($\chi^2_{red}$) less than 30). This high $\chi^2_{red}$ value was chosen since the SFR and $M_{*}$ of galaxies below this limit differ by less than 0.1~dex in comparison with \citet{bri04}. Despite this good agreement, SED with $\chi^2_{red}>5$ should be approached with caution \citep{sal16}. Therefore, we decided to exclude them (5.8\%, 5~655 galaxies) from the final analysis.
Additionally, we selected only galaxies with \textit{GALEX} FUV-NUV measurements, at least one WISE-3/WISE-4 measurement (to better constrain the IR part of the SED, see Sect.~\ref{Subsect:GSWLC_selection}) for which the derived $L_{TIR}$ is not contaminated by AGN, and which are part of the MGS. This selection was made in order to have a very good wavelength coverage and to allow for the best SED fitting and parameter estimations with CIGALE. Unfortunately, it is not possible to obtain such excellent wavelength coverage for VIPERS and to have a significant sample at the same time. It follows that the estimated parameters for VIPERS are more dependent on templates than GSWLC.

\begin{figure*}[t]
\centering
\includegraphics[angle=0,width=0.506\textwidth]{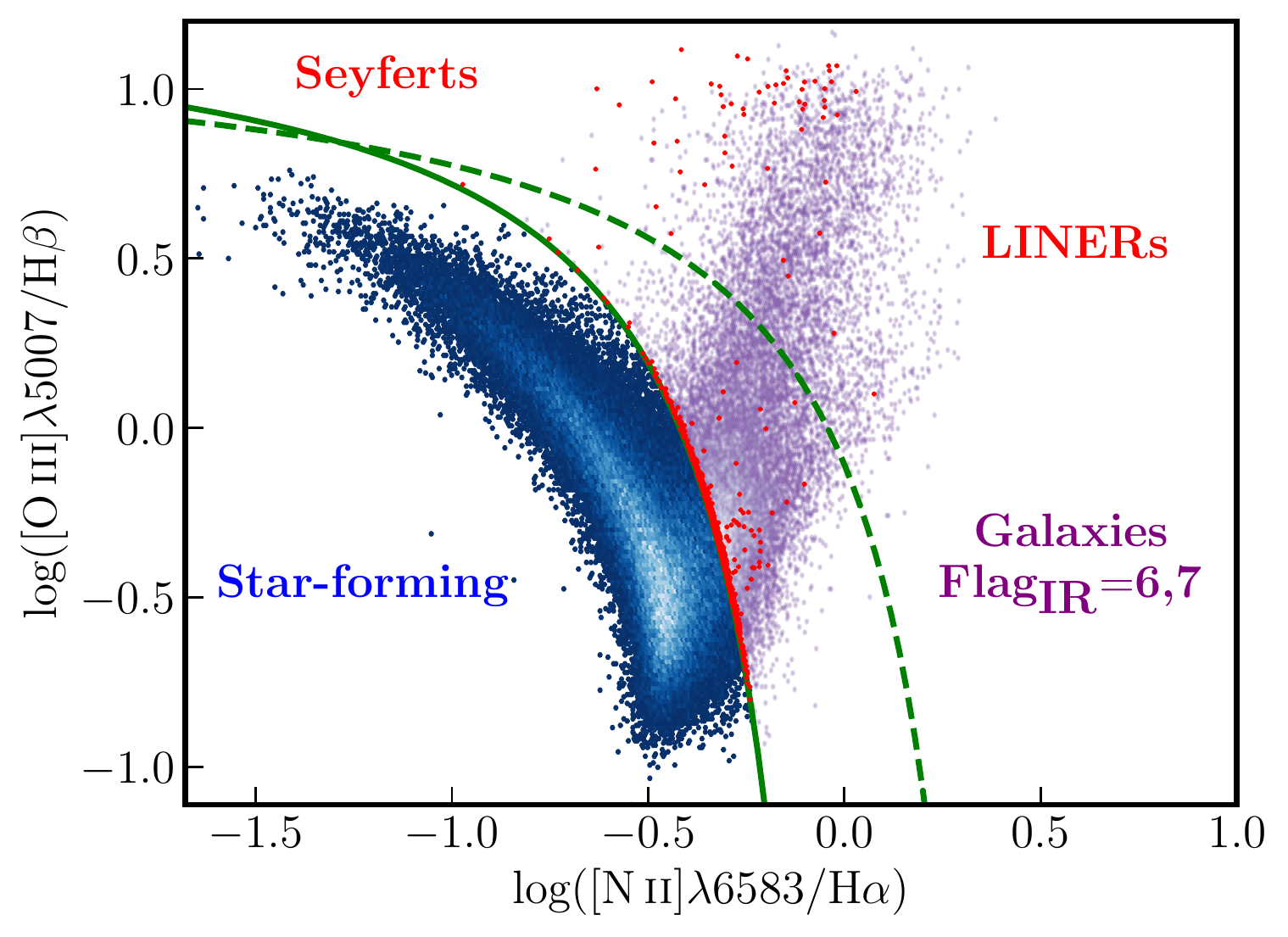}\includegraphics[angle=0,width=0.492\textwidth]{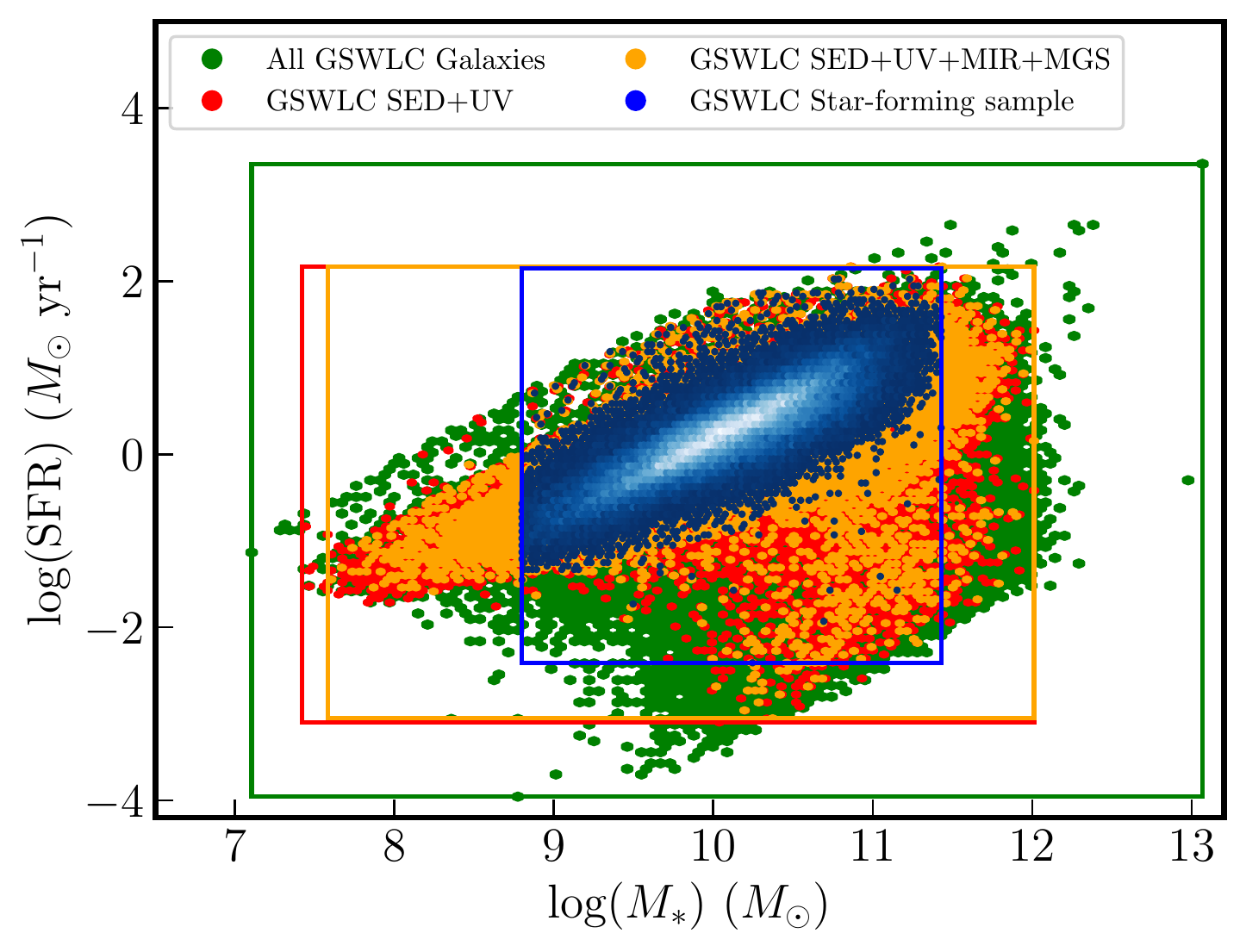}
\caption{Distribution of GSWLC galaxies throughout the selection process. Left panel: BPT diagram for the sample of GSWLC galaxies. The separation between Seyfert and LINERS (red points), and star-forming galaxies (blue to white density plot) is represented by the green curve \citep{kau03}, while the dashed-green line represents the limit from \citet{kew01}. Galaxies for which $L_{TIR}$ was corrected for AGN emission (and previously removed based on the GSWLC flags) are shown for information (purple to white density plot). Right panel: SFR versus $M_{*}$ for GSWLC galaxies for each selection: all GSWLC galaxies (green), selection on the quality of the fits, $\chi^2_r$, and availability of GALEX measurements (red), selection on MIR measurements from which $L_{TIR}$ is not contaminated by the AGN component and belonging to the MGS (orange), and the star-forming sample (blue). Visible green, red, and orange points therefore represent galaxies excluded after each step. Rectangles represent the extent in SFR and $M_{*}$ after each selection.}  
\label{Fig:BPT_GSWLC}
\end{figure*}

\noindent The lines measurements were obtained by cross-matching the GSWLC catalog with the MPA-JHU catalog\footnote{\url{https://wwwmpa.mpa-garching.mpg.de/SDSS/DR7/}}. In total, 99.8\% of the GSWLC subsample has a counterpart in the MPA-JHU catalog. The MPA-JHU catalog already accounts for foreground Galactic reddening corrections using \citet{odo94}.  To create a reliable sample of star-forming galaxies, we followed \citet{tre04} and selected galaxies with a ${\rm{S/N}}>5$ for H$\alpha$, H$\beta$ and [N\,{\sc{ii}}], and a ${\rm{S/N}}>3$ for [O\,{\sc{iii}}]$\lambda$5007 detections. We selected the star-forming galaxies using the BPT diagram \citep{bal81} and the more restrictive limit established by \citet{kau03} (see Fig.~\ref{Fig:BPT_GSWLC} left). From 93~605 galaxies with good lines measurements, 92~696 selected objects are star-forming galaxies and 909 are Seyfert galaxies or LINERs.\\

We note that only a few galaxies are classified as Seyferts or LINERs in the BPT diagram (red dots) and most of them are at the edge of the \citet{kau03} limit. This apparent lack of Seyferts and LINERs comes from the previous step when the selection of galaxies based on the WISE$-$3 and WISE-4 measurements availability was performed: galaxies for which the WISE-based $L_{TIR}$ was corrected for AGN emission (flag 6 and 7 in GSWLC) were already removed. The galaxies close to the \citet{kau03} limit were not removed, probably because they showed no or weak signs of AGN emission. These galaxies, removed in the previous selection step (17~029 galaxies), are shown in Fig.~\ref{Fig:BPT_GSWLC} (purple dots).\\

For consistency, we applied a cut in $M_{*}$ for the GSWLC star-forming sample to match the $M_{*}$ range of the VIPERS sample ($8.8<{\rm{log(}}M_{*}{\rm{)}}<11.4$~$M_{\odot}$). We remind the reader that this range of $M_{*}$ is slightly different compared to what is seen in Fig.~\ref{Fig:BPT_VIPERS} (right) because new $M_{*}$ estimations were obtained using CIGALE.\\
We ended-up with a GSWLC main sample of 91~533 star-forming galaxies. We also defined a [O\,{\sc{ii}}] sample where a selection criteria was also applied to [O\,{\sc{ii}}] (${\rm{S/N}}>5$, 48~845 galaxies, referred as the [O\,{\sc{ii}}] sample) and a second one with an additional selection for [O\,{\sc{iii}}]$\lambda 4959$ (${\rm{S/N}}>3$, 35~456 galaxies, referred as the [O\,{\sc{ii}}]+[O\,{\sc{iii}}] sample). These two subsamples will be used when estimating the SFR from [O\,{\sc{ii}}] and when using the metallicity. The detailed selection is shown in Table~\ref{tab:GSWLC_flags_selection}.\\

\begin{table*}[t]
\caption{Number of GSWLC galaxies after the different selection steps.}            
\label{tab:GSWLC_flags_selection}      
\centering
\begin{tabular}{c||c|c|c|c|c}
\hline
 \hline
Catalog flags &Initial catalog & Good SED ($\chi^2_r<5$)  & \textit{GALEX} data  &  $L_{IR}-$WISE & MGS \\
&640~659 & 595~586 & 209~628 & 154~623 & 149~712 \\
\hline
\hline
\multicolumn{6}{c}{} \\
\hline
\hline
Lines S/N&\multicolumn{3}{c|}{Cross-correlation MPA-JHU} & \multicolumn{2}{c}{S/N selection} \\
&\multicolumn{3}{c|}{149~374} & \multicolumn{2}{c}{93~605} \\
\hline
\hline
\multicolumn{6}{c}{} \\
\hline
\hline
BPT diagram&\multicolumn{3}{c|}{Star-forming galaxies} & \multicolumn{2}{c}{Seyfert - LINERs} \\
&\multicolumn{3}{c|}{92~696} & \multicolumn{2}{c}{909} \\
\hline
\hline
\multicolumn{6}{c}{\hspace{0.7cm} Mass cut ($8.8<\textrm{log(}M_{*}\textrm{)}<11.4$~$M_{\odot}$)} \\
\hline
Final samples&\multicolumn{3}{c|}{Main sample} & \multicolumn{2}{c}{[O\,{\sc{ii}}] sample / [O\,{\sc{ii}}]+[O\,{\sc{iii}}] sample} \\
&\multicolumn{3}{c|}{91~533} & \multicolumn{2}{c}{48~845 / 35~456} \\
\hline
\hline
\end{tabular}
\end{table*}

As for VIPERS, we show the effects of the different selections applied to GSWLC in Fig.~\ref{Fig:BPT_GSWLC} (right). Selection on the FUV and NUV measurements removed part of the red passive galaxies as these fluxes trace the emission from young high-mass stars. This is the selection of star-forming galaxies that removed the major part of the low-SFR galaxies. While we have a comparable mass range, the average SFR of the VIPERS sample is higher than GSWLC, as is expected from younger star-forming galaxies.

\subsection{CIGALE}\label{Sect:Cigale}

As we discussed in the previous section, the public main physical parameters obtained for VIPERS galaxies with \textit{hyperz} suffer from the low number of templates used in the SED reconstruction (see Fig.~\ref{Fig:BPT_VIPERS} right). To improve them, we used CIGALE \citep{boq19}. Based on flux measurements from UV to radio, CIGALE reconstructs the SED of galaxies based on templates built with different modules: the SFH (e.g., exponential, delayed, periodic), the stellar populations library \citep{bru03,mara05}, the dust attenuation model (\citealt{cal00,cha00} or general double power laws, see \citealt{bua11b,lof17} for instance), the dust emission model (e.g., \citealt{dra07,cas12,dal14,dra14,jon17}), the AGN contribution (e.g., \citealt{fri06,bae11,sta12,camp15,sta16}), and the radio emission.\\

Usually, the estimation of physical parameters is obtained from the best SED template. However, galaxies with different properties can have similar SEDs and the best template does not provide information about such uncertainties. To obtain more realistic values of the physical parameters, along with an estimation of the associated uncertainties, CIGALE employs a Bayesian analysis \citep{nol09,boq19}. Instead of taking the best SED only, CIGALE selects a subsample of the best templates, each of them being weighted by their likelihood (which depends on $\chi^2$). The Bayesian parameters and uncertainties are therefore the likelihood-weighted means and likelihood-weighted standard deviations of the parameters, which are more representative of the true nature of the galaxies compared to the parameter derived using the best template only.\\

The SED fitting of galaxies strongly relies on the chosen SFH \citep{pfo12,cas16} and is hard to constrain from broad-band SED fitting, even with good measurements in the UV and optical part of the spectrum. The MS SFH is well defined by a function depending on the seed mass of the galaxy \citep{cie17}. This analytical form is, however, difficult to use as the SFH parameters cannot be reliably estimated using broad-band SED fitting techniques. Modeling the variations of the SFR within the scatter of the MS would require additional free parameters in this analytical form, making it complicated to use in SED modeling. The SFR is, however, relatively well-estimated based on the SFH models used in the literature. \citet{cie17} proposed an SFH composed of a delayed SFH (for $t<t_0$), followed by a recent quench or burst (for $t>t_0$) equal to ${\rm{r}}\times {\rm{SFR}}(t=t_0$), where ${\rm{r}}$ represents the strength of the quench ($r<1$) or burst ($r>1$), giving good estimations of the SFR for several galaxy types at different redshifts. In this paper, we made use of this SFH but we discarded the possibility of a quench as our sample is composed of star-forming galaxies. Moreover, we wanted to keep the process of SED fitting as simple as possible, without modeling possible starburst galaxies during the quenching period, and limited photometric coverage of VIPERS galaxies would not allow for such a detailed study. We underline again that the choice of a particular SFH can have a certain impact on the derived properties from SED fitting, and more explanations about the difference between SFH models used in SED fitting, as well as the construction of this delayed SFH plus a quench or burst, can be found in \citet{cie17}.\\

To take into account the emission lines due to the ionization from high-mass stars radiation, we included the nebular module based on CLOUDY13.01 models \citep{ino11,fer13}, which takes into account 124 lines from He{\,{\sc{ii}}}~$\lambda$30.38~nm to N{\,{\sc{ii}}}~$\lambda$205.4~nm. Nebular emission is important for active star-forming galaxies because it also contributes to the continuum emission, and therefore has a certain impact on the galaxies' SEDs. In this module, we assumed that the Lyman continuum photons are not escaping the galaxy and are not absorbed by the dust.\\

The choice of a dust attenuation law (\citealt{cha00,cal00} and modified versions) has no significant impact on the derivation of the SFR or $L_{TIR}$ \citep{bua11b,lof17,mal18} but can influence the estimation of $M_*$. We chose the \citet{cha00} attenuation law as it is often used in other works and gives good results when compared to radiative-transfer models \citep{bua18}. We used the \citet{bru03} stellar population models and the dust emission templates of \citet{dra07} with updates from \citet{dra14}. As most of the AGNs should be excluded from our sample, we did not include any AGN modules, allowing us to save computational time. The modules and parameters used in CIGALE are shown in Table~\ref{tab:cigale_parameters}. In addition, the rest-frame luminosities, and hence the k-corrected values, were obtained using the \textit{restframe\_parameters} module of CIGALE through a Bayesian approach.\\

The optical data from CFHTLS were cross-matched with \textit{GALEX}, WIRCam-K$_s$, VIDEO-K$_s$ (W1 field only), WISE and \textit{Spitzer} data (W1 field only). A detailed description of the VIPERS multiwavelength catalog construction can be found in \citet{mou16b,mou16a}. Photometric data were corrected from reddening using the \citet{sch98} dust maps. This correction is small as $E(B-V)$ is, on average, equal to 0.03 and 0.05 for the W1 and W4 fields, respectively. In addition, we complemented the catalog with \textit{Herschel} measurements (W1 field only) from the HELP database \citep{shi19,shi21}  using a cross-matching radius of 1$^{\prime\prime}$.\\
We chose to not include the spectral lines' fluxes in the SED reconstruction of the galaxies. In this way, the line measurements remained independent from the CIGALE SFR estimations, contrary to rest-frame luminosity estimations, which depend on the templates used. All measurements located below the Lyman break ($\lambda=912$~\AA) in the rest-frame were discarded as CIGALE is not optimized to handle them: FUV measurements for galaxies at $z>0.698$ (140 galaxies, 4\%) were discarded.\\

\begin{table*}[h]
\small
\caption{Modules and parameters of the CIGALE run for VIPERS galaxies.}            
\label{tab:cigale_parameters}      
\centering                          
\begin{tabular}{c|c}
Parameters\tablefootmark{$1$} & Values \\
\hline
\hline
\multicolumn{1}{c|}{Star formation history: SFH delayed + quench/burst} & \\
\hline
 $\tau_{main}$: e-folding time of the main stellar population  (Gyr) & 4, 5, 6, 7, 8, 9, 10, 13 \\
 $age_{main}$: Age of the main stellar population in the galaxy (Gyr) & 5, 5.5, 6, 6.5, 7  \\
 $age_{burst}$: Age of the burst/quench episode (Myr) & 10, 50, 100, 150, 200, 300, 450  \\
 $r_{SFR}$: Ratio of the SFR after/before $age_{burst}$ & 1, 1.6, 2.1, 2.7, 3.3, 3.9, 4.4, 5 \\
 \hline
\multicolumn{1}{c|}{Stellar population models: \citet{bru03}} & \\
\hline
 IMF & \citet{cha03} \\
 $Z$: metallicity & $Z_{\odot}$ \\
 $age_{sep}$: Age of the separation between the young and the old star populations (Myr) & 10 \\
 \hline
\multicolumn{1}{c|}{Dust attenuation law: \citet{cha00}} &  \\
\hline
 $A_{V}$\_BC: V-band attenuation in the birth clouds (mag) & 0,0.2,0.4,0.6,0.8,1,1.2,1.4,1.6,1.8,2 \\
 BC\_to\_ISM\_factor: $A_V$\_ISM / $A_V$\_BC  & 0.8 \\
 $\alpha_{BC}$: Power law slope of the attenuation in the birth clouds & $-$0.7 \\
 $\alpha_{ISM}$: Power law slope of the attenuation in the ISM & $-$0.7 \\
 \hline
\multicolumn{1}{c|}{Dust emission: \citet{dra07,dra14}} & \\ 
\hline
 $q_{PAH}$: Mass fraction of PAH & 1.77, 2.5, 3.19 \\
 $U_{min}$: Minimum radiation field & 10, 15, 25 \\
 $\alpha$: Powerlaw slope dU/dM propto U$^{{\rm{\alpha}}}$ factor & 2 \\
 $\gamma$: Fraction illuminated from U$_{\textrm{min}}$ to U$_{\textrm{max}}$ & 0.02 \\
 \hline
 \hline
\end{tabular}
\tablefoot{
\tablefoottext{$1$}{See \citet{cie16,boq19}}
}
\end{table*}

In the case of the 351 galaxies for which a counterpart was found in the AllWISE catalog, 33\% of WISE-2, 53\% of WISE-3, and 86\% of WISE-4 detections were upper limits and were considered as such during the SED fitting. In addition, since the WISE-3 and WISE-4 conversion from magnitudes to fluxes were uncertain by 10\% \citep{wri10}, we increased the flux uncertainties by 10\% in the SED fitting process for these two bands.\\

Most of the \textit{Herschel} measurements have a $\textrm{S/N}<3$ but their associated uncertainty is large enough so that these data do not significantly impact the SED reconstruction. The comparison of SFR and $M_{*}$ for a run with and without \textit{Herschel} data is found to be on the order of $\sim0.03$~dex. All the different bands used in the CIGALE run and the number of available measurements are listed in Table~\ref{tab:flux_values}.\\

To check the reliability of the SED fitting of CIGALE, one can refer to the reduced $\chi^2$ associated with each SED (median value of 1.2). However, quantifying the quality of an SED using its $\chi^2$ is not straightforward due to the nonuniform number of bands used in the SED fitting \citep{mal18}.\\
Another way to estimate the reliability of the SED fitting is to create a mock catalog for which the true physical parameters are known and recovered using the same initial configuration (see Table~\ref{tab:cigale_parameters}). The comparison between the recovered parameters and estimated parameters in our CIGALE run characterizes the accuracy of the SED fitting. This process is automatically implemented in CIGALE \citep{gio11,boq19} and the results are presented in Fig.~\ref{Fig:Mock_CIGALE} for $L_{TIR}$, SFR, $M_{*}$, and the attenuation for H$\beta$, [O{\,\sc{ii}}], and [O{\,\sc{iii}}]. For the CIGALE run presented in this work, we observed that the recovered physical parameters are in good agreement with the input ones.\\

\begin{figure*}[t]
\centering
\includegraphics[angle=0,width=0.5\textwidth]{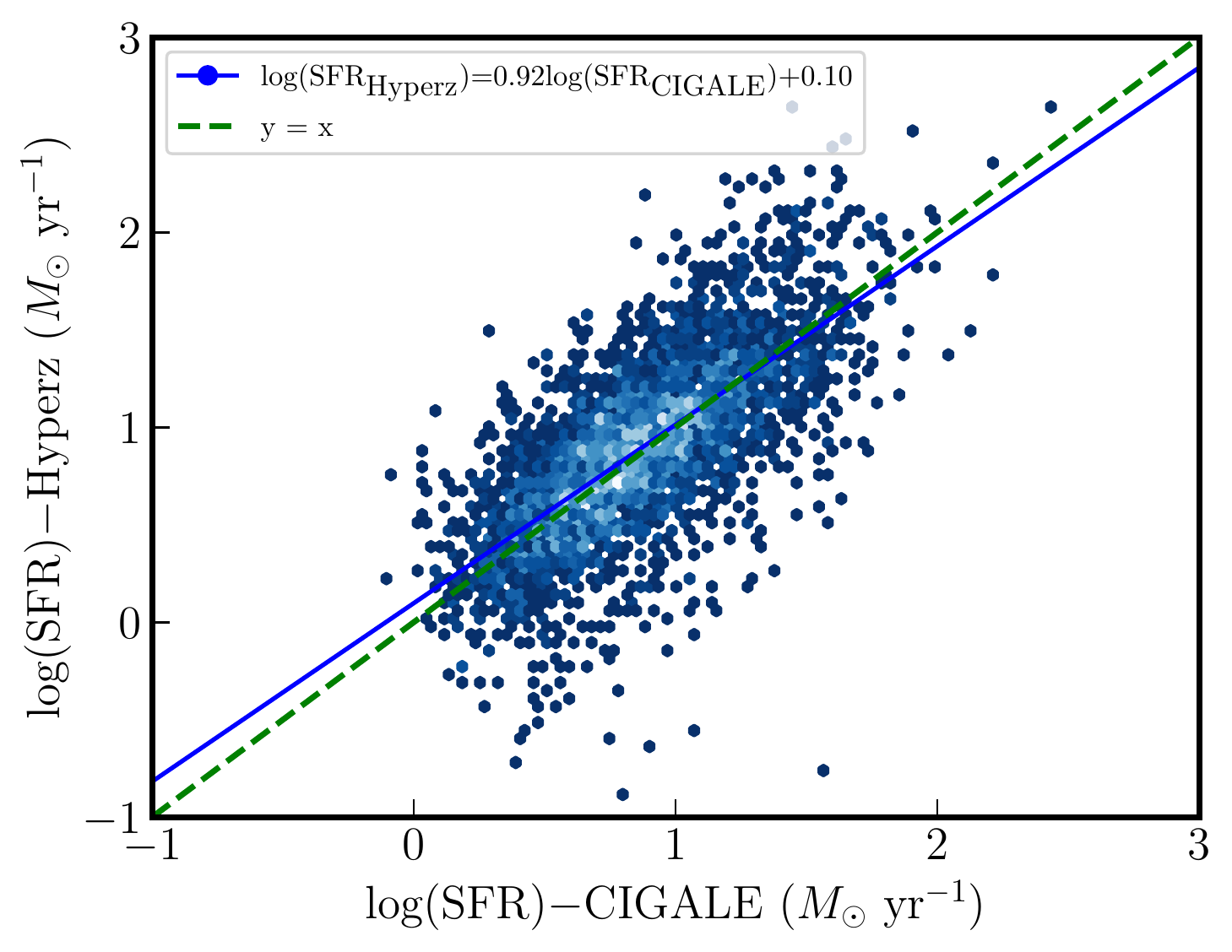}\includegraphics[angle=0,width=0.5\textwidth]{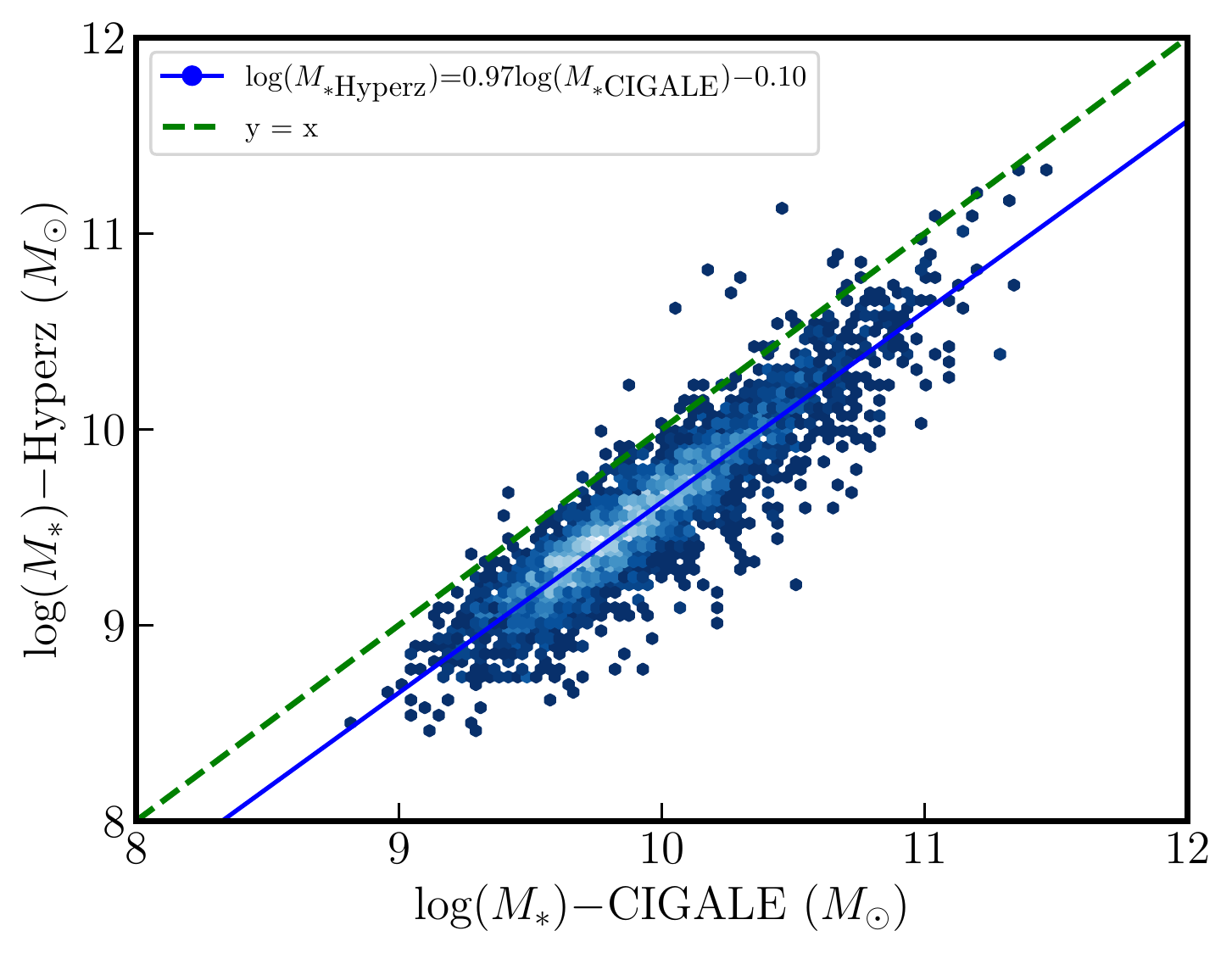}
\caption{Comparison between the main physical properties estimated using \textit{hyperz} and CIGALE tools. Left: SFR derived using \textit{hyperz} versus SFR derived using CIGALE. There is a good agreement between both estimations (0.03~dex), although there is a significant scatter (0.32~dex). Right: $M_{*}$ derived using \textit{hyperz} versus $M_{*}$ derived using CIGALE. The scatter is smaller than for the SFR (0.17~dex) but an offset of 0.37~dex is observed. Such an offset arises from the different attenuation laws used between \textit{hyperz} (\citealt{cal00}, \citealt{pre84} \& \citealt{bou85}) and CIGALE (\citealt{cha00}). The different SFHs and metallicity used may also be possible reasons for the increase in the $M_{*}$ in CIGALE.}
\label{Fig:Comparison_hyperz_CIGALE}
\end{figure*}

Figure~\ref{Fig:Comparison_hyperz_CIGALE} shows the comparison between the stellar mass and SFR estimations from CIGALE and \textit{hyperz}. While SFR estimations seem to agree, although there is a large scatter, the $M_{*}$ comparison shows an offset with an overestimation in CIGALE, or an underestimation in \textit{hyperz}. For the SED fitting of VIPERS galaxies \citep{dav13}, \textit{hyperz} uses the \citet{bru03} stellar population models, a \citet{cha03} IMF, an exponentially declining or constant SFR, \citet{cal00} or \citet{pre84,bou85} attenuation laws, and a metallicity $Z=0.2Z_{\odot}$ or $Z=Z_{\odot}$.\\

We performed a run using a \citet{cal00} attenuation law to estimate the offset in $M_{*}$ as compared to \citet{cha00}. The $M_{*}$ is systematically higher using the \citet{cha00} recipe, by 0.1$-$0.2~dex at all stellar masses. This is in agreement with \citet{mal18} and \citet{bua19}, who found that a \citet{cha00} attenuation law led to higher $M_{*}$ compared to \citet{cal00} by a factor of 1.5 and 1.7, respectively. To confirm that dust is one of the main reason for this offset, we performed an additional run based on the \citet{cha00} recipe and compared red and blue galaxies, and found that the offset in $M_{*}$ was larger for blue than for red galaxies.\\
Using exponentially declining or delayed+burst SFHs, \citet{cie17} showed that the mean error on SFR between both estimations is close to 0, which is what is observed in Fig.~\ref{Fig:Comparison_hyperz_CIGALE}. Both SFH prescriptions are found to recover the $M_{*}$ with an error lower than 25\%. Based on Fig.~11 of \citet{cie17}, we estimated that the offset could reach 0.11~dex between both estimations.\\
We also performed a run using $Z=0.2Z_{\odot}$. This does not significantly impact the SFR,  while the $M_{*}$ is slightly lower at low $M_{*}$ by $\sim$0.06~dex.\\
To summarize, these three different parameters, in particular the dust attenuation recipe, do not significantly impact the derivation of the SFR but can lead to a different $M_{*}$ estimation with a scatter of $\sim 0.3$~dex, comparable to the offset observed in Fig.~\ref{Fig:Comparison_hyperz_CIGALE} (right).\\

To check our choice of stellar absorption correction (${\rm{EW}}=2$~\AA), we compared the estimated flux of the H$\beta$ emission line from CIGALE (without stellar absorption) and from observations (stellar absorption not corrected). When a single value for the stellar absorption is used, as in this work, a correction ranging from 2 to 4~\AA in EW applied to the observed line fluxes gives a good agreement with estimated fluxes from CIGALE.

\begin{table}[h]
\small
\caption{VIPERS photometric data used in CIGALE.}            
\label{tab:flux_values}      
\centering                          
\begin{tabular}{|l|r|r|r|}
\hline
Band & Wavelength & Available  & $\textrm{S/N}>3$ \\
&  &  measurements & (\%) \\
\hline
\hline
\textit{GALEX} FUV & 1549.0~\AA & 415 & 78.8 \\
\textit{GALEX} NUV & 2304.7~\AA & 2114 & 98 \\
\hline
CFHT u  & 3884.7~\AA & 3453 & 99.9 \\
CFHT g  & 4804.1~\AA & 3457 & 100 \\
CFHT r  & 6207.9~\AA & 3457 & 100 \\
CFHT i  & 7498.3~\AA & 3457 & 100 \\
CFHT z  & 8815.0~\AA & 3457 & 99.9 \\
CFHT K$_s$  & 2133.78~nm & 3189 & 99.5 \\
\hline
VISTA K$_s$  & 2137.66~nm & 583 & 99.6 \\
\hline
WISE-1 & 3.4~$\mu$m & 351 & 100 \\
WISE-2 & 4.6~$\mu$m & 234, 117\tablefootmark{$1$} & 74.1 \\
WISE-3 & 12~$\mu$m & 163, 188\tablefootmark{$1$} & 59.5 \\
WISE-4 & 22.8~$\mu$m & 49, 302\tablefootmark{$1$} & 14.3 \\
\hline
\textit{Spitzer} IRAC1 & 3.6~$\mu$m & 533 & 100 \\
\textit{Spitzer} IRAC2 & 4.5~$\mu$m & 320 & 100 \\
\textit{Spitzer} IRAC3 & 5.8~$\mu$m & 41 & 100 \\
\textit{Spitzer} IRAC4 & 8~$\mu$m & 35 & 100 \\
\textit{Spitzer} MIPS1 & 24~$\mu$m & 150 & 100 \\
\textit{Spitzer} MIPS2 & 70~$\mu$m & 15 & 100 \\
\textit{Spitzer} MIPS3 & 160~$\mu$m & 1 & 100 \\
\hline
\textit{Herschel} PACS-Green & 100~$\mu$m & 441 & 2.9 \\
\textit{Herschel} PACS-Red & 160~$\mu$m & 441 & 6.8 \\
\textit{Herschel} SPIRE-S & 250~$\mu$m & 441 & 20.6 \\
\textit{Herschel} SPIRE-M & 350~$\mu$m & 441 & 7 \\
\textit{Herschel} SPIRE-L & 500~$\mu$m & 441 & 1.8 \\
\hline
\end{tabular}
\tablefoot{
\tablefoottext{$1$}{Detections and upper limits. The $S/N$ column takes into account detections only.}
}
\end{table}

\subsection{Dust-attenuation correction}

To correctly estimate the properties of a galaxy, dust attenuation has to be taken into account in order to recover the unattenuated flux, especially toward the bluer part of the spectrum where this effect is significant. The relation between the intrinsic (unattenuated or dust-free) and observed (attenuated) fluxes, $F_{int}$ and $F_{obs}$, respectively, is defined as:

\begin{equation}
F_{int}=F_{obs}\times 10^{0.4A(\lambda)}=F_{obs}\times 10^{0.4k(\lambda)E(B-V)}
\end{equation}

\noindent where $E(B-V)$ is the color excess between the B and V bands, $A(\lambda)$ is the attenuation at $\lambda$ (in mag), and $k(\lambda)$ is the attenuation curve. Estimations of $A(\lambda)$ for each continuum band and spectral line were obtained through CIGALE for VIPERS. From [O~{\sc{iii}}]$\lambda5007$ to [O~{\sc{ii}}]$\lambda3727$, the average attenuation applied to $F_{obs}$ ranges from 1.2 to 1.5~mag.\\

\noindent GSWLC does not provide the attenuation correction for the lines but only for the FUV and NUV bands. To estimate this correction, we used the Balmer decrement technique, which links the H$\alpha$/H$\beta$ ratio to the attenuation at different wavelengths:

\begin{equation}
A(\lambda) = k(\lambda) \times \frac{-2.5}{k(H\beta)-k(H\alpha)}\rm{log}\left(\frac{(H\alpha/H\beta)_{int}}{(H\alpha/H\beta)_{obs}}\right),
\end{equation}

\noindent where (H$\alpha$/H$\beta$)$_{\rm{int}}$ corresponds to the theoretical ratio in the dust-free case, and (H$\alpha$/H$\beta$)$_{\rm{obs}}$ is the observed ratio where H$\alpha$ and H$\beta$ are corrected for stellar absorption. For the theoretical ratio, we used the case B recombination with $T=10^{4}$~K and $n_e=10^{2}$~cm$^{-3}$ equal to 2.86. This value is widely used to determine the dust attenuation correction from the Balmer decrement, although the true value might be in between the case A and B recombination (e.g., \citealt{gro12}). For $k(\lambda)$, we used a Milky Way reddening law (e.g., \citealt{car89,odo94}) with $R_{V}=3.1$ and corrected the H$\alpha$, H$\beta$, [O\,{\sc{iii}}]$\lambda\lambda$4959,5007, and [O\,{\sc{ii}}]$\lambda$3727 emission lines, for which the average attenuation ranges from 0.8 to 1.4~mag.

\section{Analysis}\label{Sect:Analysis}

In this section, we tested different relations proposed in the literature for UV, \textit{u}, 8- and 24-$\mu$m bands, $L_{TIR}$, and spectral lines. In the following, all luminosities are in the rest-frame and are estimated from CIGALE. For each calibration, we computed the coefficients of the linear fit and the Pearson coefficient between the SFR from CIGALE and the SFR estimated using the calibration. The mean and scatter are estimated as the mean and standard deviation of the difference of the log(SFR), respectively. All this information can be found in Appendix~\ref{Appendix:Fit_parameters}.\\

We point out that the scatter derived from VIPERS fits is not completely representative. For the VIPERS sample, we do not have enough data in the MIR and FIR range of galaxy spectrum to constrain the precise value of the dust attenuation. As a consequence, dust attenuation is low ($A_V$ peaking around 0.4), which results in an almost negligible change of estimated SFR based on the optical data only or the full range of the optical part of the spectra.
We checked the impact of the wavelength coverage on the rest-frame luminosity by performing a dedicated test, described in Sect.~\ref{Subsect:absoluteluminosity}.

\subsection{SFR from continuum bands}
\subsubsection{Ultraviolet indicators}

The UV emission of galaxies is dominated by the radiation from young high-mass stars. Due to their short lifetime, their UV radiation represents an excellent tracer of star formation with a timescale of $\sim100$~Myr. The main drawback is the dust attenuation, which is particularly significant at this wavelength. Either a correction is applied based on an attenuation law \citep{cha00,cal00}, or an IR band is used as a proxy representing the re-emitted UV flux absorbed by the dust (see Sect.~\ref{Sect:composite_tracers}). In this section, we focused only on the UV bands and used the calibrations of \citet{ros02}, \citet{sal07}, \citet{dav16}, and \citet{bro17}, written in Appendix.~\ref{Appendix:Laws_UV}.\\

\begin{figure*}[h!]
\centering
\includegraphics[angle=0,width=0.5\textwidth]{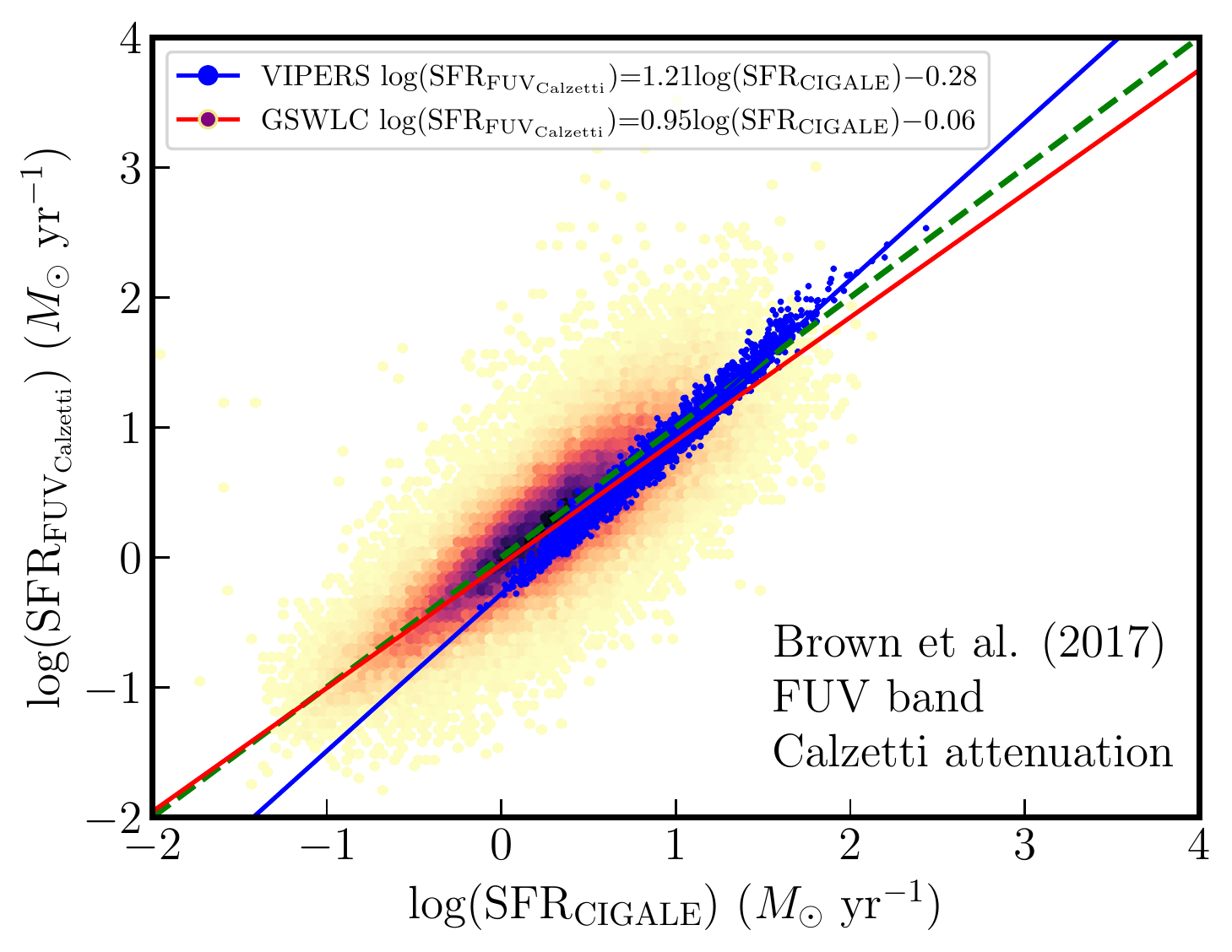}\includegraphics[angle=0,width=0.5\textwidth]{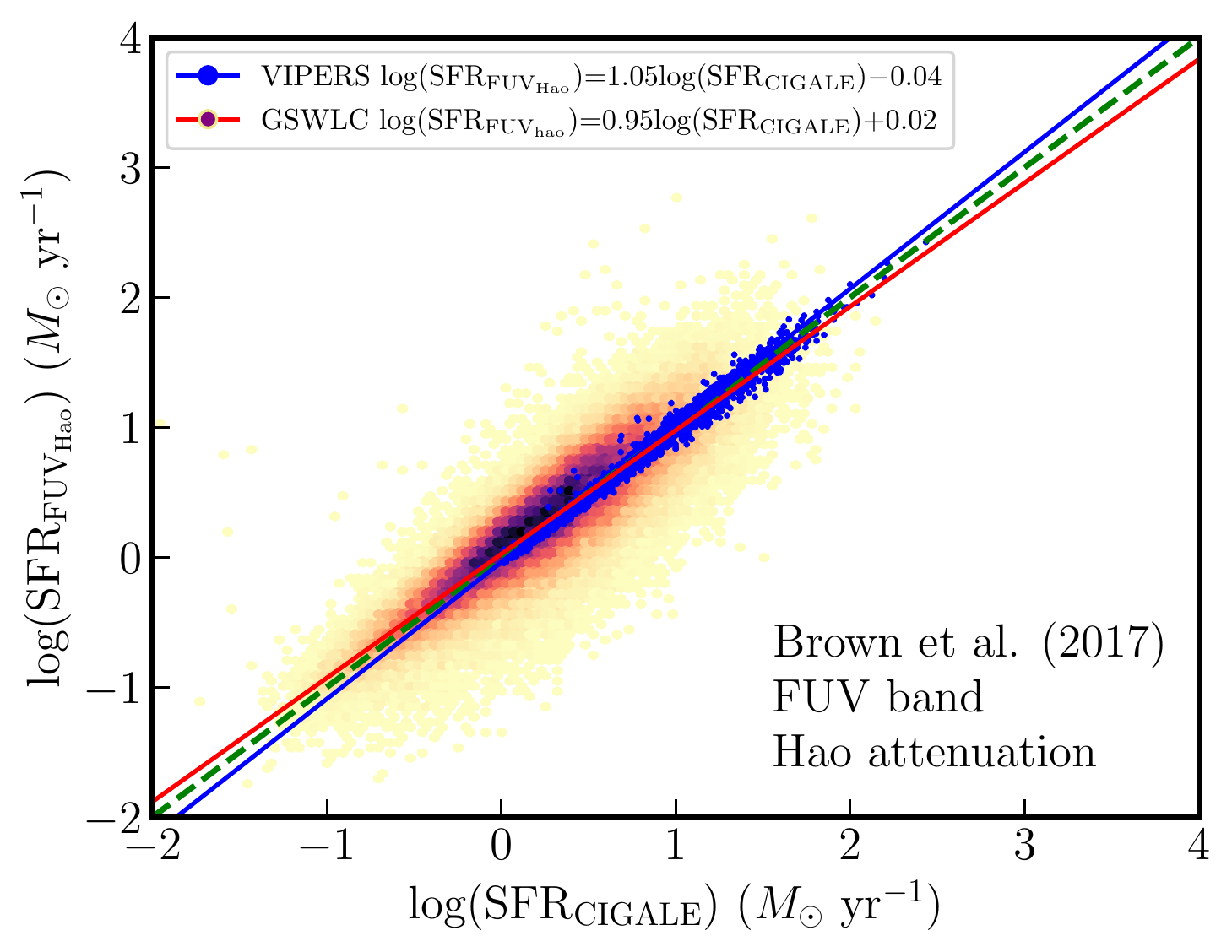}
\caption{SFR derived from the relations given in \citet{bro17} using \textit{GALEX} FUV and the attenuation law from \citet{cal00} (left) and \citet{hao11} (right). Blue points represent the VIPERS galaxies and density hexabin represent the GSWLC galaxies sample. The blue line is the fit for VIPERS galaxies, the red line is the fit for GSWLC galaxies, and the green line is the 1:1 relation. The agreement is better with the \citet{hao11} attenuation correction recipe, both for VIPERS (mean and scatter of 0.01 and 0.05~dex) and GSWLC (-0.02 and 0.25~dex), compared to \citet{cal00} (0.09 and 0.10~dex for VIPERS, 0.02 and 0.30~dex for GSWLC).} 
\label{Fig_FUVNUV_Brown17}
\end{figure*}

In \citet{bro17}, the dust attenuation is corrected following two prescriptions, both using the FUV and NUV magnitudes of the galaxies. The first is characterized by $(M_{FUV}-M_{NUV})_{{\rm{dust~free}}}=0$ and a \citet{cal00} extinction law, while the second assumes $(M_{FUV}-M_{NUV})_{{\rm{dust~free}}}=0.022$ and the prescription of \citet{hao11}. SFRs from FUV corrected with these two prescriptions are presented in Fig.~\ref{Fig_FUVNUV_Brown17}.
The SFR obtained using the \citet{cal00} attenuation law underestimates the SFR at low luminosity for VIPERS, while there is a relatively good agreement with GSWLC but with a high scatter. Such small discrepancies could originate from the use of different attenuation laws in CIGALE: \citet{cha00} for VIPERS and a modified \citet{cal00} for GSWLC. On the contrary, the \citet{hao11} attenuation recipe gives a better agreement with CIGALE SFR estimations and a smaller scatter with GSWLC.\\

\citet{dav16} computed the SFR following different bands, which were then recalibrated using the radiative transfer code of \citet{pop11}. To correct for dust attenuation, $\beta_{UV}$ is estimated from FUV and NUV data \citep{cal94} and linked to $A_{UV}$ \citep{meu99}:

\begin{equation}
A_{UV}=4.43+1.99\beta_{UV}.
\end{equation}

As seen in Fig.~\ref{Fig_FUVNUV_Davies}, the SFR for VIPERS galaxies are underestimated and the SFR for GSWLC galaxies are overestimated at low SFR. Because the calibration of the attenuation from \citet{meu99} is computed at 1600~\AA, it should naturally differ from the real attenuation suffered in the NUV. This could partially explain the strong disagreement observed for the NUV-based SFR. A smaller disagreement is also observed for the FUV-based SFR. For both bands, another source of uncertainty might come from our own derivation of $\beta_{UV}$, based on $L_{FUV}$ and $L_{NUV}$, which may be impacted by the templates, in particular for VIPERS.\\
To obtain a better estimation of the attenuation from $\beta_{UV}$, we derived new relations from the SED fitting from rest-frame FUV and NUV bands estimated by CIGALE:

\begin{figure*}[h!]
\centering
\includegraphics[angle=0,width=0.5\textwidth]{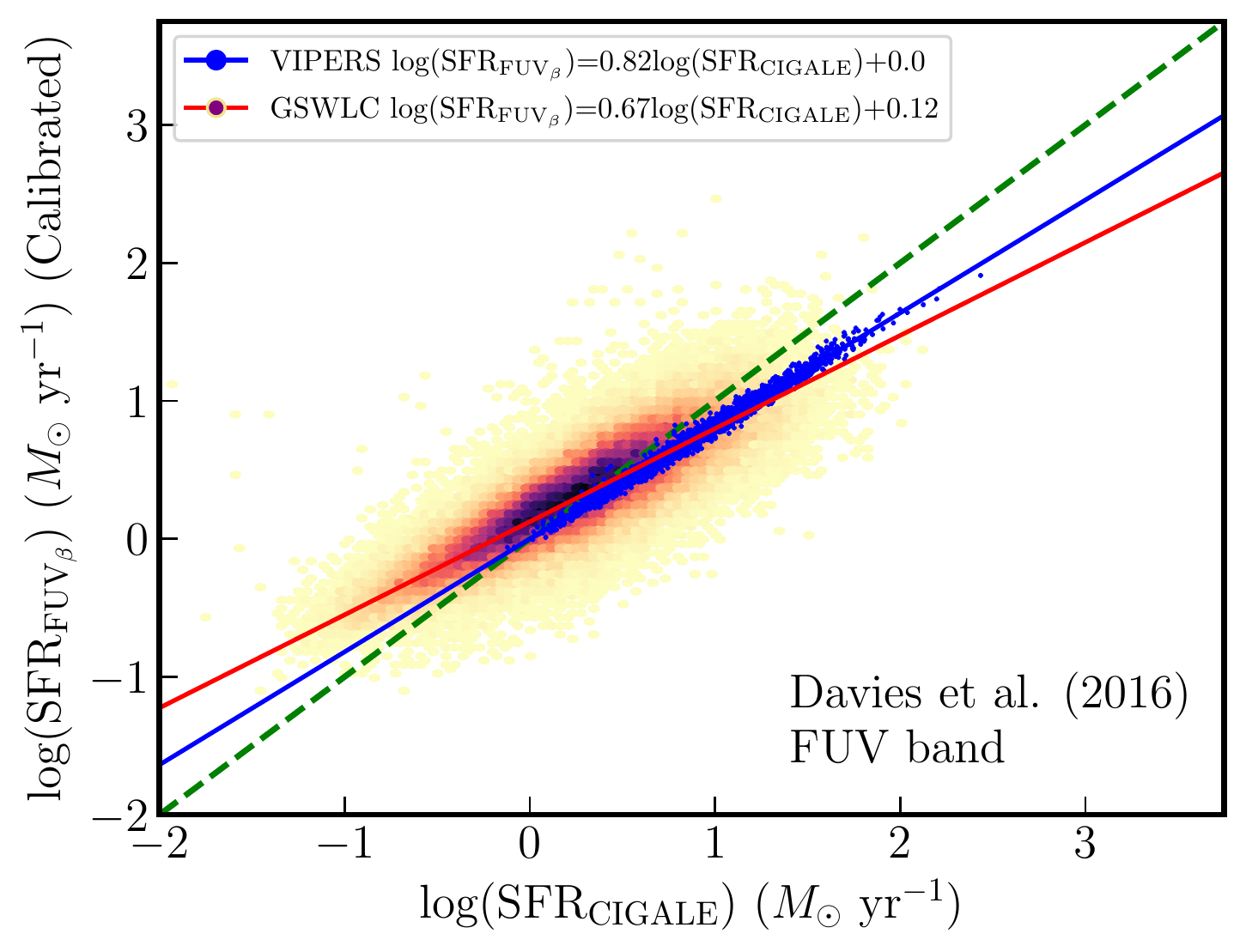}\includegraphics[angle=0,width=0.5\textwidth]{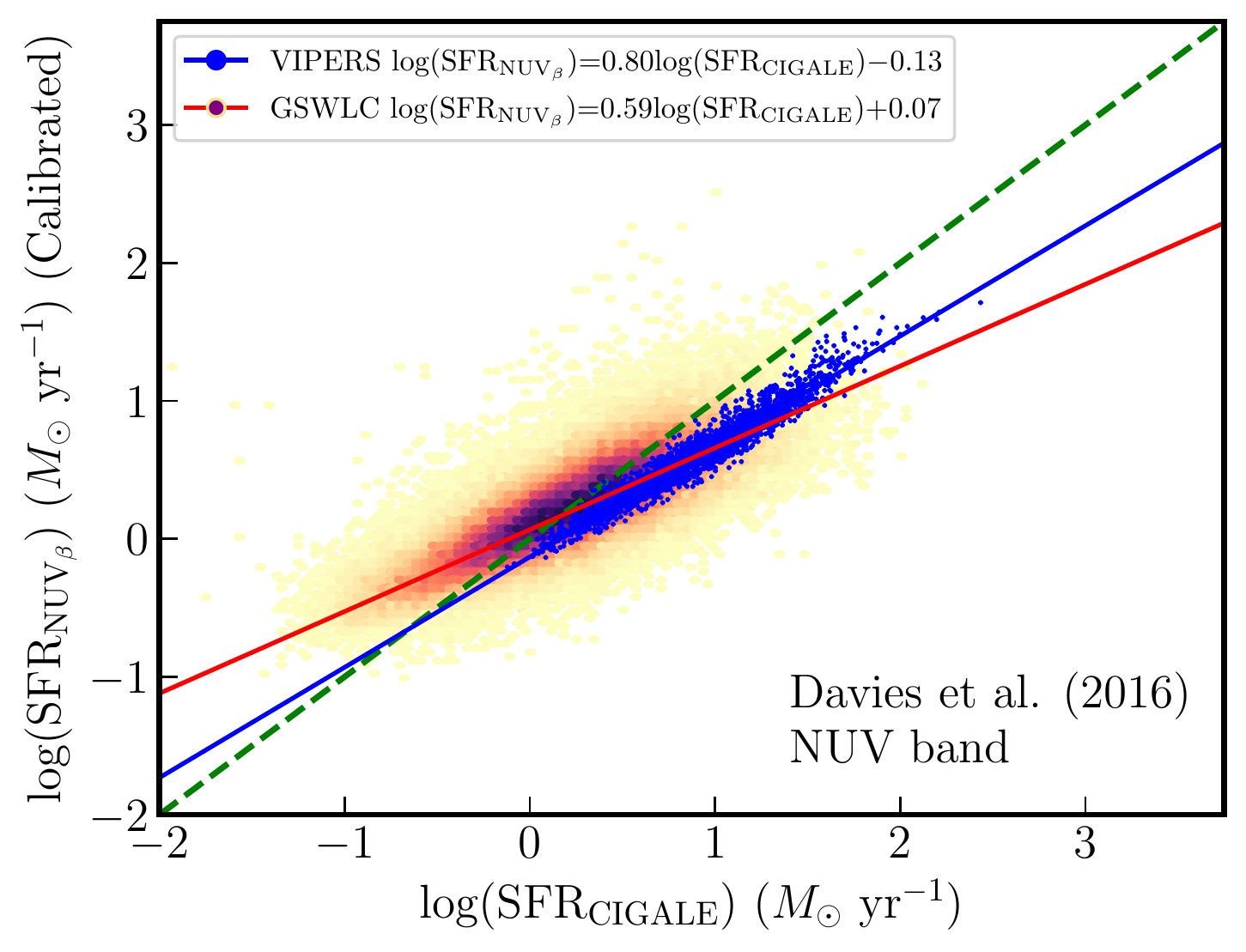}
\caption{SFR derived from the relations given in \citet{dav16} using \textit{GALEX} FUV (left) and NUV (right), and the attenuation law from the $A_{UV}-\beta_{UV}$ relation of \citet{meu99}. We observed a disagreement with CIGALE both for VIPERS and GSWLC in the FUV (mean of 0.16~dex and scatter of 0.07~dex for VIPERS, -0.07 and 0.25~dex for GSWLC), stronger in the NUV band (0.29 and 0.10~dex  for VIPERS, -0.01 and 0.29~dex for GSWLC).} 
\label{Fig_FUVNUV_Davies}
\end{figure*} 

\begin{equation}\label{Eq:SlopeBetaNew_1}
A_{FUV}=(1.78\pm0.02)\beta_{UV}+(3.71\pm0.02),
\end{equation}
\begin{equation}\label{Eq:SlopeBetaNew_2}
A_{NUV}=(1.35\pm0.01)\beta_{UV}+(2.80\pm0.02).
\end{equation}

We recall here that $\beta_{UV}$ and, by extension, $A_{FUV}$ and $A_{NUV}$, are also dependent on the assumed SFH. The SFH chosen influences the mass distribution and the dominance of high-mass stars, which directly affects the value of $\beta_{UV}$. In addition, the stellar evolution may have an impact on $\beta_{UV}$ within the SFH: after a few dozen Myr, most of the high-mass stars leave the MS and the old stellar population gains more importance relatively to the young population, leading to a decrease in $\beta_{UV}$ (i.e., it becomes redder). \citet{wil12} studied the influence of different SFHs on $\beta_{UV}$ and found an uncertainty of 0.31 for $A_{FUV}$. In addition, $\beta_{UV}$ is also affected by the metallicity, with an uncertainty of 0.7 for $A_{FUV}$.\\

\begin{figure*}[h!]
\centering
\includegraphics[angle=0,width=0.5\textwidth]{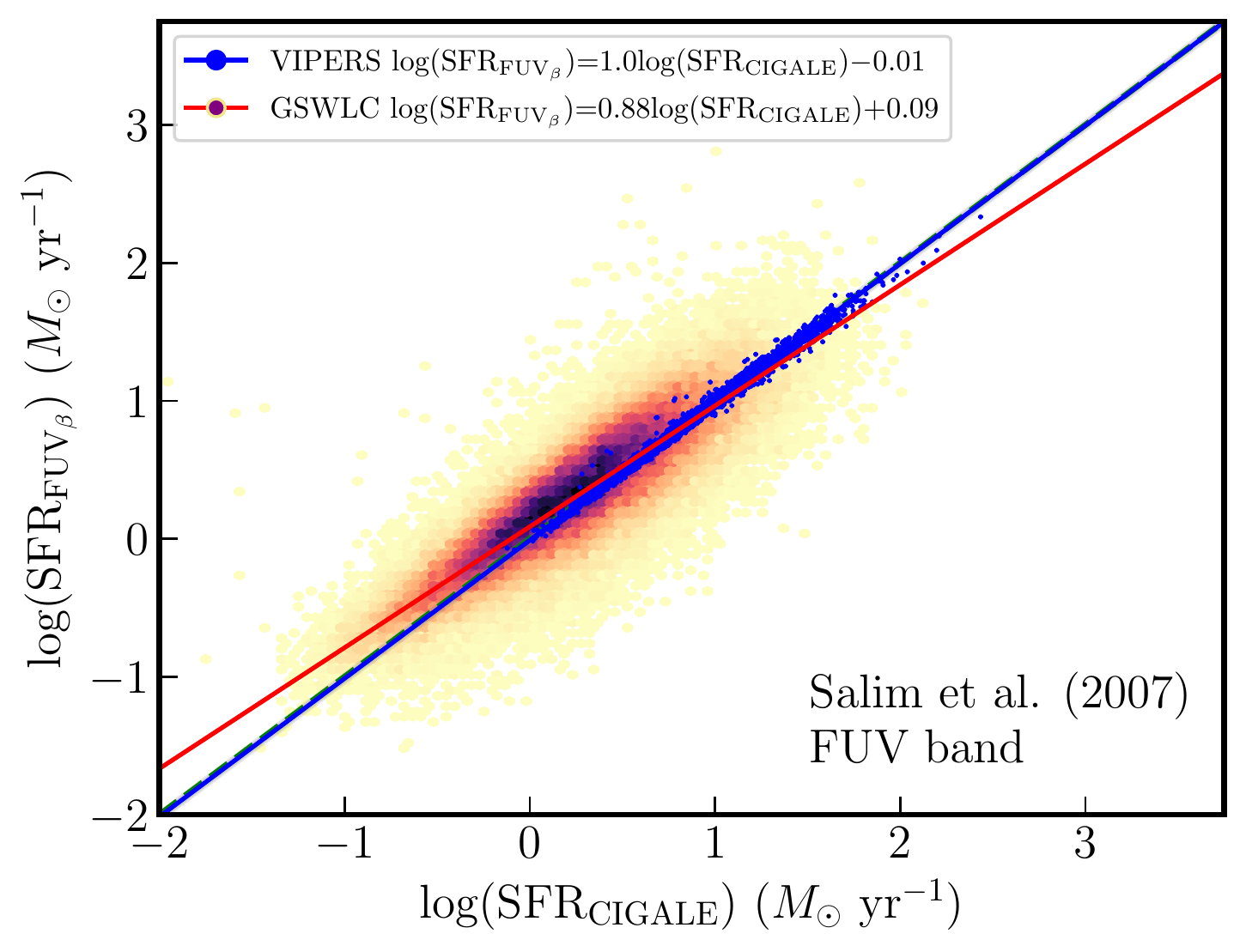}\includegraphics[angle=0,width=0.5\textwidth]{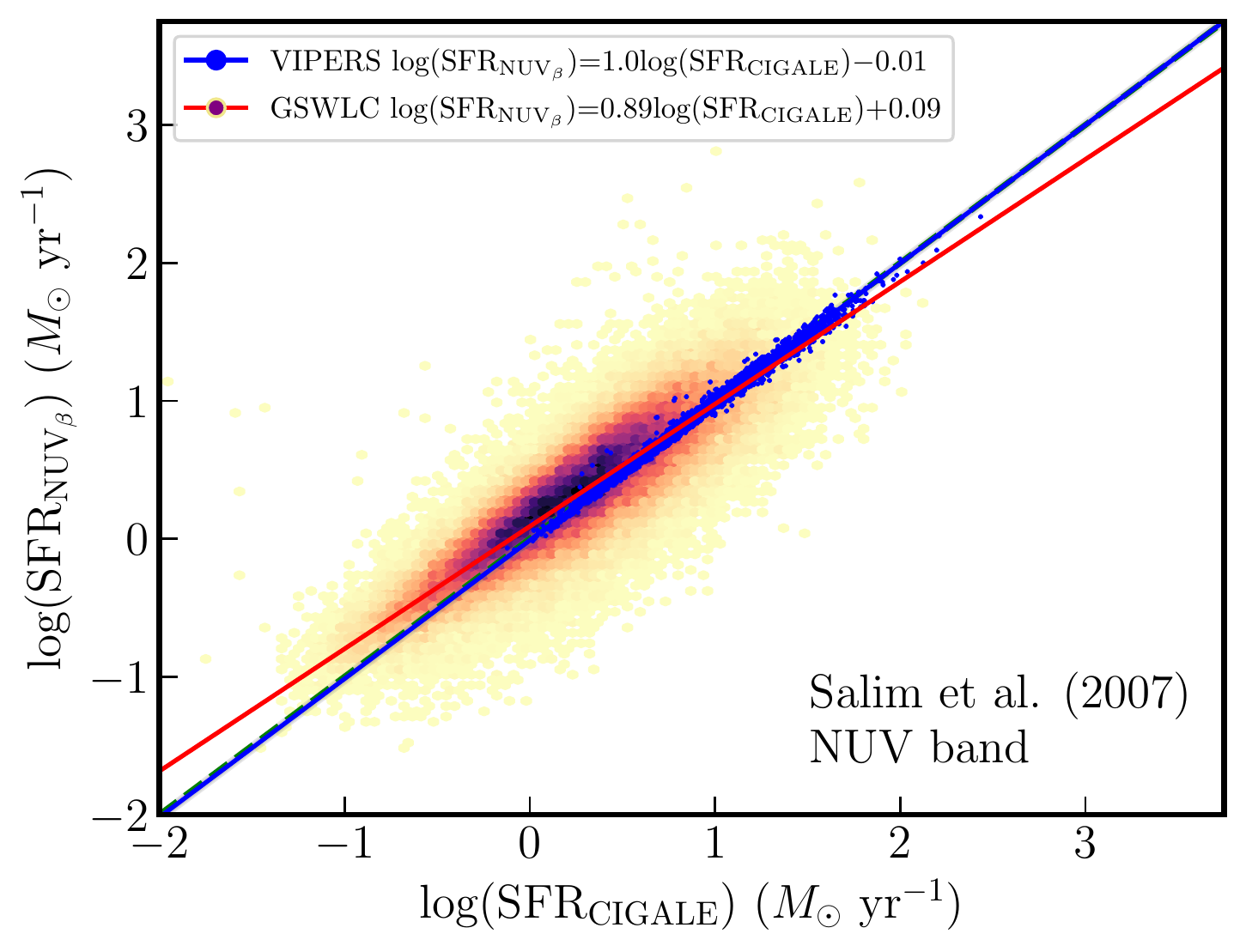}
\caption{SFR derived from the relations given in \citet{sal07} using \textit{GALEX} FUV (left) and NUV (right), and the recalibrated attenuation law from $\beta_{UV}$ (based on Eqs.~\ref{Eq:SlopeBetaNew_1} and \ref{Eq:SlopeBetaNew_2} and VIPERS galaxies). There is a much better agreement with VIPERS (mean of 0.02~dex and a scatter of 0.03~dex for FUV and NUV, respectively) but also with GSWLC (-0.08 and 0.24~dex for FUV and NUV, respectively) considering that the attenuation prescription applied is based only on VIPERS galaxies.} 
\label{Fig_FUVNUV_Davies_uncal}
\end{figure*}

Despite being recalibrated with the radiative transfer method of \citet{pop11}, the SFR calibrations of \citet{dav16} are not in good agreement with VIPERS and GSWLC for either band. We were able to recover a better agreement by using Eqs.~\ref{Eq:SlopeBetaNew_1} and \ref{Eq:SlopeBetaNew_2} and the calibration from \citet{sal07}, shown in Fig.~\ref{Fig_FUVNUV_Davies_uncal}. For GSWLC, the slight deviation may be explained by the use of a different attenuation law to the one used to derive Eqs.~\ref{Eq:SlopeBetaNew_1} and \ref{Eq:SlopeBetaNew_2}. This shows that SFR estimations heavily depend on the method chosen to retrieve the intrinsic luminosity in the UV. Nonetheless, the attenuation given by Eqs.~\ref{Eq:SlopeBetaNew_1} and \ref{Eq:SlopeBetaNew_2} and based on VIPERS, gives reliable corrections at low redshifts, giving good SFR estimations.\\

Because the dust attenuation correction might be difficult to derive, \citet{ros02} proposed an SFR calibration based on NUV luminosity and for which the attenuation correction \citet{cal00} was already taken into account. As observed in Fig.~\ref{Fig_NUV_Ros02} (top-left), using a calibration for which the dust attenuation correction is already included might lead to a high scatter when used on another sample of galaxies. When the average SFR of the VIPERS and GSWLC samples is compared to the estimation from \citet{ros02}, the SFR differ by 0.11 and 0.20~dex, respectively.\\

In Figs.~\ref{Fig_FUVNUV_Brown17}, \ref{Fig_FUVNUV_Davies}, and \ref{Fig_FUVNUV_Davies_uncal}, for instance, and throughout this work, we note that the fits for VIPERS galaxies are steeper than the fits for GSWLC galaxies. Although we made sure to select the same type of galaxies at low and intermediate redshifts, the wavelength coverage for GSWLC was better than that of the VIPERS galaxies in the IR bands. This in turn affects the SFRs when estimating them from the panchromatic coverage using the balanced SED. In such SED fitting tools, SFRs are sensitive to IR emission in particular. That is why we focused our efforts on modeling as best as possible the short wavelength fluxes, using carefully chosen SFHs and attenuation, to ensure a more reliable fit in the IR where the photometry was not as good as the short wavelength counterparts. While this method ensures a reliable SFR estimation using the full SEDs, attenuation plays a bigger role in estimating the total SFRs, assuming a more important hidden SFR fraction.\\
The fact that GSWLC have observations in UV wavelengths, directly tracing young stellar populations in those galaxies, may also partially explain the difference  in slopes in Fig.~\ref{Fig_FUVNUV_Davies}, which relies more on templates. However, more direct measurements do not necessarily imply a better agreement with known calibrations, as seen in Fig.~\ref{Fig_FUVNUV_Davies} for the GSWLC sample.

\subsubsection{The u-band indicator}

The \textit{u}-band can also be used as an SFR tracer but, in addition to significant dust attenuation, it can be more strongly affected by the old stellar population than the FUV and NUV bands \citep{bel03}. Moreover, the \textit{u}-band luminosity can vary significantly during the stellar evolution, making it less reliable as an SFR tracer \citep{hop03}. In this subsection, we used the calibrations of \citet{hop03}, \citet{mou06b}, \citet{dav16}, and \citet{zho17}, written in Appendix~\ref{Appendix:Laws_u}.\\

Due to our selection criteria, VIPERS and GSWLC are predominantly composed of blue galaxies and the red population should not significantly contaminate the star-forming sample. Using a criterion based on the NUVrK$_{s}$ diagram \citep{arn13}, we found no passive galaxies in VIPERS and only 158 in GSWLC (0.2\% of the sample). This number of passive galaxies should be taken with caution for GSWLC as this method is not well-constrained at $z=0$.\\

However, even for star-forming galaxies, part of the stellar population is composed of old stars that can contaminate the \textit{u}-band. Using CIGALE, we estimated this contamination as the old stellar population ($\geq$100~Myr) to the total stellar luminosity ratio through the \textit{u} filter \footnote{The old stellar population contamination in the \textit{u}-band and $L_{TIR}$, and the stellar contamination at 8~$\mu$m (in Sect.~\ref{Sect:MidFar}) were estimated from a Bayesian analysis, but were not implemented in CIGALE.}. The contamination ranges from 16\%  to 53\%, with an average of 36\%. This is consistent with the hydrodynamical simulations at $1<z<2$ of \citet{boq14}, who found that SFR estimations based on the \textit{u}-band overestimates the SFR by $\sim39$\%.\\

\begin{figure*}[h!]
\centering 
\includegraphics[angle=0,width=0.5\textwidth]{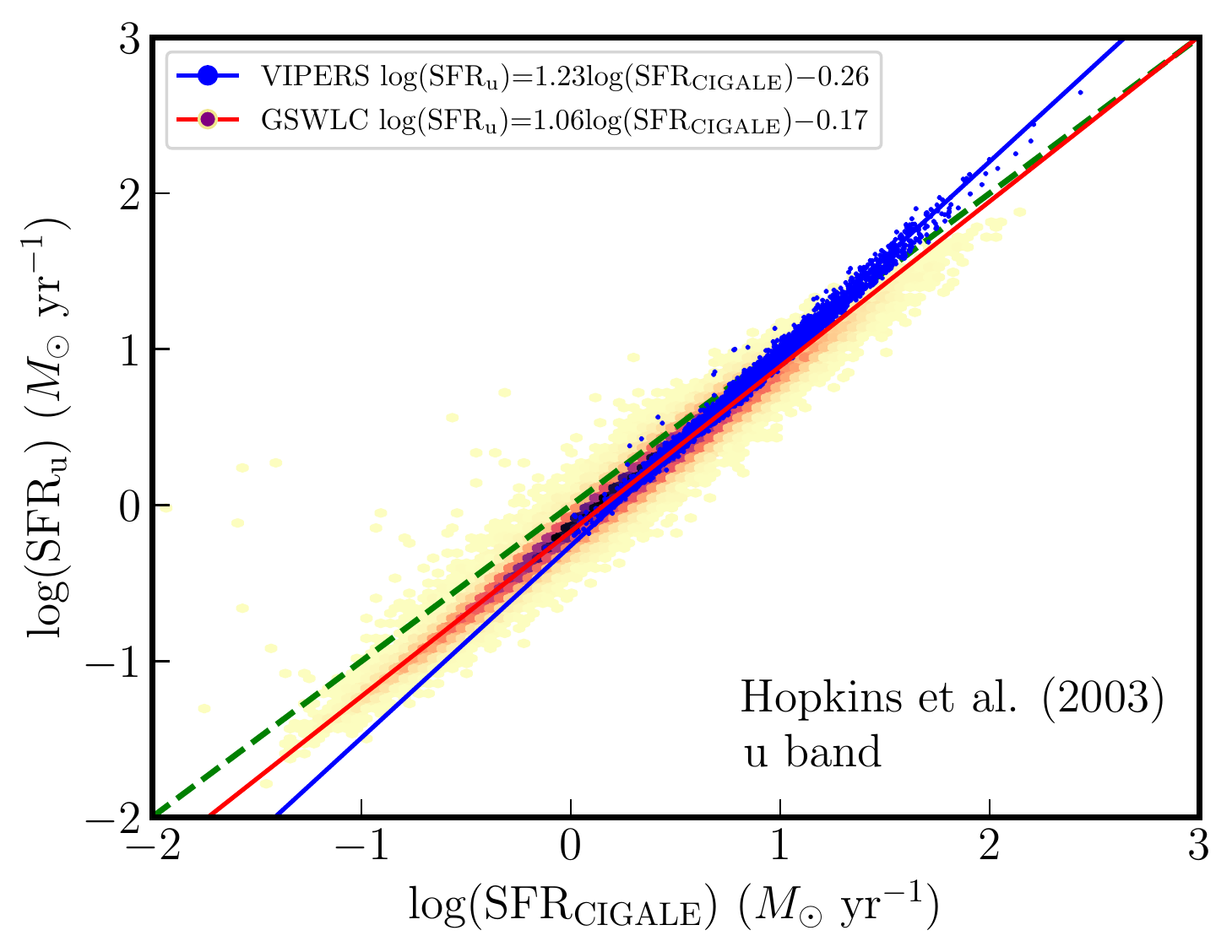}\includegraphics[angle=0,width=0.5\textwidth]{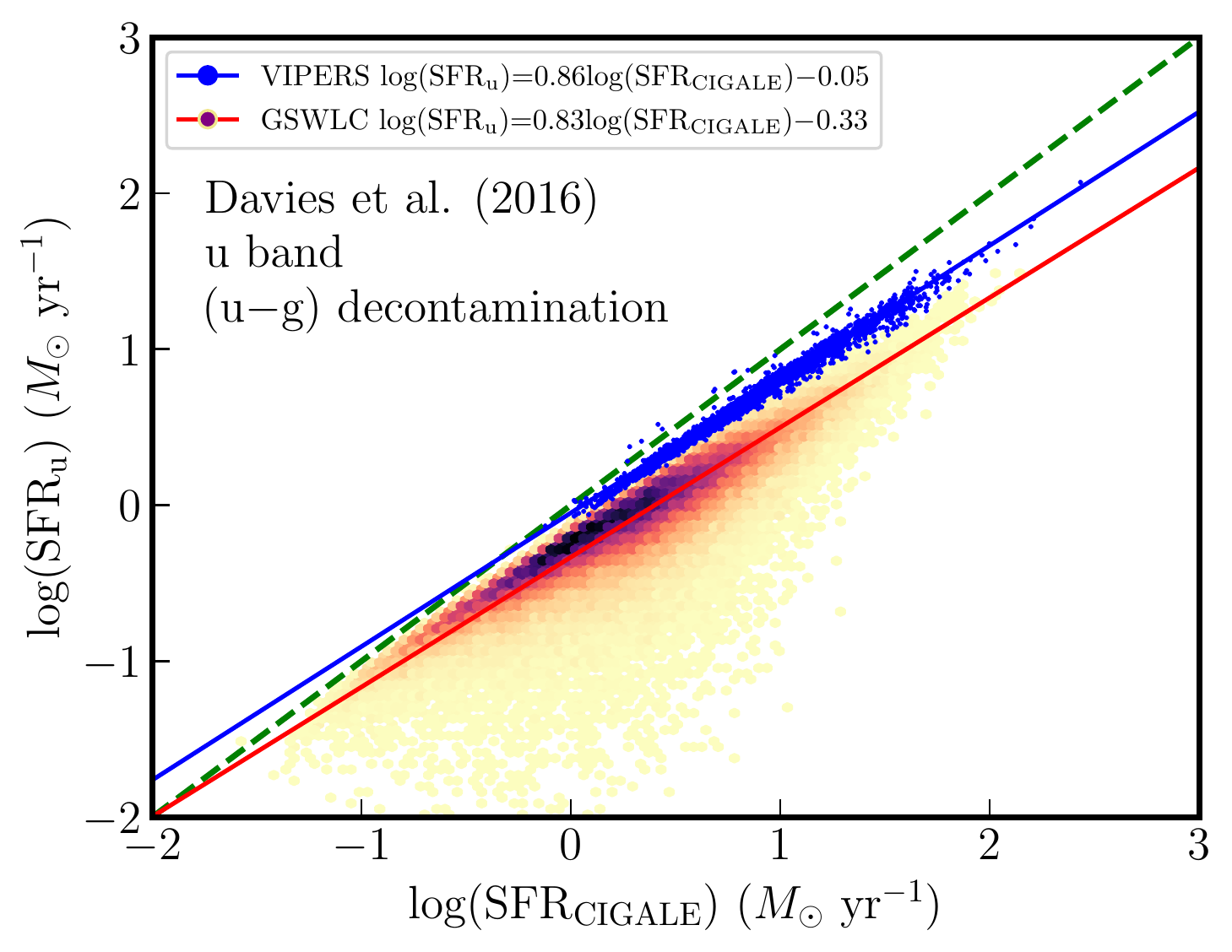}
\caption{Comparisons between the SFR from CIGALE and the literature using the \textit{u}-band. Left panel: SFR derived from the relation given in \citet{hop03} using the \textit{u}-band. Right panel: SFR derived from the calibrated relation given in \citet{dav16} using the \textit{u}-band and the old population decontamination from the \textit{u-g} color. While the calibration of \citet{hop03} is in good agreement with CIGALE (mean of 0.07~dex and scatter of 0.08~dex for VIPERS, 0.15 and 0.09~dex for GSWLC), the decontamination from the old stellar population luminosity proposed by \citet{dav16} leads to an underestimation of the SFR for VIPERS (0.18 and 0.06~dex) at high SFR, and to a strong underestimation for GSWLC (0.42 and 0.22~dex).}
\label{Fig_optical_Davies}
\end{figure*} 

Based on a sample of galaxies with \textit{u} and H$\alpha$ flux measurements, \citet{hop03} derived a nonlinear calibration that reflects the change in star formation history and old-stellar population contamination (Fig.~\ref{Fig_optical_Davies} left). Despite the assumption of an equivalent old stellar contamination between the sample of \citet{hop03}, GSWLC, and VIPERS, which may not be entirely true, particularly for VIPERS due to the different redshift range, the nonlinear calibration of \citet{hop03} gives a good estimation of the SFR for VIPERS and GSWLC, without having to perform any additional decontamination.\\

\citet{dav16} derived a linear calibration for the \textit{u}-band where the luminosity is scaled based on the $u-g$ color for $u-g>0.55$ (reddest galaxies) to account for this old stellar population contamination. This correction concerns 34\% of VIPERS galaxies, with an average contamination of 11\%, and 99\% of GSWLC galaxies, with a much higher average contamination equal to 51\%. For some galaxies that are part of GSWLC, this correction leads to negative luminosity and they are excluded when performing the linear fit. No correlation is found between the $u-g$ color and the contamination in the \textit{u}-band for VIPERS, as previously estimated with CIGALE. \\
This decontamination method associated with the calibration of \citet{dav16} leads to a significant underestimation of the SFR, especially for GSWLC, compared to CIGALE (Fig.~\ref{Fig_optical_Davies} right). We also estimated the SFR based on the calibration of \citet{dav16} without performing any decontamination. As seen in Fig.~\ref{Fig_NUV_Ros02} (top right), it leads to a better agreement with CIGALE values.\\

\citet{zho17} also found that a nonlinear calibration was in better agreement with the SFR derived from H$\alpha$ and that the residuals were weakly correlated with the old stellar population. This SFR calibration is presented in Fig.~\ref{Fig_NUV_Ros02} (middle left) and is roughly equivalent to the calibration of \citet{hop03}. At $z\sim0.1$, \citet{mou06b} showed that the scatter in the \textit{u}-band luminosity to SFR calibration with respect to the $D_{4000}$ break was mostly caused by the reddening effect. Therefore, a simple linear calibration based on the \textit{u}-band should give a rather good estimation of the SFR. This calibration (see Fig.~\ref{Fig_NUV_Ros02}, middle right) shows a high scatter  and an underestimation of the SFR compared to CIGALE. However, the dust attenuation correction is already accounted for in their calibration, and such scatter and low SFR estimations may originate from the different treatement of dust reddening between the samples, rather than from a contamination from the old stellar population.\\

A proper calibration of the SFR based on the \textit{u}-band should account for the old-stellar population contamination, assuming that the contribution to the luminosity of the old stars can be accurately removed. Because uncertainties due to dust attenuation dominate, a decontamination is not mandatory and the calibration of \citet{hop03}, for instance, is able to give a reliable estimation of the SFR. This implies, however, that galaxies should not be significantly different from the sample from which the calibration is derived. Extrapolation of these calibrations to galaxies at higher redshifts is also subject to a larger scatter.

\subsubsection{MIR and FIR indicators}\label{Sect:MidFar}

The IR bands are often used as SFR tracers because their flux is related to high-mass stars through the emission of PAHs found around H{\,\sc{ii}} regions, and to the emission of dust. We used the calibrations of \citet{ken98a}, \citet{wu05}, \citet{per06}, \citet{rel07}, \citet{zhu08}, \citet{rie09}, \citet{gil10}, \citet{you14}, \citet{dav16} and \citet{bro17}, written in Appendix~\ref{Appendix:Laws_IR}.\\

\begin{figure*}[h!]
\centering
\includegraphics[angle=0,width=0.5\textwidth]{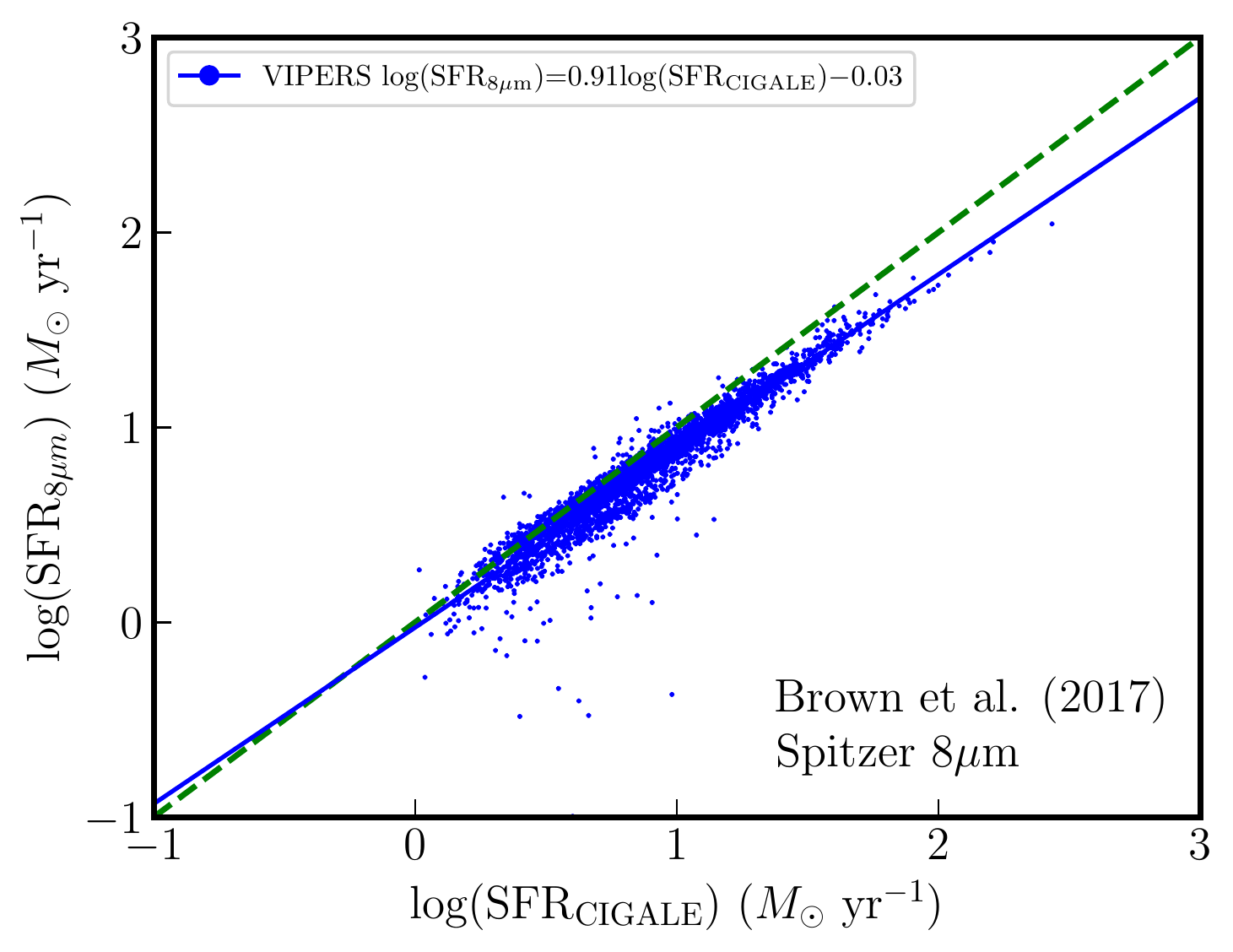}\includegraphics[angle=0,width=0.5\textwidth]{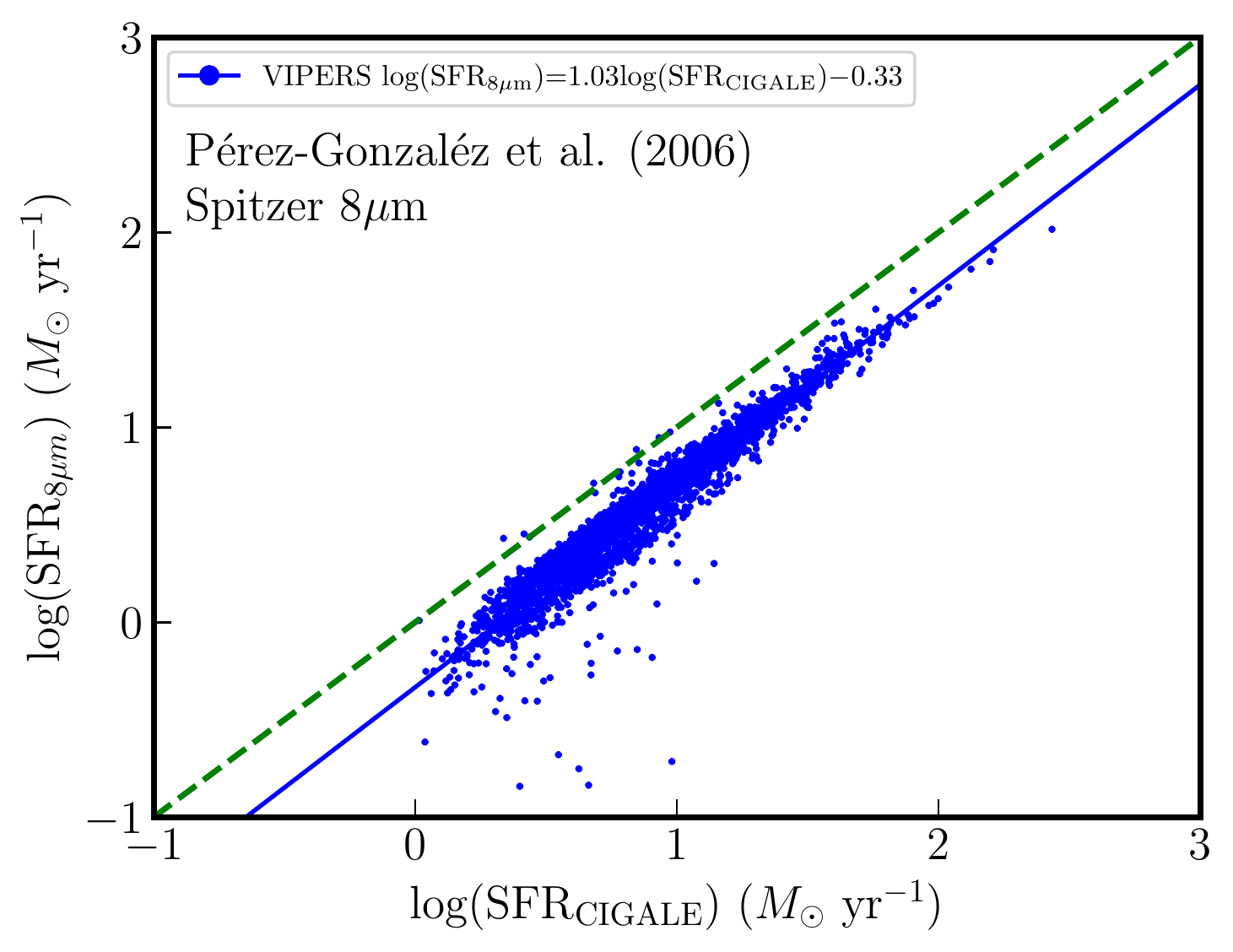}
\caption{Comparisons between the SFR from CIGALE and the literature at 8~$\mu$m. Left panel: SFR derived from the relations given in \citet{bro17} using \textit{Spitzer} 8~$\mu$m from CIGALE (blue). Right panel: SFR derived from the relations given in \citet{per06} using \textit{Spitzer} 8~$\mu$m from CIGALE. Both SFR calibrations show a disagreement with SFR from CIGALE (mean of 0.10~dex and scatter of 0.10~dex for \citealt{bro17}, 0.29 and 0.11~dex for \citealt{per06}), which could originate from the sample from which the calibration was established and from the different PAH properties of the galaxies.} 
\label{Fig_8_Brown17}
\end{figure*}

At $z=0$, the emission of PAHs excited by the UV radiation from OB stars is the main contributor to the luminosity at $\lambda=8$~$\mu$m, followed by the emission from very small grains (VSGs) and stellar emission. Even if the PAH emission is correlated with metallicity \citep{sch18}, the luminosity at 8~$\mu$m remains roughly correlated with the SFR and useful due to its strength in star-forming galaxies. A better comprehension of PAH physics is crucial for future observations with JWST \citep{shi16}, whose MIR bands from 5.6 to 25.5~$\mu$m were calibrated as SFR indicators \citep{bat15,sen18}.\\

Before calibrating the 8-$\mu$m emission as an SFR tracer, we removed the stellar component from $L_{8\mu m}$. Using CIGALE, we estimated this contamination as the ratio of the flux from the old and young stellar populations to the total luminosity in the \textit{Spitzer} 8~$\mu$m filter. Galaxies with stellar contamination higher than 50\% (72 galaxies) were found to have unreliable SED in the IR part of the spectrum and were not taken into account when dealing with SFR tracers at $\lambda\geq8$~$\mu$m.\\
The average stellar contamination in the 8~$\mu$m is 5\%. This value is lower for the VIPERS sample than other samples such as \citet{eng05} or \citet{wu05}, which have a contamination of 10\% for a sample of 34 nearby galaxies and a sample of 91 galaxies, respectively. In comparison, the estimated stellar contamination at 24~$\mu$m is, on average, equal to 0.2\%.\\

Using different calibrations at 8~$\mu$m \citep{wu05,per06,you14,bro17}, we observed an underestimation of the SFR for each of them (Figs.~\ref{Fig_8_Brown17}, \ref{Fig_NUV_Ros02} bottom). We note that \citet{bro17} did not remove the stellar component at 8~$\mu$m as it did not significantly change the fit parameters, contrary to the other calibrations. In addition to the stellar contamination, this discrepancy could also come from the sample of galaxies from which each SFR calibration was derived. For instance, the calibration of \citet{bro17} is based on 66 nearby star-forming galaxies for which the H$\alpha$ flux ranges from $\sim$10$^{38}$ to $\sim$10$^{43}$~erg~s$^{-1}$. Based on H$\beta$ measurements for VIPERS galaxies, H$\alpha$ could be as high as $\sim$10$^{44}$~erg~s$^{-1}$, a luminosity at which the calibration of \citet{bro17} could be out of the range of applicability. The metallicity of galaxies could also impact the emission of PAHs, directly affecting the derived calibrations. An ideal SFR calibration at this wavelength requires a very good understanding of the PAHs' behavior with respect to the properties of the galaxy such as the metallicity, which is not yet achieved. This is underlined by the recent work of \citet{gre22}, who showed that the scatter of SFR-8~$\mu$m relation is related to the metallicity distribution.\\

In the mid-IR range, the 24-$\mu$m emission represents the dust heated by VSGs with a stable temperature. It is a reliable SFR tracer as this emission traces the interior of H{\,{\sc{ii}}} regions powered by high-mass stars. Two advantages of this band are: (1) it does not require any stellar subtraction, and (2) the attenuation is low enough to be omitted.\\
In GSWLC, the \textit{Spitzer} data were not included, but the similarity between the 24-$\mu$m and WISE-4 filters makes it possible to find a relation between the luminosity in both bands \citep{jar13}. We compared the rest-frame luminosities and showed that $L_{24\mu m}$ is higher than $L_{W4}$ by $\sim$8\% for star-forming galaxies. We applied this 8\% correction to $L_{W4}$ when calibrations based on $L_{24\mu m}$ were used, bearing in mind that the impact on the SFR remains negligible.\\
 
\begin{figure*}[h!]
\centering
\includegraphics[angle=0,width=0.5\textwidth]{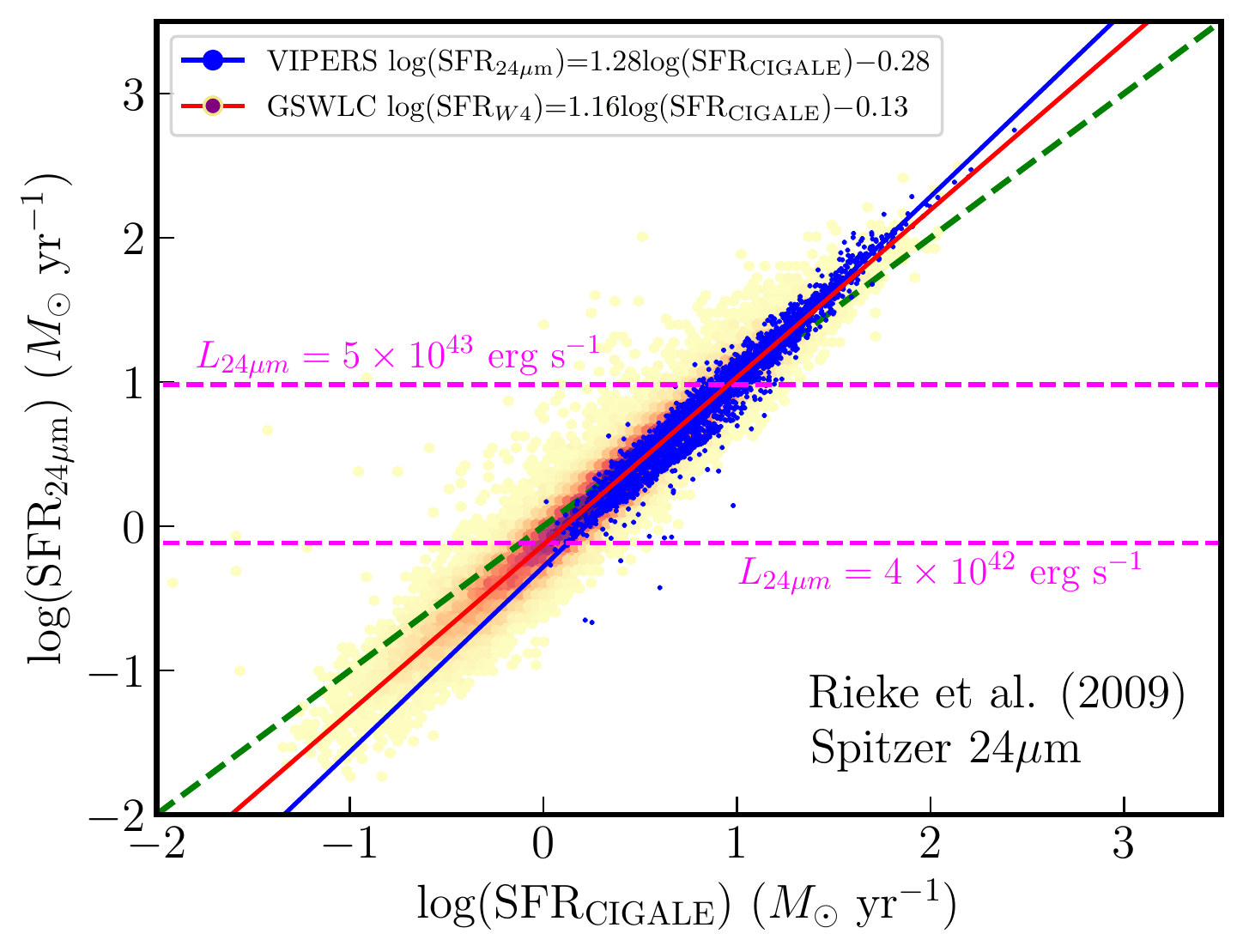}\includegraphics[angle=0,width=0.5\textwidth]{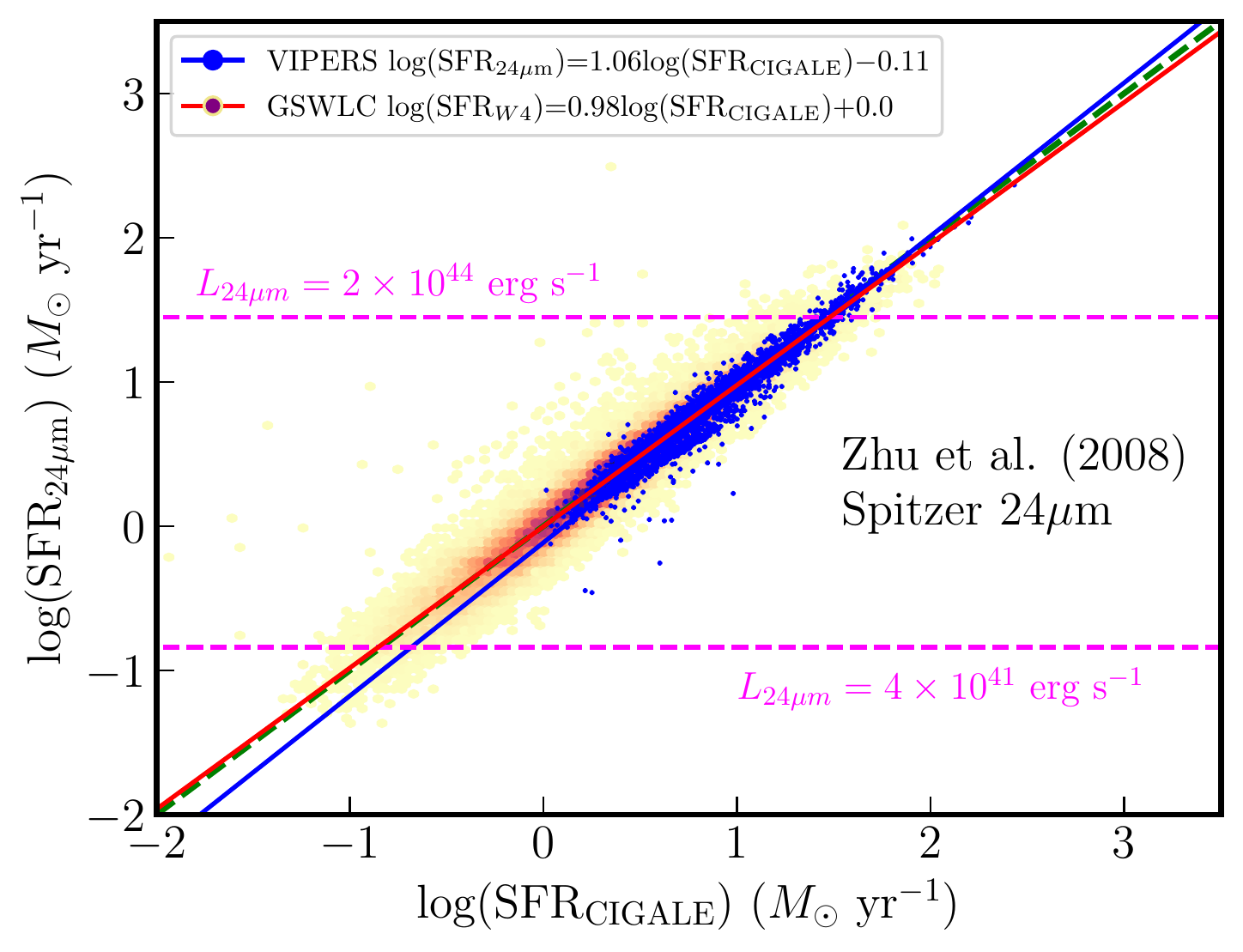}
\caption{Comparisons between the SFR from CIGALE and the literature at 24~$\mu$m. Left panel: SFR derived from the relations given in \citet{rie09} using \textit{Spitzer} 24~$\mu$m from CIGALE (blue). The continuous magenta line represents the limit above which a nonlinear correction is used. Right panel: SFR derived from the relations given in \citet{zhu08} using \textit{Spitzer} 24~$\mu$m. The dashed magenta lines represent the domain of applicability, as defined by \citet{cal10}. The small discrepancy observed using \citet{rie09} (mean of 0.004~dex and scatter of 0.13~dex for VIPERS, 0.07 and 0.16~dex for GSWLC) may originate from the hypothesis of linearity (the nonlinearity correction for \citet{rie09} is small), while a nonlinear calibration such as \citet{zhu08} gives a better agreement (mean of 0.04~dex and scatter of 0.08~dex for VIPERS, 0.01 and 0.12~dex for GSWLC).} 
\label{Fig_24_R09}
\end{figure*}

\citet{cal10} listed several calibrations at 24~$\mu$m that have been proposed in the literature. Relations assuming a linear behavior work relatively well for VIPERS and GSWLC \citep{wu05,zhu08,rie09}. The calibration giving the smallest scatter is given by \citet{rie09}, which, however, assumes a nonlinear behavior at $L_{24\mu m}>5\times 10^{43}$~erg~s$^{-1}$ (Fig.~\ref{Fig_24_R09} left). The assumption of linearity induces an overestimation at high SFR and an underestimation at low SFR, for GSWLC.\\

Several works have shown that $L_{24\mu m}$ may not be directly proportional to the SFR. It could originate from different dust attenuation corrections at high SFR \citep{bua07}, a higher dust temperature inside H{\,\textsc{ii}} regions leading to higher $L_{24\mu m}$, an underestimation of Pa$\alpha$ from which SFR$_{24\mu m}$ may be calibrated, or self-absorption at 24~$\mu$m \citep{rie09}. At low luminosity, galaxies become transparent and $L_{24\mu m}$ cannot reliably trace the SFR as a significant part comes from unabsorbed UV emission \citep{cal10}. Several authors calibrated their SFR in such a way \citep{wu05,per06,alo06,cal07,rel07,zhu08}. For the VIPERS and GSWLC samples, a better agreement is indeed found when nonlinear relations are used; the best fit is given by the calibration of \citet{zhu08}, which holds even outside the luminosity range for which their relation was calibrated (Fig.~\ref{Fig_24_R09}, right).\\

We also note that calibrations based on the observations of entire galaxies give better SFR estimations compared to those based on H{\,\sc{ii}} regions. This might be due to different dust temperatures between H{\,\sc{ii}} regions and galaxies, leading to a higher calibration constant in the case of entire galaxies \citep{cal10}. To test this, we used a subsample of 441 VIPERS galaxies having \textit{Herschel} counterparts and we modeled the SED as a modified black-body (MBB, \citealt{cas12}). Taking into account fits for which the uncertainty on $T_{dust}$ is smaller than 50\% (70 galaxies), $T_{dust}$ ranges from 14.0$\pm$1.4 to 46.5$\pm$16.2~K, with an average equal to 27.7$\pm$1.4~K. In comparison, the dust temperature of H{\,{\sc{ii}}} regions in the Milky Way is similar \citep{and14,fig17}. Therefore, the difference between entire-galaxy- and H{\,{\sc{ii}}}-region-based calibrations may not come entirely from a difference in temperature. An example is shown in Fig.~\ref{Fig_Cal_Brown} (top left) where the \citet{rel07} calibration based on H{\,{\sc{ii}}} regions shows an underestimation of the SFR for VIPERS galaxies. The calibration of \citet{bro17} shown in Fig.~\ref{Fig_Cal_Brown} (right) also underestimates the SFR of VIPERS galaxies at high luminosity, despite being calibrated on entire galaxies.\\

\begin{figure}[h!]
\centering
\includegraphics[angle=0,width=0.5\textwidth]{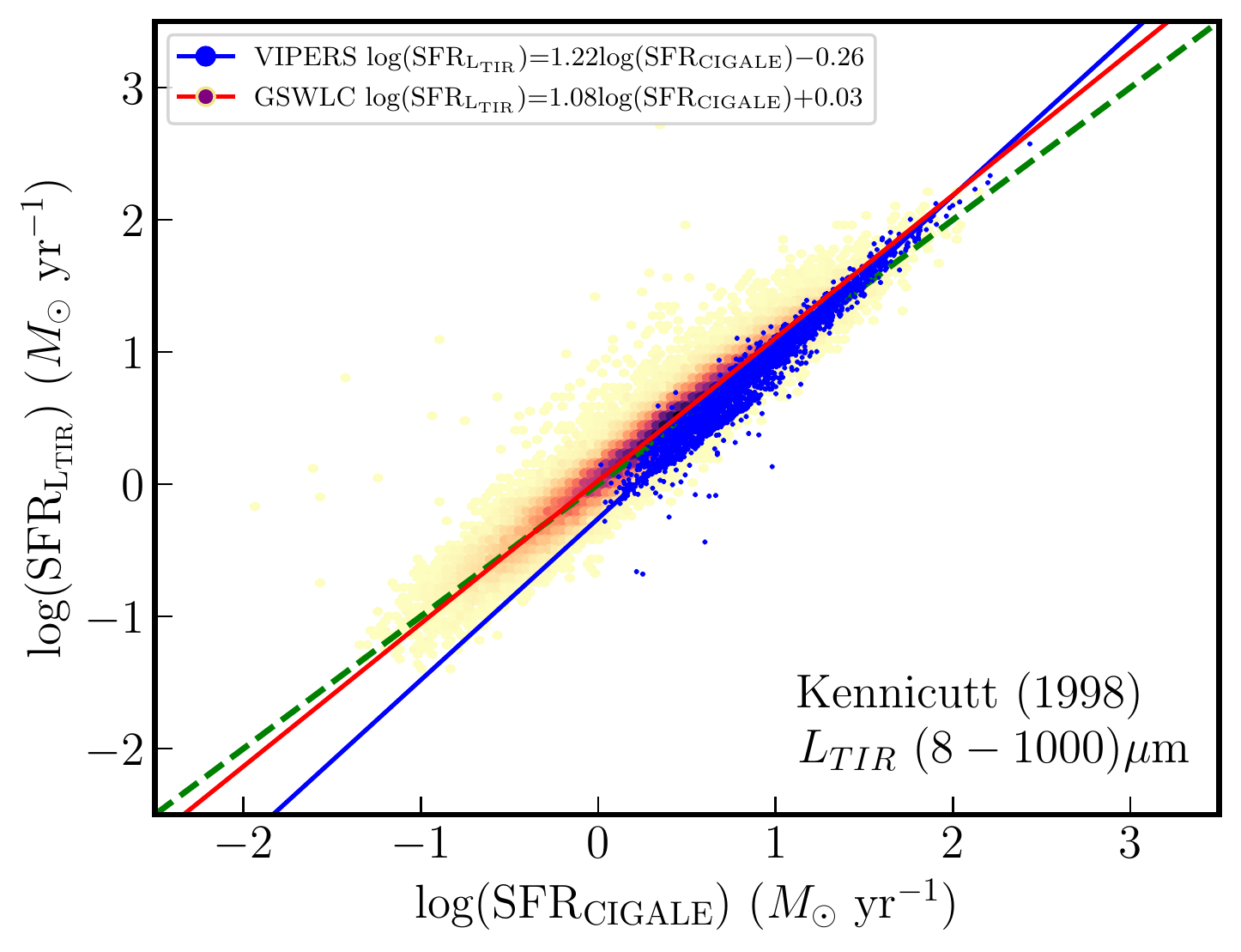}
\caption{SFR derived from the relations given in \citet{ken98a} using CIGALE for VIPERS (blue) and WISE-4 for GSWLC (density plot). Blue lines and red lines are the fits for VIPERS and GSWLC, respectively. The agreement between the calibration and CIGALE SFRs is good (mean of 0.04~dex and scatter of 0.11~dex for VIPERS, 0.04 and 0.13~dex for GSWLC).} 
\label{Fig_LTIR_Salim}
\end{figure}

We estimated the SFR from $L_{TIR}$ (measured with CIGALE for VIPERS and using the WISE-4 band for GSWLC, with the BOSA templates, \citealt{boq21}) following the calibration of \citet{ken98a}, and Fig~\ref{Fig_LTIR_Salim} shows that there is a rather good agreement with the SFR from CIGALE. At this redshift, this correlation is expected since our samples, both at low and intermediate z, are composed of star-forming galaxies and should not contain outliers. In Sect.~\ref{Subsect:absoluteluminosity}, we show that the scatter in $L_{TIR}$ for VIPERS should be around $\sim$0.2~dex, and a similar scatter be added to the SFR derived from $L_{TIR}$. For GSWLC, the scatter in $L_{TIR}$ and SFR from BOSA templates is around 0.15 and 0.17~dex \citep{boq21}. In the next section, we show that part of the dust emission is due to the old stellar population but, as for the \textit{u}-band, calibrations usually do not take this into account. This could explain why the SFR is slightly overestimated using the calibration of \citet{ken98a}, which assumes a population of young stars.\\

\subsubsection{Composite SFR indicators}\label{Sect:composite_tracers}

Previously, we used SFR calibrations based on a single band only (two close bands in the case of attenuation correction in the UV). While this method is practical, since just one band is needed, it requires correction for dust attenuation in the UV part of the spectrum. Such corrections depend on the dust attenuation recipe used, which also depends on galaxies' properties, such as inclination \citep{bat17}, axis ratio \citep{wil11}, optical opacity \citep{sal18}, dust luminosity \citep{deb16}, or redshift \citep{bog20}. In addition, these laws are averaged over samples of galaxies that might be different from the studied sample.\\ A popular method to correct for dust attenuation without relying on attenuation laws is to use a direct tracer of young high-mass stars (UV, H$\alpha$, [O{\,\sc{ii}}]) and an indirect tracer that takes into account the flux absorbed and re-emitted in the infrared ($L_{TIR}$, 8~$\mu$m, 24~$\mu$m). Appendix~\ref{Appendix:Laws_Composites} includes the laws used in this subsection: \citet{bel05}, \citet{ken09}, \citet{arn13}, \citet{boq14}, and \citet{cla15}.\\

In \citet{bel05}, the SFR was calibrated with the PEGASE population synthesis model \citep{fio97} using the FUV band and $L_{TIR}$. While slightly overestimated, the SFRs show a very good agreement with CIGALE values (Fig.~\ref{Fig_Cal_Brown} middle-left). \citet{cla15} calibrated the SFR using the FUV and 24-$\mu$m bands from \citet{hir03}, \citet{bua08}, and \citet{jar13}, but took into account that, in addition to OB stars, the old stellar population can also heat the dust, overestimating the SFR estimation.\\
To estimate this quantity for VIPERS galaxies, we computed the absorbed luminosity of the young and old stellar population relatively to $L_{dust}$. The fraction of dust emission due to the young or old stellar population, depending on the component that is considered, is estimated following

\begin{equation}\label{Eq:dust_heat}
R_{heat-dust}=\frac{\int_{91.2~nm}^{+\infty}L_{stellar}^{non-att}d\lambda}{\int_{91.2~nm}^{+\infty}L_{dust}d\lambda}.
\end{equation}

\noindent The integration is done from 91.2~nm onward, corresponding to the range in which the absorption is due to dust.\\ For GSWLC galaxies, we estimated $R_{heat-dust}^{old}$ using the relation between $sSFR$ and the $R_{heat-dust}^{young}$ parameter established by \citet{ner19} for 814 nearby galaxies. The fraction of dust emission due to the old stellar population is

\begin{equation}\label{eq:Percentage_dust_heat}
{R_{heat-dust}^{old}} = \left\{
    \begin{array}{l}
       11\%\hspace*{0.5cm} {\rm{(VIPERS) }}\\
       45\%\hspace*{0.5cm} {\rm{(GSWLC) }}\\
    \end{array}
\right..
\end{equation}

\begin{figure}[h!]
\centering
\includegraphics[angle=0,width=0.5\textwidth]{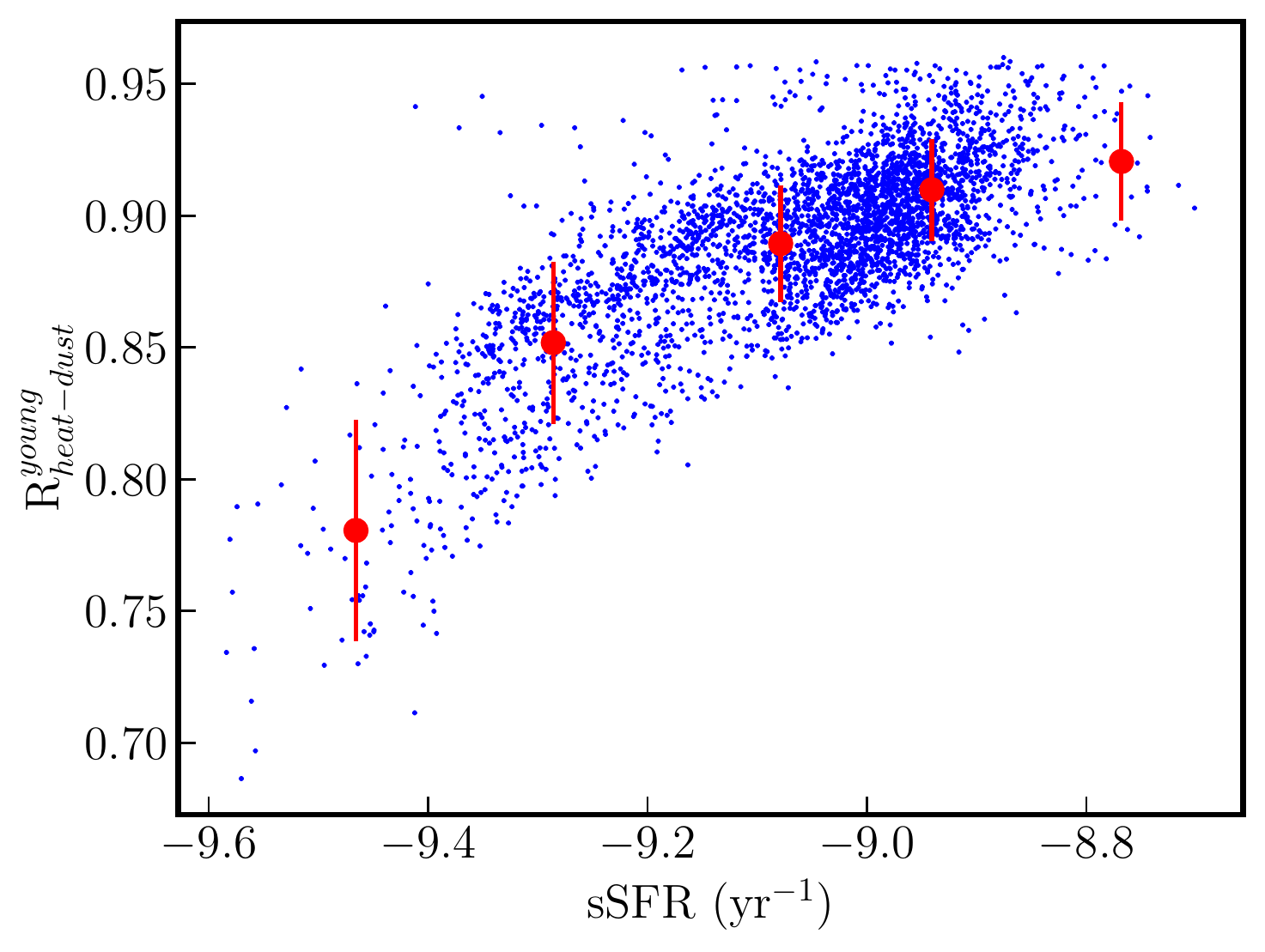}
\caption{Fraction of dust heated by the young stellar population with respect to the sSFR for VIPERS, where each bin (red points) represents the median for $\sim$420 galaxies and uncertainties are estimated as the standard deviation. The fraction of dust heated by the young stellar population increases as the stellar activity of the galaxy increases.} 
\label{Fig_Rdust}
\end{figure} 

As expected, the fraction of dust heated by the old stellar population is lower for intermediate redshift galaxies due to the higher number of young stars compared to local galaxies. This is also what is observed in \citet{ner19}, where more than 80\% of dust emission in early-type galaxies (E to S0) is due to old stellar population and decreases to $\sim$40\% for late-type and irregular galaxies. In Fig.~\ref{Fig_Rdust} we show how the fraction of dust emission due to the young stellar population increases with the specific star formation rate (sSFR) for the VIPERS sample.\\

\begin{figure*}[h!]
\centering
\includegraphics[angle=0,width=0.5\textwidth]{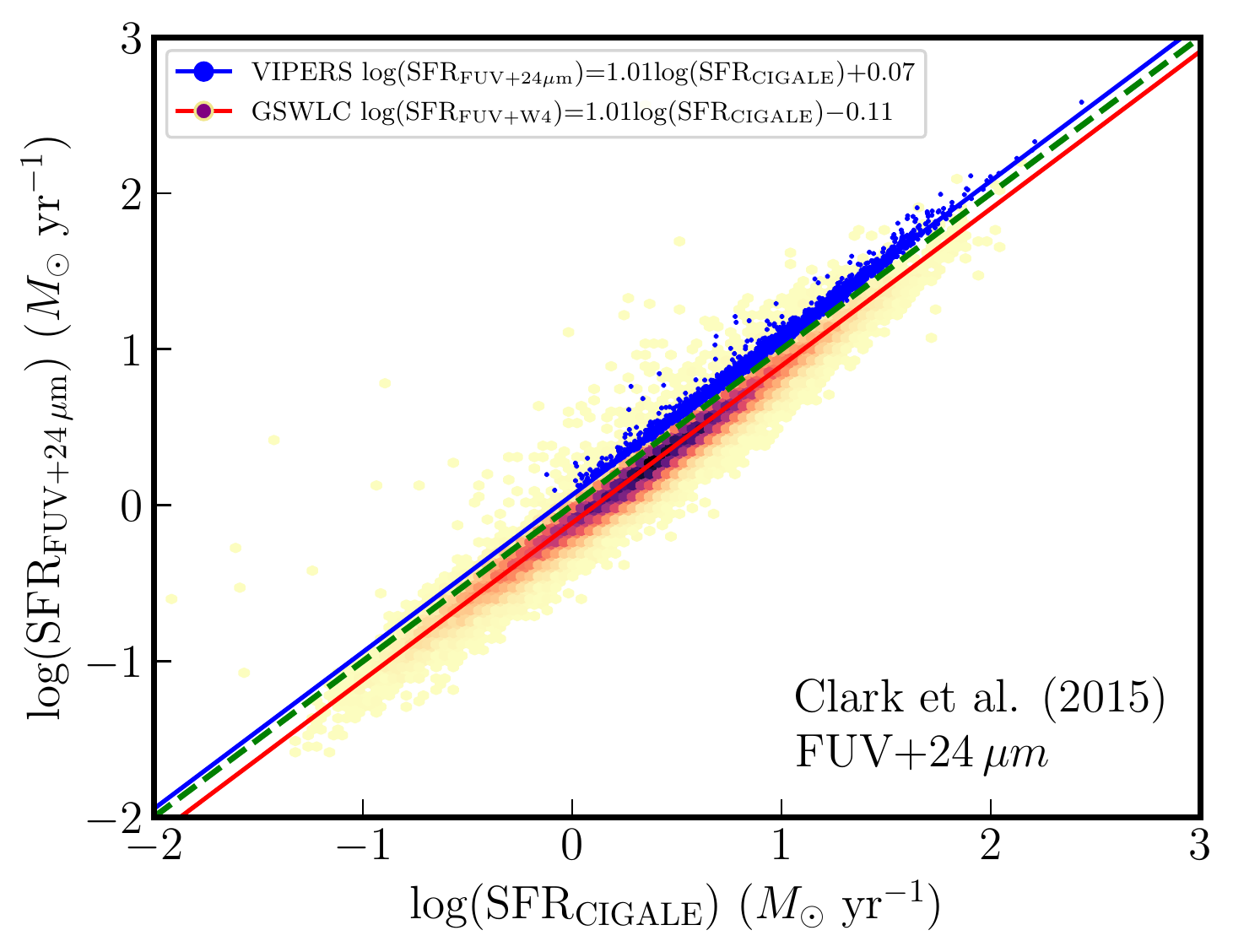}\includegraphics[angle=0,width=0.5\textwidth]{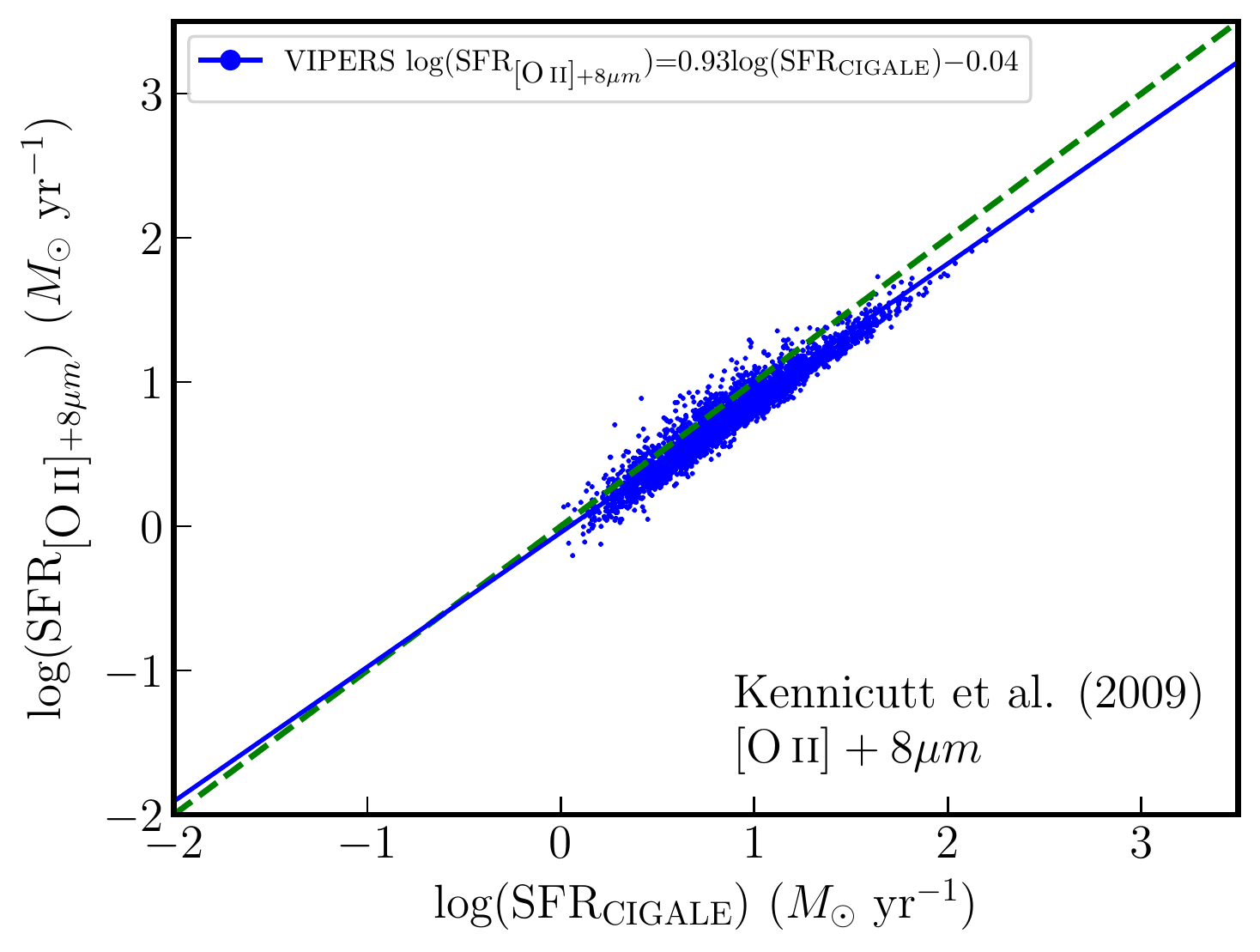}
\includegraphics[angle=0,width=0.5\textwidth]{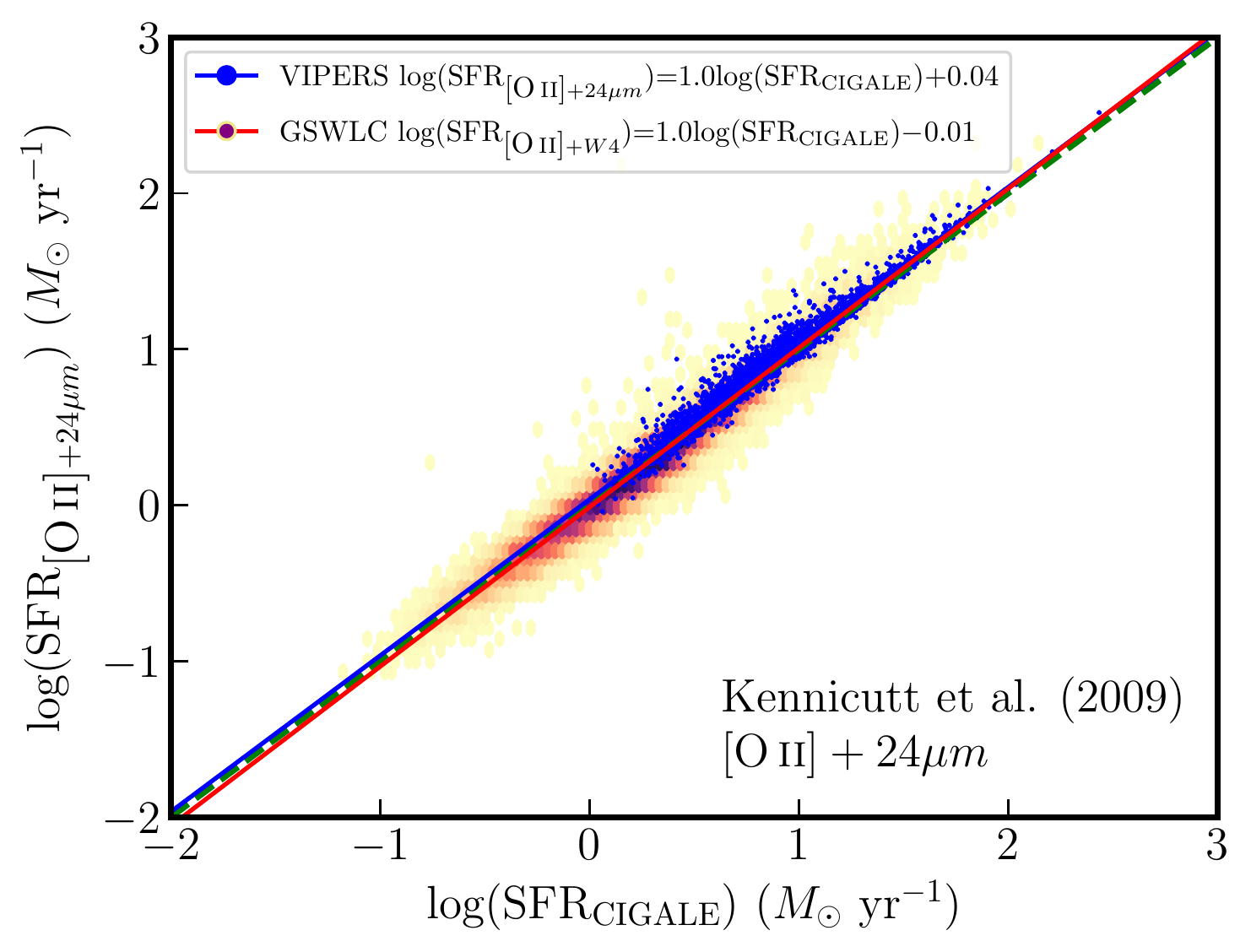}\includegraphics[angle=0,width=0.5\textwidth]{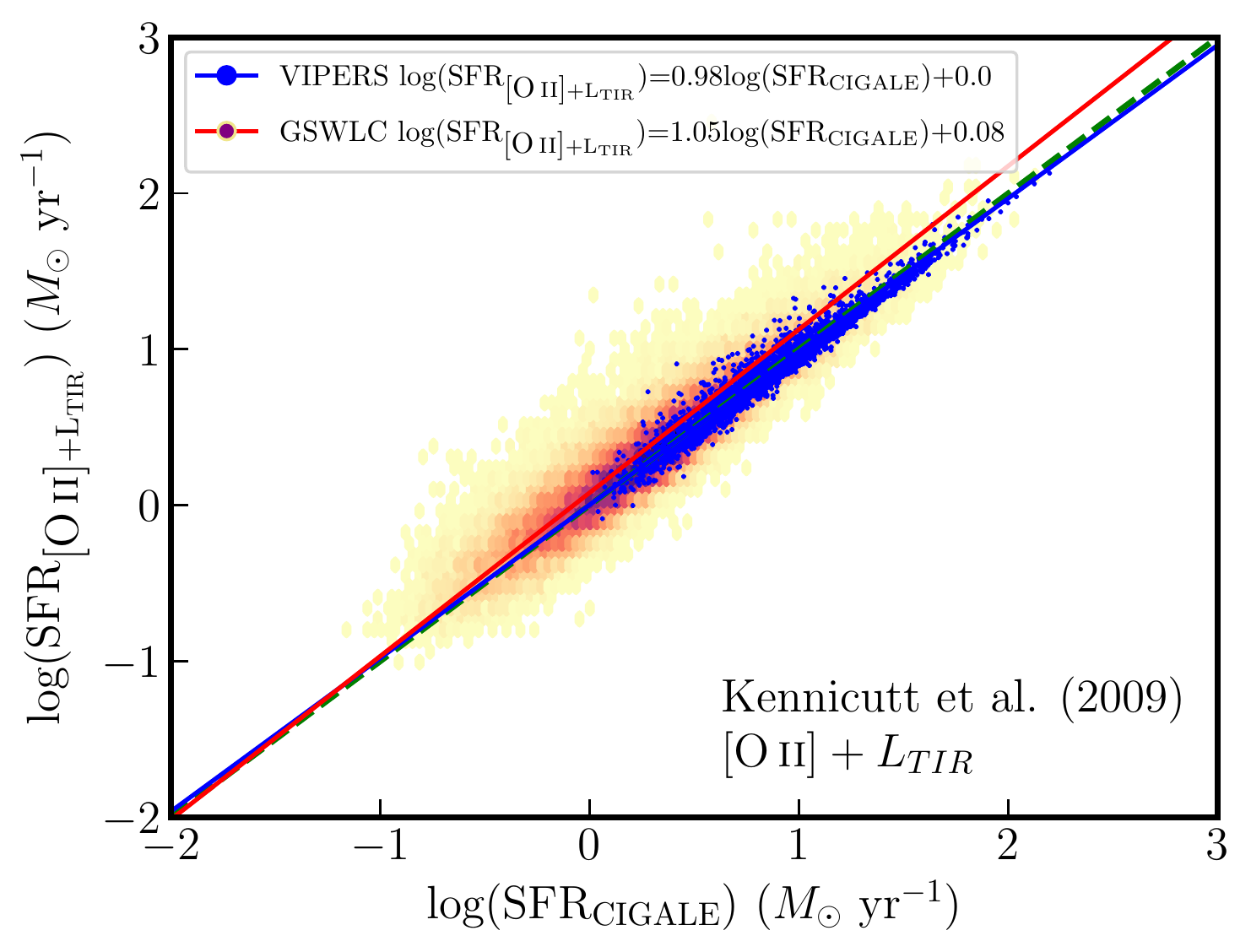}
\caption{SFR derived from the relations given in \citet{cla15} using \textit{GALEX} FUV and $L_{24\mu m}$ or $L_{W4}$ using the old stellar population contamination (Eq.~\ref{eq:Percentage_dust_heat}) (top left, mean of -0.08~dex and scatter of 0.03~dex for VIPERS, 0.12 and 0.10~dex for GSWLC). SFR derived from the relations given in \citet{ken09} using [O\,{\sc{ii}}] associated with 8~$\mu$m (top right, mean of 0.12~dex and scatter of 0.09~dex for VIPERS), 24~$\mu$m (bottom left, mean of -0.04~dex and scatter of 0.06~dex for VIPERS, 0.02 and 0.10~dex for GSWLC), and $L_{TIR}$ (bottom right, {mean of 0.01~dex and scatter of 0.07~dex for VIPERS, -0.06 and 0.18~dex for GSWLC)}.} 
\label{Fig_Ken09}
\end{figure*}

Other studies such as \citet{del14} and \citet{via16} derived such a relation between $R_{heat-dust}^{young}$ and sSFR, whose shape resembles the relation of \citet{ner19} but gives a higher $R_{heat-dust}^{young}$ at a lower sSFR. Applied to GSWLC, these relations would cause the fraction of dust heated by young stars to increase by $\sim$20\%, giving a contamination on the order of 25\%. \citet{ner19} attributes this offset to the different methods used to retrieve this fraction. Such low contamination was observed by \citet{bua11a} for a sample of local galaxies from AKARI with $R_{heat-dust}^{old}$=17\%, but the contamination reaches $\sim$10\% for sSFR similar to VIPERS.\\ Other works (e.g., \citealt{hir03,bel03,hao11}) found values on the order of 30\%\ to 50\%. This underlines the fact that an accurate derivation is not trivial and depends on the method and stellar population templates used to derive this fraction. The main conclusion, in agreement with most of the studies, is that the contamination of dust emission by the old stellar population can be significant, even for blue star-forming galaxies. Using the values of contamination that we found (Eq.~\ref{Eq:dust_heat}) in the calibration proposed by \citet{cla15}, we obtain a good agreement with CIGALE (Fig.~\ref{Fig_Ken09} top left).

Instead of UV, \citet{ken09} used the [O{\,\sc{ii}}] spectral line as a direct tracer of young high-mass stars and different bands to represent the PAH or dust emission (8, 24~$\mu$m and $L_{TIR}$, see Fig.~\ref{Fig_Ken09}). Using the 8-$\mu$m band as a complement of the blue band leads to a similar underestimation of the SFR as in \citet{bro17}, which may originate from the inadequacy of the 8-$\mu$m emission as a reliable tracer of SFR due to its dependence on metallicity. There is a good agreement when using $L_{TIR}$ as a tracer of obscured SFR for the VIPERS galaxies but the best agreement, both for VIPERS and GSWLC, is reached when using 24~$\mu$m as a tracer of dust emission.\\

Similarly to \citet{bel05}, \citet{arn13} used a relation between $L_{NUV}$ and $L_{TIR}$ to compute the SFR where $L_{TIR}$ is estimated from the NUVrK$_{\rm{s}}$ diagram. Briefly, the infrared excess, defined as IRX=$L_{TIR}/L_{NUV}$, is found to be constant along with stripes in the $NUVrK$ diagram, and can be parametrized using the vector perpendicular to those stripes (the NRK vector) and $z$. Prior to estimating the SFR, we only kept galaxies at $z>0.1$ (19~441 galaxies, 21\% of the GSWLC sample) because the evolution of IRX with respect to $z$ is not constrained around $z\sim0$. We also removed the passive galaxies from this GSWLC subsample ($NRK_{sSFR}>1.9$, 3 galaxies).\\
The SFR given by the NRK method for VIPERS galaxies has a scatter of 0.14~dex and is slightly overestimated at low SFR by around 0.1~dex (Fig.~\ref{Fig_Cal_Brown} middle right). In addition, 265 VIPERS galaxies (8\% of the sample) are found to have catastrophic SFR estimations ($|\textrm{SFR}_{\textrm{CIGALE}}-$SFR$_{\textrm{NrK}}|>3\sigma$, \citealt{arn13}). For GSWLC, the SFR estimation based on the NUVrK diagram underestimates the SFR by $\sim$0.21~dex, with catastrophic estimations for 1558 GSWLC galaxies (8\% of the sample). For both samples, the offsets are less than the accuracy of the method derived by \citet{arn13} to recover $L_{TIR}$, which is close to 0.3~dex for the low-z sample and 0.21~dex for $0.2<z<1.3$. Additionally, \citet{arn13} showed that the observed stripes (with constant IRX) can be only reproduced using a two-component dust attenuation law (birth clouds + ISM), whereas the dust attenuation correction for GSWLC is based on a modified Calzetti attenuation law. \\

\begin{figure}[h!]
\centering
\includegraphics[angle=0,width=0.43\textwidth]{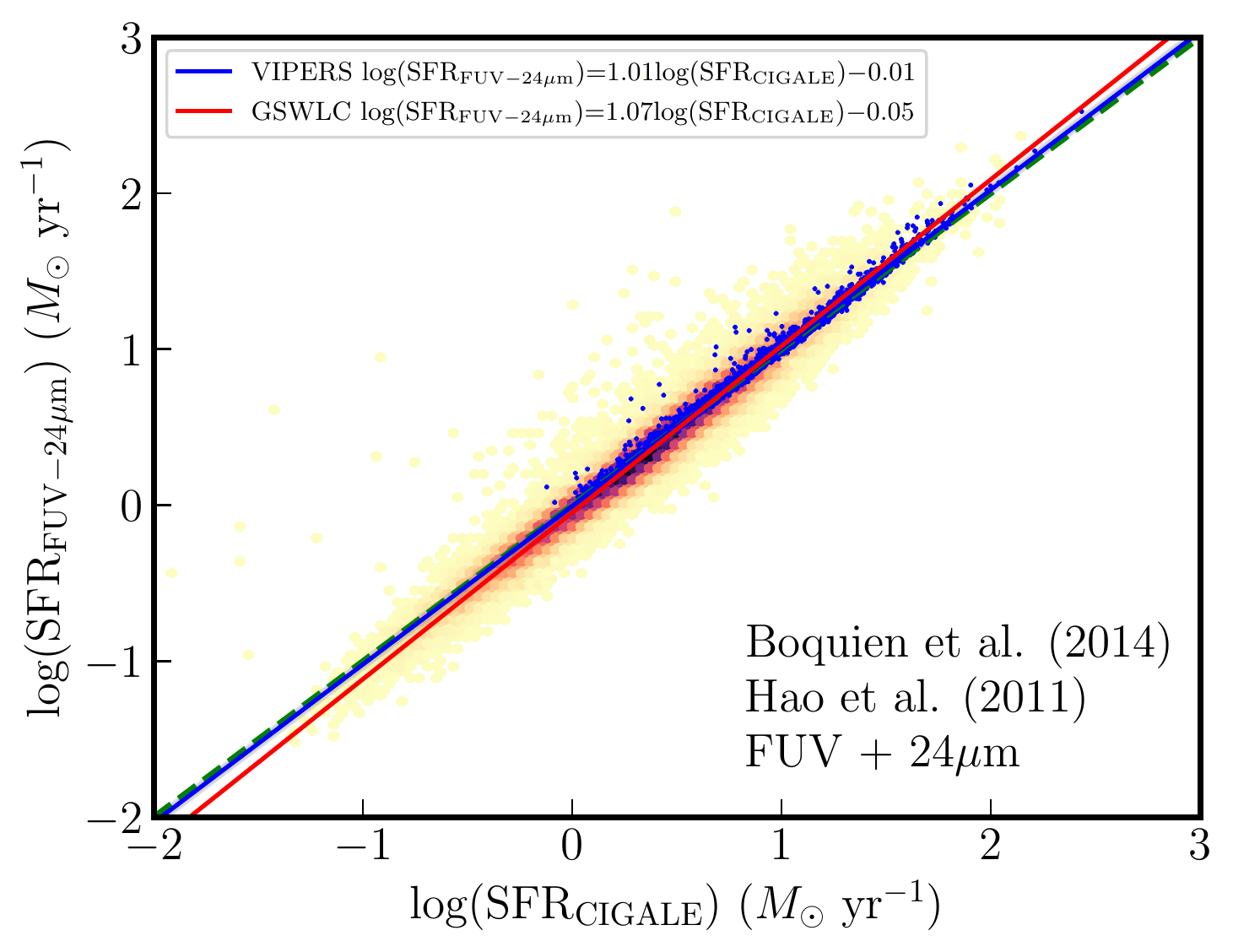}
\caption{SFR derived from the relation given in \citet{boq14} using the coefficient of \citet{hao11}. There is an excellent agreement with SFR from CIGALE for VIPERS (mean of -0.01~dex and scatter of 0.03~dex) and GSWLC (0.02 and 0.10~dex)} 
\label{Fig_Boq14}
\end{figure} 

Using eight nearby galaxies from the KINGFISH sample, \citet{boq16} derived the fraction of IR luminosity that has to be added to the attenuated $L_{FUV}$ to obtain a reliable estimation of the SFR. Based on CIGALE estimations, they found that the coefficient that should be applied to $L_{24\mu m}$ depends on the sSFR and can vary by an order of magnitude. This shows that it is generally not correct to use the same single value for all the galaxies in a sample. However, without resolved observations at NIR wavelength, we are forced to use a simple value representing the whole sample. Using the relation of \citet{hao11} to scale $L_{24\mu m}$ and the conversion factor from FUV to SFR from \citet{boq14}, we found a good agreement with the CIGALE SFR (see Fig.~\ref{Fig_Boq14} middle).\\

Composite calibrations offer a way to estimate the total SFR (obscured + unobscured), which is very a convenient method as dust attenuation in the blue part of the spectrum can carry significant uncertainties. Nonetheless, two bands or more are required to estimate the SFR, which are not necessarily obtainable for every galaxy. In addition, the monochromatic data must be k-corrected when $z>0$, which involves SED fitting from which we can directly obtain an estimation of the SFR. As a consequence, the usefulness of this method really depends on the availability of multiwavelength data for the galaxies considered.

\subsection{SFR from spectral lines}

\subsubsection{H$\beta$ line}

Toward local galaxies, the H$\beta$ line is mostly used to correct for dust attenuation through the Balmer decrement but generally not as an SFR indicator, as H$\alpha$ is available. When the redshift increases, H$\alpha$ is shifted out the optical window and is more difficult to observe. Making use of H$\beta$ as an SFR tracer relies on the assumption that it can be related to H$\alpha$ through the intrinsic Balmer decrement. While this value can be different based on the temperature and electronic density of the H{\,\sc{ii}} region, it is generally assumed to be equal to 2.86.

\begin{figure}[h!]
\centering
\includegraphics[angle=0,width=0.5\textwidth]{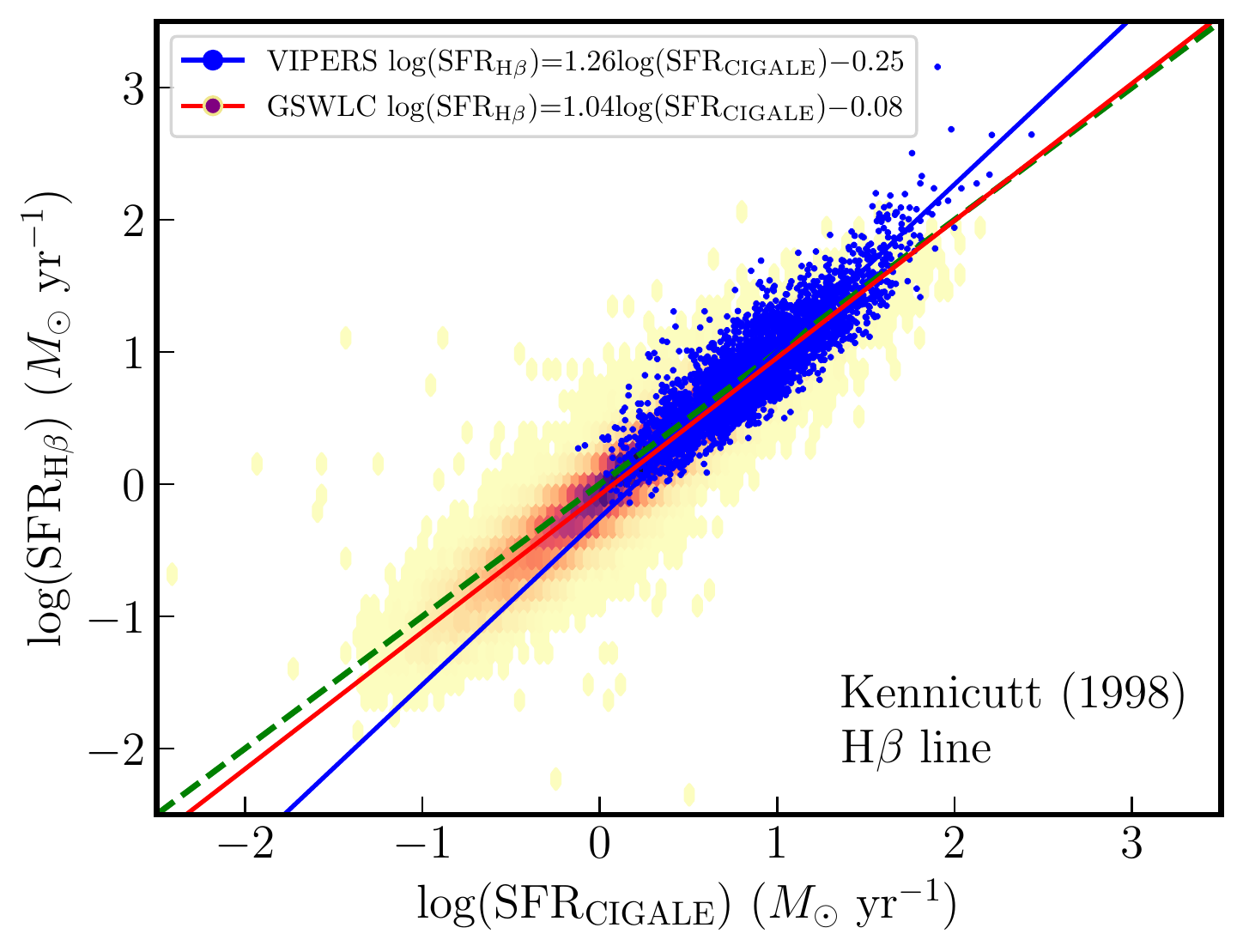}
\caption{SFR derived from H$\beta$ \citep{ken98a} assuming (H$\alpha$/H$\beta$)$_{\rm{int}}=2.86$ vs. SFR derived from CIGALE. A good agreement is observed for GSWLC (mean of 0.07~dex and scatter of 0.19~dex), while there is an overestimation at high SFR for VIPERS (-0.01 and 0.19~dex).}
\label{Fig:Hb_Kennicutt}
\end{figure}

For VIPERS, the SFR obtained from H$\beta$ and the H$\alpha$ calibration from \citet{ken98a}, assuming a Balmer decrement, reproduces the SFR given by CIGALE with a scatter of 0.19~dex, but an overestimation of the SFR can be observed around $\sim100$~$M_{\odot}$~yr$^{-1}$ (Fig.~\ref{Fig:Hb_Kennicutt}). Different electron density ($n_e$) and electron temperature ($T_e$) values for H{\,{\sc{ii}}} regions translates into different values of the Balmer decrement. For $n_e=10^2$~cm$^{-3}$ and $T_e=5000,10~000,20~000$~K, the Balmer decrement ranges from 3.03 to 2.74 \citep{hum87} but cannot explain the deviation observed. We checked if these highly active galaxies could be Seyfert 2 but most of them are located in the star-forming part of the BPT diagram (Fig.~\ref{Fig:BPT_VIPERS}). We checked the evolution of the S/N ratio of fluxes and EWs of the VIPERS star-forming galaxies with respect to the redshift but no change is observed, indicating that a selection bias at high luminosity cannot explain the observed overestimation. Such an overestimation is also seen when using a stellar absorption correction of 3 and 4~\AA. A more detailed study would require H$\alpha$ measurements in order to see if this overestimation is also observed. For GSWLC, the agreement with CIGALE SFR is excellent (scatter of 0.19~dex).

\subsubsection{$\mbox{[O\,\sc{ii}]}\lambda 3727$ line}\label{Subsect:OII}

The [O{\,\sc{ii}}] line is often used as an SFR tracer as it originates from the same location as H$\alpha$, with a similar timescale. At $z\sim0.7$, \citet{mai15} consider the SFR calibration based on [O\,{\sc{ii}}] to be the second best one after H$\alpha$ using the calibration of \citet{gil10}. However, [O{\,\sc{ii}}] suffers from two main disadvantages: high dust attenuation and a dependence on metallicity. While dust attenuation can be corrected easily in this work as CIGALE estimates the attenuation in specific bands, the metallicity is more difficult to constrain. Several studies have been performed at low redshifts to estimate the metallicity using the H$\alpha$, H$\beta$, [N{\,\sc{ii}}], [S{\,\sc{ii}}], [O{\,\sc{ii}}], and [O{\,\sc{iii}}] lines (e.g., \citealt{kew02,zhu19}). Since only the three bluest lines (H$\beta$, [O{\,\sc{ii}}] and [O{\,\sc{iii}}]) are available in VIPERS, we based our metallicity estimation on $R_{23}$ \citep{pag79}, defined as:

\begin{equation}
R_{23}=\frac{[\rm{O\mbox{\sc{ii}}}]\lambda3727+\rm{[O\mbox{\sc{iii}}}]\lambda\lambda4959,5007}{\rm{H}\beta},
\end{equation}

\noindent and the calibration of \citet{zar94} (hereafter Z94):

\begin{equation}\label{eq:metal_zar94}
{\rm{log(O/H)+12}} = \left\{
    \begin{array}{l}
        9.265-0.33x - 0.202x^{2}\\
       \hspace*{2.05cm} -\,0.207x^{3}-0.333x^{4},\\
        x={\rm{log(}}R_{23}), \\
        \\
         {\mbox{Valid for }\rm{log(O/H)+12}}>8.4.\\
    \end{array}
\right.
\end{equation}

\begin{figure}[h!]
\includegraphics[width=\linewidth]{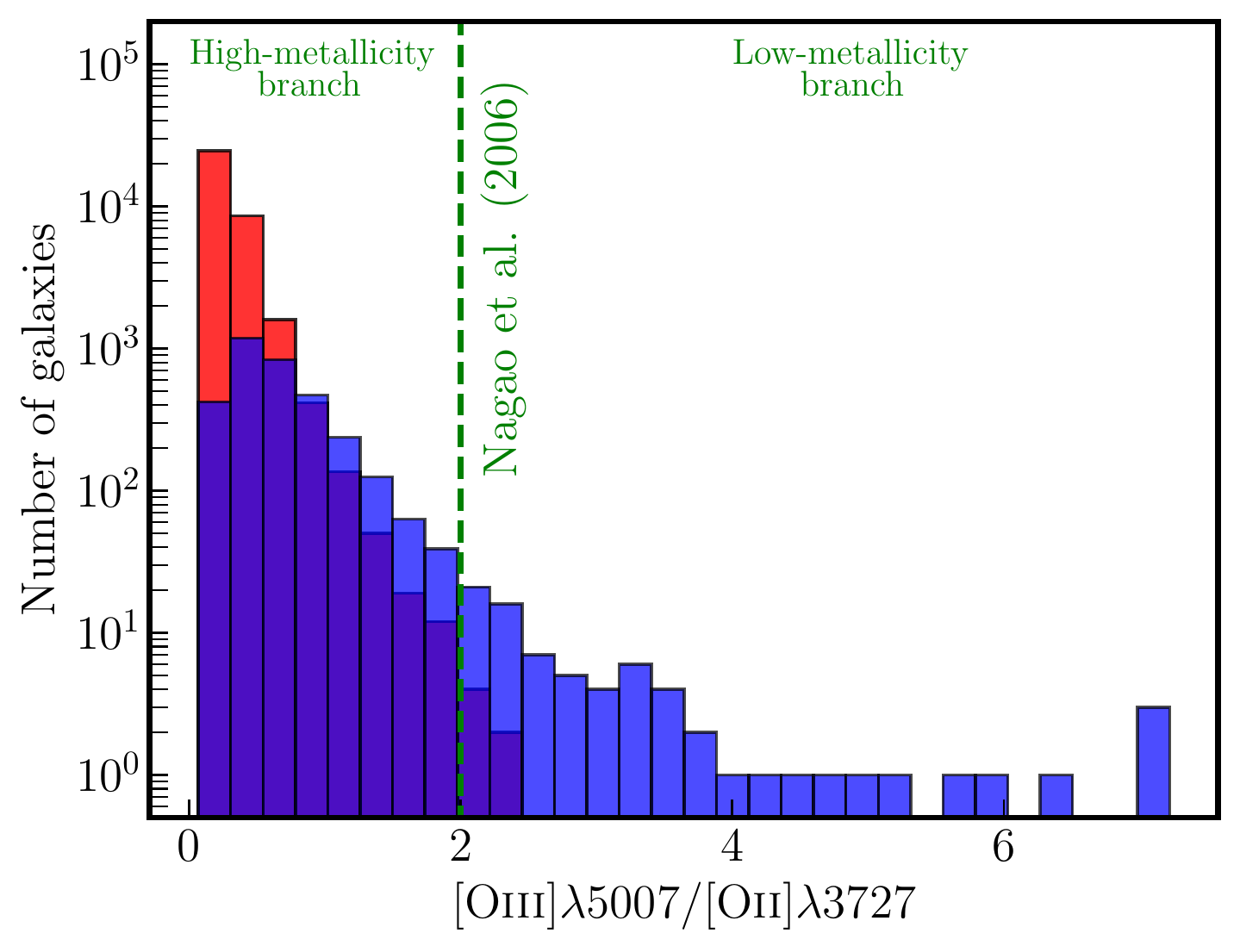}
\caption{Histogram of [O\,{\sc{iii}}]$\lambda$5007/[O\,{\sc{ii}}]$\lambda$3727 for the GSWLC (red) and the VIPERS star-forming (blue) samples. The boundary between the lower and upper branch in the $R_{23}$ diagnostic from \citet{nag06} is shown as a dashed green line. Most of the galaxies are found on the high-metallicity branch (99.98\% for GSWLC and 98\% for VIPERS).} 
\label{Fig:OIIIOII}
\end{figure}

\noindent Diagnostics based on $R_{23}$ are problematic because this parameter is double-valued. Therefore, this method is often used in combination with a first estimation of the metallicity to break the degeneracy \citep{kew02}. Since we did not have another reliable method to make the first estimation, we based our choice of the $R_{23}$ branch on the [O\,{\sc{iii}}]$\lambda$5007/[O\,{\sc{ii}}]$\lambda$3727 ratio. In \citet{nag06}, this ratio was proposed to distinguish between the two branches, where the high-metallicity branch is associated with [O\,{\sc{iii}}]$\lambda$5007/[O\,{\sc{ii}}]$\lambda 3727<2$ (Fig.~\ref{Fig:OIIIOII}). We remind the reader that the estimation of metallicity for GSWLC is based on the [O\,{\sc{ii}}]+[O\,{\sc{iii}}] sample (see Table~\ref{tab:GSWLC_flags_selection}). Based on this criterion, we found that five GSWLC (0.02\%) and 76 VIPERS (2\%) galaxies are part of the lower metallicity branch. These galaxies were excluded from SFR comparisons and we focused on high-metallicity galaxies, which are predominant in our sample.\\

In addition, the $R_{23}$ parameter depends on the [O{\sc{iii}}]$\lambda5007$/H$\beta$ ratio, which increase with redshift (Sect.~\ref{subsubsect:VIPERS_BPT}). For increasing ionization parameters ranging from $-$3.2 (typical value for SDSS galaxies, \citealt{lia06}) to $-$2.3, the metallicity is overestimated by 0.1~dex at most \citep{cul16}, which corresponds to the uncertainty of the $R_{23}$ method. This dependence is, therefore, not problematic at $z<1,$ but should be considered when estimating and comparing the metallicity over a higher and wider range of redshift.\\

\begin{figure}[h!]
\centering
\includegraphics[angle=0,width=0.5\textwidth]{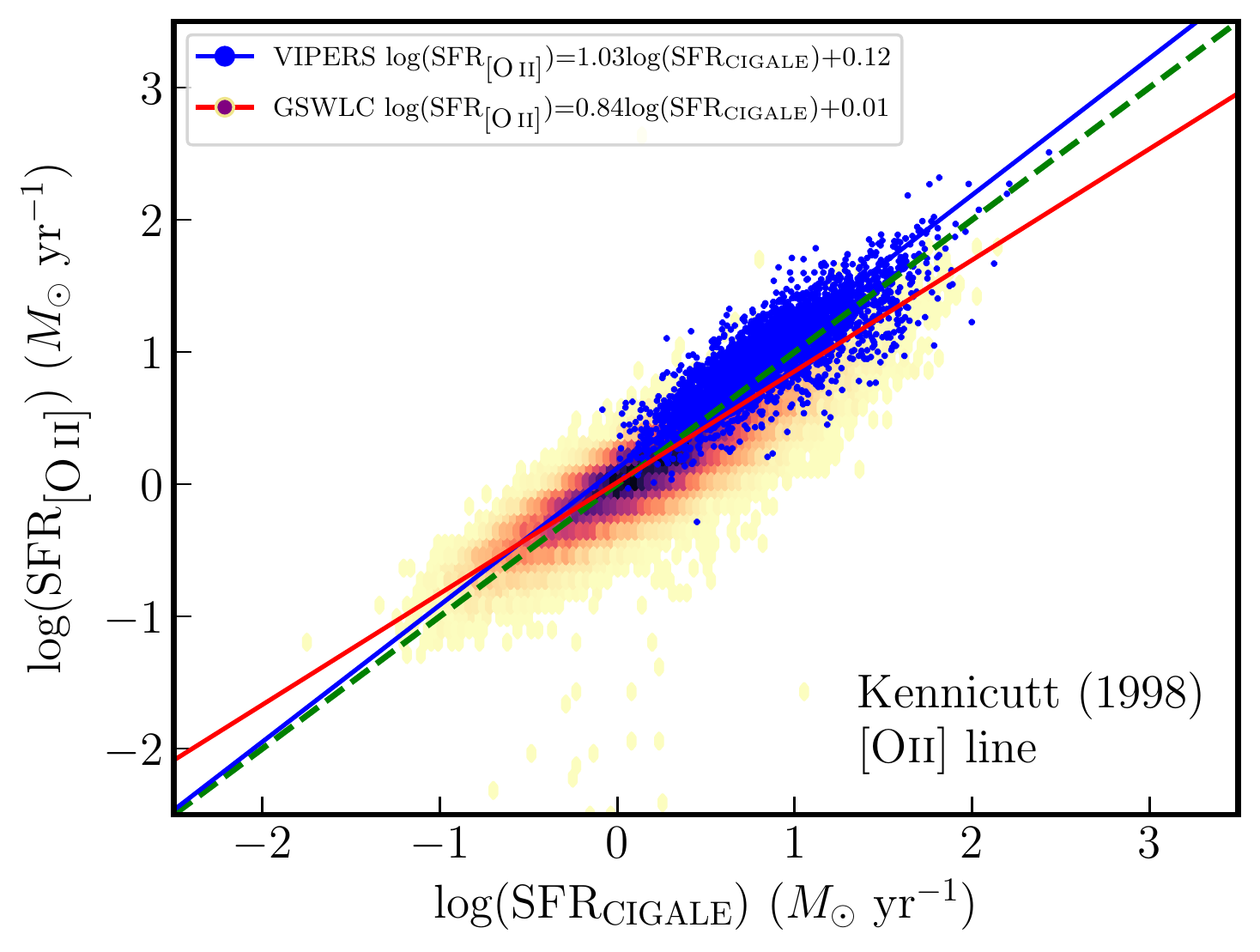}
\caption{SFR derived from [O{\,\sc{ii}}] using the direct calibration from \citet{ken98a}. A good agreement is observed for VIPERS (mean of -0.11~dex and scatter of 0.19~dex), whereas the SFR for GSWLC is underestimated at high SFR (0.08 and 0.23~dex).} 
\label{Fig:OII_Kennicutt98}
\end{figure}

To estimate the SFR from the [O{\,\sc{ii}}] line, we first used the initial calibration of \citet{ken98a}. Despite assuming an average reddening based on the sample used by \citet{ken98a}, the agreement is good with a scatter of 0.19~dex for VIPERS and a mean difference of 0.11~dex (Fig.~\ref{Fig:OII_Kennicutt98}), while the SFR is slightly underestimated at high SFR for GSWLC.\\

\begin{figure*}[h!]
\includegraphics[width=0.5\linewidth]{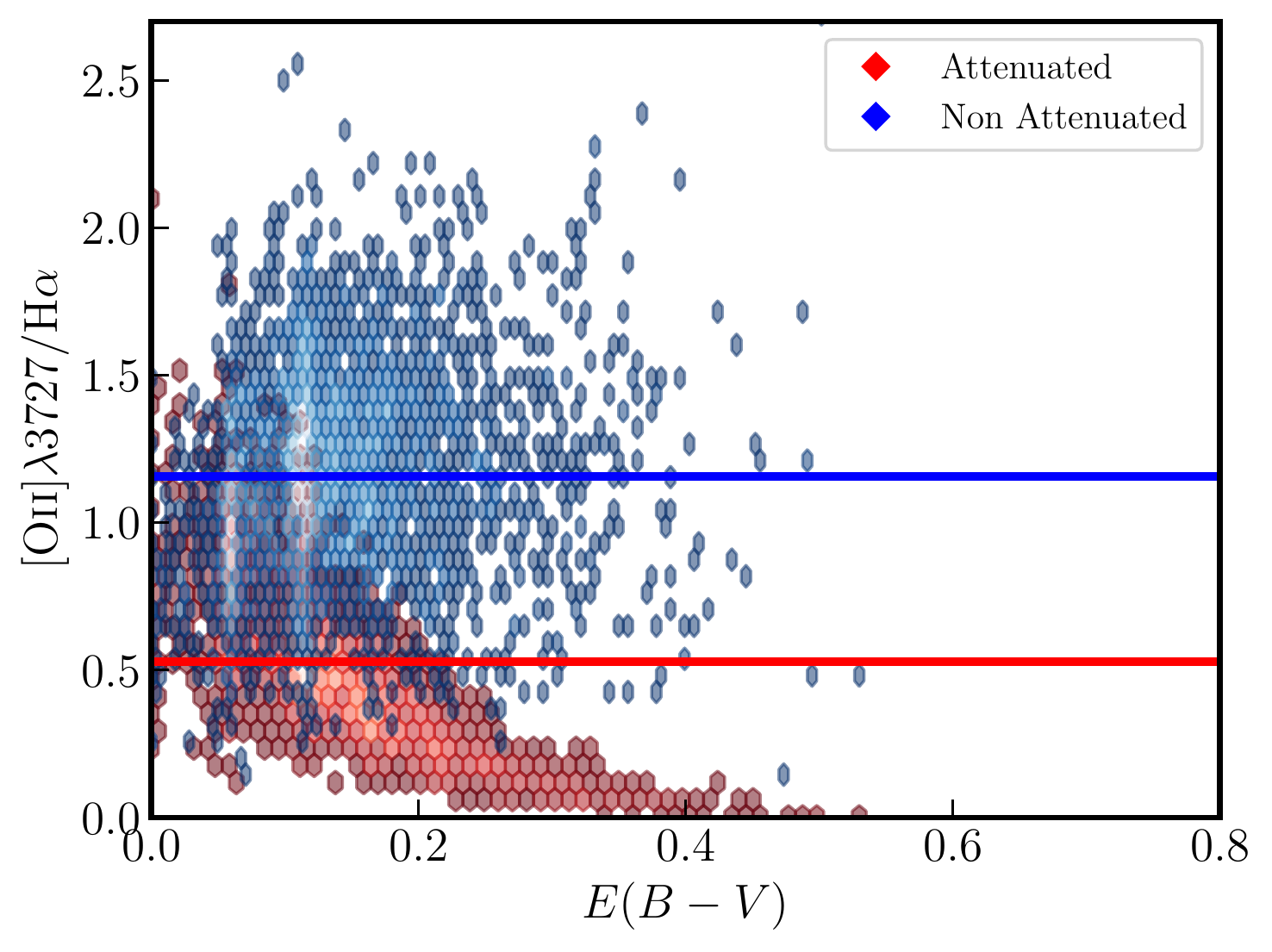}\includegraphics[angle=0,width=0.5\textwidth]{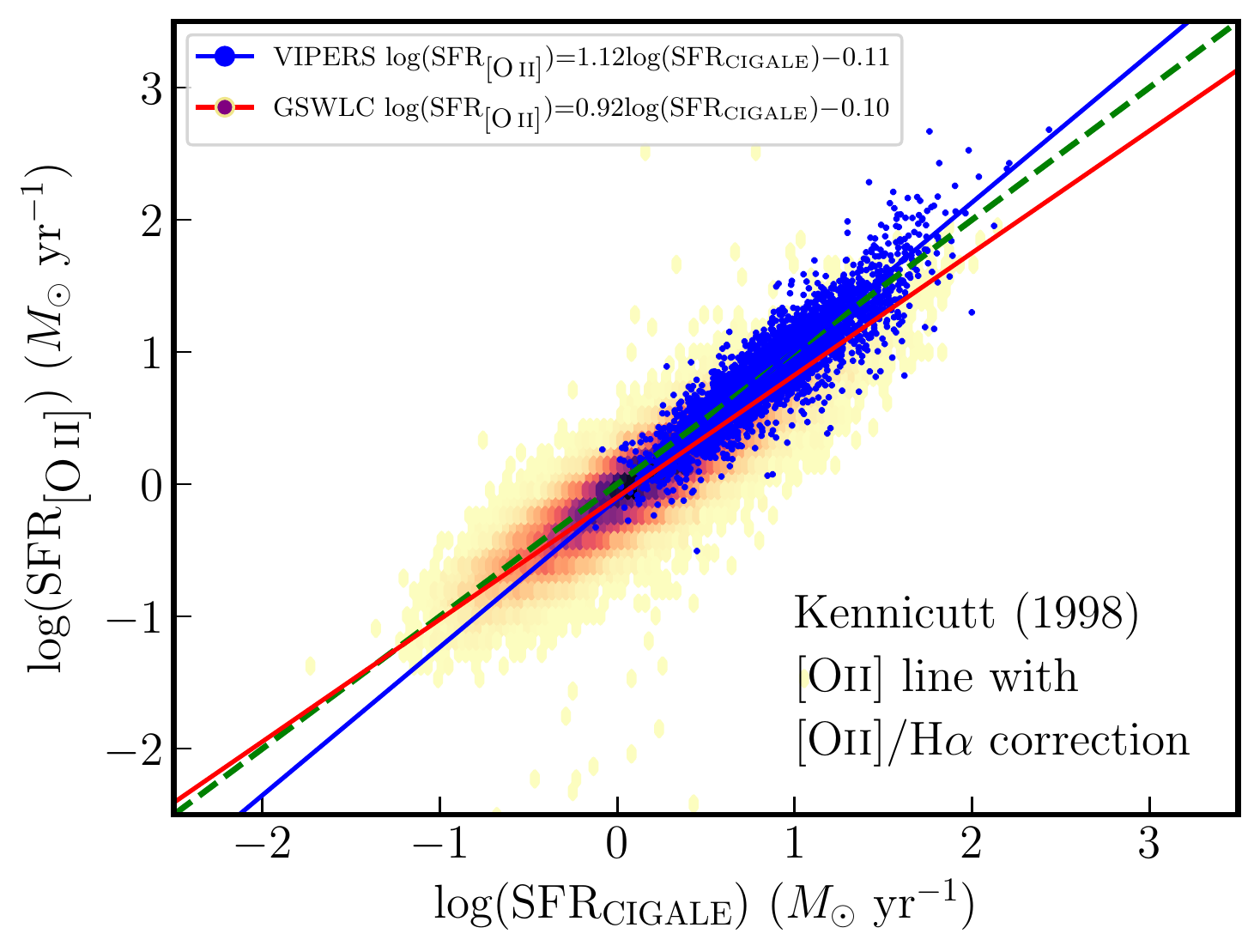}
\caption{SFR derivation from [O{\,\sc{ii}}] using \citet{ken98a} and a reddening correction. Left panel: [O{\,\sc{ii}}]/H$\alpha$ ratio (assuming H$\alpha=2.86$H$\beta$) as a function of the color excess, without (red) and with (blue) dust attenuation correction; the continuous lines represent the average ratio. Right panel: SFR derived from [O{\,\sc{ii}}] using the H$\alpha$ calibration of \citet{ken98a} and the attenuation-corrected [O{\,\sc{ii}}]/H$\alpha$ ratio. For VIPERS and GSWLC, the slope increases in comparison with the direct [O{\,\sc{ii}}] calibration, but the scatter decreases (mean of 0.01~dex and scatter of 0.17~dex for VIPERS, 0.12 and 0.22~dex for GSWLC).}
\label{Fig:OIIHa}
\end{figure*}

To improve this relation, we used the H$\alpha$ calibration of \citet{ken98a} and converted it to a [O{\,\sc{ii}}] SFR calibration,taking the correction-attenuated [O\,{\sc{ii}}]/H$\alpha$ ratio as a conversion factor between the H$\alpha$-SFR and [O\,{\sc{ii}}]-SFR relations. Here, the H$\alpha$ flux comes from the H$\beta$ flux, scaled by the Balmer decrement for both samples for consistency, even if H$\alpha$ is available for GSWLC. The dependence of the attenuated [O\,{\sc{ii}}]/H$\alpha$ with the color excess E(B$-$V), estimated with CIGALE, is shown in Fig.\ref{Fig:OIIHa} and disappears when the dust attenuation is correctly taken into account, with an average [O\,{\sc{ii}}]/H$\alpha$ equal to 1.16. Using this new calibration, the scatter is reduced from 0.19 to 0.17~dex at the cost of an increase in the slope (Fig.~\ref{Fig:OIIHa} right). We note that H$\alpha$ for VIPERS is derived from H$\beta,$ and so any deviation, as observed in Fig.~\ref{Fig:Hb_Kennicutt}, would increase the uncertainty of this new [O{\,\sc{ii}}] SFR calibration. For GSWLC, the scatter slightly increases but the slope gets closer to one when the proper attenuation is taken into account.\\

\begin{figure*}[h!]
\includegraphics[width=0.5\linewidth]{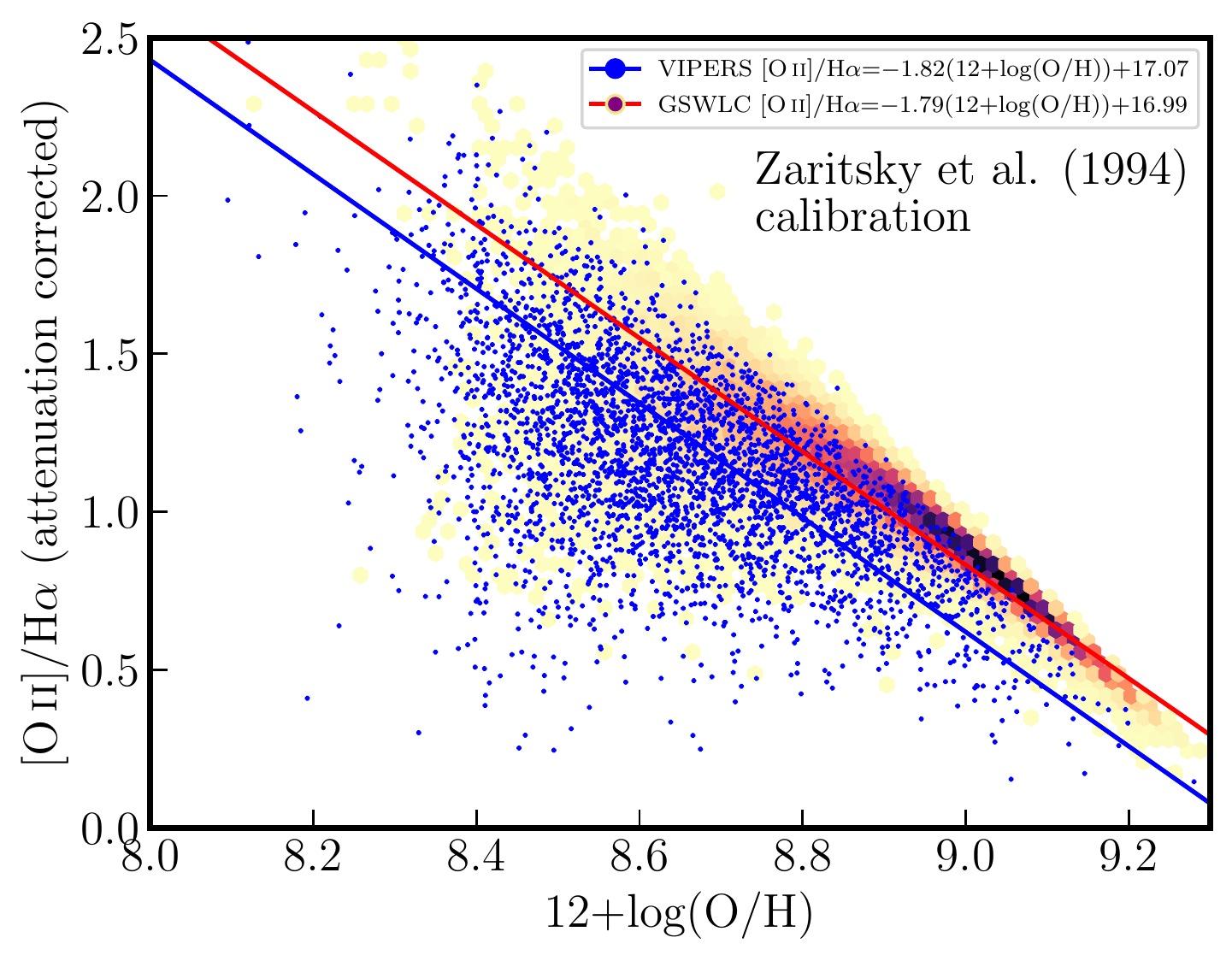}\includegraphics[angle=0,width=0.5\textwidth]{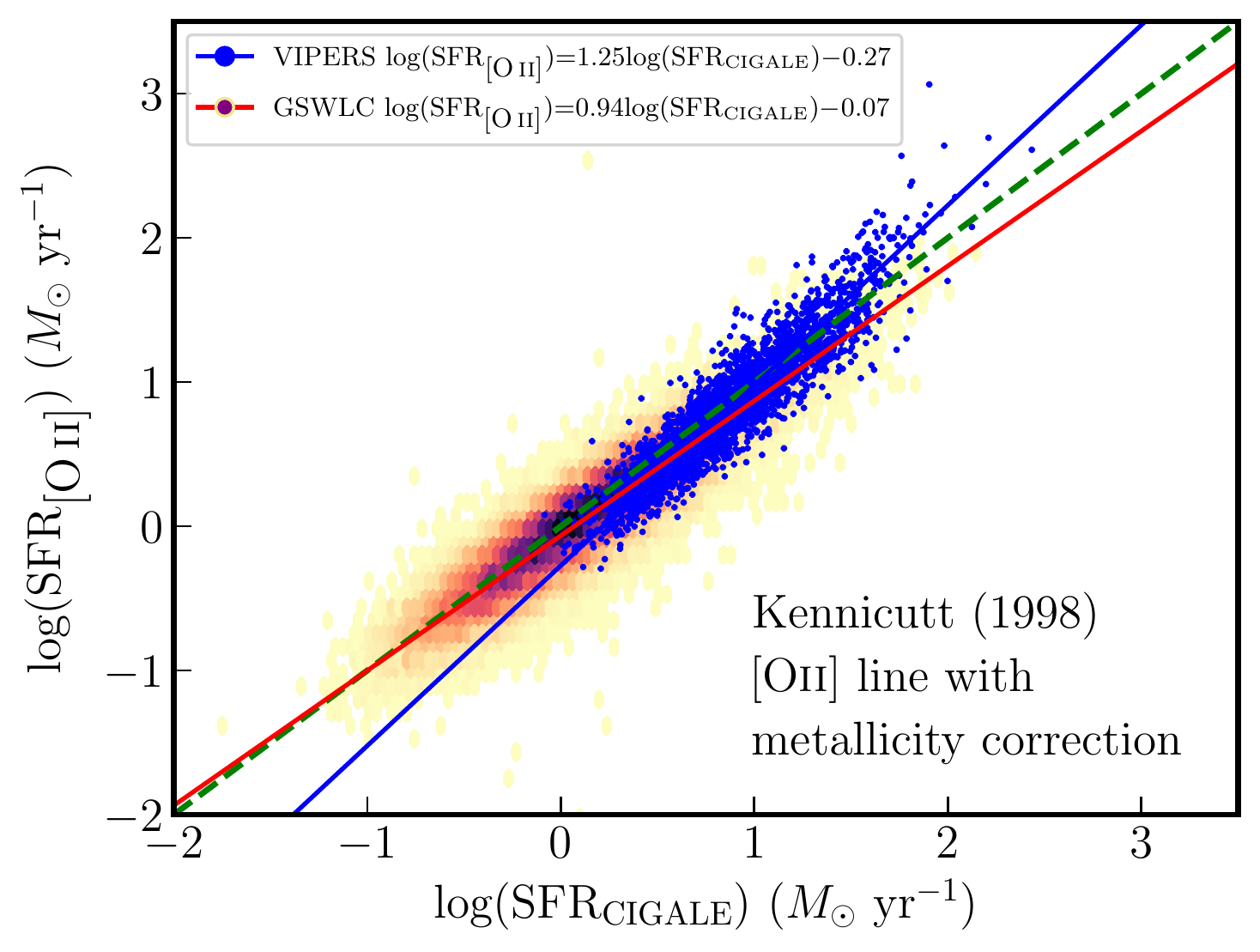}
\caption{SFR derivation from [O{\,\sc{ii}}] using \citet{ken98a} and a metallicity correction. Left panel: [O{\,\sc{ii}}]/H$\alpha$ ratio (assuming H$\alpha=2.86$H$\beta$) corrected for dust attenuation versus metallicity estimated using the \citet{zar94} calibration for the VIPERS (blue) and GSWLC (density) samples, with their associated linear fit. Right panel: SFR derived from [O{\,\sc{ii}}] using the H$\alpha$ calibration of \citet{ken98a} and the attenuation-corrected [O{\,\sc{ii}}]/H$\alpha$ expressed as a function of metallicity. The slope increases for both samples and the scatter decreases compared to the direct [O{\,\sc{ii}}] calibration (mean of 0.05~dex and scatter of 0.17~dex for VIPERS, 0.06 and 0.20~dex for GSWLC).}
\label{Fig:OIIHa_Metal}
\end{figure*}

Based on the Z94 calibration, we estimated the metallicity dependence on the [O\,{\sc{ii}}]/H$\alpha$ ratio. Figure~\ref{Fig:OIIHa_Metal} (left) shows this dependence for VIPERS and GSWLC. For GSWLC, the observed distribution is similar to the one found by \citet{zhu19} using the Z94 and \citet{kew02} methods. The higher dispersion observed for VIPERS galaxies could originate from a higher dispersion of ionization parameters, modifying the shape of the [O\,{\sc{ii}}]/H$\alpha$ ratio versus log(O/H)+12 established from theoretical models \citep{kew04}.
The fit of [O\,{\sc{ii}}]/H$\alpha$ versus 12+log(O/H) is given by:

\begin{equation}\label{eq:fit_zar94_OII}
{ [\mbox{O{\,\sc{ii}}]/H}\alpha} = \left\{
    \begin{array}{l}
        (-1.82\pm 0.08)x+(17.07\pm 0.70) \mbox{ (VIPERS)}, \\
        (-1.79\pm 0.07)x+(16.99\pm 0.65) \mbox{ (GSWLC)}, \\
        (-1.75\pm 0.25)x+(16.73\pm 2.23),\\
         \mbox{\citep{kew04}}. \\
       \\
        x=12+{\rm{log(O/H)}}\\
    \end{array}
\right.
\end{equation}

The fit for GSWLC and the relation for VIPERS are found to be in agreement within the uncertainties when compared with \citet{kew04}. We applied this metallicity calibration for galaxies that are part of the upper $R_{23}$ branch. Compared to the original law of \citet{ken98a}, we observe a slight decrease in the scatter, as well as an increase in the slope, when a metallicity correction is applied (Fig.~\ref{Fig:OIIHa_Metal} right). We observed an improvement of the SFR estimation for GSWLC when the metallicity correction is taken into account.\\

\begin{figure*}[t!]
\includegraphics[width=0.33\linewidth]{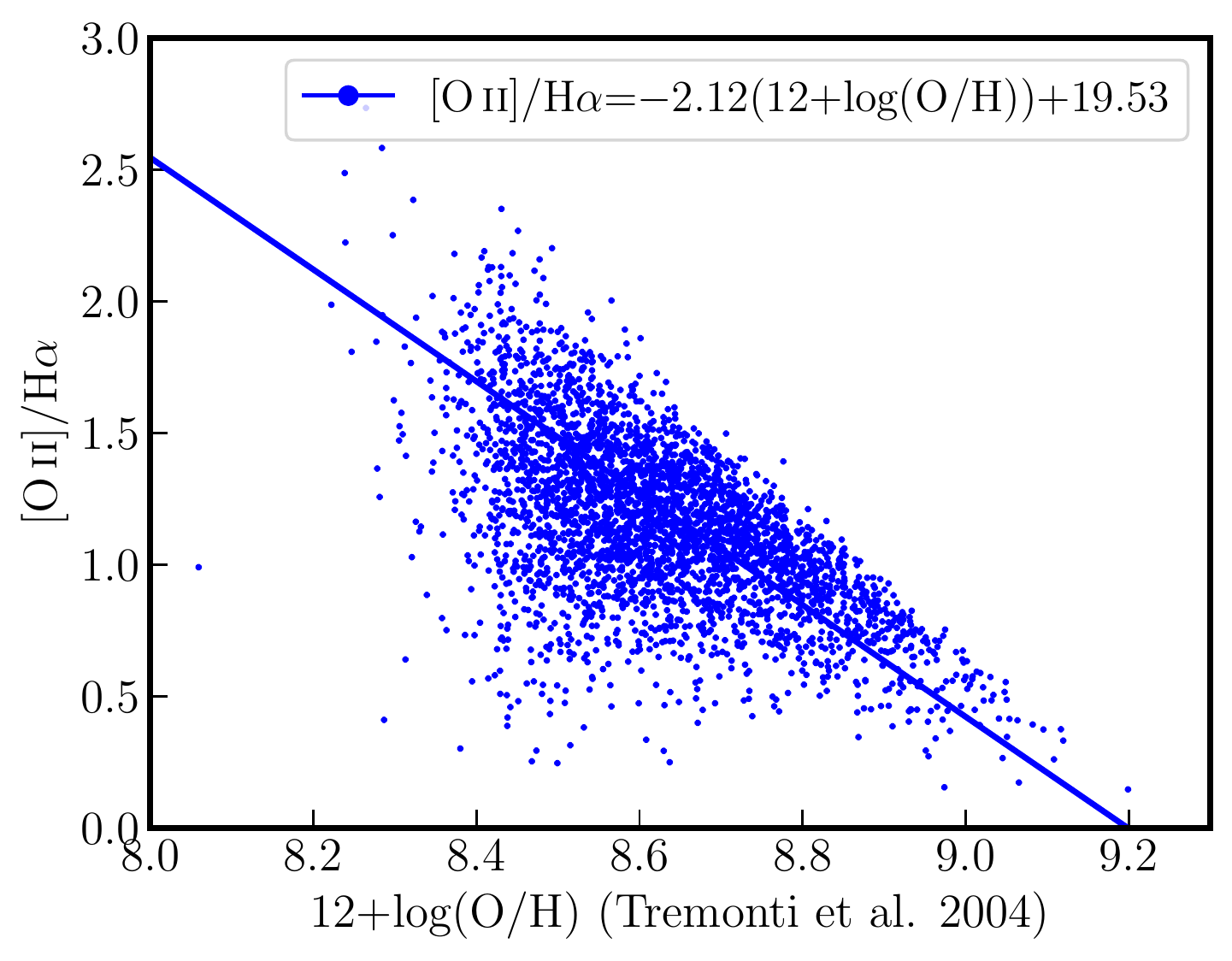}\includegraphics[width=0.33\linewidth]{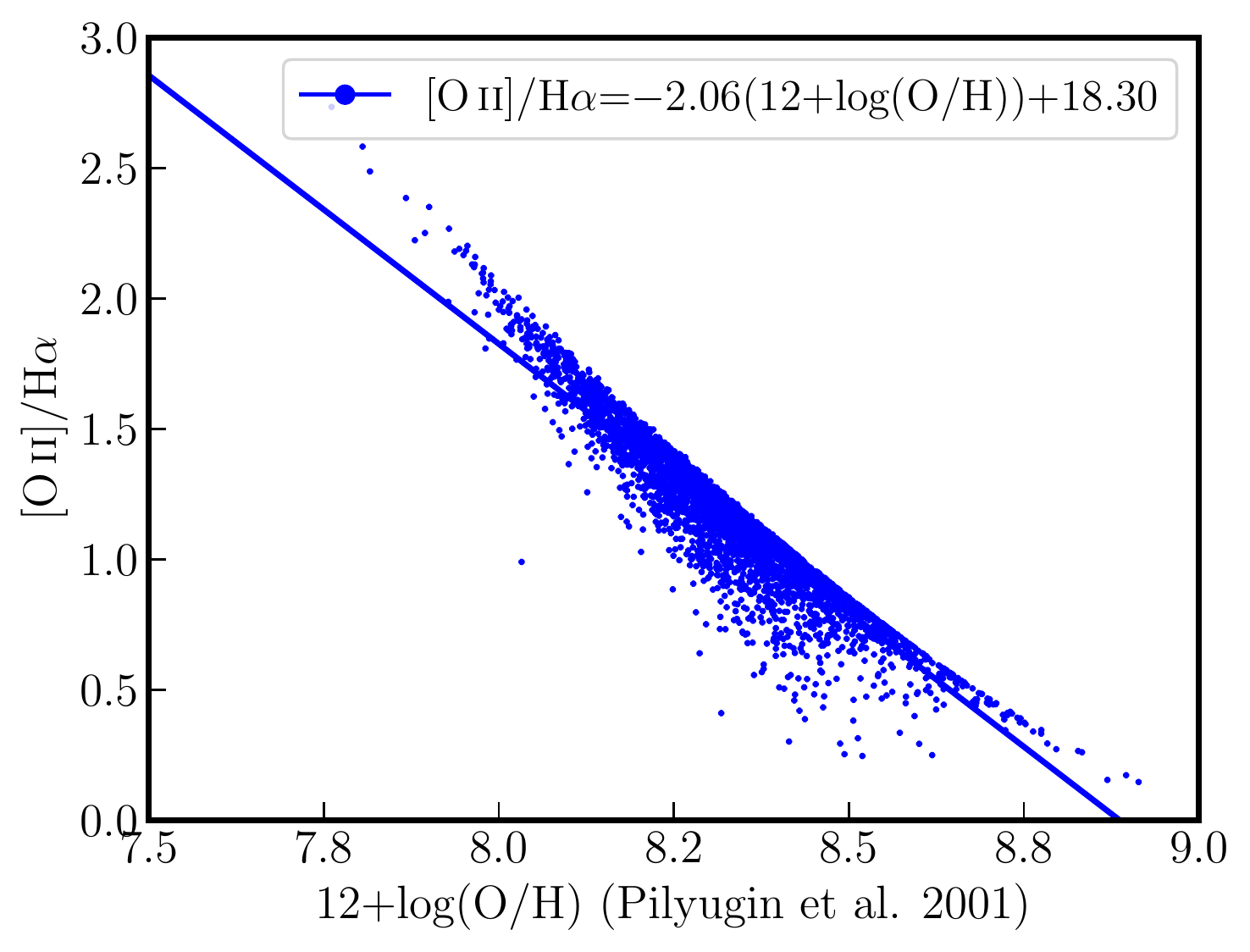}\includegraphics[width=0.33\linewidth]{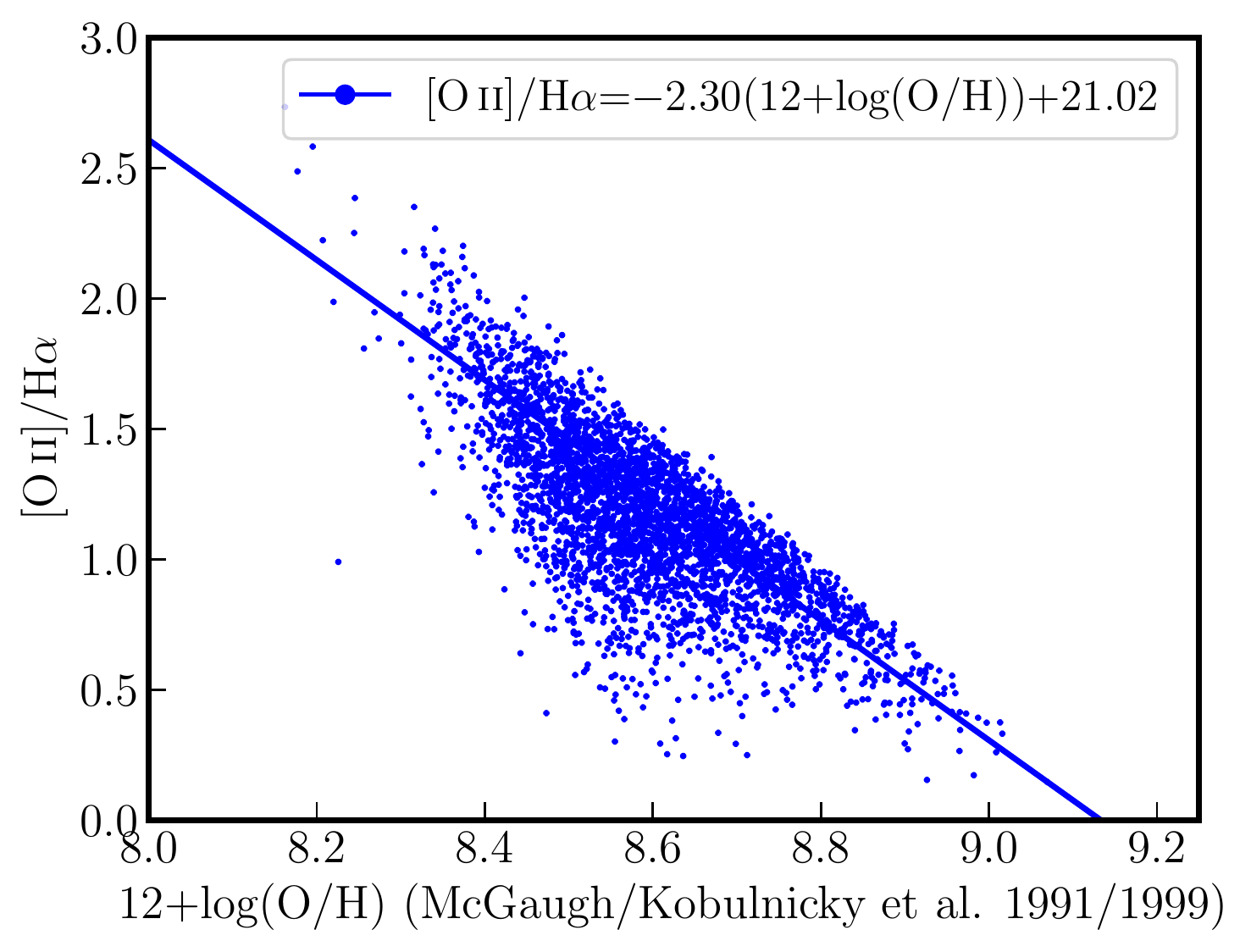}
\caption{[O{\,\sc{ii}}]/H$\alpha$ versus metallicity for VIPERS, estimated with the $R_{23}$ calibrations of \citet{tre04} (left), \citet{pil01} (middle), and \citet{mcg91} (right). Compared to Z94, there is a good agreement with \citet{tre04} and \citet{mcg91} (mean difference of 0.03 and 0.06~dex) but above the typical uncertainty for metallicity based on $R_{23}$ (0.1~dex) for \citet{mcg91} (mean difference of 0.3~dex).}
\label{Fig:OIIHa_metallicity_estimators}
\end{figure*}

Known methods to estimate the metallicity are mainly based on studies performed in local galaxies due to the availability of good estimators such as H$\alpha$ and [N\,{\sc{ii}}]. Following \citet{kew04}, we estimated the metallicity using other calibrations: \citet{mcg91}, \citet{kob99}, \citet{pil01}, and \citet{tre04}, all based on R$_{23}$.\\
We did not use the calibration of \citet{char01} because more than 92\% of the galaxies in our sample are characterized by [O{\,\sc{ii}}]/[O{\,\sc{iii}}]$\lambda 5007>0.8$. This means that the metallicity will be estimated through the [O{\,\sc{iii}}]/H$\beta$ ratio, which is more sensitive to the ionization parameter rather than the metallicity. The resulting metallicity shows a very high dispersion, in addition to being uncorrelated with the abundance, contrary to the previous calibrations of \citet{kew04} and \citet{zhu19}.\\
Figure~\ref{Fig:OIIHa_metallicity_estimators} shows the relation between [O{\,\sc{ii}}]/H$\alpha$ and metallicity from \citet{mcg91}, \citet{kob99}, \citet{pil01}, and \citet{tre04}. The shape of the distribution of galaxies is in agreement with the one based on the [N\,{\sc{ii}}]/[O\,{\sc{ii}}] from \citet{kew02}, and the one from Z94 (Fig.~\ref{Fig:OIIHa_Metal}). The lower slope with \citet{mcg91} is attributed to the different stellar atmosphere and models used for the calibration \citep{kew02}. We applied these different metallicity corrections but none of them were found to significantly improve the estimation of the SFR compared to Z94.\\

Similarly, as for the NUV band, \citet{ros02} proposed a calibration based on the [O\,{\sc{ii}}] luminosity, where the attenuation and metallicity are directly taken into account. This calibration overestimates the SFR by an offset of $\sim$0.45~dex for both samples (Fig.~\ref{Fig_Cal_Brown} bottom left). It is not surprising to observe a large discrepancy as the dust and metallicity corrections might be significantly different between VIPERS and the sample used in \citet{ros02}.\\
\citet{gil10} estimates the SFR from [O\,{\sc{ii}}], assuming a constant attenuation of 1~mag at H$\alpha$ and a general correction that takes into account the metallicity from the mass metallicity relation. Despite these strong assumptions, we found a surprisingly good agreement (Fig.~\ref{Fig_Cal_Brown} bottom right) and note that this relation gives a better agreement with CIGALE SFR for GSWLC than the metallicity-corrected [O\,{\sc{ii}}] calibration. The attenuation correction is an average value that is often used in the local Universe and we note that the metallicity correction for the calibration of \citet{gil10} is based on the SDSS DR4. This could explain why the agreement does not strongly deviate from CIGALE SFR for GSWLC, as for the calibration of \citet{ros02}, but gives a worse agreement for VIPERS.

\subsubsection{$\mbox{[O\,\sc{iii}]}\lambda 5007$ line}

\begin{figure}[h!]
\includegraphics[angle=0,width=0.5\textwidth]{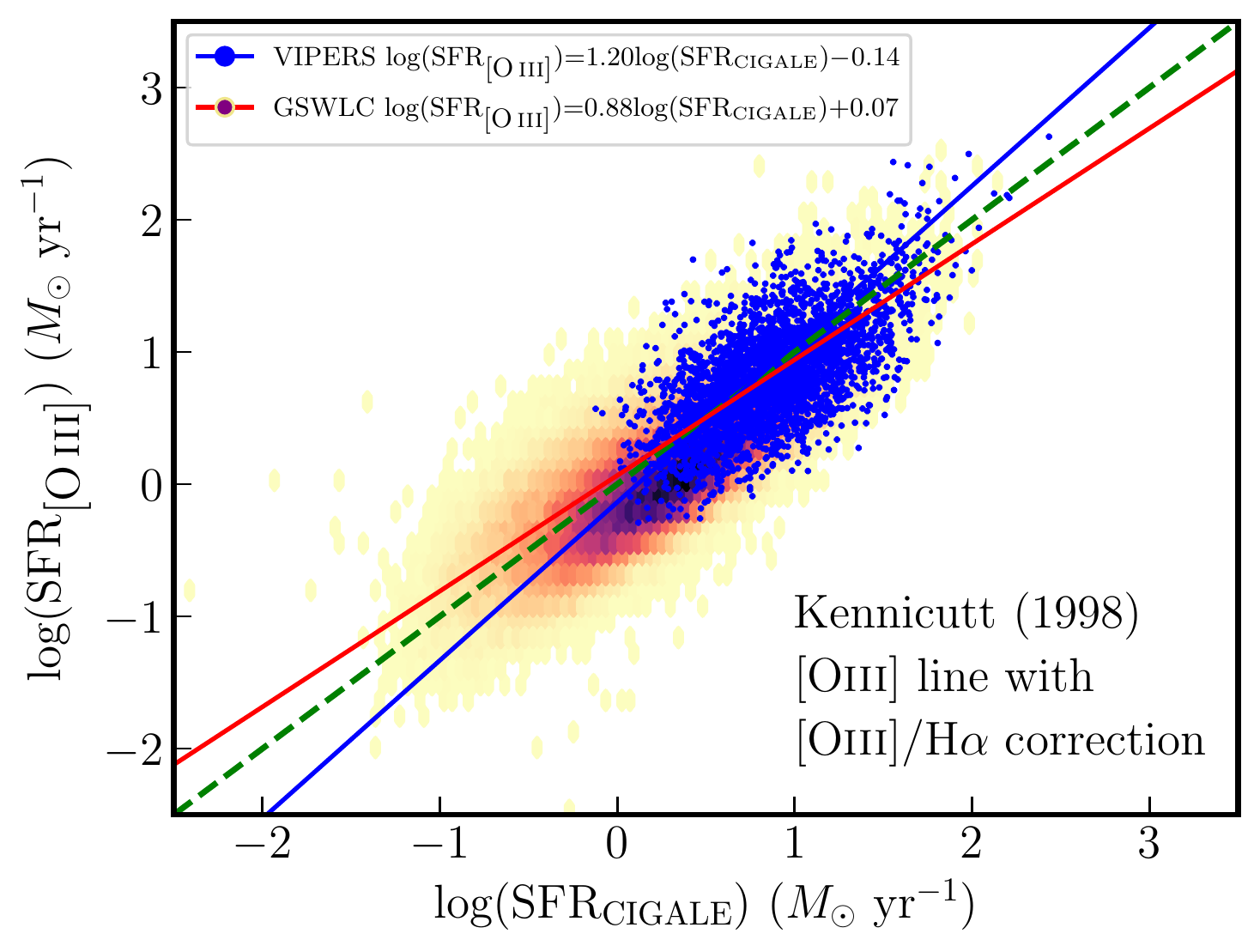}
\caption{SFR derived from H$\alpha$ SFR calibration of \citet{ken98a} and the attenuation-corrected [O{\,\sc{iii}}]/H$\alpha$ ratio. Estimation of SFR with [O{\,\sc{iii}}] leads to a significant scatter for VIPERS (mean of -0.04~dex and scatter of 0.31~dex) and GSWLC (-0.17 and 0.32~dex).}
\label{Fig:OIIIHa}
\end{figure}

The [O{\,\sc{iii}}]$\lambda$5007 line is not widely used as an SFR indicator, as only a rough estimation of the SFR can be obtained \citep{tep00,ken09}. Similarly as for [O{\,\sc{ii}}], we used the \citet{ken98a} calibration and converted it into a [O{\,\sc{iii}}]-SFR calibration using the average value [O{\,\sc{iii}}]/H$\alpha$ corrected for attenuation. Compared to other line calibrators, the SFR estimation using [O{\,\sc{iii}}] shows a large scatter of 0.3~dex for VIPERS and GSWLC (Fig.~\ref{Fig:OIIIHa}). We also estimated the SFR using the general calibration from \citet{vil21} but we found a similar scatter for VIPERS and a mean difference of 0.5~dex for GSWLC.\\
The [O{\,\sc{iii}}]/H$\alpha$ ratio is equal to 0.72 and 0.21 for VIPERS and GSWLC, respectively, but others works found different ratios at different redshifts: 1 (\citealt{tep00}, $3.3<z<3.4$), 0.79 (\citealt{hip03}, $0.4<z<0.64$), 1.05 (\citealt{ly07}, $0.07<z<1.47$), 1.23 (\citealt{str09}, $0.1<z<1.1$), and 0.86 (\citealt{vil21}, $1.4<z<1.68$). Due to this wide range of values, a direct calibration (i.e., an SFR only proportional to the luminosity) of the SFR encompassing a large range of redshift,  such as for [O{\,\sc{ii}}], cannot be obtained.\\

\begin{figure*}[h!]
\centering
\includegraphics[width=0.5\linewidth]{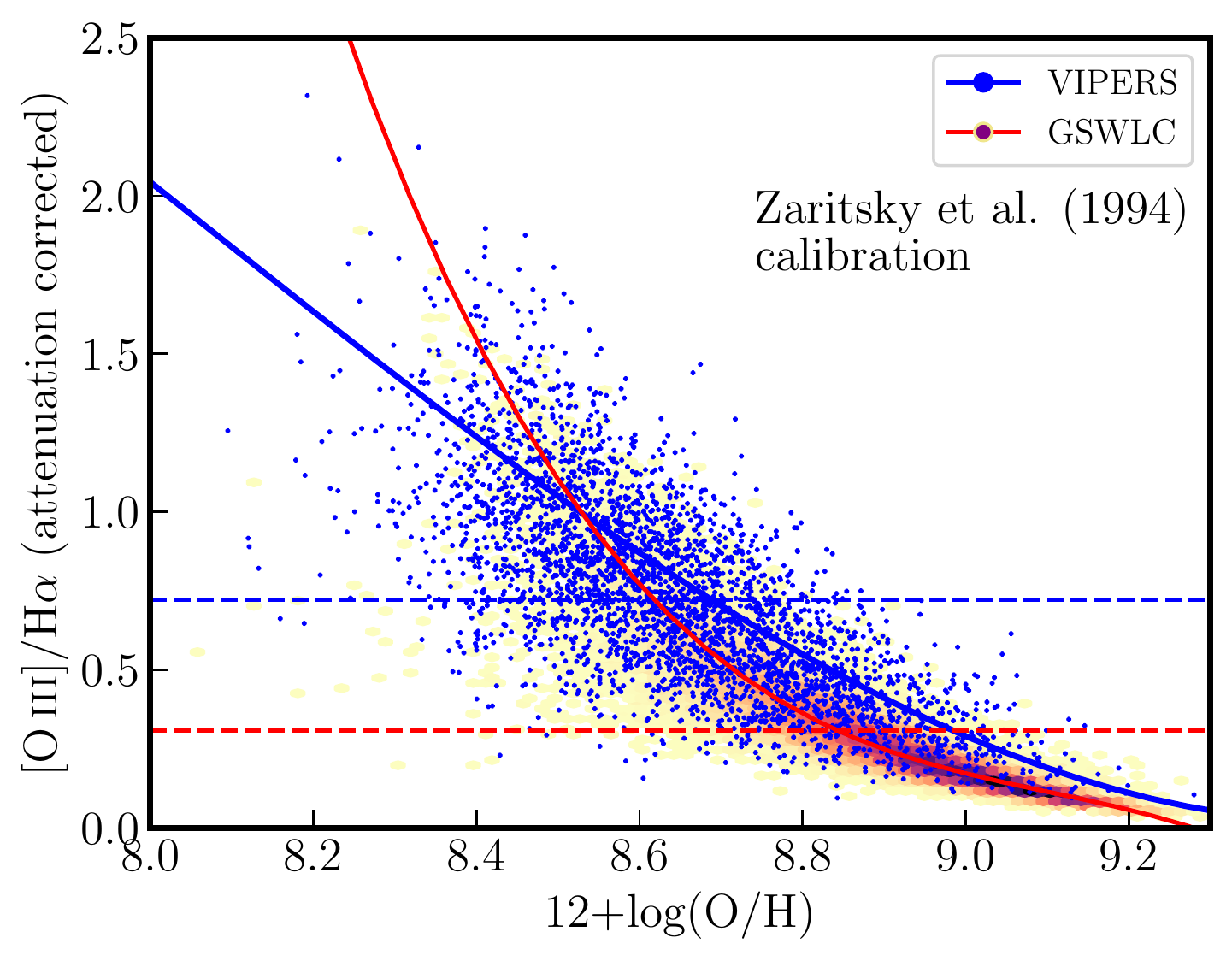}\includegraphics[angle=0,width=0.5\textwidth]{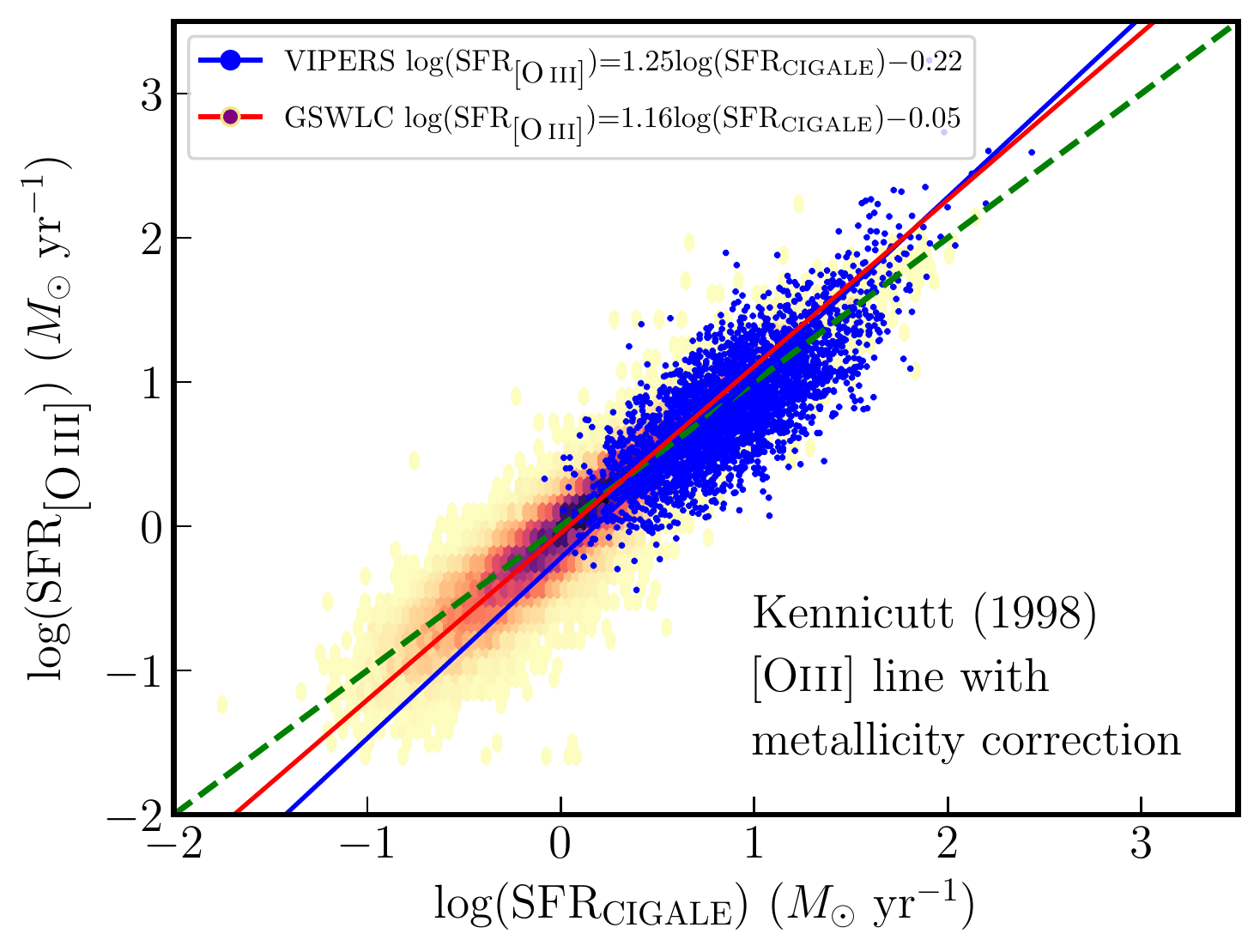}
\caption{SFR derived from H$\alpha$ SFR calibration of \citet{ken98a} and a metallicity correction. Left: [O{\,\sc{iii}}]/H$\alpha$ ratio (assuming H$\alpha=2.86$H$\beta$) corrected for dust attenuation versus metallicity estimated using the \citet{zar94} calibration for the VIPERS (blue) and GSWLC (density) samples, with their associated linear fit. Right panel: SFR derived from [O{\,\sc{iii}}] using the H$\alpha$ calibration of \citet{ken98a} and the attenuation-corrected [O{\,\sc{iii}}]/H$\alpha$ expressed as a function of metallicity (Eq.~\ref{eq:metal_zar94_OIII}). Taking into account the metallicity decreases the scatter for both samples (mean of 0.05~dex and scatter of 0.26~dex for VIPERS, 0.04 and 0.20~dex for GSWLC).} 
\label{Fig:OIII_Kennicutt}
\end{figure*}

To derive a calibration over a wide range of redshift, a metallicity correction should be included. Figure~\ref{Fig:OIII_Kennicutt} (left) shows the dependence of [O{\,\sc{iii}}]/H$\alpha$ with respect to the metallicity, using the calibration of Z94. The results of the fit are

\begin{equation}\label{eq:metal_zar94_OIII}
{ [\mbox{O{\,\sc{iii}}]/H}\alpha} = \left\{
    \begin{array}{l}
 -140.6304 + 57.9913x-7.5446x^2 \\
\hspace*{3.05cm}+\,0.3157x^3 \mbox{ (VIPERS)},\\
 2281.0376-750.5980x+82.3943x^2 \\
\hspace*{3.05cm}-\,3.0170x^3 \mbox{ (GSWLC)}\\
       \\
        \hspace*{1.4cm} x=12+{\rm{log(O/H)}}.\\
    \end{array}
\right.    
\end{equation}

We applied this metallicity correction to the \citet{ken98a} relation and the resulting SFR estimation can be seen in Fig.~\ref{Fig:OIII_Kennicutt} (right). The scatter is reduced to 0.26 and 0.20~dex for VIPERS and GSWLC, respectively, compared to 0.3~dex when no correction is applied.\\

\begin{figure}[h!]
\centering
\includegraphics[angle=0,width=0.5\textwidth]{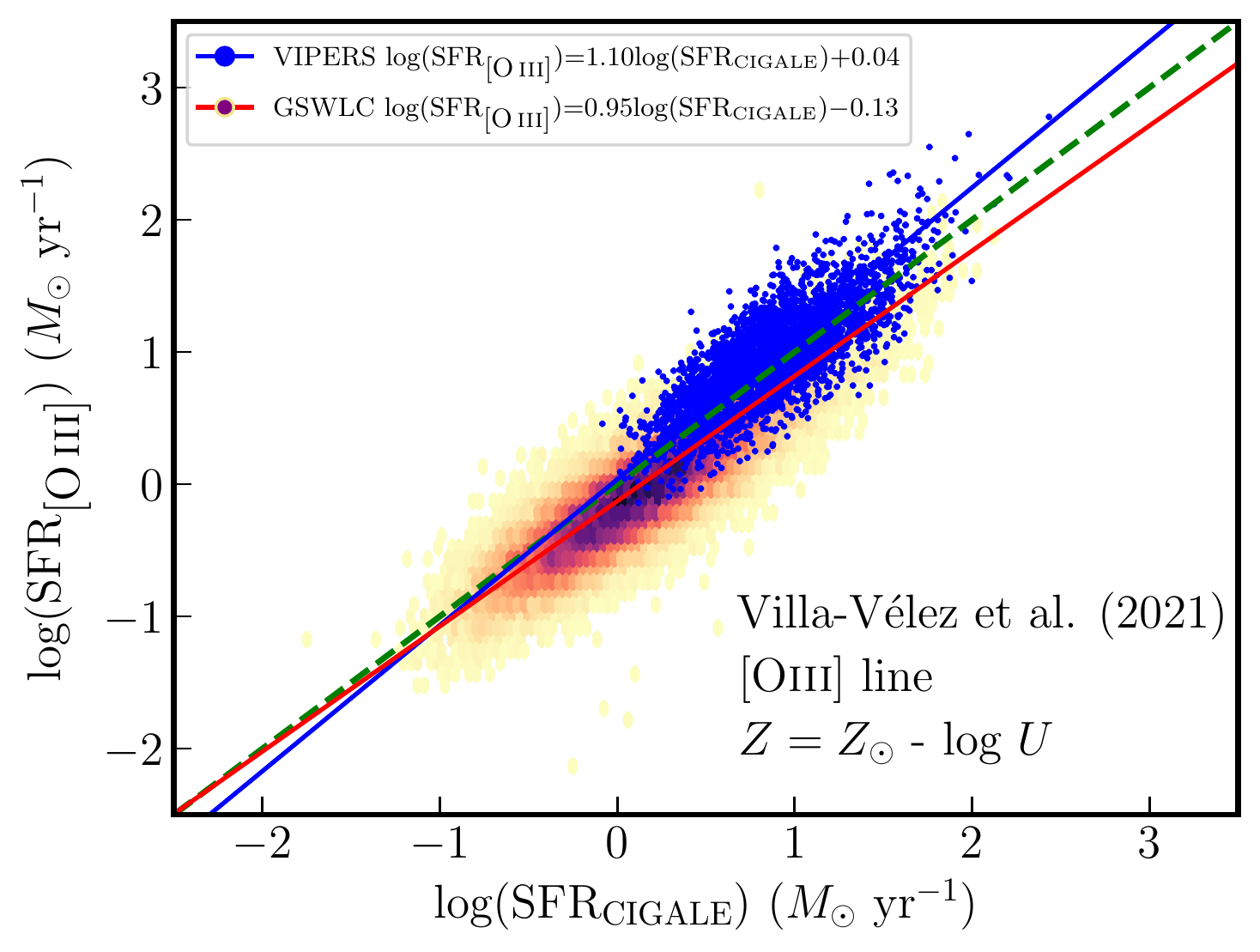}
\caption{SFR derived from [O{\,\sc{iii}}] using the calibration factors at $Z=Z_{\odot}$ and depending on U proposed by \citet{vil21}. U was estimated based on [O{\,\sc{iii}}]/[O{\,\sc{ii}}] and the model of \citet{kew02} at $Z=Z_{\odot}$. The agreement is good with CIGALE SFR (mean of 0.09~dex and scatter of 0.24~dex for VIPERS, 0.19 and 0.24~dex for GSWLC).} 
\label{Fig:OIII_Kennicutt_VV}
\end{figure}

In addition to the metallicity, the coefficient used to convert L$_{\textrm{[O\,\sc{iii]}}}$ to SFR also depends on the ionization parameter. In addition to a general [O{\,\sc{iii}}]-SFR calibration, \citet{vil21} established a grid where linear calibrations are given depending on the metallicity and ionization parameter.
Based on the [O{\,\sc{iii}}]/[O{\,\sc{ii}}] ratio and the [O{\,\sc{iii}}]/[O{\,\sc{ii}}] versus the ionization parameter q for $Z=Z_{\odot}$ in \citet{kew02}, we estimated a dimensionless ionization parameter $U$ ($U=q/c$) ranging from -3.5 to -2.2. Using the Z94 calibration, the metallicity ranges from 0.8 to 1.04~$Z_{\odot}$ so we used the grid values from \citet{vil21} (shown in their Table~5) at $Z=Z_{\odot}$ to estimate the SFR based on the estimated U for each galaxy. The SFR comparison is presented in Fig.~\ref{Fig:OIII_Kennicutt_VV} and they show a good agreement, with a scatter of $\sim$0.24~dex for VIPERS and GSWLC, despite the fact that we considered the model of \citet{kew02} at $Z=Z_{\odot}$ for every galaxy in VIPERS and GSWLC. We note that the scatter for VIPERS is lower when using the grid of \citet{vil21} compared to the calibration including a metallicity correction. This confirms that metallicity and the ionization parameter have to be included in a calibration based on [O{\,\sc{iii}}].\\

Some other works have also shown that the [O{\,\sc{iii}}] line can trace the SFR of galaxies, but with a significant scatter \citep{mou06b,suz16}, as in this work where [O{\,\sc{iii}}] is one of the poorest SFR tracers. It is fundamental to better understand what impacts the SFR estimation based on [O{\,\sc{iii}}] regarding future observations, as JSWT will be able to detect the [O{\,\sc{iii}}] line with $\textrm{S/N}>20$ at redshifts similar to those of the Hubble Deep Field \citep{che19}.

\section{Discussion}\label{Sect:discussion}

\subsection{Derivation of rest-frame luminosity}\label{Subsect:absoluteluminosity}

In this paper, we used the CIGALE SED fitting tool to derive the rest-frame luminosities used for the SFR calibrations. Because SFR comparisons depend on templates and parameters used for CIGALE to model the spectra, we performed an additional test using SEDs reconstructed previously with the \textit{hyperz} tool. Based on the two sets of SED results, we estimated the impact of different templates on the rest-frame luminosity estimations.\\

The scatter calculated between the absolute magnitudes from \textit{hyperz} and CIGALE is on the order of 0.29~mag for the \textit{ugriz} bands, which corresponds to a scatter of 0.12~dex for the rest-frame luminosity. This low value is mostly due to the excellent coverage of the optical part of the spectra. For the K$_s$ and NUV band, there is a slight increase to 0.34~mag, corresponding to a scatter of 0.14~dex. For the FUV band, the scatter increases to 0.62~mag, corresponding to 0.25~dex in luminosity. This results from the partial and poor coverage at NUV and FUV wavelengths (39\% and 84\% of galaxies have no NUV and no FUV measurements, respectively), increasing the uncertainty in the UV. This increase is also seen in a mock analysis where the scatter between the true and recovered FUV rest-frame luminosity is higher  compared to the other bands (0.06~dex for $L_{FUV}$ and 0.02~dex for the $g$-band, for instance). The test shows that the choice of templates can significantly influence the absolute magnitude estimation, especially with a poor coverage of photometric measurements, causing the recovery of the SED from the broad band photometry to be less constrained. However, we are more confident with estimations obtained using the CIGALE tool due to the use of Bayesian analysis, rather than \textit{hyperz} where only the best SED is taken into account.

Due to the lack of coverage in the MIR and FIR ranges of wavelength, the dust attenuations' true values are unknown. The low values of dust attenuation used in our fitting process impact the estimation of the blue bands' luminosities, affecting the estimation of IR luminosities due to the principle of energy-balance. We use a subsample of 441 galaxies for which \textit{Herschel} data are available to quantify this influence. We performed two runs similar to the one discussed in Sect.~\ref{Sect:Cigale} using input data from FUV to \textit{Herschel}-500~$\mu$m on one side and from \textit{u} to $K_{\mathrm{s}}$ band on the other side. Comparing the rest-frame luminosity between both runs, we find that the scatter decreases from 0.08 to 0.02~dex (FUV to $r$-band), then increases from 0.02 to 0.06~dex up to $K_{\mathrm{s}}$, and finally drastically increases for the 8- and 24-$\mu$m bands, and $L_{\mathrm{TIR}}$, with a scatter of 0.2~dex. This is the consequence of the energy-balance principle used in CIGALE, where energy attenuated by dust is re-emitted in the infrared wavelengths. In practice, by increasing or decreasing the attenuation in the bluer bands, the observed energy emitted by the stellar population becomes lower or higher, respectively, and finally the luminosity emitted by the dust increases or decreases, respectively, affecting the luminosity in the MIR.

\begin{figure}[h!]
\centering
\includegraphics[angle=0,width=0.5\textwidth]{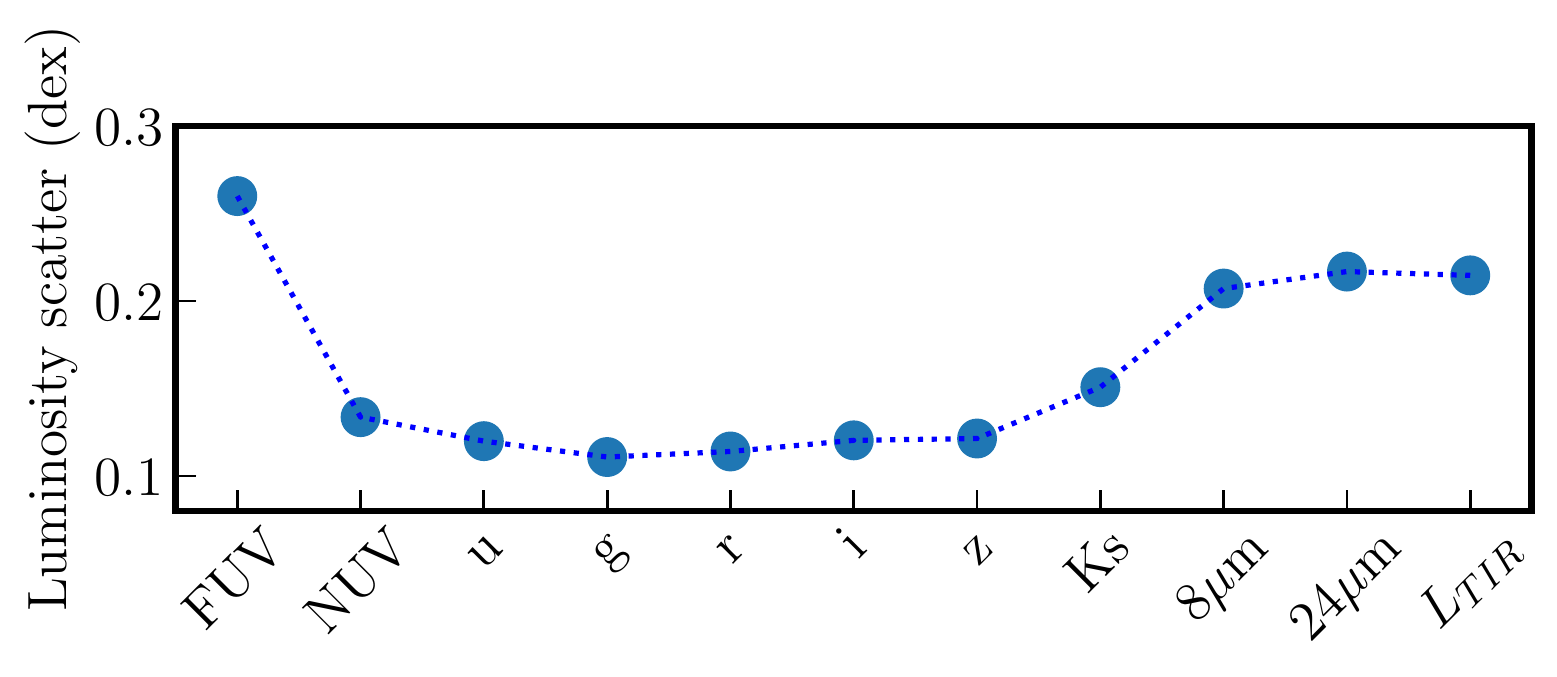}
\caption{Evolution of the scatter along different wavelengths ranging from FUV to $L_{TIR}$}
\label{Fig:scatter}
\end{figure}

The scatter estimated for the different calibrations, shown in Table~\ref{Appendix:Fit_parameters}, does not take into account this effect on templates, and is systematically small for calibrations based on photometric bands and spectral lines coming directly from observations, and is smaller compared to GSWLC for which k-corrections are smaller. We suggest that a lower limit on the scatter from FUV to 24~$\mu$m should take this factor into account, which, as a first estimation from the scatter from CIGALE versus \textit{hyperz} and CIGALE (FUV-500~$\mu$m) versus CIGALE(u-K$_{\mathrm{s}}$), is equal to 0.26 (FUV), 0.14 (NUV), 0.12 (ugriz), 0.15 (K$_{\mathrm{s}}$), and 0.20~dex (8, 24~$\mu$m and $L_{\mathrm{TIR}}$). The variation of this scatter from FUV to $L_{TIR}$ is shown in Fig.~\ref{Fig:scatter}

\subsection{SFR calibrations for star-forming galaxies from $z=0$ to $z=0.9$}

\begin{table*}[t!]
\caption{SFR calibrations, based on the \citet{cha03} IMF, for each band and line.}
\label{tab:SFR_calibration_Miguel}      
\centering                          
\begin{tabular}{c|r|r|c|c}
\multicolumn{5}{c}{${\rm{log[SFR_{\mbox{band}}}}~(M_{\odot}~{\rm{yr^{-1}}})]=A\times{\rm{log[}L_{\mbox{band}}(units)]}+B$}\\
\multicolumn{5}{c}{}\\
Rest-frame band\tablefootmark{$1$} & A & B & Luminosity range & Unit \\
\hline
\hline
FUV&1.04$\pm$0.01 & $-$21.99$\pm$0.02& $2.1\times10^{19}<L<4.7\times10^{23}$ &  W~Hz$^{-1}$  \\
NUV&1.03$\pm$0.01 & $-$21.81$\pm$0.01& $3.9\times10^{19}<L<4.3\times10^{23}$ & W~Hz$^{-1}$  \\
\textit{u}-band&1.11$\pm$0.0 & $-$23.62$\pm$0.01& $8.3\times10^{19}<L<4.5\times10^{23}$ & W~Hz$^{-1}$  \\
8~$\mu$m&0.85$\pm$0.01 & $-$18.53$\pm$0.14& $3.9\times10^{21}<L<4.4\times10^{24}$ & W~Hz$^{-1}$  \\
24~$\mu$m&0.81$\pm$0.0 & $-$18.22$\pm$0.01& $7.3\times10^{20}<L<2.6\times10^{25}$ & W~Hz$^{-1}$ \\
$L_{TIR}$&0.99$\pm$0.01 & $-$9.97$\pm$0.03& $3.7\times10^{8}<L<4.8\times10^{12}$ & $L_{\odot}$ \\
H$\beta$&0.94$\pm$0.01 & $-$38.34$\pm$0.04& $9.3\times10^{38}<L<1.0\times10^{44}$ & erg~s$^{-1}$ \\\
[O\,{\sc{ii}}]$\lambda$3727&0.96$\pm$0.01 & $-$39.69$\pm$0.07& $6.4\times10^{39}<L<1.1\times10^{44}$ & erg~s$^{-1}$ \\\
[O\,{\sc{iii}}]$\lambda$5007&0.89$\pm$0.01 & $-$35.94$\pm$0.35& $4.4\times10^{38}<L<6.1\times10^{43}$ & erg~s$^{-1}$ \\
\hline
\hline
\end{tabular}
\tablefoot{
\tablefoottext{$1$}{The rest-frame fluxes are corrected for attenuation.}
}
\end{table*}

Comparing two estimates of the same physical parameter given by two different methods can be done using the mean and the scatter between these two estimates. However, it might be difficult to estimate which estimate is closest to the true value when one method gives a significant scatter and a small mean while the other gives a small scatter and a significant mean. In addition, a small value for the mean difference does not necessarily translate as a small offset if the estimates are rotated around $y=x$ (Fig.~\ref{Fig_FUVNUV_Davies}, right). Therefore, we chose to use a test giving a unique number, measuring the precision and accuracy of a method, instead of separately comparing the mean and scatter.\\

To estimate the reliability of SFR calibrations, we computed the concordance correlation coefficient (CCC,  \citealt{lin89}). This coefficient is used to compare gold standard measurements (the true SFR estimated with CIGALE in this work) and a new set of measurements obtained with another method (here: another SFR calibration from the literature). The CCC is a combination of the Pearson coefficient, measuring the dispersion from a straight line, and another factor that measures how much the relationship is close to the identity. For the Pearson coefficient, the CCC ranges from -1 to 1. We estimated the CCC for each relation with VIPERS and GSWLC separately.\\

To estimate which relation works the best over the entire redshift range ($0\le z \le 0.9$), we employed the CCC, using both samples at low and intermediate redshifts (referred to as CCC$_{GV}$). Because of the different sample sizes between GSWLC and VIPERS, we constructed a sample of GSWLC galaxies with the same number of galaxies as VIPERS by performing 1000 samplings without replacement, where the median SFR of each sampling should not differ by more than $\mathrm{log(SFR)=0.1}$ compared to the total GSWLC sample. The CCC and CCC$_{GV}$ are shown in Table~\ref{Appendix:Fit_parameters}, where the best relation for VIPERS ($0.5\le z \le 0.9$) and GSWLC ($z \le 0.3$) are underlined in blue, and in magenta for VIPERS and GSWLC simultaneously ($0\le z \le 0.9$). Given that VIPERS measurements are more affected by extrapolations compared to GSWLC, we performed a comparison per band only.

\subsubsection{Calibration from the UV bands and the \textit{u}-band}
For the FUV band, the relation of \citet{bro17} using the \citet{hao11} attenuation law gives the best agreement for VIPERS ($\mathrm{CCC}=0.99$), GSWLC ($\mathrm{CCC}=0.88$), and both ($\mathrm{CCC}_{GV}=0.94$). The relation given by \citet{sal07}, despite giving a worse agreement for GSWLC (0.86), is equally good for VIPERS (0.99) and both samples (0.94). It is reassuring to find a good agreement with another calibration from the literature given the higher uncertainty of the FUV and NUV luminosity. For the \textit{u}-band, a good agreement is found for VIPERS with \citet{zho17} (0.97), and for GSWLC (0.94) with \citet{hop03} and \citet{dav16}. However, the relation of \citet{hop03} gives the best agreement for $0\le z \le 0.9$ with a CCC of 0.97.\\

\subsubsection{Calibration from MIR bands}
In agreement with the literature (e.g., \citealt{sch18, gre22}), 8~$\mu$m only gives a rough estimation of the SFR, the best calibration being given by \citet{bro17} with a CCC of 0.91, while the others relations give a poor CCC when applied to the VIPERS sample ($<0.78$).
All the calibrations tested at 24~$\mu$m give a high and close CCC$_{GV}$, the highest being given by the relation of \citet{zhu08}, which is when the nonlinearity of $L_{24}$ versus SFR is taken into account and the calibration is established for the whole galaxy.

\subsubsection{Calibrations from multiple bands}
Between all the calibrations, composite tracers give the best CCC$_{GV}$ with the calibration of \citet{boq14}, with the weighting coefficient for $L_{24}$ from \citet{hao11} giving the highest one (CCC$_{GV}$=0.99). This is not surprising as no attenuation law has to be used, and these composite calibrations retrieve the unobscured and obscured SFR of galaxies.

\subsubsection{Calibrations from spectral lines}
For spectral lines, H$\beta$ is a good tracer, in particular for GSWLC ($\mathrm{CCC}=0.92$) given its relation with H$\alpha$. For VIPERS, the lower CCC of 0.87 for VIPERS might come from our treatment of the stellar absorption, taken to be equal to 2~\AA\ for each galaxy, while this correction was done automatically for GSWLC. Given the assumptions on H$\beta$, we would need H$\alpha$ measurements to understand if this lower CCC is due to a deviation from the \citet{ken98a} calibration or due to the corrections and conversion factor with H$\beta$.

The [O\,{\sc{ii}}] line gives a reliable estimation of SFR, and the agreement improves when the samples are corrected for reddening and metallicity. When this last correction is applied, the [O\,{\sc{ii}}] line has the same CCC$_{GV}$ as H$\beta$. We note that the calibration of \citet{gil10} gives a slightly better estimation compared to the one corrected for metallicity for GSWLC galaxies ($\mathrm{CCC}=0.91$). The assumptions for the attenuation ($A_{H\alpha}$=1~mag) and metallicity corrections (derived from SDSS DR4) in the calibration of \citet{gil10} naturally explains this good agreement and, at the same time, the worse agreement with VIPERS, for which $A_{H\alpha}$ and metallicity are found to be different. In agreement with our work and \citet{gil10}, \citet{kew02} tested their [O\,{\sc{ii}}] calibration based on the Nearby Field Galaxies Survey (NFGS) from \citet{hic02} and \citet{tre02} at $0.5<z<1.6$, and showed that the [O\,{\sc{ii}}] line is a reliable tracer of SFR ($\sim$0.17~dex) when the metallicity is taken into account.

The same applies to the SFR estimated with [O\,{\sc{iii}}] where the metallicity correction improves the estimation, increasing CCC$_{GV}$ from 0.81 to 0.91. The relation of \citet{vil21} gives a slightly better estimation for VIPERS when the metallicity and the ionization parameter are included. A more extended grid associated with models to derive the ionization parameter would be useful, with the aim of giving a better estimation of the SFR from the [O\,{\sc{iii}}] spectral line.\\

\subsubsection{Calibrations from GSWLC and VIPERS}
Using both GSWLC and VIPERS, we derived our own set of SFR calibrations for single-band tracers. We remind the reader that these calibrations were obtained for star-forming galaxies selected through the BPT diagram, and whose SFR and attenuation-corrected rest-frame luminosities were estimated using CIGALE, assuming a \citet{cha03} IMF (Table~\ref{tab:SFR_calibration_Miguel}).

The FUV and NUV calibrations are close to the those established by \citet{sal07} at low redshifts. For the \textit{u}-band, we did not apply any correction for the old stellar population contamination as it is difficult to accurately remove. However, it implies that the \textit{u}-band SFR calibration is restricted to star-forming galaxies that have similar contamination. If using a \textit{u}-band luminosity that had been decontaminated of the old stellar population, we found that a factor of 0.64 should be applied to $L_{u}$ in order to recover the luminosity contribution from young stars. For 8~$\mu$m, we removed the stellar contamination ($\sim$5\% in average) as explained in Sect. \ref{Sect:MidFar}. Contrary to the other bands, this calibration is obtained using VIPERS data only, and extrapolation to the local Universe should be considered with caution. At 24~$\mu$m, the WISE-4 luminosity from GSWLC is converted into a 24-$\mu$m luminosity (Sect.~\ref{Sect:MidFar}) before performing the fit.

Calibrations based on the \textit{u}-band, and 8 and 24~$\mu$m are found to be nonlinear as the luminosity does not directly trace the SFR, and some other effects have to be taken into account: old-stellar population contamination and variation due to stellar evolution for the \textit{u}-band, metallicity dependence of the PAHs at 8~$\mu$m, and self-absorption and transparency for the 24-$\mu$m band. On the contrary, the linearity of the SFR calibration based on $L_{TIR}$ shows that this property is a reliable and direct tracer of the SFR up to $z\sim0.9$. For the [O\,{\sc{ii}}] and [O\,{\sc{iii}}] lines, the metallicity is already included, so no additional corrections are needed in order to use them. The strong nonlinearity of the SFR-[O\,{\sc{iii}}] calibration may reflect the dependence on metallicity and the ionization parameter suffered by this spectral line.
 
\section{Conclusions}\label{Sect:conclusion}

We used VIPERS, a spectroscopic survey based on CFHTLS galaxies, with $0.5<z<0.9$ and measurements from FUV to FIR bands including H$\beta$, [O\,{\sc{ii}}]$\lambda$3727 and [O\,{\sc{iii}}]$\lambda \lambda$4959,5007 spectral lines. We tested different SFR recipes from the literature, mainly calibrated on local galaxies, to check how consistent those calibrations are at intermediate redshifts. For comparison purposes, we used GSWLC, a catalog of local galaxies based on SDSS data merged with \textit{GALEX} and WISE ($z<0.3$) data. For both the VIPERS and GSWLC catalogs, the physical parameters were derived using the CIGALE code, in particular the SFR, which is considered to be the reference SFR in this work. The main conclusions are:\\

$-$ In the blue part of the spectrum, the FUV and NUV bands can be excellent tracers of SFR up to $z\sim0.9$, using the calibrations of \citet{bro17} (scatter of 0.05 and 0.25~dex for VIPERS and GSWLC, respectively for the FUV band) and \citet{sal07} (scatter of 0.03 and 0.24~dex for VIPERS and GSWLC, respectively for the NUV band). The attenuation should be adequately corrected and we calibrated the $A_{UV}-\beta_{UV}$ relation for VIPERS for the FUV and NUV bands.\\

$-$ We estimated the contamination by the old stellar population in the \textit{u}-band to be on the order of 36\%. Up to $z\sim0.9$, the calibration of \citet{hop03}, taking into account the contamination and stellar evolution through a nonlinear calibrations, gives the best SFR estimations (scatter of 0.08 and 0.09~dex for VIPERS and GSWLC, respectively) up to $z\sim0.9$. However, such calibrations should be limited to galaxies suffering roughly from the same old stellar population contamination.\\

$-$ The luminosity at 8~$\mu$m, dominated by PAH emission, is correlated with the SFR, but calibrations tested in this work underestimate the SFR compared to the one obtained with CIGALE. The calibration giving the best agreement with CIGALE is given by \citet{bro17} (scatter of 0.10~dex with a mean difference of 0.10~dex). More work is needed to understand PAH physics and how PAHs are impacted by metallicity to derive, if possible, more accurate calibrations. This is particularly important in the context of future JWST observations.\\

$-$ 24~$\mu$m is an excellent SFR tracer, and previous calibrations are found to work very well, in particular when the calibrations are based on whole galaxies and take into account the nonlinear behavior of $L_{24\mu m}$ with respect to the SFR. The calibration of \citet{zhu08} gives the best agreement (scatter of 0.08~dex with a mean difference of 0.12~dex). If WISE-4 observations are able to reliably estimate the SFR, this is not true beyond the local Universe, as they are not sensitive enough and can only be considered as upper limits.\\

$-$ Composite tracers are part of the best SFR indicators as they do not rely on attenuation prescriptions and are able to associate unobscured and obscured SFRs. Up to $z\sim0.9$, the calibration of \citet{boq14}/\citet{hao11} is in good agreement with CIGALE (scatter of 0.03 and 0.10~dex for VIPERS and GSWLC). However, the need for two bands, or more, make them less practical than single-band SFR calibrations. For VIPERS and GSWLC, we estimated the dust emission to be contaminated by the old stellar population by 11\%\ and 45\% for VIPERS and GSWLC, respectively.\\

$-$ Estimation of rest-frame luminosity from CIGALE is impacted from the choice of templates and wavelength coverage of the SED. By comparing our CIGALE luminosity estimations with those from \textit{hyperz} and CIGALE runs, with and without IR coverage, the impact ranges from 0.12 to 0.26~dex depending on the band considered.\\

$-$ The spectral line H$\beta$ gives a reliable estimation of the SFR up to $z\sim0.9$ (scatter of 0.19~dex for VIPERS and GSWLC), and is therefore a good alternative when H$\alpha$ is not observed.\\
In GSWLC and VIPERS, the [O\,{\sc{ii}}] and [O\,{\sc{iii}}] lines' strengths are higher and easier to observe, making an SFR calibration based on them particularly important for future surveys. The [O\,{\sc{ii}}] spectral line gives also a good estimation of the SFR, reaching an equivalent accuracy compared to H$\beta$ when the metallicity dependence, through the $R_{23}$ parameter in this work, is taken into account (scatter of 0.17 and 0.20~dex for VIPERS and GSWLC, respectively). The [O\,{\sc{iii}}] is one of the least accurate SFR indicators, the calibrations being different between local and intermediate redshift galaxies in addition to showing a high scatter (scatter of 0.31 and 0.32~dex for VIPERS and GSWLC, respectively). A correction for metallicity is mandatory to reliably recover the SFR (scatter of 0.26 and 0.20~dex for VIPERS and GSWLC, respectively). In addition, this calibration also depends on the ionization parameter of the galaxies considered. Calibrations using the metallicity-ionization parameter grid proposed by \citet{vil21} gives good SFR estimations (scatter of 0.24~dex for VIPERS and GSWLC).\\

$-$ We calibrated the FUV, NUV, \textit{u-}, 8-$\mu$m, and 24~$\mu$m bands, $L_{TIR}$, and the H$\beta$, [O\,{\sc{ii}}]$\lambda$3727 and [O\,{\sc{iii}}]$\lambda$,5007 spectral lines, based on the SFR derived with CIGALE in the range $0<z<0.9$ for star-forming galaxies. Those relations are in the form ${\rm{log[SFR_{\mbox{band}}}}~(M_{\odot}~{\rm{yr^{-1}}})]=A\times{\rm{log[}L_{\mbox{band}}]}+B$ where the luminosity in the band $L_{band}$ has to be corrected for dust attenuation. We did not account for old stellar population correction for the \textit{u}-band and $L_{TIR}$, but this contribution can be removed based on the average values given in this work. The [O\,{\sc{ii}}]$\lambda$3727 and [O\,{\sc{iii}}]$\lambda$5007 calibration are not corrected for metallicity.

\begin{acknowledgements}
We thank the anonymous referee for the useful comments and suggestions. This research was supported by the Polish National Science Centre via grants UMO-2018/30/M/ST9/00757 and UMO-2018/30/E/ST9/00082, and and by Polish Ministry of Science and Higher Education grant DIR/WK/2018/12. M.F. acknowledges support from the First TEAM grant of the Foundation for Polish Science No. POIR.04.04.00-00-5D21/18-00 (PI: A. Karska). This paper uses data from the VIMOS Public Extragalactic Redshift Survey (VIPERS).
VIPERS has been performed using the ESO Very Large Telescope, under the
“Large Programme” 182.A-0886. The participating institutions and funding
agencies are listed at \url{http://vipers.inaf.it}. We thank B. Garilli for the
line measurements for the VIPERS sample.
\end{acknowledgements}

%
   \bibliographystyle{aa} 
   \bibliography{biblio}

\begin{thebibliography}{200}
\expandafter\ifx\csname natexlab\endcsname\relax\def\natexlab#1{#1}\fi

\bibitem[{{Ahn} {et~al.}(2014){Ahn}, {Alexandroff}, {Allende Prieto}, {Anders},
  {Anderson}, {Anderton}, {Andrews}, {Aubourg}, {Bailey}, {Bastien},
  {Bautista}, {Beers}, {Beifiori}, {Bender}, {Berlind}, {Beutler}, {Bhardwaj},
  {Bird}, {Bizyaev}, {Blake}, {Blanton}, {Blomqvist}, {Bochanski}, {Bolton},
  {Borde}, {Bovy}, {Shelden Bradley}, {Brandt}, {Brauer}, {Brinkmann},
  {Brownstein}, {Busca}, {Carithers}, {Carlberg}, {Carnero}, {Carr},
  {Chiappini}, {Chojnowski}, {Chuang}, {Comparat}, {Crepp}, {Cristiani},
  {Croft}, {Cuesta}, {Cunha}, {da Costa}, {Dawson}, {De Lee}, {Dean},
  {Delubac}, {Deshpande}, {Dhital}, {Ealet}, {Ebelke}, {Edmondson},
  {Eisenstein}, {Epstein}, {Escoffier}, {Esposito}, {Evans}, {Fabbian}, {Fan},
  {Favole}, {Femen{\'\i}a Castell{\'a}}, {Fern{\'a}ndez Alvar}, {Feuillet},
  {Filiz Ak}, {Finley}, {Fleming}, {Font-Ribera}, {Frinchaboy},
  {Galbraith-Frew}, {Garc{\'\i}a-Hern{\'a}ndez}, {Garc{\'\i}a P{\'e}rez}, {Ge},
  {G{\'e}nova-Santos}, {Gillespie}, {Girardi}, {Gonz{\'a}lez Hern{\'a}ndez},
  {Gott}, {Gunn}, {Guo}, {Halverson}, {Harding}, {Harris}, {Hasselquist},
  {Hawley}, {Hayden}, {Hearty}, {Herrero Dav{\'o}}, {Ho}, {Hogg}, {Holtzman},
  {Honscheid}, {Huehnerhoff}, {Ivans}, {Jackson}, {Jiang}, {Johnson},
  {Kinemuchi}, {Kirkby}, {Klaene}, {Kneib}, {Koesterke}, {Lan}, {Lang}, {Le
  Goff}, {Leauthaud}, {Lee}, {Lee}, {Long}, {Loomis}, {Lucatello}, {Lupton},
  {Ma}, {Mack}, {Mahadevan}, {Maia}, {Majewski}, {Malanushenko},
  {Malanushenko}, {Manchado}, {Manera}, {Maraston}, {Margala}, {Martell},
  {Masters}, {McBride}, {McGreer}, {McMahon}, {M{\'e}nard}, {M{\'e}sz{\'a}ros},
  {Miralda-Escud{\'e}}, {Miyatake}, {Montero-Dorta}, {Montesano}, {More},
  {Morrison}, {Muna}, {Munn}, {Myers}, {Nguyen}, {Nichol}, {Nidever},
  {Noterdaeme}, {Nuza}, {O'Connell}, {O'Connell}, {O'Connell}, {Olmstead},
  {Oravetz}, {Owen}, {Padmanabhan}, {Palanque-Delabrouille}, {Pan}, {Parejko},
  {Parihar}, {P{\^a}ris}, {Pepper}, {Percival}, {P{\'e}rez-R{\`a}fols}, {Dotto
  Perottoni}, {Petitjean}, {Pieri}, {Pinsonneault}, {Prada}, {Price-Whelan},
  {Raddick}, {Rahman}, {Rebolo}, {Reid}, {Richards}, {Riffel}, {Robin},
  {Rocha-Pinto}, {Rockosi}, {Roe}, {Ross}, {Ross}, {Rossi}, {Roy},
  {Rubi{\~n}o-Martin}, {Sabiu}, {S{\'a}nchez}, {Santiago}, {Sayres},
  {Schiavon}, {Schlegel}, {Schlesinger}, {Schmidt}, {Schneider}, {Schultheis},
  {Sellgren}, {Seo}, {Shen}, {Shetrone}, {Shu}, {Simmons}, {Skrutskie},
  {Slosar}, {Smith}, {Snedden}, {Sobeck}, {Sobreira}, {Stassun}, {Steinmetz},
  {Strauss}, {Streblyanska}, {Suzuki}, {Swanson}, {Terrien}, {Thakar},
  {Thomas}, {Thompson}, {Tinker}, {Tojeiro}, {Troup}, {Vandenberg}, {Vargas
  Maga{\~n}a}, {Viel}, {Vogt}, {Wake}, {Weaver}, {Weinberg}, {Weiner}, {White},
  {White}, {Wilson}, {Wisniewski}, {Wood-Vasey}, {Y{\`e}che}, {York}, {Zamora},
  {Zasowski}, {Zehavi}, {Zhao}, {Zheng}, \& {Zhu}}]{ahn14}
{Ahn}, C.~P., {Alexandroff}, R., {Allende Prieto}, C., {et~al.} 2014, \apjs,
  211, 17

\bibitem[{{Alonso-Herrero} {et~al.}(2006){Alonso-Herrero}, {Rieke}, {Rieke},
  {Colina}, {P{\'e}rez-Gonz{\'a}lez}, \& {Ryder}}]{alo06}
{Alonso-Herrero}, A., {Rieke}, G.~H., {Rieke}, M.~J., {et~al.} 2006, \apj, 650,
  835

\bibitem[{{Anderson} {et~al.}(2014){Anderson}, {Bania}, {Balser}, {Cunningham},
  {Wenger}, {Johnstone}, \& {Armentrout}}]{and14}
{Anderson}, L.~D., {Bania}, T.~M., {Balser}, D.~S., {et~al.} 2014, \apjs, 212,
  1

\bibitem[{{Arnouts} {et~al.}(2013){Arnouts}, {Le Floc'h}, {Chevallard},
  {Johnson}, {Ilbert}, {Treyer}, {Aussel}, {Capak}, {Sanders}, {Scoville},
  {McCracken}, {Milliard}, {Pozzetti}, \& {Salvato}}]{arn13}
{Arnouts}, S., {Le Floc'h}, E., {Chevallard}, J., {et~al.} 2013, \aap, 558, A67

\bibitem[{{Baes} {et~al.}(2011){Baes}, {Verstappen}, {De Looze}, {Fritz},
  {Saftly}, {Vidal P{\'e}rez}, {Stalevski}, \& {Valcke}}]{bae11}
{Baes}, M., {Verstappen}, J., {De Looze}, I., {et~al.} 2011, \apjs, 196, 22

\bibitem[{{Baldwin} {et~al.}(1981){Baldwin}, {Phillips}, \&
  {Terlevich}}]{bal81}
{Baldwin}, J.~A., {Phillips}, M.~M., \& {Terlevich}, R. 1981, \pasp, 93, 5

\bibitem[{{Battisti} {et~al.}(2017){Battisti}, {Calzetti}, \& {Chary}}]{bat17}
{Battisti}, A.~J., {Calzetti}, D., \& {Chary}, R.~R. 2017, \apj, 851, 90

\bibitem[{{Battisti} {et~al.}(2015){Battisti}, {Calzetti}, {Johnson}, \&
  {Elbaz}}]{bat15}
{Battisti}, A.~J., {Calzetti}, D., {Johnson}, B.~D., \& {Elbaz}, D. 2015, \apj,
  800, 143

\bibitem[{{Bell}(2003)}]{bel03}
{Bell}, E.~F. 2003, \apj, 586, 794

\bibitem[{{Bell} {et~al.}(2005){Bell}, {Papovich}, {Wolf}, {Le Floc'h},
  {Caldwell}, {Barden}, {Egami}, {McIntosh}, {Meisenheimer},
  {P{\'e}rez-Gonz{\'a}lez}, {Rieke}, {Rieke}, {Rigby}, \& {Rix}}]{bel05}
{Bell}, E.~F., {Papovich}, C., {Wolf}, C., {et~al.} 2005, \apj, 625, 23

\bibitem[{{Berman}(1936)}]{ber36}
{Berman}, L. 1936, \mnras, 96, 890

\bibitem[{{Berta} {et~al.}(2013){Berta}, {Lutz}, {Santini}, {Wuyts}, {Rosario},
  {Brisbin}, {Cooray}, {Franceschini}, {Gruppioni}, {Hatziminaoglou}, {Hwang},
  {Le Floc'h}, {Magnelli}, {Nordon}, {Oliver}, {Page}, {Popesso}, {Pozzetti},
  {Pozzi}, {Riguccini}, {Rodighiero}, {Roseboom}, {Scott}, {Symeonidis},
  {Valtchanov}, {Viero}, \& {Wang}}]{ber13}
{Berta}, S., {Lutz}, D., {Santini}, P., {et~al.} 2013, \aap, 551, A100

\bibitem[{{Bianchi} {et~al.}(2014){Bianchi}, {Conti}, \& {Shiao}}]{bia14}
{Bianchi}, L., {Conti}, A., \& {Shiao}, B. 2014, Advances in Space Research,
  53, 900

\bibitem[{{Blanton} {et~al.}(2017){Blanton}, {Bershady}, {Abolfathi},
  {Albareti}, {Allende Prieto}, {Almeida}, {Alonso-Garc{\'{\i}}a}, {Anders},
  {Anderson}, {Andrews}, \& et~al.}]{bla17}
{Blanton}, M.~R., {Bershady}, M.~A., {Abolfathi}, B., {et~al.} 2017, \aj, 154,
  28

\bibitem[{{Bogdanoska} \& {Burgarella}(2020)}]{bog20}
{Bogdanoska}, J. \& {Burgarella}, D. 2020, \mnras, 496, 5341

\bibitem[{{Boissier}(2013)}]{boi13}
{Boissier}, S. 2013, {Star Formation in Galaxies}, ed. T.~D. {Oswalt} \& W.~C.
  {Keel}, Vol.~6, 141

\bibitem[{{Boissier} {et~al.}(2003){Boissier}, {Prantzos}, {Boselli}, \&
  {Gavazzi}}]{boi03}
{Boissier}, S., {Prantzos}, N., {Boselli}, A., \& {Gavazzi}, G. 2003, \mnras,
  346, 1215

\bibitem[{{Bolzonella} {et~al.}(2010){Bolzonella}, {Kova{\v{c}}}, {Pozzetti},
  {Zucca}, {Cucciati}, {Lilly}, {Peng}, {Iovino}, {Zamorani}, {Vergani},
  {Tasca}, {Lamareille}, {Oesch}, {Caputi}, {Kampczyk}, {Bardelli}, {Maier},
  {Abbas}, {Knobel}, {Scodeggio}, {Carollo}, {Contini}, {Kneib}, {Le
  F{\`e}vre}, {Mainieri}, {Renzini}, {Bongiorno}, {Coppa}, {de la Torre}, {de
  Ravel}, {Franzetti}, {Garilli}, {Le Borgne}, {Le Brun}, {Mignoli},
  {Pell{\'o}}, {Perez-Montero}, {Ricciardelli}, {Silverman}, {Tanaka},
  {Tresse}, {Bottini}, {Cappi}, {Cassata}, {Cimatti}, {Guzzo}, {Koekemoer},
  {Leauthaud}, {Maccagni}, {Marinoni}, {McCracken}, {Memeo}, {Meneux},
  {Porciani}, {Scaramella}, {Aussel}, {Capak}, {Halliday}, {Ilbert},
  {Kartaltepe}, {Salvato}, {Sanders}, {Scarlata}, {Scoville}, {Taniguchi}, \&
  {Thompson}}]{bol10}
{Bolzonella}, M., {Kova{\v{c}}}, K., {Pozzetti}, L., {et~al.} 2010, \aap, 524,
  A76

\bibitem[{{Bolzonella} {et~al.}(2000){Bolzonella}, {Miralles}, \&
  {Pell{\'o}}}]{bol00}
{Bolzonella}, M., {Miralles}, J.~M., \& {Pell{\'o}}, R. 2000, \aap, 363, 476

\bibitem[{{Bongiorno} {et~al.}(2010){Bongiorno}, {Mignoli}, {Zamorani},
  {Lamareille}, {Lanzuisi}, {Miyaji}, {Bolzonella}, {Carollo}, {Contini},
  {Kneib}, {Le F{\`e}vre}, {Lilly}, {Mainieri}, {Renzini}, {Scodeggio},
  {Bardelli}, {Brusa}, {Caputi}, {Civano}, {Coppa}, {Cucciati}, {de la Torre},
  {de Ravel}, {Franzetti}, {Garilli}, {Halliday}, {Hasinger}, {Koekemoer},
  {Iovino}, {Kampczyk}, {Knobel}, {Kova{\v{c}}}, {Le Borgne}, {Le Brun},
  {Maier}, {Merloni}, {Nair}, {Pello}, {Peng}, {Perez Montero}, {Ricciardelli},
  {Salvato}, {Silverman}, {Tanaka}, {Tasca}, {Tresse}, {Vergani}, {Zucca},
  {Abbas}, {Bottini}, {Cappi}, {Cassata}, {Cimatti}, {Guzzo}, {Leauthaud},
  {Maccagni}, {Marinoni}, {McCracken}, {Memeo}, {Meneux}, {Oesch}, {Porciani},
  {Pozzetti}, \& {Scaramella}}]{bong10}
{Bongiorno}, A., {Mignoli}, M., {Zamorani}, G., {et~al.} 2010, \aap, 510, A56

\bibitem[{{Boquien} {et~al.}(2014){Boquien}, {Buat}, \& {Perret}}]{boq14}
{Boquien}, M., {Buat}, V., \& {Perret}, V. 2014, \aap, 571, A72

\bibitem[{{Boquien} {et~al.}(2019){Boquien}, {Burgarella}, {Roehlly}, {Buat},
  {Ciesla}, {Corre}, {Inoue}, \& {Salas}}]{boq19}
{Boquien}, M., {Burgarella}, D., {Roehlly}, Y., {et~al.} 2019, \aap, 622, A103

\bibitem[{{Boquien} {et~al.}(2016){Boquien}, {Kennicutt}, {Calzetti}, {Dale},
  {Galametz}, {Sauvage}, {Croxall}, {Draine}, {Kirkpatrick}, {Kumari}, {Hunt},
  {De Looze}, {Pellegrini}, {Rela{\~n}o}, {Smith}, \& {Tabatabaei}}]{boq16}
{Boquien}, M., {Kennicutt}, R., {Calzetti}, D., {et~al.} 2016, \aap, 591, A6

\bibitem[{{Boquien} \& {Salim}(2021)}]{boq21}
{Boquien}, M. \& {Salim}, S. 2021, \aap, 653, A149

\bibitem[{{Bouchet} {et~al.}(1985){Bouchet}, {Lequeux}, {Maurice}, {Prevot}, \&
  {Prevot-Burnichon}}]{bou85}
{Bouchet}, P., {Lequeux}, J., {Maurice}, E., {Prevot}, L., \&
  {Prevot-Burnichon}, M.~L. 1985, \aap, 149, 330

\bibitem[{{Brinchmann} {et~al.}(2004){Brinchmann}, {Charlot}, {White},
  {Tremonti}, {Kauffmann}, {Heckman}, \& {Brinkmann}}]{bri04}
{Brinchmann}, J., {Charlot}, S., {White}, S.~D.~M., {et~al.} 2004, \mnras, 351,
  1151

\bibitem[{{Brown} {et~al.}(2014{\natexlab{a}}){Brown}, {Jarrett}, \&
  {Cluver}}]{bro14a}
{Brown}, M.~J.~I., {Jarrett}, T.~H., \& {Cluver}, M.~E. 2014{\natexlab{a}},
  \pasa, 31, e049

\bibitem[{{Brown} {et~al.}(2017){Brown}, {Moustakas}, {Kennicutt}, {Bonne},
  {Intema}, {de Gasperin}, {Boquien}, {Jarrett}, {Cluver}, {Smith}, {da Cunha},
  {Imanishi}, {Armus}, {Brandl}, \& {Peek}}]{bro17}
{Brown}, M. J.~I., {Moustakas}, J., {Kennicutt}, R.~C., {et~al.} 2017, \apj,
  847, 136

\bibitem[{{Brown} {et~al.}(2014{\natexlab{b}}){Brown}, {Moustakas}, {Smith},
  {da Cunha}, {Jarrett}, {Imanishi}, {Armus}, {Brandl}, \& {Peek}}]{bro14b}
{Brown}, M. J.~I., {Moustakas}, J., {Smith}, J. D.~T., {et~al.}
  2014{\natexlab{b}}, \apjs, 212, 18

\bibitem[{{Bruzual} \& {Charlot}(2003)}]{bru03}
{Bruzual}, G. \& {Charlot}, S. 2003, \mnras, 344, 1000

\bibitem[{{Buat}(2015)}]{bua15}
{Buat}, V. 2015, in Introduction to Cosmology, ed. M.~{Biernacka}, K.~{Bajan},
  G.~{Stachowski}, \& P.~{Flin}, 134--141

\bibitem[{{Buat} {et~al.}(2008){Buat}, {Boissier}, {Burgarella}, {Takeuchi},
  {Le Floc'h}, {Marcillac}, {Huang}, {Nagashima}, \& {Enoki}}]{bua08}
{Buat}, V., {Boissier}, S., {Burgarella}, D., {et~al.} 2008, \aap, 483, 107

\bibitem[{{Buat} {et~al.}(2018){Buat}, {Boquien}, {Ma{\l}ek}, {Corre}, {Salas},
  {Roehlly}, {Shirley}, \& {Efstathiou}}]{bua18}
{Buat}, V., {Boquien}, M., {Ma{\l}ek}, K., {et~al.} 2018, \aap, 619, A135

\bibitem[{{Buat} {et~al.}(2019){Buat}, {Ciesla}, {Boquien}, {Ma{\l}ek}, \&
  {Burgarella}}]{bua19}
{Buat}, V., {Ciesla}, L., {Boquien}, M., {Ma{\l}ek}, K., \& {Burgarella}, D.
  2019, \aap, 632, A79

\bibitem[{{Buat} {et~al.}(2011{\natexlab{a}}){Buat}, {Giovannoli}, {Heinis},
  {Charmandaris}, {Coia}, {Daddi}, {Dickinson}, {Elbaz}, {Hwang}, {Morrison},
  {Dasyra}, {Aussel}, {Altieri}, {Dannerbauer}, {Kartaltepe}, {Leiton},
  {Magdis}, {Magnelli}, \& {Popesso}}]{bua11b}
{Buat}, V., {Giovannoli}, E., {Heinis}, S., {et~al.} 2011{\natexlab{a}}, \aap,
  533, A93

\bibitem[{{Buat} {et~al.}(2011{\natexlab{b}}){Buat}, {Giovannoli}, {Takeuchi},
  {Heinis}, {Yuan}, {Burgarella}, {Noll}, \& {Iglesias-P{\'a}ramo}}]{bua11a}
{Buat}, V., {Giovannoli}, E., {Takeuchi}, T.~T., {et~al.} 2011{\natexlab{b}},
  \aap, 529, A22

\bibitem[{{Buat} {et~al.}(2007){Buat}, {Takeuchi}, {Iglesias-P{\'a}ramo}, {Xu},
  {Burgarella}, {Boselli}, {Barlow}, {Bianchi}, {Donas}, {Forster}, {Friedman},
  {Heckman}, {Lee}, {Madore}, {Martin}, {Milliard}, {Morissey}, {Neff}, {Rich},
  {Schiminovich}, {Seibert}, {Small}, {Szalay}, {Welsh}, {Wyder}, \&
  {Yi}}]{bua07}
{Buat}, V., {Takeuchi}, T.~T., {Iglesias-P{\'a}ramo}, J., {et~al.} 2007, \apjs,
  173, 404

\bibitem[{{Burgarella} {et~al.}(2005){Burgarella}, {Buat}, \&
  {Iglesias-P{\'a}ramo}}]{bur05}
{Burgarella}, D., {Buat}, V., \& {Iglesias-P{\'a}ramo}, J. 2005, \mnras, 360,
  1413

\bibitem[{{Calzetti}(2013)}]{cal13}
{Calzetti}, D. 2013, {Star Formation Rate Indicators}, ed.
  J.~{Falc{\'o}n-Barroso} \& J.~H. {Knapen}, 419

\bibitem[{{Calzetti} {et~al.}(2000){Calzetti}, {Armus}, {Bohlin}, {Kinney},
  {Koornneef}, \& {Storchi-Bergmann}}]{cal00}
{Calzetti}, D., {Armus}, L., {Bohlin}, R.~C., {et~al.} 2000, \apj, 533, 682

\bibitem[{{Calzetti} {et~al.}(2007){Calzetti}, {Kennicutt}, {Engelbracht},
  {Leitherer}, {Draine}, {Kewley}, {Moustakas}, {Sosey}, {Dale}, {Gordon},
  {Helou}, {Hollenbach}, {Armus}, {Bendo}, {Bot}, {Buckalew}, {Jarrett}, {Li},
  {Meyer}, {Murphy}, {Prescott}, {Regan}, {Rieke}, {Roussel}, {Sheth}, {Smith},
  {Thornley}, \& {Walter}}]{cal07}
{Calzetti}, D., {Kennicutt}, R.~C., {Engelbracht}, C.~W., {et~al.} 2007, \apj,
  666, 870

\bibitem[{{Calzetti} {et~al.}(1994){Calzetti}, {Kinney}, \&
  {Storchi-Bergmann}}]{cal94}
{Calzetti}, D., {Kinney}, A.~L., \& {Storchi-Bergmann}, T. 1994, \apj, 429, 582

\bibitem[{{Calzetti} {et~al.}(2010){Calzetti}, {Wu}, {Hong}, {Kennicutt},
  {Lee}, {Dale}, {Engelbracht}, {van Zee}, {Draine}, {Hao}, {Gordon},
  {Moustakas}, {Murphy}, {Regan}, {Begum}, {Block}, {Dalcanton}, {Funes}, {Gil
  de Paz}, {Johnson}, {Sakai}, {Skillman}, {Walter}, {Weisz}, {Williams}, \&
  {Wu}}]{cal10}
{Calzetti}, D., {Wu}, S.~Y., {Hong}, S., {et~al.} 2010, \apj, 714, 1256

\bibitem[{{Camps} \& {Baes}(2015)}]{camp15}
{Camps}, P. \& {Baes}, M. 2015, Astronomy and Computing, 9, 20

\bibitem[{{Cardelli} {et~al.}(1989){Cardelli}, {Clayton}, \& {Mathis}}]{car89}
{Cardelli}, J.~A., {Clayton}, G.~C., \& {Mathis}, J.~S. 1989, \apj, 345, 245

\bibitem[{{Casey}(2012)}]{cas12}
{Casey}, C.~M. 2012, \mnras, 425, 3094

\bibitem[{{Cassar{\`a}} {et~al.}(2016){Cassar{\`a}}, {Maccagni}, {Garilli},
  {Scodeggio}, {Thomas}, {Le F{\`e}vre}, {Zamorani}, {Schaerer}, {Lemaux},
  {Cassata}, {Le Brun}, {Pentericci}, {Tasca}, {Vanzella}, {Zucca},
  {Amor{\'\i}n}, {Bardelli}, {Castellano}, {Cimatti}, {Cucciati}, {Durkalec},
  {Fontana}, {Giavalisco}, {Grazian}, {Hathi}, {Ilbert}, {Paltani}, {Ribeiro},
  {Sommariva}, {Talia}, {Tresse}, {Vergani}, {Capak}, {Charlot}, {Contini}, {de
  la Torre}, {Dunlop}, {Fotopoulou}, {Guaita}, {Koekemoer},
  {L{\'o}pez-Sanjuan}, {Mellier}, {Pforr}, {Salvato}, {Scoville}, {Taniguchi},
  \& {Wang}}]{cas16}
{Cassar{\`a}}, L.~P., {Maccagni}, D., {Garilli}, B., {et~al.} 2016, \aap, 593,
  A9

\bibitem[{{Chabrier}(2003)}]{cha03}
{Chabrier}, G. 2003, \pasp, 115, 763

\bibitem[{{Charlot} \& {Fall}(2000)}]{cha00}
{Charlot}, S. \& {Fall}, S.~M. 2000, \apj, 539, 718

\bibitem[{{Charlot} \& {Longhetti}(2001)}]{char01}
{Charlot}, S. \& {Longhetti}, M. 2001, \mnras, 323, 887

\bibitem[{{Chary} \& {Elbaz}(2001)}]{cha01}
{Chary}, R. \& {Elbaz}, D. 2001, \apj, 556, 562

\bibitem[{{Chevallard} {et~al.}(2019){Chevallard}, {Curtis-Lake}, {Charlot},
  {Ferruit}, {Giardino}, {Franx}, {Maseda}, {Amorin}, {Arribas}, {Bunker},
  {Carniani}, {Husemann}, {Jakobsen}, {Maiolino}, {Pforr}, {Rawle}, {Rix},
  {Smit}, \& {Willott}}]{che19}
{Chevallard}, J., {Curtis-Lake}, E., {Charlot}, S., {et~al.} 2019, \mnras, 483,
  2621

\bibitem[{{Ciesla} {et~al.}(2014){Ciesla}, {Boquien}, {Boselli}, {Buat},
  {Cortese}, {Bendo}, {Heinis}, {Galametz}, {Eales}, {Smith}, {Baes},
  {Bianchi}, {De Looze}, {di Serego Alighieri}, {Galliano}, {Hughes}, {Madden},
  {Pierini}, {R{\'e}my-Ruyer}, {Spinoglio}, {Vaccari}, {Viaene}, \&
  {Vlahakis}}]{cie14}
{Ciesla}, L., {Boquien}, M., {Boselli}, A., {et~al.} 2014, \aap, 565, A128

\bibitem[{{Ciesla} {et~al.}(2016){Ciesla}, {Boselli}, {Elbaz}, {Boissier},
  {Buat}, {Charmandaris}, {Schreiber}, {B{\'e}thermin}, {Baes}, {Boquien}, {De
  Looze}, {Fern{\'a}ndez-Ontiveros}, {Pappalardo}, {Spinoglio}, \&
  {Viaene}}]{cie16}
{Ciesla}, L., {Boselli}, A., {Elbaz}, D., {et~al.} 2016, \aap, 585, A43

\bibitem[{{Ciesla} {et~al.}(2017){Ciesla}, {Elbaz}, \& {Fensch}}]{cie17}
{Ciesla}, L., {Elbaz}, D., \& {Fensch}, J. 2017, \aap, 608, A41

\bibitem[{{Clark} {et~al.}(2015){Clark}, {Dunne}, {Gomez}, {Maddox}, {De Vis},
  {Smith}, {Eales}, {Baes}, {Bendo}, {Bourne}, {Driver}, {Dye}, {Furlanetto},
  {Grootes}, {Ivison}, {Schofield}, {Robotham}, {Rowlands}, {Valiante},
  {Vlahakis}, {van der Werf}, {Wright}, \& {de Zotti}}]{cla15}
{Clark}, C.~J.~R., {Dunne}, L., {Gomez}, H.~L., {et~al.} 2015, \mnras, 452, 397

\bibitem[{{Cluver} {et~al.}(2017){Cluver}, {Jarrett}, {Dale}, {Smith},
  {August}, \& {Brown}}]{clu17}
{Cluver}, M.~E., {Jarrett}, T.~H., {Dale}, D.~A., {et~al.} 2017, \apj, 850, 68

\bibitem[{{Cluver} {et~al.}(2014){Cluver}, {Jarrett}, {Hopkins}, {Driver},
  {Liske}, {Gunawardhana}, {Taylor}, {Robotham}, {Alpaslan}, {Baldry}, {Brown},
  {Peacock}, {Popescu}, {Tuffs}, {Bauer}, {Bland -Hawthorn}, {Colless},
  {Holwerda}, {Lara-L{\'o}pez}, {Leschinski}, {L{\'o}pez-S{\'a}nchez},
  {Norberg}, {Owers}, {Wang}, \& {Wilkins}}]{clu14}
{Cluver}, M.~E., {Jarrett}, T.~H., {Hopkins}, A.~M., {et~al.} 2014, \apj, 782,
  90

\bibitem[{{Colless} {et~al.}(2001){Colless}, {Dalton}, {Maddox}, {Sutherland },
  {Norberg}, {Cole}, {Bland -Hawthorn}, {Bridges}, {Cannon}, {Collins},
  {Couch}, {Cross}, {Deeley}, {De Propris}, {Driver}, {Efstathiou}, {Ellis},
  {Frenk}, {Glazebrook}, {Jackson}, {Lahav}, {Lewis}, {Lumsden}, {Madgwick},
  {Peacock}, {Peterson}, {Price}, {Seaborne}, \& {Taylor}}]{col01}
{Colless}, M., {Dalton}, G., {Maddox}, S., {et~al.} 2001, \mnras, 328, 1039

\bibitem[{{Colless} {et~al.}(2003){Colless}, {Peterson}, {Jackson}, {Peacock},
  {Cole}, {Norberg}, {Baldry}, {Baugh}, {Bland-Hawthorn}, {Bridges}, {Cannon},
  {Collins}, {Couch}, {Cross}, {Dalton}, {De Propris}, {Driver}, {Efstathiou},
  {Ellis}, {Frenk}, {Glazebrook}, {Lahav}, {Lewis}, {Lumsden}, {Maddox},
  {Madgwick}, {Sutherland}, \& {Taylor}}]{col03}
{Colless}, M., {Peterson}, B.~A., {Jackson}, C., {et~al.} 2003, arXiv e-prints,
  astro

\bibitem[{{Cullen} {et~al.}(2016){Cullen}, {Cirasuolo}, {Kewley}, {McLure},
  {Dunlop}, \& {Bowler}}]{cul16}
{Cullen}, F., {Cirasuolo}, M., {Kewley}, L.~J., {et~al.} 2016, \mnras, 460,
  3002

\bibitem[{{da Cunha} {et~al.}(2012){da Cunha}, {Charlot}, {Dunne}, {Smith}, \&
  {Rowlands}}]{dac12}
{da Cunha}, E., {Charlot}, S., {Dunne}, L., {Smith}, D., \& {Rowlands}, K.
  2012, in IAU Symposium, Vol. 284, The Spectral Energy Distribution of
  Galaxies - SED 2011, ed. R.~J. {Tuffs} \& C.~C. {Popescu}, 292--296

\bibitem[{{Dale} \& {Helou}(2002)}]{dal02}
{Dale}, D.~A. \& {Helou}, G. 2002, \apj, 576, 159

\bibitem[{{Dale} {et~al.}(2014){Dale}, {Ngoumou}, {Ercolano}, \&
  {Bonnell}}]{dal14}
{Dale}, J.~E., {Ngoumou}, J., {Ercolano}, B., \& {Bonnell}, I.~A. 2014, \mnras,
  442, 694

\bibitem[{{Davidzon} {et~al.}(2013){Davidzon}, {Bolzonella}, {Coupon},
  {Ilbert}, {Arnouts}, {de la Torre}, {Fritz}, {De Lucia}, {Iovino}, {Granett},
  {Zamorani}, {Guzzo}, {Abbas}, {Adami}, {Bel}, {Bottini}, {Branchini},
  {Cappi}, {Cucciati}, {Franzetti}, {Fumana}, {Garilli}, {Krywult}, {Le Brun},
  {Le F{\`e}vre}, {Maccagni}, {Ma{\l}ek}, {Marulli}, {McCracken}, {Paioro},
  {Peacock}, {Polletta}, {Pollo}, {Schlagenhaufer}, {Scodeggio}, {Tasca},
  {Tojeiro}, {Vergani}, {Zanichelli}, {Burden}, {Di Porto}, {Marchetti},
  {Marinoni}, {Mellier}, {Moscardini}, {Moutard}, {Nichol}, {Percival},
  {Phleps}, \& {Wolk}}]{dav13}
{Davidzon}, I., {Bolzonella}, M., {Coupon}, J., {et~al.} 2013, \aap, 558, A23

\bibitem[{{Davies} {et~al.}(2016){Davies}, {Driver}, {Robotham}, {Grootes},
  {Popescu}, {Tuffs}, {Hopkins}, {Alpaslan}, {Andrews}, {Bland -Hawthorn},
  {Bremer}, {Brough}, {Brown}, {Cluver}, {Croom}, {da Cunha}, {Dunne},
  {Lara-L{\'o}pez}, {Liske}, {Loveday}, {Moffett}, {Owers}, {Phillipps},
  {Sansom}, {Taylor}, {Michalowski}, {Ibar}, {Smith}, \& {Bourne}}]{dav16}
{Davies}, L.~J.~M., {Driver}, S.~P., {Robotham}, A.~S.~G., {et~al.} 2016,
  \mnras, 461, 458

\bibitem[{{Davies} {et~al.}(2018){Davies}, {Robotham}, {Driver}, {Lagos},
  {Cortese}, {Mannering}, {Foster}, {Lidman}, {Hashemizadeh}, {Koushan},
  {O'Toole}, {Baldry}, {Bilicki}, {Bland-Hawthorn}, {Bremer}, {Brown},
  {Bryant}, {Catinella}, {Croom}, {Grootes}, {Holwerda}, {Jarvis}, {Maddox},
  {Meyer}, {Moffett}, {Phillipps}, {Taylor}, {Windhorst}, \& {Wolf}}]{dav18}
{Davies}, L.~J.~M., {Robotham}, A.~S.~G., {Driver}, S.~P., {et~al.} 2018,
  \mnras, 480, 768

\bibitem[{{De Barros} {et~al.}(2016){De Barros}, {Reddy}, \& {Shivaei}}]{deb16}
{De Barros}, S., {Reddy}, N., \& {Shivaei}, I. 2016, \apj, 820, 96

\bibitem[{{De Looze} {et~al.}(2014){De Looze}, {Fritz}, {Baes}, {Bendo},
  {Cortese}, {Boquien}, {Boselli}, {Camps}, {Cooray}, {Cormier}, {Davies}, {De
  Geyter}, {Hughes}, {Jones}, {Karczewski}, {Lebouteiller}, {Lu}, {Madden},
  {R{\'e}my-Ruyer}, {Spinoglio}, {Smith}, {Viaene}, \& {Wilson}}]{del14}
{De Looze}, I., {Fritz}, J., {Baes}, M., {et~al.} 2014, \aap, 571, A69

\bibitem[{{Donas} \& {Deharveng}(1984)}]{don84}
{Donas}, J. \& {Deharveng}, J.~M. 1984, \aap, 140, 325

\bibitem[{{Draine} {et~al.}(2014){Draine}, {Aniano}, {Krause}, {Groves},
  {Sandstrom}, {Braun}, {Leroy}, {Klaas}, {Linz}, {Rix}, {Schinnerer},
  {Schmiedeke}, \& {Walter}}]{dra14}
{Draine}, B.~T., {Aniano}, G., {Krause}, O., {et~al.} 2014, \apj, 780, 172

\bibitem[{{Draine} {et~al.}(2007){Draine}, {Dale}, {Bendo}, {Gordon}, {Smith},
  {Armus}, {Engelbracht}, {Helou}, {Kennicutt}, {Li}, {Roussel}, {Walter},
  {Calzetti}, {Moustakas}, {Murphy}, {Rieke}, {Bot}, {Hollenbach}, {Sheth}, \&
  {Teplitz}}]{dra07}
{Draine}, B.~T., {Dale}, D.~A., {Bendo}, G., {et~al.} 2007, \apj, 663, 866

\bibitem[{{Driver} {et~al.}(2018){Driver}, {Andrews}, {da Cunha}, {Davies},
  {Lagos}, {Robotham}, {Vinsen}, {Wright}, {Alpaslan}, {Bland -Hawthorn},
  {Bourne}, {Brough}, {Bremer}, {Cluver}, {Colless}, {Conselice}, {Dunne},
  {Eales}, {Gomez}, {Holwerda}, {Hopkins}, {Kafle}, {Kelvin}, {Loveday},
  {Liske}, {Maddox}, {Phillipps}, {Pimbblet}, {Rowlands}, {Sansom}, {Taylor},
  {Wang}, \& {Wilkins}}]{dri18}
{Driver}, S.~P., {Andrews}, S.~K., {da Cunha}, E., {et~al.} 2018, \mnras, 475,
  2891

\bibitem[{{Driver} {et~al.}(2016){Driver}, {Wright}, {Andrews}, {Davies},
  {Kafle}, {Lange}, {Moffett}, {Mannering}, {Robotham}, {Vinsen}, {Alpaslan},
  {Andrae}, {Baldry}, {Bauer}, {Bamford}, {Bland-Hawthorn}, {Bourne}, {Brough},
  {Brown}, {Cluver}, {Croom}, {Colless}, {Conselice}, {da Cunha}, {De Propris},
  {Drinkwater}, {Dunne}, {Eales}, {Edge}, {Frenk}, {Graham}, {Grootes},
  {Holwerda}, {Hopkins}, {Ibar}, {van Kampen}, {Kelvin}, {Jarrett}, {Jones},
  {Lara-Lopez}, {Liske}, {Lopez-Sanchez}, {Loveday}, {Maddox}, {Madore},
  {Mahajan}, {Meyer}, {Norberg}, {Penny}, {Phillipps}, {Popescu}, {Tuffs},
  {Peacock}, {Pimbblet}, {Prescott}, {Rowlands}, {Sansom}, {Seibert}, {Smith},
  {Sutherland}, {Taylor}, {Valiante}, {Vazquez-Mata}, {Wang}, {Wilkins}, \&
  {Williams}}]{dri16}
{Driver}, S.~P., {Wright}, A.~H., {Andrews}, S.~K., {et~al.} 2016, \mnras, 455,
  3911

\bibitem[{{Elbaz} {et~al.}(2010){Elbaz}, {Hwang}, {Magnelli}, {Daddi},
  {Aussel}, {Altieri}, {Amblard}, {Andreani}, {Arumugam}, {Auld}, {Babbedge},
  {Berta}, {Blain}, {Bock}, {Bongiovanni}, {Boselli}, {Buat}, {Burgarella},
  {Castro-Rodriguez}, {Cava}, {Cepa}, {Chanial}, {Chary}, {Cimatti},
  {Clements}, {Conley}, {Conversi}, {Cooray}, {Dickinson}, {Dominguez},
  {Dowell}, {Dunlop}, {Dwek}, {Eales}, {Farrah}, {F{\"o}rster Schreiber},
  {Fox}, {Franceschini}, {Gear}, {Genzel}, {Glenn}, {Griffin}, {Gruppioni},
  {Halpern}, {Hatziminaoglou}, {Ibar}, {Isaak}, {Ivison}, {Lagache}, {Le
  Borgne}, {Le Floc'h}, {Levenson}, {Lu}, {Lutz}, {Madden}, {Maffei}, {Magdis},
  {Mainetti}, {Maiolino}, {Marchetti}, {Mortier}, {Nguyen}, {Nordon},
  {O'Halloran}, {Okumura}, {Oliver}, {Omont}, {Page}, {Panuzzo},
  {Papageorgiou}, {Pearson}, {Perez Fournon}, {P{\'e}rez Garc{\'\i}a},
  {Poglitsch}, {Pohlen}, {Popesso}, {Pozzi}, {Rawlings}, {Rigopoulou},
  {Riguccini}, {Rizzo}, {Rodighiero}, {Roseboom}, {Rowan-Robinson},
  {Saintonge}, {Sanchez Portal}, {Santini}, {Sauvage}, {Schulz}, {Scott},
  {Seymour}, {Shao}, {Shupe}, {Smith}, {Stevens}, {Sturm}, {Symeonidis},
  {Tacconi}, {Trichas}, {Tugwell}, {Vaccari}, {Valtchanov}, {Vieira},
  {Vigroux}, {Wang}, {Ward}, {Wright}, {Xu}, \& {Zemcov}}]{elb10}
{Elbaz}, D., {Hwang}, H.~S., {Magnelli}, B., {et~al.} 2010, \aap, 518, L29

\bibitem[{{Engelbracht} {et~al.}(2005){Engelbracht}, {Gordon}, {Rieke},
  {Werner}, {Dale}, \& {Latter}}]{eng05}
{Engelbracht}, C.~W., {Gordon}, K.~D., {Rieke}, G.~H., {et~al.} 2005, \apjl,
  628, L29

\bibitem[{{Ferland} {et~al.}(2013){Ferland}, {Porter}, {van Hoof}, {Williams},
  {Abel}, {Lykins}, {Shaw}, {Henney}, \& {Stancil}}]{fer13}
{Ferland}, G.~J., {Porter}, R.~L., {van Hoof}, P.~A.~M., {et~al.} 2013, \rmxaa,
  49, 137

\bibitem[{{Figueira} {et~al.}(2017){Figueira}, {Zavagno}, {Deharveng},
  {Russeil}, {Anderson}, {Men'shchikov}, {Schneider}, {Hill}, {Motte},
  {M{\`e}ge}, {LeLeu}, {Roussel}, {Bernard}, {Traficante}, {Paradis},
  {Tig{\'e}}, {Andr{\'e}}, {Bontemps}, \& {Abergel}}]{fig17}
{Figueira}, M., {Zavagno}, A., {Deharveng}, L., {et~al.} 2017, \aap, 600, A93

\bibitem[{{Fioc} \& {Rocca-Volmerange}(1997)}]{fio97}
{Fioc}, M. \& {Rocca-Volmerange}, B. 1997, \aap, 500, 507

\bibitem[{{Fritz} {et~al.}(2006){Fritz}, {Franceschini}, \&
  {Hatziminaoglou}}]{fri06}
{Fritz}, J., {Franceschini}, A., \& {Hatziminaoglou}, E. 2006, \mnras, 366, 767

\bibitem[{{Fukugita} {et~al.}(1996){Fukugita}, {Ichikawa}, {Gunn}, {Doi},
  {Shimasaku}, \& {Schneider}}]{fuk96}
{Fukugita}, M., {Ichikawa}, T., {Gunn}, J.~E., {et~al.} 1996, \aj, 111, 1748

\bibitem[{{Garc{\'{\i}}a} {et~al.}(2014){Garc{\'{\i}}a}, {Bronfman}, {Nyman},
  {Dame}, \& {Luna}}]{gar14}
{Garc{\'{\i}}a}, P., {Bronfman}, L., {Nyman}, L.-{\AA}., {Dame}, T.~M., \&
  {Luna}, A. 2014, \apjs, 212, 2

\bibitem[{{Garilli} {et~al.}(2010){Garilli}, {Fumana}, {Franzetti}, {Paioro},
  {Scodeggio}, {Le F{\`e}vre}, {Paltani}, \& {Scaramella}}]{gar10}
{Garilli}, B., {Fumana}, M., {Franzetti}, P., {et~al.} 2010, \pasp, 122, 827

\bibitem[{{Garilli} {et~al.}(2014){Garilli}, {Guzzo}, {Scodeggio},
  {Bolzonella}, {Abbas}, {Adami}, {Arnouts}, {Bel}, {Bottini}, {Branchini},
  {Cappi}, {Coupon}, {Cucciati}, {Davidzon}, {De Lucia}, {de la Torre},
  {Franzetti}, {Fritz}, {Fumana}, {Granett}, {Ilbert}, {Iovino}, {Krywult}, {Le
  Brun}, {Le F{\`e}vre}, {Maccagni}, {Ma{\l}ek}, {Marulli}, {McCracken},
  {Paioro}, {Polletta}, {Pollo}, {Schlagenhaufer}, {Tasca}, {Tojeiro},
  {Vergani}, {Zamorani}, {Zanichelli}, {Burden}, {Di Porto}, {Marchetti},
  {Marinoni}, {Mellier}, {Moscardini}, {Nichol}, {Peacock}, {Percival},
  {Phleps}, \& {Wolk}}]{gari14}
{Garilli}, B., {Guzzo}, L., {Scodeggio}, M., {et~al.} 2014, \aap, 562, A23

\bibitem[{{Garilli} {et~al.}(2008){Garilli}, {Le F{\`e}vre}, {Guzzo},
  {Maccagni}, {Le Brun}, {de la Torre}, {Meneux}, {Tresse}, {Franzetti},
  {Zamorani}, {Zanichelli}, {Gregorini}, {Vergani}, {Bottini}, {Scaramella},
  {Scodeggio}, {Vettolani}, {Adami}, {Arnouts}, {Bardelli}, {Bolzonella},
  {Cappi}, {Charlot}, {Ciliegi}, {Contini}, {Foucaud}, {Gavignaud}, {Ilbert},
  {Iovino}, {Lamareille}, {McCracken}, {Marano}, {Marinoni}, {Mazure},
  {Merighi}, {Paltani}, {Pell{\`o}}, {Pollo}, {Pozzetti}, {Radovich}, {Zucca},
  {Blaizot}, {Bongiorno}, {Cucciati}, {Mellier}, {Moreau}, \& {Paioro}}]{gar08}
{Garilli}, B., {Le F{\`e}vre}, O., {Guzzo}, L., {et~al.} 2008, \aap, 486, 683

\bibitem[{{Gilbank} {et~al.}(2010){Gilbank}, {Baldry}, {Balogh}, {Glazebrook},
  \& {Bower}}]{gil10}
{Gilbank}, D.~G., {Baldry}, I.~K., {Balogh}, M.~L., {Glazebrook}, K., \&
  {Bower}, R.~G. 2010, \mnras, 405, 2594

\bibitem[{{Giovannoli} {et~al.}(2011){Giovannoli}, {Buat}, {Noll},
  {Burgarella}, \& {Magnelli}}]{gio11}
{Giovannoli}, E., {Buat}, V., {Noll}, S., {Burgarella}, D., \& {Magnelli}, B.
  2011, \aap, 525, A150

\bibitem[{{Goranova} {et~al.}(2009){Goranova}, {Hudelot}, {Magnard}, \&
  {al}}]{gor09}
{Goranova}, Y., {Hudelot}, P., {Magnard}, \& {al}. 2009, ~,
  \url{http://terapix.iap.fr/cplt/table_syn_T0006.html}

\bibitem[{{Goto} {et~al.}(2003){Goto}, {Okamura}, {Sekiguchi}, {Bernardi},
  {Brinkmann}, {G{\'o}mez}, {Harvanek}, {Kleinman}, {Krzesinski}, {Long},
  {Loveday}, {Miller}, {Neilsen}, {Newman}, {Nitta}, {Sheth}, {Snedden}, \&
  {Yamauchi}}]{got03}
{Goto}, T., {Okamura}, S., {Sekiguchi}, M., {et~al.} 2003, \pasj, 55, 757

\bibitem[{{Gregg} {et~al.}(2022){Gregg}, {Calzetti}, \& {Heyer}}]{gre22}
{Gregg}, B., {Calzetti}, D., \& {Heyer}, M. 2022, \apj, 928, 120

\bibitem[{{Groves} {et~al.}(2012{\natexlab{a}}){Groves}, {Brinchmann}, \&
  {Walcher}}]{gro06}
{Groves}, B., {Brinchmann}, J., \& {Walcher}, C.~J. 2012{\natexlab{a}}, \mnras,
  419, 1402

\bibitem[{{Groves} {et~al.}(2012{\natexlab{b}}){Groves}, {Brinchmann}, \&
  {Walcher}}]{gro12}
{Groves}, B., {Brinchmann}, J., \& {Walcher}, C.~J. 2012{\natexlab{b}}, \mnras,
  419, 1402

\bibitem[{{Guzzo} \& {VIPERS Team}(2013)}]{guz13}
{Guzzo}, L. \& {VIPERS Team}. 2013, The Messenger, 151, 41

\bibitem[{{Hao} {et~al.}(2011){Hao}, {Kennicutt}, {Johnson}, {Calzetti},
  {Dale}, \& {Moustakas}}]{hao11}
{Hao}, C.-N., {Kennicutt}, R.~C., {Johnson}, B.~D., {et~al.} 2011, \apj, 741,
  124

\bibitem[{{Hicks} {et~al.}(2002){Hicks}, {Malkan}, {Teplitz}, {McCarthy}, \&
  {Yan}}]{hic02}
{Hicks}, E. K.~S., {Malkan}, M.~A., {Teplitz}, H.~I., {McCarthy}, P.~J., \&
  {Yan}, L. 2002, \apj, 581, 205

\bibitem[{{Hippelein} {et~al.}(2003){Hippelein}, {Maier}, {Meisenheimer},
  {Wolf}, {Fried}, {von Kuhlmann}, {K{\"u}mmel}, {Phleps}, \&
  {R{\"o}ser}}]{hip03}
{Hippelein}, H., {Maier}, C., {Meisenheimer}, K., {et~al.} 2003, \aap, 402, 65

\bibitem[{{Hirashita} {et~al.}(2003){Hirashita}, {Buat}, \& {Inoue}}]{hir03}
{Hirashita}, H., {Buat}, V., \& {Inoue}, A.~K. 2003, \aap, 410, 83

\bibitem[{{Holden} {et~al.}(2016){Holden}, {Oesch}, {Gonz{\'a}lez},
  {Illingworth}, {Labb{\'e}}, {Bouwens}, {Franx}, {van Dokkum}, \&
  {Spitler}}]{hol16}
{Holden}, B.~P., {Oesch}, P.~A., {Gonz{\'a}lez}, V.~G., {et~al.} 2016, \apj,
  820, 73

\bibitem[{{Hopkins} {et~al.}(2003){Hopkins}, {Miller}, {Nichol}, {Connolly},
  {Bernardi}, {G{\'o}mez}, {Goto}, {Tremonti}, {Brinkmann}, {Ivezi{\'c}}, \&
  {Lamb}}]{hop03}
{Hopkins}, A.~M., {Miller}, C.~J., {Nichol}, R.~C., {et~al.} 2003, \apj, 599,
  971

\bibitem[{{Hummer} \& {Storey}(1987)}]{hum87}
{Hummer}, D.~G. \& {Storey}, P.~J. 1987, \mnras, 224, 801

\bibitem[{{Inoue}(2011)}]{ino11}
{Inoue}, A.~K. 2011, \mnras, 415, 2920

\bibitem[{{Jarrett} {et~al.}(2000){Jarrett}, {Chester}, {Cutri}, {Schneider},
  {Skrutskie}, \& {Huchra}}]{jar00}
{Jarrett}, T.~H., {Chester}, T., {Cutri}, R., {et~al.} 2000, \aj, 119, 2498

\bibitem[{{Jarrett} {et~al.}(2013){Jarrett}, {Masci}, {Tsai}, {Petty},
  {Cluver}, {Assef}, {Benford}, {Blain}, {Bridge}, {Donoso}, {Eisenhardt},
  {Koribalski}, {Lake}, {Neill}, {Seibert}, {Sheth}, {Stanford}, \&
  {Wright}}]{jar13}
{Jarrett}, T.~H., {Masci}, F., {Tsai}, C.~W., {et~al.} 2013, \aj, 145, 6

\bibitem[{{Jones} {et~al.}(2017){Jones}, {K{\"o}hler}, {Ysard}, {Bocchio}, \&
  {Verstraete}}]{jon17}
{Jones}, A.~P., {K{\"o}hler}, M., {Ysard}, N., {Bocchio}, M., \& {Verstraete},
  L. 2017, \aap, 602, A46

\bibitem[{{Kauffmann} {et~al.}(2003){Kauffmann}, {Heckman}, {Tremonti},
  {Brinchmann}, {Charlot}, {White}, {Ridgway}, {Brinkmann}, {Fukugita}, {Hall},
  {Ivezi{\'c}}, {Richards}, \& {Schneider}}]{kau03}
{Kauffmann}, G., {Heckman}, T.~M., {Tremonti}, C., {et~al.} 2003, \mnras, 346,
  1055

\bibitem[{{Kennicutt}(1983)}]{ken83}
{Kennicutt}, R.~C., J. 1983, \apj, 272, 54

\bibitem[{{Kennicutt}(1998)}]{ken98a}
{Kennicutt}, Robert~C., J. 1998, \araa, 36, 189

\bibitem[{{Kennicutt} {et~al.}(2009){Kennicutt}, {Hao}, {Calzetti},
  {Moustakas}, {Dale}, {Bendo}, {Engelbracht}, {Johnson}, \& {Lee}}]{ken09}
{Kennicutt}, Robert~C., J., {Hao}, C.-N., {Calzetti}, D., {et~al.} 2009, \apj,
  703, 1672

\bibitem[{{Kennicutt} {et~al.}(1994){Kennicutt}, {Tamblyn}, \&
  {Congdon}}]{ken94}
{Kennicutt}, Robert~C., J., {Tamblyn}, P., \& {Congdon}, C.~E. 1994, \apj, 435,
  22

\bibitem[{{Kennicutt} \& {Evans}(2012)}]{kenn12}
{Kennicutt}, R.~C. \& {Evans}, N.~J. 2012, \araa, 50, 531

\bibitem[{{Kewley} \& {Dopita}(2002)}]{kew02}
{Kewley}, L.~J. \& {Dopita}, M.~A. 2002, \apjs, 142, 35

\bibitem[{{Kewley} {et~al.}(2001){Kewley}, {Dopita}, {Sutherland}, {Heisler},
  \& {Trevena}}]{kew01}
{Kewley}, L.~J., {Dopita}, M.~A., {Sutherland}, R.~S., {Heisler}, C.~A., \&
  {Trevena}, J. 2001, \apj, 556, 121

\bibitem[{{Kewley} {et~al.}(2004){Kewley}, {Geller}, \& {Jansen}}]{kew04}
{Kewley}, L.~J., {Geller}, M.~J., \& {Jansen}, R.~A. 2004, \aj, 127, 2002

\bibitem[{{Kewley} {et~al.}(2013){Kewley}, {Maier}, {Yabe}, {Ohta}, {Akiyama},
  {Dopita}, \& {Yuan}}]{kew13a}
{Kewley}, L.~J., {Maier}, C., {Yabe}, K., {et~al.} 2013, \apjl, 774, L10

\bibitem[{{Kewley} {et~al.}(2015){Kewley}, {Zahid}, {Geller}, {Dopita},
  {Hwang}, \& {Fabricant}}]{kew15}
{Kewley}, L.~J., {Zahid}, H.~J., {Geller}, M.~J., {et~al.} 2015, \apjl, 812,
  L20

\bibitem[{{Kobulnicky} {et~al.}(1999){Kobulnicky}, {Kennicutt}, \&
  {Pizagno}}]{kob99}
{Kobulnicky}, H.~A., {Kennicutt}, Robert~C., J., \& {Pizagno}, J.~L. 1999,
  \apj, 514, 544

\bibitem[{{Komatsu} {et~al.}(2011){Komatsu}, {Smith}, {Dunkley}, {Bennett},
  {Gold}, {Hinshaw}, {Jarosik}, {Larson}, {Nolta}, {Page}, {Spergel},
  {Halpern}, {Hill}, {Kogut}, {Limon}, {Meyer}, {Odegard}, {Tucker}, {Weiland},
  {Wollack}, \& {Wright}}]{kom11}
{Komatsu}, E., {Smith}, K.~M., {Dunkley}, J., {et~al.} 2011, \apjs, 192, 18

\bibitem[{{Kroupa}(2002)}]{kro02}
{Kroupa}, P. 2002, Science, 295, 82

\bibitem[{{Lamareille}(2010)}]{lam10}
{Lamareille}, F. 2010, \aap, 509, A53

\bibitem[{{Lamareille} {et~al.}(2009){Lamareille}, {Brinchmann}, {Contini},
  {Walcher}, {Charlot}, {P{\'e}rez-Montero}, {Zamorani}, {Pozzetti},
  {Bolzonella}, {Garilli}, {Paltani}, {Bongiorno}, {Le F{\`e}vre}, {Bottini},
  {Le Brun}, {Maccagni}, {Scaramella}, {Scodeggio}, {Tresse}, {Vettolani},
  {Zanichelli}, {Adami}, {Arnouts}, {Bardelli}, {Cappi}, {Ciliegi}, {Foucaud},
  {Franzetti}, {Gavignaud}, {Guzzo}, {Ilbert}, {Iovino}, {McCracken}, {Marano},
  {Marinoni}, {Mazure}, {Meneux}, {Merighi}, {Pell{\`o}}, {Pollo}, {Radovich},
  {Vergani}, {Zucca}, {Romano}, {Grado}, \& {Limatola}}]{lam09}
{Lamareille}, F., {Brinchmann}, J., {Contini}, T., {et~al.} 2009, \aap, 495, 53

\bibitem[{{Lang} {et~al.}(2016){Lang}, {Hogg}, \& {Schlegel}}]{lan16}
{Lang}, D., {Hogg}, D.~W., \& {Schlegel}, D.~J. 2016, \aj, 151, 36

\bibitem[{{Le F{\`e}vre} {et~al.}(2003){Le F{\`e}vre}, {Saisse}, {Mancini},
  {Brau-Nogue}, {Caputi}, {Castinel}, {D'Odorico}, {Garilli}, {Kissler-Patig},
  {Lucuix}, {Mancini}, {Pauget}, {Sciarretta}, {Scodeggio}, {Tresse}, \&
  {Vettolani}}]{lef03}
{Le F{\`e}vre}, O., {Saisse}, M., {Mancini}, D., {et~al.} 2003, in Society of
  Photo-Optical Instrumentation Engineers (SPIE) Conference Series, Vol. 4841,
  \procspie, ed. M.~{Iye} \& A.~F.~M. {Moorwood}, 1670--1681

\bibitem[{{Le F{\`e}vre} {et~al.}(2005){Le F{\`e}vre}, {Vettolani}, {Garilli},
  {Tresse}, {Bottini}, {Le Brun}, {Maccagni}, {Picat}, {Scaramella},
  {Scodeggio}, {Zanichelli}, {Adami}, {Arnaboldi}, {Arnouts}, {Bardelli},
  {Bolzonella}, {Cappi}, {Charlot}, {Ciliegi}, {Contini}, {Foucaud},
  {Franzetti}, {Gavignaud}, {Guzzo}, {Ilbert}, {Iovino}, {McCracken}, {Marano},
  {Marinoni}, {Mathez}, {Mazure}, {Meneux}, {Merighi}, {Paltani}, {Pell{\`o}},
  {Pollo}, {Pozzetti}, {Radovich}, {Zamorani}, {Zucca}, {Bondi}, {Bongiorno},
  {Busarello}, {Lamareille}, {Mellier}, {Merluzzi}, {Ripepi}, \&
  {Rizzo}}]{lef05}
{Le F{\`e}vre}, O., {Vettolani}, G., {Garilli}, B., {et~al.} 2005, \aap, 439,
  845

\bibitem[{{Liang} {et~al.}(2006){Liang}, {Yin}, {Hammer}, {Deng}, {Flores}, \&
  {Zhang}}]{lia06}
{Liang}, Y.~C., {Yin}, S.~Y., {Hammer}, F., {et~al.} 2006, \apj, 652, 257

\bibitem[{{Lilly} {et~al.}(1995){Lilly}, {Tresse}, {Hammer}, {Crampton}, \& {Le
  Fevre}}]{lil95}
{Lilly}, S.~J., {Tresse}, L., {Hammer}, F., {Crampton}, D., \& {Le Fevre}, O.
  1995, \apj, 455, 108

\bibitem[{Lin(1989)}]{lin89}
Lin, L. I.-K. 1989, Biometrics, 45, 255

\bibitem[{{Lo Faro} {et~al.}(2017){Lo Faro}, {Buat}, {Roehlly},
  {Alvarez-Marquez}, {Burgarella}, {Silva}, \& {Efstathiou}}]{lof17}
{Lo Faro}, B., {Buat}, V., {Roehlly}, Y., {et~al.} 2017, \mnras, 472, 1372

\bibitem[{{Ly} {et~al.}(2007){Ly}, {Malkan}, {Kashikawa}, {Shimasaku}, {Doi},
  {Nagao}, {Iye}, {Kodama}, {Morokuma}, \& {Motohara}}]{ly07}
{Ly}, C., {Malkan}, M.~A., {Kashikawa}, N., {et~al.} 2007, \apj, 657, 738

\bibitem[{{Madau} \& {Dickinson}(2014)}]{mad14}
{Madau}, P. \& {Dickinson}, M. 2014, \araa, 52, 415

\bibitem[{{Madau} {et~al.}(1996){Madau}, {Ferguson}, {Dickinson}, {Giavalisco},
  {Steidel}, \& {Fruchter}}]{mad96}
{Madau}, P., {Ferguson}, H.~C., {Dickinson}, M.~E., {et~al.} 1996, \mnras, 283,
  1388

\bibitem[{{Madden} {et~al.}(2006){Madden}, {Galliano}, {Jones}, \&
  {Sauvage}}]{mad06}
{Madden}, S.~C., {Galliano}, F., {Jones}, A.~P., \& {Sauvage}, M. 2006, \aap,
  446, 877

\bibitem[{{Magdis} {et~al.}(2012){Magdis}, {Daddi}, {B{\'e}thermin}, {Sargent},
  {Elbaz}, {Pannella}, {Dickinson}, {Dannerbauer}, {da Cunha}, {Walter},
  {Rigopoulou}, {Charmandaris}, {Hwang}, \& {Kartaltepe}}]{mag12}
{Magdis}, G.~E., {Daddi}, E., {B{\'e}thermin}, M., {et~al.} 2012, \apj, 760, 6

\bibitem[{{Magnelli} {et~al.}(2010){Magnelli}, {Lutz}, {Berta}, {Altieri},
  {Andreani}, {Aussel}, {Casta{\~n}eda}, {Cava}, {Cepa}, {Cimatti}, {Daddi},
  {Dannerbauer}, {Dominguez}, {Elbaz}, {F{\"o}rster Schreiber}, {Genzel},
  {Grazian}, {Gruppioni}, {Magdis}, {Maiolino}, {Nordon}, {P{\'e}rez Fournon},
  {P{\'e}rez Garc{\'\i}a}, {Poglitsch}, {Popesso}, {Pozzi}, {Riguccini},
  {Rodighiero}, {Saintonge}, {Santini}, {Sanchez-Portal}, {Shao}, {Sturm},
  {Tacconi}, {Valtchanov}, {Wieprecht}, \& {Wiezorrek}}]{mag10}
{Magnelli}, B., {Lutz}, D., {Berta}, S., {et~al.} 2010, \aap, 518, L28

\bibitem[{{Maier} {et~al.}(2015){Maier}, {Ziegler}, {Lilly}, {Contini},
  {P{\'e}rez-Montero}, {Lamareille}, {Bolzonella}, \& {Le Floc'h}}]{mai15}
{Maier}, C., {Ziegler}, B.~L., {Lilly}, S.~J., {et~al.} 2015, \aap, 577, A14

\bibitem[{{Ma{\l}ek} {et~al.}(2018){Ma{\l}ek}, {Buat}, {Roehlly}, {Burgarella},
  {Hurley}, {Shirley}, {Duncan}, {Efstathiou}, {Papadopoulos}, {Vaccari},
  {Farrah}, {Marchetti}, \& {Oliver}}]{mal18}
{Ma{\l}ek}, K., {Buat}, V., {Roehlly}, Y., {et~al.} 2018, \aap, 620, A50

\bibitem[{{Maraston}(2005)}]{mara05}
{Maraston}, C. 2005, \mnras, 362, 799

\bibitem[{{Martin} {et~al.}(2005){Martin}, {Fanson}, {Schiminovich},
  {Morrissey}, {Friedman}, {Barlow}, {Conrow}, {Grange}, {Jelinsky},
  {Milliard}, {Siegmund}, {Bianchi}, {Byun}, {Donas}, {Forster}, {Heckman},
  {Lee}, {Madore}, {Malina}, {Neff}, {Rich}, {Small}, {Surber}, {Szalay},
  {Welsh}, \& {Wyder}}]{mart05}
{Martin}, D.~C., {Fanson}, J., {Schiminovich}, D., {et~al.} 2005, \apjl, 619,
  L1

\bibitem[{{McCarthy} {et~al.}(1999){McCarthy}, {Yan}, {Freudling}, {Teplitz},
  {Malumuth}, {Weymann}, {Malkan}, {Fosbury}, {Gardner}, {Storrie-Lombardi},
  {Thompson}, {Williams}, \& {Heap}}]{mcc99}
{McCarthy}, P.~J., {Yan}, L., {Freudling}, W., {et~al.} 1999, \apj, 520, 548

\bibitem[{{McGaugh}(1991)}]{mcg91}
{McGaugh}, S.~S. 1991, \apj, 380, 140

\bibitem[{{Meurer} {et~al.}(1999){Meurer}, {Heckman}, \& {Calzetti}}]{meu99}
{Meurer}, G.~R., {Heckman}, T.~M., \& {Calzetti}, D. 1999, \apj, 521, 64

\bibitem[{{Miller} \& {Owen}(2002)}]{mil02}
{Miller}, N.~A. \& {Owen}, F.~N. 2002, \aj, 124, 2453

\bibitem[{{Moustakas} \& {Kennicutt}(2006)}]{mou06a}
{Moustakas}, J. \& {Kennicutt}, Robert~C., J. 2006, \apjs, 164, 81

\bibitem[{{Moustakas} {et~al.}(2006){Moustakas}, {Kennicutt}, \&
  {Tremonti}}]{mou06b}
{Moustakas}, J., {Kennicutt}, Robert~C., J., \& {Tremonti}, C.~A. 2006, \apj,
  642, 775

\bibitem[{{Moutard} {et~al.}(2016{\natexlab{a}}){Moutard}, {Arnouts}, {Ilbert},
  {Coupon}, {Davidzon}, {Guzzo}, {Hudelot}, {McCracken}, {Van Waerbeke},
  {Morrison}, {Le F{\`e}vre}, {Comte}, {Bolzonella}, {Fritz}, {Garilli}, \&
  {Scodeggio}}]{mou16b}
{Moutard}, T., {Arnouts}, S., {Ilbert}, O., {et~al.} 2016{\natexlab{a}}, \aap,
  590, A103

\bibitem[{{Moutard} {et~al.}(2016{\natexlab{b}}){Moutard}, {Arnouts}, {Ilbert},
  {Coupon}, {Hudelot}, {Vibert}, {Comte}, {Conseil}, {Davidzon}, {Guzzo},
  {Llebaria}, {Martin}, {McCracken}, {Milliard}, {Morrison}, {Schiminovich},
  {Treyer}, \& {Van Werbaeke}}]{mou16a}
{Moutard}, T., {Arnouts}, S., {Ilbert}, O., {et~al.} 2016{\natexlab{b}}, \aap,
  590, A102

\bibitem[{{Nagao} {et~al.}(2006){Nagao}, {Maiolino}, \& {Marconi}}]{nag06}
{Nagao}, T., {Maiolino}, R., \& {Marconi}, A. 2006, \aap, 459, 85

\bibitem[{{Nersesian} {et~al.}(2019){Nersesian}, {Xilouris}, {Bianchi},
  {Galliano}, {Jones}, {Baes}, {Casasola}, {Cassar{\`a}}, {Clark}, {Davies},
  {Decleir}, {Dobbels}, {De Looze}, {De Vis}, {Fritz}, {Galametz}, {Madden},
  {Mosenkov}, {Tr{\v{c}}ka}, {Verstocken}, {Viaene}, \& {Lianou}}]{ner19}
{Nersesian}, A., {Xilouris}, E.~M., {Bianchi}, S., {et~al.} 2019, \aap, 624,
  A80

\bibitem[{{Noll} {et~al.}(2009){Noll}, {Burgarella}, {Giovannoli}, {Buat},
  {Marcillac}, \& {Mu{\~n}oz-Mateos}}]{nol09}
{Noll}, S., {Burgarella}, D., {Giovannoli}, E., {et~al.} 2009, \aap, 507, 1793

\bibitem[{{Nordon} {et~al.}(2012){Nordon}, {Lutz}, {Genzel}, {Berta}, {Wuyts},
  {Magnelli}, {Altieri}, {Andreani}, {Aussel}, {Bongiovanni}, {Cepa},
  {Cimatti}, {Daddi}, {Fadda}, {F{\"o}rster Schreiber}, {Lagache}, {Maiolino},
  {P{\'e}rez Garc{\'\i}a}, {Poglitsch}, {Popesso}, {Pozzi}, {Rodighiero},
  {Rosario}, {Saintonge}, {Sanchez-Portal}, {Santini}, {Sturm}, {Tacconi},
  {Valtchanov}, \& {Yan}}]{nor12}
{Nordon}, R., {Lutz}, D., {Genzel}, R., {et~al.} 2012, \apj, 745, 182

\bibitem[{{Nordon} {et~al.}(2010){Nordon}, {Lutz}, {Shao}, {Magnelli}, {Berta},
  {Altieri}, {Andreani}, {Aussel}, {Bongiovanni}, {Cava}, {Cepa}, {Cimatti},
  {Daddi}, {Dominguez}, {Elbaz}, {F{\"o}rster Schreiber}, {Genzel}, {Grazian},
  {Magdis}, {Maiolino}, {P{\'e}rez Garc{\'\i}a}, {Poglitsch}, {Popesso},
  {Pozzi}, {Riguccini}, {Rodighiero}, {Saintonge}, {Sanchez-Portal}, {Santini},
  {Sturm}, {Tacconi}, {Valtchanov}, {Wetzstein}, \& {Wieprecht}}]{nor10}
{Nordon}, R., {Lutz}, D., {Shao}, L., {et~al.} 2010, \aap, 518, L24

\bibitem[{{O'Donnell}(1994)}]{odo94}
{O'Donnell}, J.~E. 1994, \apj, 422, 158

\bibitem[{{Oke}(1974)}]{oke74}
{Oke}, J.~B. 1974, \apjs, 27, 21

\bibitem[{{Pagel} {et~al.}(1979){Pagel}, {Edmunds}, {Blackwell}, {Chun}, \&
  {Smith}}]{pag79}
{Pagel}, B.~E.~J., {Edmunds}, M.~G., {Blackwell}, D.~E., {Chun}, M.~S., \&
  {Smith}, G. 1979, \mnras, 189, 95

\bibitem[{{Peek} \& {Schiminovich}(2013)}]{pee13}
{Peek}, J.~E.~G. \& {Schiminovich}, D. 2013, \apj, 771, 68

\bibitem[{{P{\'e}rez-Gonz{\'a}lez} {et~al.}(2006){P{\'e}rez-Gonz{\'a}lez},
  {Kennicutt}, {Gordon}, {Misselt}, {Gil de Paz}, {Engelbracht}, {Rieke},
  {Bendo}, {Bianchi}, {Boissier}, {Calzetti}, {Dale}, {Draine}, {Jarrett},
  {Hollenbach}, \& {Prescott}}]{per06}
{P{\'e}rez-Gonz{\'a}lez}, P.~G., {Kennicutt}, Robert~C., J., {Gordon}, K.~D.,
  {et~al.} 2006, \apj, 648, 987

\bibitem[{{Pforr} {et~al.}(2012){Pforr}, {Maraston}, \& {Tonini}}]{pfo12}
{Pforr}, J., {Maraston}, C., \& {Tonini}, C. 2012, \mnras, 422, 3285

\bibitem[{{Pilyugin}(2001)}]{pil01}
{Pilyugin}, L.~S. 2001, \aap, 369, 594

\bibitem[{{Pope} {et~al.}(2008){Pope}, {Chary}, {Alexander}, {Armus},
  {Dickinson}, {Elbaz}, {Frayer}, {Scott}, \& {Teplitz}}]{pop08}
{Pope}, A., {Chary}, R.-R., {Alexander}, D.~M., {et~al.} 2008, \apj, 675, 1171

\bibitem[{{Popescu} {et~al.}(2011){Popescu}, {Tuffs}, {Dopita}, {Fischera},
  {Kylafis}, \& {Madore}}]{pop11}
{Popescu}, C.~C., {Tuffs}, R.~J., {Dopita}, M.~A., {et~al.} 2011, \aap, 527,
  A109

\bibitem[{{Prescott} {et~al.}(2009){Prescott}, {Baldry}, \& {James}}]{pre09}
{Prescott}, M., {Baldry}, I.~K., \& {James}, P.~A. 2009, \mnras, 397, 90

\bibitem[{{Prevot} {et~al.}(1984){Prevot}, {Lequeux}, {Maurice}, {Prevot}, \&
  {Rocca-Volmerange}}]{pre84}
{Prevot}, M.~L., {Lequeux}, J., {Maurice}, E., {Prevot}, L., \&
  {Rocca-Volmerange}, B. 1984, \aap, 132, 389

\bibitem[{{Rela{\~n}o} {et~al.}(2007){Rela{\~n}o}, {Lisenfeld},
  {P{\'e}rez-Gonz{\'a}lez}, {V{\'\i}lchez}, \& {Battaner}}]{rel07}
{Rela{\~n}o}, M., {Lisenfeld}, U., {P{\'e}rez-Gonz{\'a}lez}, P.~G.,
  {V{\'\i}lchez}, J.~M., \& {Battaner}, E. 2007, \apjl, 667, L141

\bibitem[{{Rieke} {et~al.}(2009){Rieke}, {Alonso-Herrero}, {Weiner},
  {P{\'e}rez-Gonz{\'a}lez}, {Blaylock}, {Donley}, \& {Marcillac}}]{rie09}
{Rieke}, G.~H., {Alonso-Herrero}, A., {Weiner}, B.~J., {et~al.} 2009, \apj,
  692, 556

\bibitem[{{Rosa-Gonz{\'a}lez} {et~al.}(2002){Rosa-Gonz{\'a}lez}, {Terlevich},
  \& {Terlevich}}]{ros02}
{Rosa-Gonz{\'a}lez}, D., {Terlevich}, E., \& {Terlevich}, R. 2002, \mnras, 332,
  283

\bibitem[{{Salim} {et~al.}(2018){Salim}, {Boquien}, \& {Lee}}]{sal18}
{Salim}, S., {Boquien}, M., \& {Lee}, J.~C. 2018, \apj, 859, 11

\bibitem[{{Salim} {et~al.}(2016){Salim}, {Lee}, {Janowiecki}, {da Cunha},
  {Dickinson}, {Boquien}, {Burgarella}, {Salzer}, \& {Charlot}}]{sal16}
{Salim}, S., {Lee}, J.~C., {Janowiecki}, S., {et~al.} 2016, \apjs, 227, 2

\bibitem[{{Salim} {et~al.}(2007){Salim}, {Rich}, {Charlot}, {Brinchmann},
  {Johnson}, {Schiminovich}, {Seibert}, {Mallery}, {Heckman}, {Forster},
  {Friedman}, {Martin}, {Morrissey}, {Neff}, {Small}, {Wyder}, {Bianchi},
  {Donas}, {Lee}, {Madore}, {Milliard}, {Szalay}, {Welsh}, \& {Yi}}]{sal07}
{Salim}, S., {Rich}, R.~M., {Charlot}, S., {et~al.} 2007, \apjs, 173, 267

\bibitem[{{Salpeter}(1955)}]{sal55}
{Salpeter}, E.~E. 1955, \apj, 121, 161

\bibitem[{{Schlegel} {et~al.}(1998){Schlegel}, {Finkbeiner}, \&
  {Davis}}]{sch98}
{Schlegel}, D.~J., {Finkbeiner}, D.~P., \& {Davis}, M. 1998, \apj, 500, 525

\bibitem[{{Schreiber} {et~al.}(2018){Schreiber}, {Elbaz}, {Pannella}, {Ciesla},
  {Wang}, \& {Franco}}]{sch18}
{Schreiber}, C., {Elbaz}, D., {Pannella}, M., {et~al.} 2018, \aap, 609, A30

\bibitem[{{Scodeggio} {et~al.}(2018){Scodeggio}, {Guzzo}, {Garilli}, {Granett},
  {Bolzonella}, {de la Torre}, {Abbas}, {Adami}, {Arnouts}, {Bottini}, {Cappi},
  {Coupon}, {Cucciati}, {Davidzon}, {Franzetti}, {Fritz}, {Iovino}, {Krywult},
  {Le Brun}, {Le F{\`e}vre}, {Maccagni}, {Ma{\l}ek}, {Marchetti}, {Marulli},
  {Polletta}, {Pollo}, {Tasca}, {Tojeiro}, {Vergani}, {Zanichelli}, {Bel},
  {Branchini}, {De Lucia}, {Ilbert}, {McCracken}, {Moutard}, {Peacock},
  {Zamorani}, {Burden}, {Fumana}, {Jullo}, {Marinoni}, {Mellier}, {Moscardini},
  \& {Percival}}]{sco18}
{Scodeggio}, M., {Guzzo}, L., {Garilli}, B., {et~al.} 2018, \aap, 609, A84

\bibitem[{{Searle} {et~al.}(1973){Searle}, {Sargent}, \& {Bagnuolo}}]{sea73}
{Searle}, L., {Sargent}, W.~L.~W., \& {Bagnuolo}, W.~G. 1973, \apj, 179, 427

\bibitem[{{Senarath} {et~al.}(2018){Senarath}, {Brown}, {Cluver}, {Moustakas},
  {Armus}, \& {Jarrett}}]{sen18}
{Senarath}, M.~R., {Brown}, M. J.~I., {Cluver}, M.~E., {et~al.} 2018, \apjl,
  869, L26

\bibitem[{{Shipley} {et~al.}(2016){Shipley}, {Papovich}, {Rieke}, {Brown}, \&
  {Moustakas}}]{shi16}
{Shipley}, H.~V., {Papovich}, C., {Rieke}, G.~H., {Brown}, M. J.~I., \&
  {Moustakas}, J. 2016, \apj, 818, 60

\bibitem[{{Shirley} {et~al.}(2021){Shirley}, {Duncan}, {Campos Varillas},
  {Hurley}, {Ma{\l}ek}, {Roehlly}, {Smith}, {Aussel}, {Bakx}, {Buat},
  {Burgarella}, {Christopher}, {Duivenvoorden}, {Eales}, {Efstathiou},
  {Gonz{\'a}lez Solares}, {Griffin}, {Jarvis}, {Faro}, {Marchetti}, {McCheyne},
  {Papadopoulos}, {Penner}, {Pons}, {Prescott}, {Rigby}, {Rottgering},
  {Saxena}, {Scudder}, {Vaccari}, {Wang}, \& {Oliver}}]{shi21}
{Shirley}, R., {Duncan}, K., {Campos Varillas}, M.~C., {et~al.} 2021, \mnras,
  507, 129

\bibitem[{{Shirley} {et~al.}(2019){Shirley}, {Roehlly}, {Hurley}, {Buat},
  {Campos Varillas}, {Duivenvoorden}, {Duncan}, {Efstathiou}, {Farrah},
  {Gonz{\'a}lez Solares}, {Malek}, {Marchetti}, {McCheyne}, {Papadopoulos},
  {Pons}, {Scipioni}, {Vaccari}, \& {Oliver}}]{shi19}
{Shirley}, R., {Roehlly}, Y., {Hurley}, P.~D., {et~al.} 2019, \mnras, 490, 634

\bibitem[{{Smol{\v{c}}i{\'c}} {et~al.}(2009){Smol{\v{c}}i{\'c}}, {Schinnerer},
  {Zamorani}, {Bell}, {Bondi}, {Carilli}, {Ciliegi}, {Mobasher}, {Paglione},
  {Scodeggio}, \& {Scoville}}]{smo09}
{Smol{\v{c}}i{\'c}}, V., {Schinnerer}, E., {Zamorani}, G., {et~al.} 2009, \apj,
  690, 610

\bibitem[{{Stalevski} {et~al.}(2012){Stalevski}, {Fritz}, {Baes}, {Nakos}, \&
  {Popovi{\'c}}}]{sta12}
{Stalevski}, M., {Fritz}, J., {Baes}, M., {Nakos}, T., \& {Popovi{\'c}},
  L.~{\v{C}}. 2012, \mnras, 420, 2756

\bibitem[{{Stalevski} {et~al.}(2016){Stalevski}, {Ricci}, {Ueda}, {Lira},
  {Fritz}, \& {Baes}}]{sta16}
{Stalevski}, M., {Ricci}, C., {Ueda}, Y., {et~al.} 2016, \mnras, 458, 2288

\bibitem[{{Steidel} {et~al.}(2014){Steidel}, {Rudie}, {Strom}, {Pettini},
  {Reddy}, {Shapley}, {Trainor}, {Erb}, {Turner}, {Konidaris}, {Kulas}, {Mace},
  {Matthews}, \& {McLean}}]{ste14}
{Steidel}, C.~C., {Rudie}, G.~C., {Strom}, A.~L., {et~al.} 2014, \apj, 795, 165

\bibitem[{{Straughn} {et~al.}(2009){Straughn}, {Pirzkal}, {Meurer}, {Cohen},
  {Windhorst}, {Malhotra}, {Rhoads}, {Gardner}, {Hathi}, {Jansen}, {Grogin},
  {Panagia}, {di Serego Alighieri}, {Gronwall}, {Walsh}, {Pasquali}, \&
  {Xu}}]{str09}
{Straughn}, A.~N., {Pirzkal}, N., {Meurer}, G.~R., {et~al.} 2009, \aj, 138,
  1022

\bibitem[{{Suzuki} {et~al.}(2016){Suzuki}, {Kodama}, {Sobral}, {Khostovan},
  {Hayashi}, {Shimakawa}, {Koyama}, {Tadaki}, {Tanaka}, {Minowa}, {Yamamoto},
  {Smail}, \& {Best}}]{suz16}
{Suzuki}, T.~L., {Kodama}, T., {Sobral}, D., {et~al.} 2016, \mnras, 462, 181

\bibitem[{{Talia} {et~al.}(2015){Talia}, {Cimatti}, {Pozzetti}, {Rodighiero},
  {Gruppioni}, {Pozzi}, {Daddi}, {Maraston}, {Mignoli}, \& {Kurk}}]{tal15}
{Talia}, M., {Cimatti}, A., {Pozzetti}, L., {et~al.} 2015, \aap, 582, A80

\bibitem[{{Teplitz} {et~al.}(2000){Teplitz}, {Malkan}, {Steidel}, {McLean},
  {Becklin}, {Figer}, {Gilbert}, {Graham}, {Larkin}, {Levenson}, \&
  {Wilcox}}]{tep00}
{Teplitz}, H.~I., {Malkan}, M.~A., {Steidel}, C.~C., {et~al.} 2000, \apj, 542,
  18

\bibitem[{{Tremonti} {et~al.}(2004){Tremonti}, {Heckman}, {Kauffmann},
  {Brinchmann}, {Charlot}, {White}, {Seibert}, {Peng}, {Schlegel}, {Uomoto},
  {Fukugita}, \& {Brinkmann}}]{tre04}
{Tremonti}, C.~A., {Heckman}, T.~M., {Kauffmann}, G., {et~al.} 2004, \apj, 613,
  898

\bibitem[{{Tresse} {et~al.}(2002){Tresse}, {Maddox}, {Le F{\`e}vre}, \&
  {Cuby}}]{tre02}
{Tresse}, L., {Maddox}, S.~J., {Le F{\`e}vre}, O., \& {Cuby}, J.~G. 2002,
  \mnras, 337, 369

\bibitem[{{Viaene} {et~al.}(2016){Viaene}, {Baes}, {Bendo}, {Boquien},
  {Boselli}, {Ciesla}, {Cortese}, {De Looze}, {Eales}, {Fritz}, {Karczewski},
  {Madden}, {Smith}, \& {Spinoglio}}]{via16}
{Viaene}, S., {Baes}, M., {Bendo}, G., {et~al.} 2016, \aap, 586, A13

\bibitem[{{Villa-V{\'e}lez} {et~al.}(2021){Villa-V{\'e}lez}, {Buat},
  {Theul{\'e}}, {Boquien}, \& {Burgarella}}]{vil21}
{Villa-V{\'e}lez}, J.~A., {Buat}, V., {Theul{\'e}}, P., {Boquien}, M., \&
  {Burgarella}, D. 2021, \aap, 654, A153

\bibitem[{{Wild} {et~al.}(2011){Wild}, {Charlot}, {Brinchmann}, {Heckman},
  {Vince}, {Pacifici}, \& {Chevallard}}]{wil11}
{Wild}, V., {Charlot}, S., {Brinchmann}, J., {et~al.} 2011, \mnras, 417, 1760

\bibitem[{{Wilkins} {et~al.}(2012){Wilkins}, {Gonzalez-Perez}, {Lacey}, \&
  {Baugh}}]{wil12}
{Wilkins}, S.~M., {Gonzalez-Perez}, V., {Lacey}, C.~G., \& {Baugh}, C.~M. 2012,
  \mnras, 424, 1522

\bibitem[{{Wright} {et~al.}(2010){Wright}, {Eisenhardt}, {Mainzer}, {Ressler},
  {Cutri}, {Jarrett}, {Kirkpatrick}, {Padgett}, {McMillan}, {Skrutskie},
  {Stanford}, {Cohen}, {Walker}, {Mather}, {Leisawitz}, {Gautier}, {McLean},
  {Benford}, {Lonsdale}, {Blain}, {Mendez}, {Irace}, {Duval}, {Liu}, {Royer},
  {Heinrichsen}, {Howard}, {Shannon}, {Kendall}, {Walsh}, {Larsen}, {Cardon},
  {Schick}, {Schwalm}, {Abid}, {Fabinsky}, {Naes}, \& {Tsai}}]{wri10}
{Wright}, E.~L., {Eisenhardt}, P.~R.~M., {Mainzer}, A.~K., {et~al.} 2010, \aj,
  140, 1868

\bibitem[{{Wu} {et~al.}(2005){Wu}, {Cao}, {Hao}, {Liu}, {Wang}, {Xia}, {Deng},
  \& {Young}}]{wu05}
{Wu}, H., {Cao}, C., {Hao}, C.-N., {et~al.} 2005, \apjl, 632, L79

\bibitem[{{Wuyts} {et~al.}(2008){Wuyts}, {Labb{\'e}}, {F{\"o}rster Schreiber},
  {Franx}, {Rudnick}, {Brammer}, \& {van Dokkum}}]{wuy08}
{Wuyts}, S., {Labb{\'e}}, I., {F{\"o}rster Schreiber}, N.~M., {et~al.} 2008,
  \apj, 682, 985

\bibitem[{{York} {et~al.}(2000){York}, {Adelman}, {Anderson}, {Anderson},
  {Annis}, {Bahcall}, {Bakken}, {Barkhouser}, {Bastian}, {Berman}, {Boroski},
  {Bracker}, {Briegel}, {Briggs}, {Brinkmann}, {Brunner}, {Burles}, {Carey},
  {Carr}, {Castander}, {Chen}, {Colestock}, {Connolly}, {Crocker}, {Csabai},
  {Czarapata}, {Davis}, {Doi}, {Dombeck}, {Eisenstein}, {Ellman}, {Elms},
  {Evans}, {Fan}, {Federwitz}, {Fiscelli}, {Friedman}, {Frieman}, {Fukugita},
  {Gillespie}, {Gunn}, {Gurbani}, {de Haas}, {Haldeman}, {Harris}, {Hayes},
  {Heckman}, {Hennessy}, {Hindsley}, {Holm}, {Holmgren}, {Huang}, {Hull},
  {Husby}, {Ichikawa}, {Ichikawa}, {Ivezi{\'c}}, {Kent}, {Kim}, {Kinney},
  {Klaene}, {Kleinman}, {Kleinman}, {Knapp}, {Korienek}, {Kron}, {Kunszt},
  {Lamb}, {Lee}, {Leger}, {Limmongkol}, {Lindenmeyer}, {Long}, {Loomis},
  {Loveday}, {Lucinio}, {Lupton}, {MacKinnon}, {Mannery}, {Mantsch}, {Margon},
  {McGehee}, {McKay}, {Meiksin}, {Merelli}, {Monet}, {Munn}, {Narayanan},
  {Nash}, {Neilsen}, {Neswold}, {Newberg}, {Nichol}, {Nicinski}, {Nonino},
  {Okada}, {Okamura}, {Ostriker}, {Owen}, {Pauls}, {Peoples}, {Peterson},
  {Petravick}, {Pier}, {Pope}, {Pordes}, {Prosapio}, {Rechenmacher}, {Quinn},
  {Richards}, {Richmond}, {Rivetta}, {Rockosi}, {Ruthmansdorfer}, {Sandford},
  {Schlegel}, {Schneider}, {Sekiguchi}, {Sergey}, {Shimasaku}, {Siegmund},
  {Smee}, {Smith}, {Snedden}, {Stone}, {Stoughton}, {Strauss}, {Stubbs},
  {SubbaRao}, {Szalay}, {Szapudi}, {Szokoly}, {Thakar}, {Tremonti}, {Tucker},
  {Uomoto}, {Vanden Berk}, {Vogeley}, {Waddell}, {Wang}, {Watanabe},
  {Weinberg}, {Yanny}, {Yasuda}, \& {SDSS Collaboration}}]{yor00}
{York}, D.~G., {Adelman}, J., {Anderson}, Jr., J.~E., {et~al.} 2000, \aj, 120,
  1579

\bibitem[{{Young} {et~al.}(2014){Young}, {Gronwall}, {Salzer}, \&
  {Rosenberg}}]{you14}
{Young}, J.~E., {Gronwall}, C., {Salzer}, J.~J., \& {Rosenberg}, J.~L. 2014,
  \mnras, 443, 2711

\bibitem[{{Yuan} {et~al.}(2013){Yuan}, {Liu}, \& {Xiang}}]{yua13}
{Yuan}, H.~B., {Liu}, X.~W., \& {Xiang}, M.~S. 2013, \mnras, 430, 2188

\bibitem[{{Zaritsky} {et~al.}(1994){Zaritsky}, {Kennicutt}, \&
  {Huchra}}]{zar94}
{Zaritsky}, D., {Kennicutt}, Robert~C., J., \& {Huchra}, J.~P. 1994, \apj, 420,
  87

\bibitem[{{Zhou} {et~al.}(2017){Zhou}, {Zhou}, {Wu}, {Fan}, {Fan}, {Jiang},
  {Jing}, {Li}, {Lesser}, {Jiang}, {Ma}, {Nie}, {Shen}, {Wang}, {Wu}, {Zhang},
  \& {Zou}}]{zho17}
{Zhou}, Z., {Zhou}, X., {Wu}, H., {et~al.} 2017, \apj, 835, 70

\bibitem[{{Zhu} {et~al.}(2008){Zhu}, {Wu}, {Cao}, \& {Li}}]{zhu08}
{Zhu}, Y.-N., {Wu}, H., {Cao}, C., \& {Li}, H.-N. 2008, \apj, 686, 155

\bibitem[{{Zhuang} \& {Ho}(2019)}]{zhu19}
{Zhuang}, M.-Y. \& {Ho}, L.~C. 2019, \apj, 882, 89

\end{thebibliography}
   
%
\clearpage
\onecolumn

\begin{appendix} 

\section{CIGALE results}

\subsection{Mock analysis}\label{Appendix:mocks}

\begin{figure*}[h!]
\centering
\includegraphics[angle=0,width=0.33\textwidth]{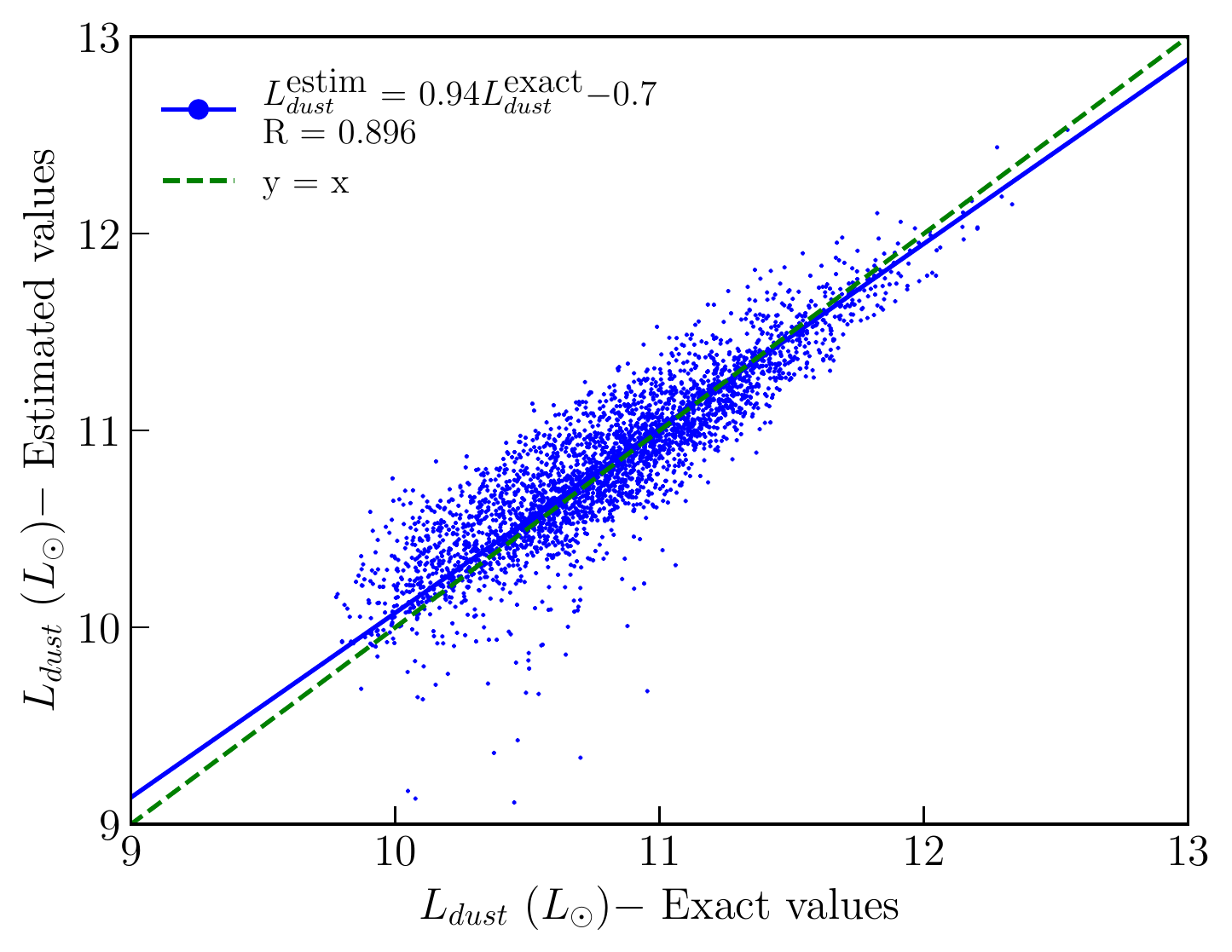}\includegraphics[angle=0,width=0.33\textwidth]{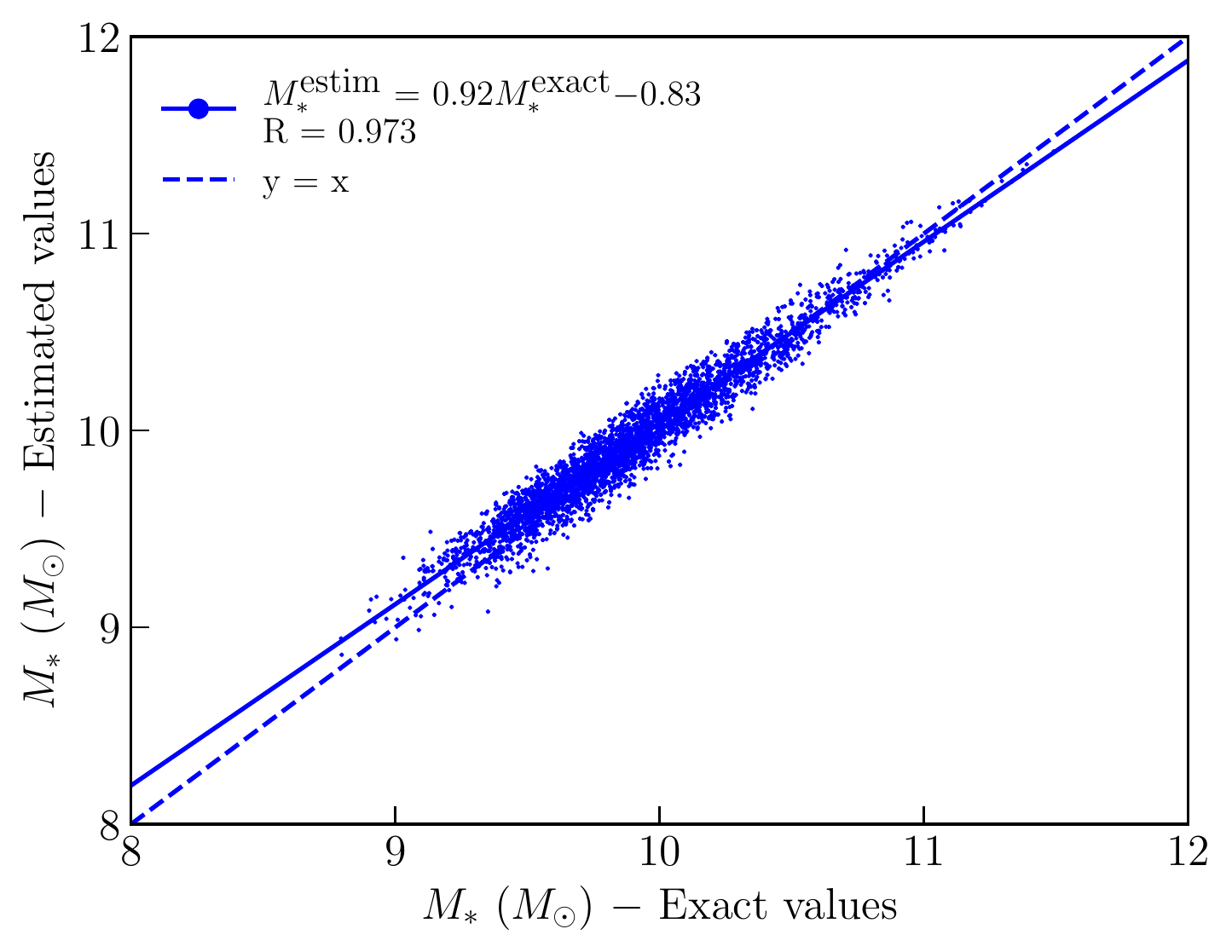}\includegraphics[angle=0,width=0.33\textwidth]{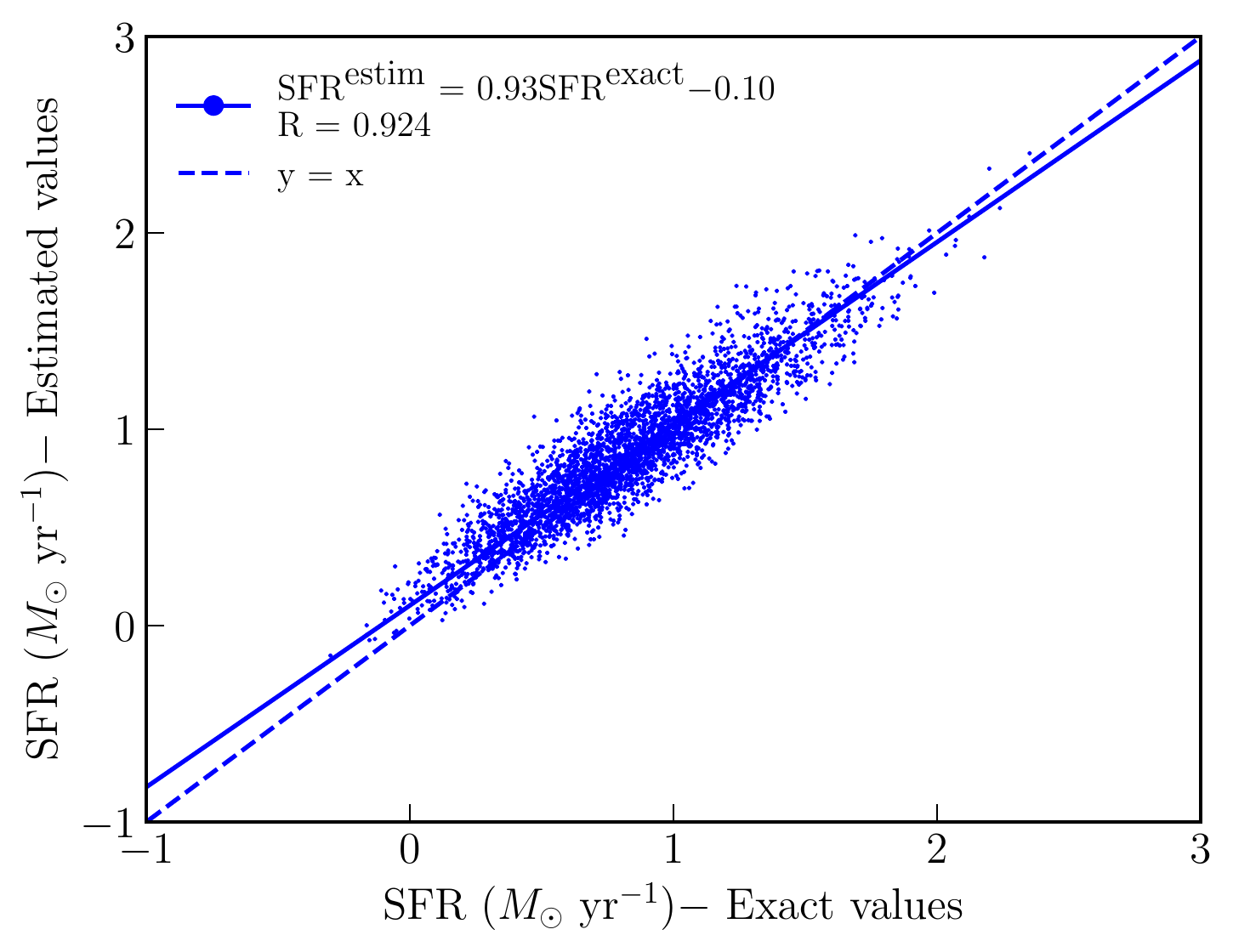}
\includegraphics[angle=0,width=0.33\textwidth]{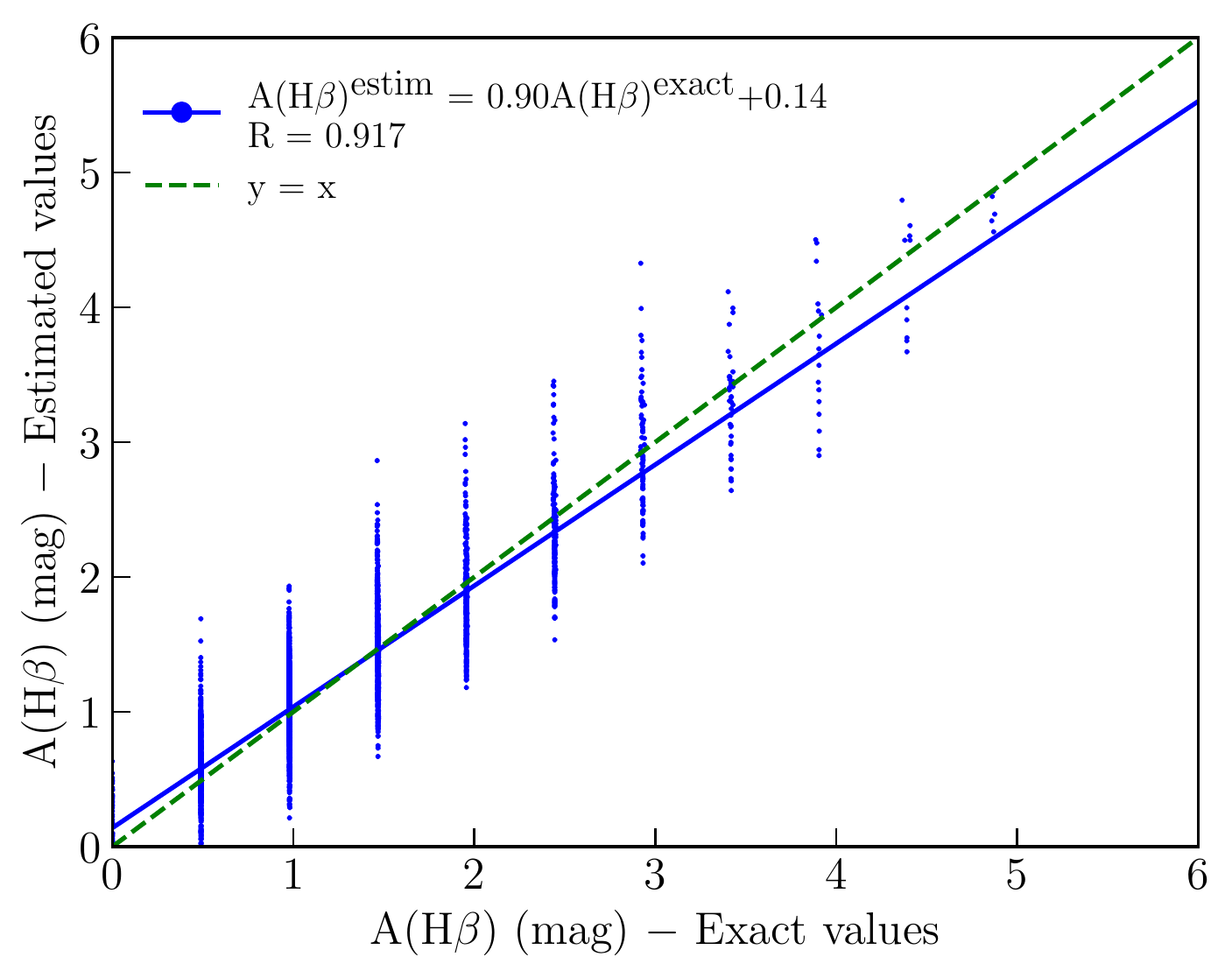}\includegraphics[angle=0,width=0.33\textwidth]{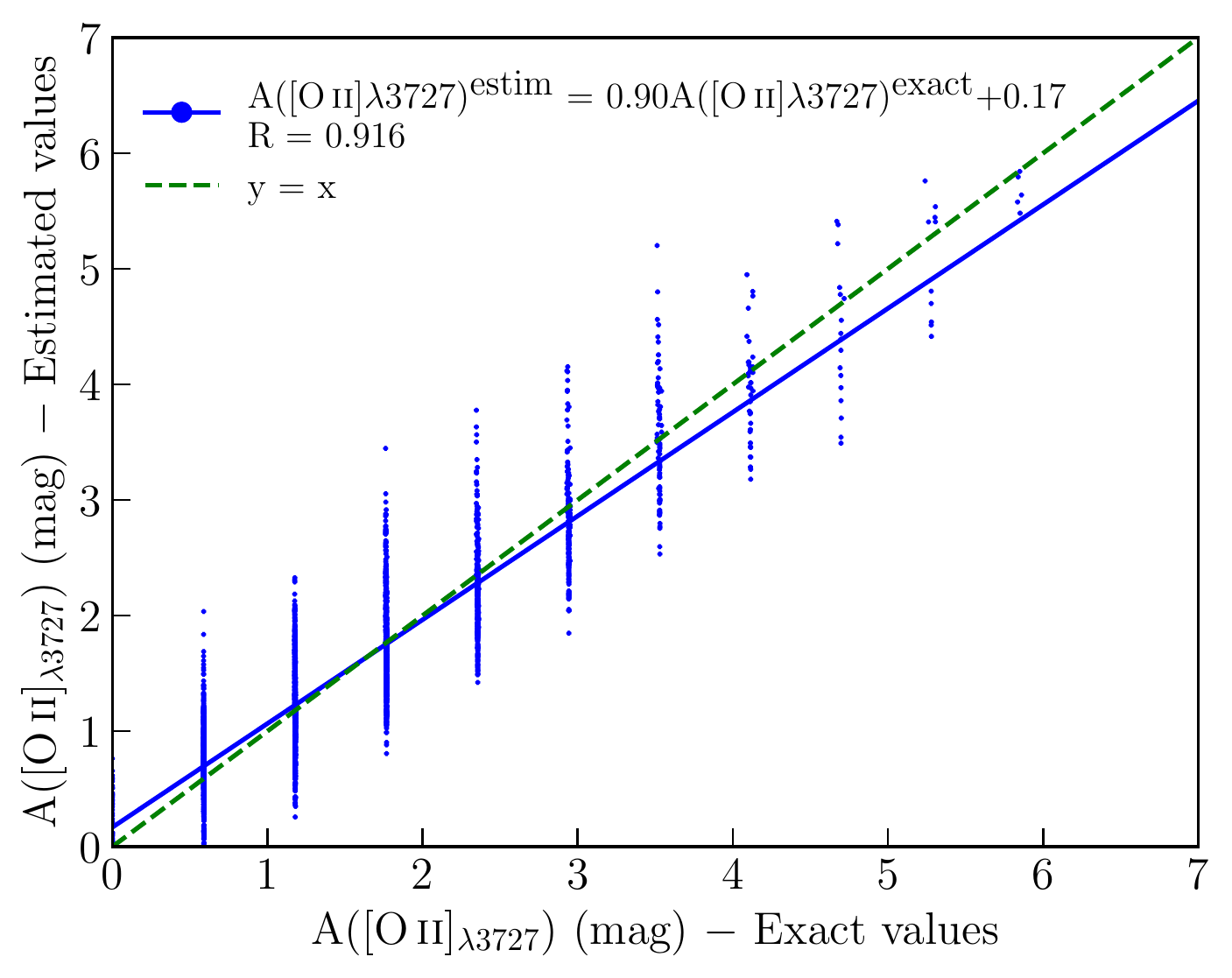}\includegraphics[angle=0,width=0.33\textwidth]{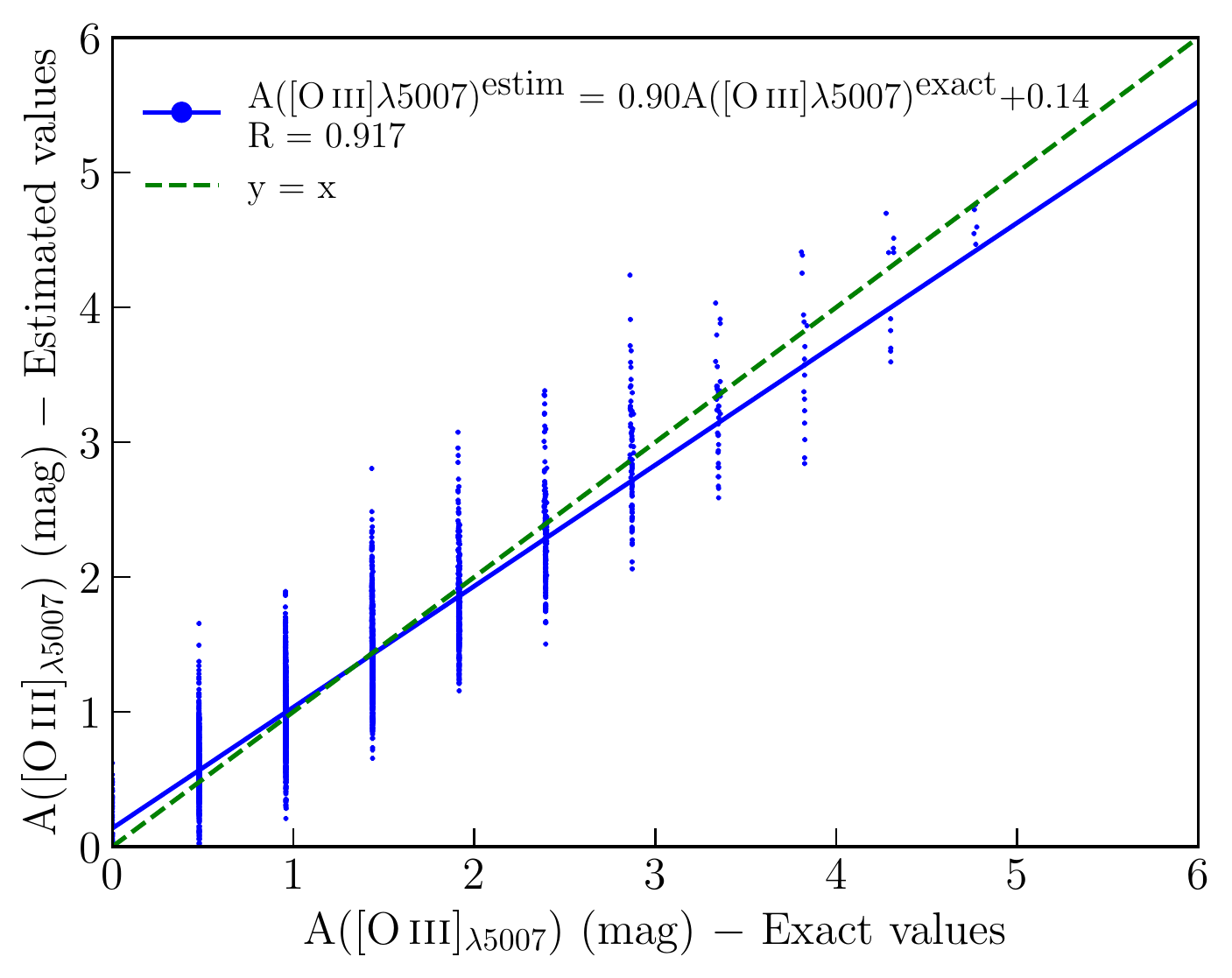}
\caption{Mock analysis of the CIGALE run: parameters estimated by CIGALE versus exact parameters (top left: $L_{dust}$, top middle: $M_{*}$, top right: SFR, bottom left: H$\beta$ attenuation, bottom middle: [O{\,\sc{ii}}]$\lambda$3727 attenuation, bottom right: [O{\,\sc{iii}}]$\lambda$5007 attenuation. The blue lines and green lines represent the fit to the data and the one-to-one relation, respectively. The equation of the fit and the Pearson correlation coefficient R are indicated in each plot.} 
\label{Fig:Mock_CIGALE}
\end{figure*}
\subsection{Examples of SEDs}
\begin{figure*}[h!]
\centering
\includegraphics[angle=0,width=0.5\textwidth]{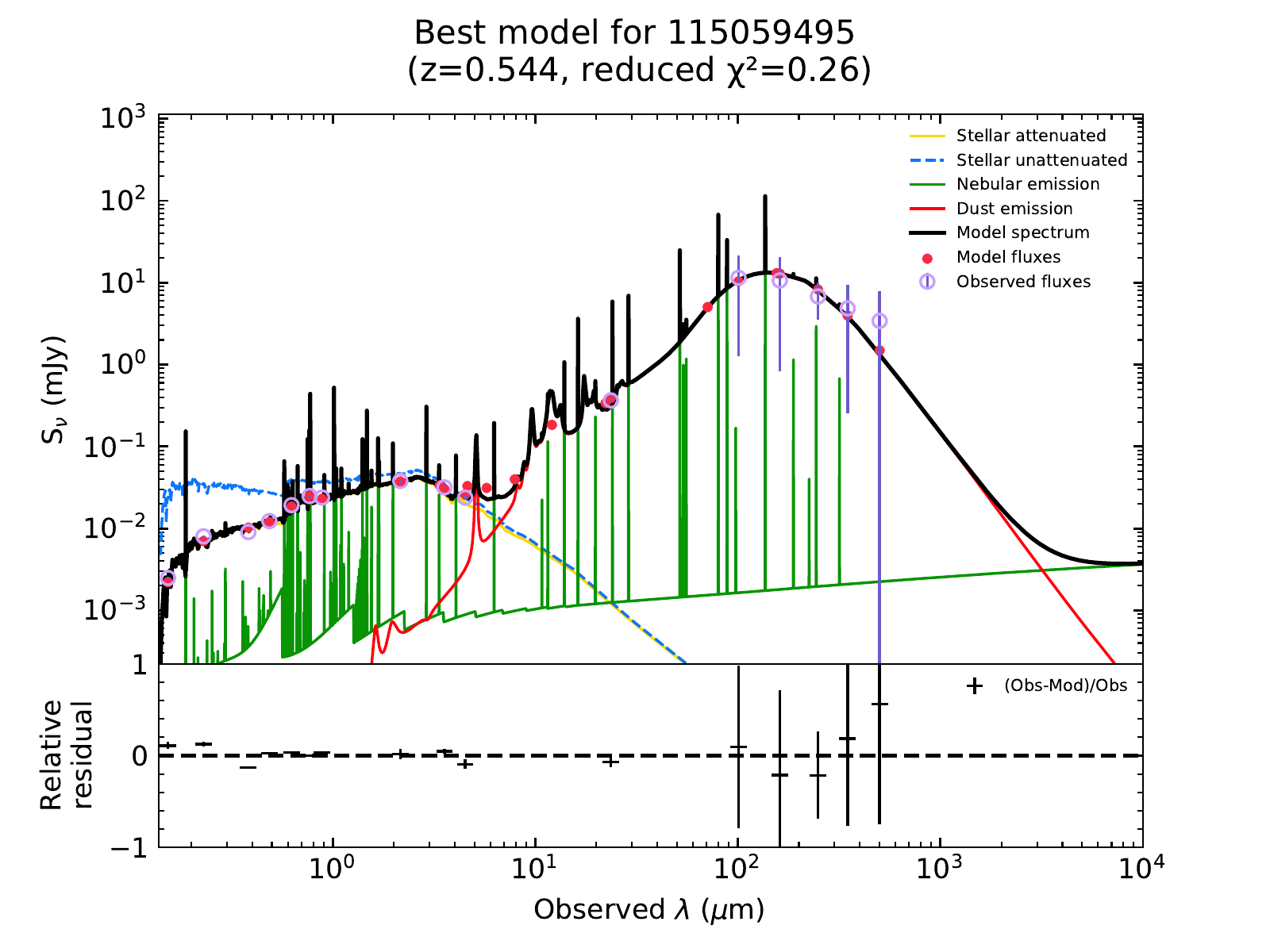}\includegraphics[angle=0,width=0.5\textwidth]{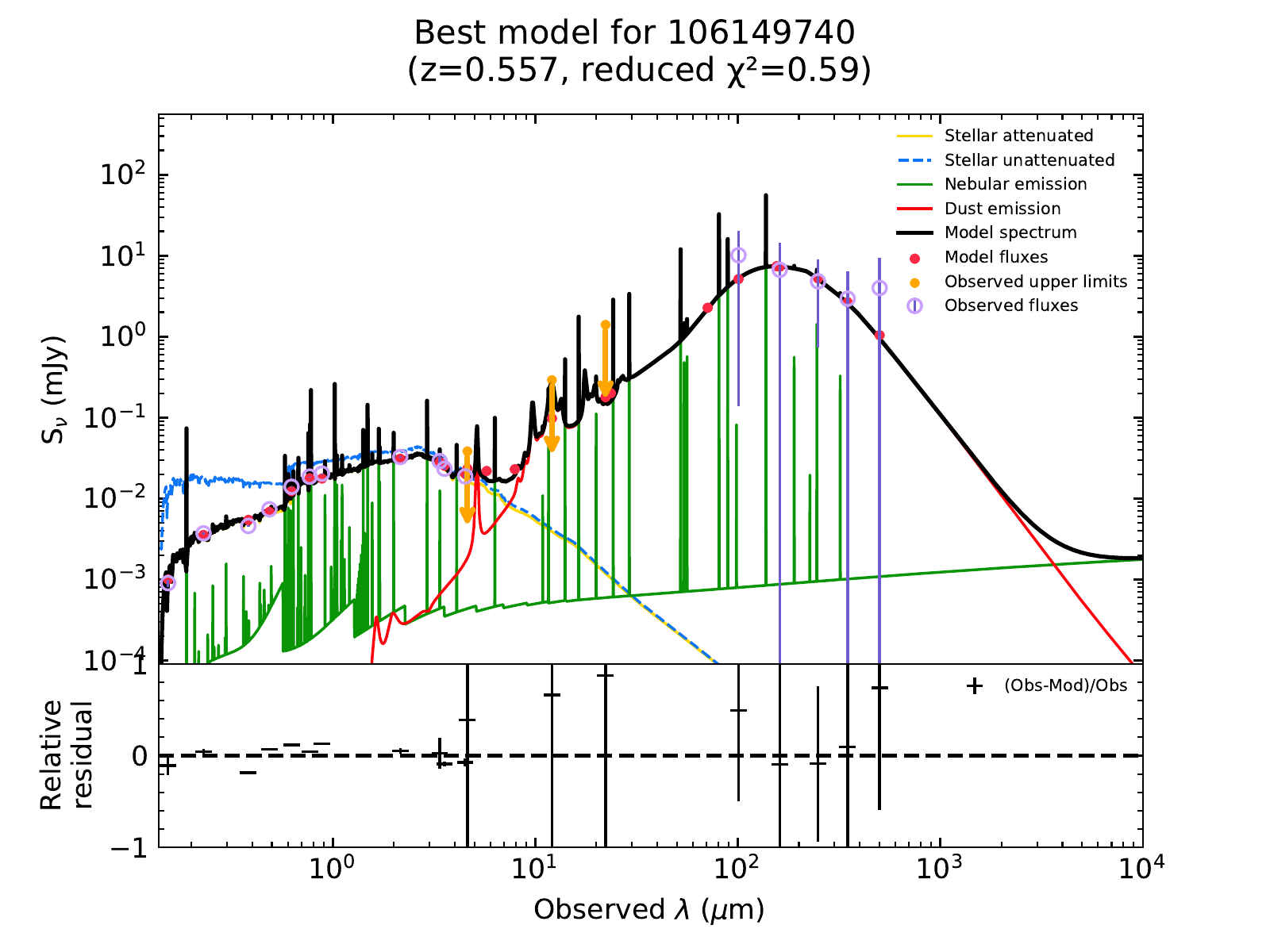}
\caption{Best SEDs for the VIPERS 115059495 and VIPERS 106149740 galaxies reconstructed with CIGALE. VIPERS 106149740 is an example of a galaxy for which the WISE-2, WISE-3, and WISE-4 measurements are upper limits (orange arrows).}
\label{Fig:VIPERS_SEDs}
\end{figure*}
\newpage
\section{List of the calibrations}\label{Appendix:Laws}

\subsection{UV bands}\label{Appendix:Laws_UV}

\textbf{\citet{bro17} $-$ \textit{GALEX}-SDSS DR3 (66 nearby galaxies at $D<10$~Mpc)  $-$ Kroupa IMF}
\begin{equation}\label{Eq:SFR_Brown}
{\rm{log(SFR)}}(M_{\odot}~{\rm{yr^{-1}}})=\frac{f-B}{A}-1.26
\end{equation}

\begin{align}\label{Eq:SFR_BrownFUVCal}
&f={\rm{log}}(L_{FUV})+2(M_{FUV}-M_{NUV})\mbox{ with }A=0.9\mbox{ and }B=42.25 \mbox{\hspace{2.5cm}[\citet{cal00} attenuation law]}\\
&f={\rm{log}}(L_{FUV})+1.532(M_{FUV}-M_{NUV})-0.033\mbox{ with }A=0.96\mbox{ and }B=42.42 \mbox{\hspace{0.535cm}[\citet{hao11} attenuation law]}
\end{align}

\noindent\textbf{\citet{dav16} $-$ GAMA II survey (3749 galaxies at $z<0.13$) $-$ Chabrier IMF}
\begin{align}\label{Eq:SFR_Davies_NUV}
&{\rm{log}(SFR_{NUV})}~(M_{\odot}~{\rm{yr^{-1}}})=0.62\times ({\rm{log}[}L_{NUV}({\rm{W~Hz}^{-1}})]-21.5)+0.014\\
&{\rm{log}(SFR_{FUV})}~(M_{\odot}~{\rm{yr^{-1}}})=0.75\times ({\rm{log}[}L_{FUV}({\rm{W~Hz}^{-1}})]-21.5)+0.17
\end{align}

\noindent \textbf{\citet{sal07} $-$ \textit{GALEX}-SDSS DR3 (48~295 galaxies at $0.005<z<0.22$) $-$ Chabrier IMF}
\begin{align}\label{Eq:SFR_Salim_NUV}
&{\rm{log}}(SFR_{NUV})~(M_{\odot}~{\rm{yr^{-1}}})={\rm{log}}[L_{NUV}({\rm{W~Hz}^{-1}})]-21.14\\
&{\rm{log}}(SFR_{FUV})~(M_{\odot}~{\rm{yr^{-1}}})={\rm{log}}[L_{NUV}({\rm{W~Hz}^{-1}})]-21.16
\end{align}

\noindent \textbf{\citet{ros02} $-$ 31 nearby star-forming galaxies $-$ Salpeter IMF $-$ Attenuation correction included}
\begin{equation}\label{Eq:Rosa_gonz_UV}
{\rm{log(SFR}_{UV})}~(M_{\odot}~{\rm{yr^{-1}}})={\rm{log[}}L_{UV}{({\rm{W~Hz}^{-1}})}]-20.19\\
\end{equation}

\subsection{u band}\label{Appendix:Laws_u}

\noindent\textbf{\citet{dav16} $-$ GAMA II survey (3749 galaxies at $z<0.13$) $-$ Chabrier IMF}
\begin{equation}\label{Eq:SFR_Davies_u}
{\rm{log}(SFR_{u})}~(M_{\odot}~{\rm{yr^{-1}}})=0.92\times ({\rm{log}[}L_{u}({\rm{W~Hz}^{-1}})]-21.25)-0.079
\end{equation}

\noindent \textbf{\citet{mou06b} $-$ SDSS (120~846 galaxies at $z\sim0.1$) $-$ Salpeter IMF $-$ Attenuation correction included}
\begin{equation}\label{Eq:Mou06}
{\rm{log(SFR_u)}}~(M_{\odot}~{\rm{yr^{-1}}})={\rm{log[}}L_{u}{\rm{(erg~s^{-1})]}}]-42.85\\
\end{equation}

\noindent \textbf{\citet{hop03} $-$ SDSS (2625 galaxies at $z<0.36$) $-$ Salpeter IMF }
\begin{equation}\label{Eq:Hop03}
{\rm{log(SFR_u)}}~(M_{\odot}~{\rm{yr^{-1}}})=1.186\times{\rm{log[}}L_{u}{({\rm{W~Hz}^{-1}})}]-21.25\\
\end{equation}

\noindent \textbf{\citet{zho17} $-$ South Galactic Cap \textit{u}-band Sky Survey (9596 galaxies at $z<0.4$) $-$ Salpeter IMF }
\begin{equation}\label{Eq:zho17}
{\rm{log(SFR_u)}}~(M_{\odot}~{\rm{yr^{-1}}})=1.25\times{\rm{log[}}L_{u}{({\rm{W~Hz}^{-1}})}]-54.17\\
\end{equation}

\subsection{Infrared bands}\label{Appendix:Laws_IR}

\textbf{\citet{bro17} $-$ Kroupa IMF}
\begin{align}\label{Eq:SFR_Brown8}
&f={\rm{log}}(L_{8{\mu m}})\mbox{ with }A=1.3\mbox{ and }B=40.88\\
&f={\rm{log}}(L_{24{\mu m}})\mbox{ with }A=1.3\mbox{ and }B=40.93
\end{align}

\noindent \textbf{\citet{wu05} $-$ \textit{Spitzer} First Look Survey (91 galaxies + lines from SDSS) $-$ Salpeter IMF}
\begin{equation}\label{Eq:Wu05_8}
{\rm{log(SFR_{8{\mu m}})}}~(M_{\odot}~{\rm{yr^{-1}}})={\rm{log[}}L_{8\mu m}{(L_{\odot})}]-9.14\\
\end{equation}

\noindent \textbf{\citet{you14} $-$ NOAO Deep Wide Field Survey (91 galaxies, $z<0.1$) $-$ Salpeter IMF}
\begin{equation}\label{Eq:You14_8}
{\rm{log(SFR_{8{\mu m}})}}~(M_{\odot}~{\rm{yr^{-1}}})=0.93\times{\rm{log[}}L_{8\mu m}{(L_{\odot})}]-8.5\\
\end{equation}

\noindent \textbf{\citet{per06} $-$ Galaxy M81 $-$ Salpeter IMF}
\begin{equation}\label{Eq:Per06_8}
{\rm{log(SFR_{8{\mu m}})}}~(M_{\odot}~{\rm{yr^{-1}}})=0.87\times{\rm{log[}}L_{8\mu m}{(L_{\odot})}]-7.9\\
\end{equation}

\noindent \textbf{\citet{rie09} $-$ SED templates based on local galaxies $-$ Kroupa IMF}
\begin{align}\label{Eq:SFR_rie09}
&{\rm{log(SFR_{24\mu m})}}~(M_{\odot}~{\rm{yr^{-1}}})={\rm{log[}}L_{24\mu m}{{\rm{(erg~s^{-1})]}}}-42.69\mbox{ if }L_{24\mu m}>4\times10^{42}\mbox{ erg s}^{-1}\\
&{\rm{log(SFR_{24\mu m})}}~(M_{\odot}~{\rm{yr^{-1}}})=1.05\times{\rm{log[}}L_{24\mu m}{{\rm{(erg~s^{-1})]}}}-44.79\mbox{ if }L_{24\mu m}>5\times10^{43}\mbox{ erg s}^{-1}
\end{align}

\noindent \textbf{\citet{zhu08} $-$ SDSS-IRAC/MIPS24$\mu$m (413 galaxies) $-$ Kroupa IMF}
\begin{equation}\label{Eq:zhu08_24}
{\rm{log(SFR_{24\mu m})}}~(M_{\odot}~{\rm{yr^{-1}}})=0.848\times{\rm{log[}}L_{24\mu m}{{\rm{(erg~s^{-1})]}}}-36.09\\
\end{equation}

\noindent \textbf{\citet{rel07} $-$ \textit{Spitzer}-HST (41 nearby galaxies) $-$ Kroupa IMF}
\begin{equation}\label{Eq:rel07_24}
{\rm{log(SFR_{24\mu m})}}~(M_{\odot}~{\rm{yr^{-1}}})=0.826\times{\rm{log[}}L_{24\mu m}{{\rm{(erg~s^{-1})]}}}-35.25\\
\end{equation}

\subsection{$L_{TIR}$}\label{Appendix:Laws_Ltir}

\noindent \textbf{\citet{ken98a} $-$ Salpeter IMF}
\begin{equation}\label{Eq:Alo_24}
{\rm{log(SFR_{TIR})}}~(M_{\odot}~\rm{yr^{-1}})={\rm{log[}}L_{TIR}{\rm{(erg~s^{-1})]}}-43.35
\end{equation}

\subsection{Composite tracers}\label{Appendix:Laws_Composites}

\textbf{\citet{bel05} $-$ PEGASE calibration $-$ Kroupa IMF}
\begin{equation}
{\rm{log(SFR_{TIR+2800\AA})}}~(M_{\odot}~{\rm{yr^{-1}}})={\rm{log[(}}L_{TIR}+3.3L_{2800\AA}{\rm{)(}}L_{\odot})]-10.0
\end{equation}\label{Eq:Bel_TIR_NUV}
\hspace*{-0.1cm}\textbf{\citet{ken09} $-$ SINGS and \citet{mou06a} survey (171 galaxies at $D<154$~Mpc) $-$ Kroupa IMF}
\begin{align}
&{\rm{log(SFR_{\textrm{[O\,\sc{ii}]}+\lambda}})}~(M_{\odot}~{\rm{yr^{-1}}}) = 42.75+\textrm{log[(}L_{\textrm{[O\,\sc{ii}}]}+a_{\lambda}L_{\lambda})({\rm{erg~s^{-1}}})]\\
&\rm{where }~a_{\lambda}=0.016-0.029-0.0036~for~\lambda=~8~\mu m,~24~\mu m~and~L_{TIR},~\rm{respectively}.
\end{align}
\noindent \textbf{\citet{cla15} $-$ H ATLAS (Phase-I Version-3)-SDSS-GAMA (42 galaxies at $D<46$~Mpc) $-$ Kroupa (UV) / Chabrier (IR) IMF}
\begin{equation}
{\rm{log(SFR_{FUV+W4})}}~(M_{\odot}~{\rm{yr^{-1}}}) = \textrm{log[(}10^{-9.69}L_{FUV} + (1-\eta)10^{-9.125}L_{W4})(L_{\odot})\textrm{]}
\end{equation}
\noindent \textbf{\citet{arn13} $-$ Deep 24~$\mu$m-COSMOS-SWIRE ($\sim$17~600 galaxies at $0.2<z<1.3$) $-$ Chabrier IMF}
\begin{align}\label{Eq:clark2015}
&NRK=0.31(NUV-r)+0.958(r-K_{\rm{s}})\\
&\textrm{log}(IRX)=-0.69+3.43z-3.49z^2+1.22z^3+0.63\times NRK\\
&{\rm{log(SFR_{NrK}}})~(M_{\odot}~{\rm{yr^{-1}}})=\textrm{log[}(L_{TIR}+2.3L_{NUV})(L_{\odot})\textrm{]}-10.07
\end{align}\label{Eq:Arn2013a}

\noindent \textbf{\citet{boq14} $-$ For $Z=0.02$ and $\tau=100$~Myr $-$ Chabrier IMF}
\begin{equation}
{\rm{log(SFR_{FUV+24\mu m})}}~(M_{\odot}~{\rm{yr^{-1}}}) = -21.09+\textrm{log[(}L_{FUV} + 3.89L_{24\mu m}\textrm{)(W~Hz}^{-1}\textrm{)]}
\end{equation}

\subsection{Lines}\label{Appendix:Laws_Lines}

\textbf{\citet{ken98a} $-$ Salpeter IMF}
\begin{align}\label{Eq:Kennicutt_Ha}
&{\rm{log(SFR_{H\alpha})}}~(M_{\odot}~{\rm{yr^{-1}}})={\rm{log[}}L_{\rm{H\alpha}}{\rm{(erg~s^{-1})]}}-41.10\\
&{\rm{log(SFR_{\textrm{[O\,\sc{ii}}]})}}~(M_{\odot}~{\rm{yr^{-1}}})={\rm{log[}}L_{\textrm{[O\,\sc{ii}}]}{\rm{(erg~s^{-1})]}}-41.20
\end{align}

\noindent \textbf{\citet{vil21} $-$ 182 galaxies of COSMOS ($z\sim1.6$) $-$ Chabrier IMF}
\begin{align}
&{\rm{log(SFR_{\textrm{[O\,\sc{iii}}]})}}~(M_{\odot}~{\rm{yr^{-1}}})={\rm{log[}}L_{\textrm{[O\,\sc{iii}}]}{\rm{(erg~s^{-1})]}}-A\\
&\rm{where }~A=41.72,~41.35,~41.24,~41.01~and~40.76~for~log(U)\sim -1.25,~-1.75,~-2.25,~-2.75~and~-3.25,~\rm{respectively}.
\end{align}

\noindent \textbf{\citet{ros02} $-$ Salpeter IMF $-$ Attenuation correction included}
\begin{equation}\label{Eq:Rosa_gonz_OII}
{\rm{log(SFR_{\textrm{[O\,\sc{ii}}]})}}~(M_{\odot}~{\rm{yr^{-1}}})={\rm{log[}}L_{\textrm{[O\,\sc{ii}}]}{\rm{(erg~s^{-1})]}}-40.08\\
\end{equation}

\noindent \textbf{\citet{gil10} $-$ SDSS (43~155 galaxies at $0.032<z<0.2$) $-$ Salpeter IMF $-$ Attenuation correction included ($A_{{\textrm{H}}\alpha}=1$~mag)}
\begin{align}\label{Eq:gil_OII}
&{\rm{log(SFR_{\textrm{[O\,\sc{ii}}]})}}~(M_{\odot}~{\rm{yr^{-1}}})={\rm{log[}}L_{\textrm{[O\,\sc{ii}}]}{\rm{(erg~s^{-1})]}}-{\rm{log[}}a\times {\rm{tanh}}((x-b)/c)+d{\rm{]}}-40.40\\
&{\rm{where }}~x={\textrm{log}}(M_{*}/M_{\odot}),~a=-1.424,~b=9.827,~c=0.572,~d=1.7
\end{align}

\clearpage

\section{SFRs from recipes versus SFRs from CIGALE}\label{Appendix:Plots}

\begin{figure*}[h!]
\centering
\includegraphics[angle=0,width=0.43\textwidth]{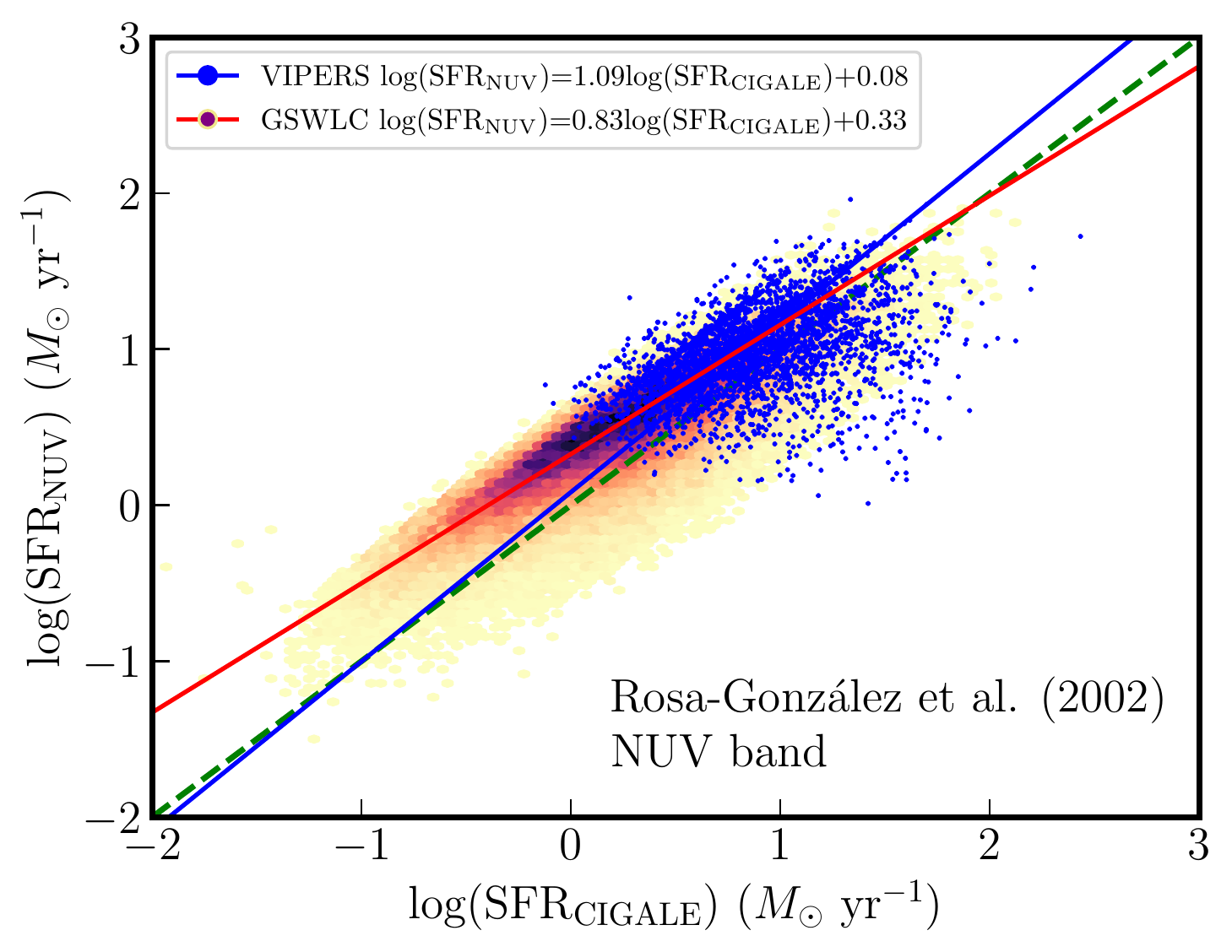}\includegraphics[angle=0,width=0.43\textwidth]{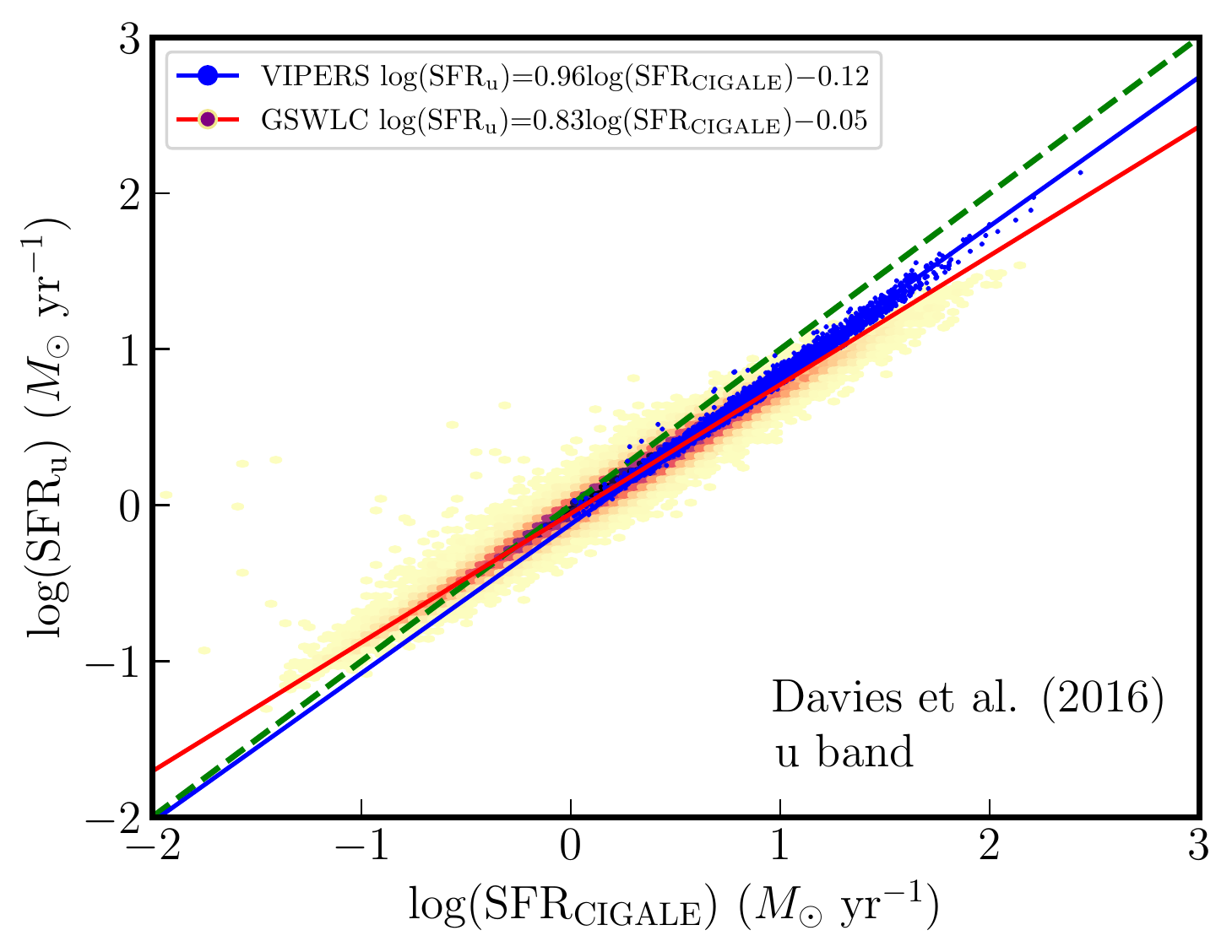}
\includegraphics[angle=0,width=0.43\textwidth]{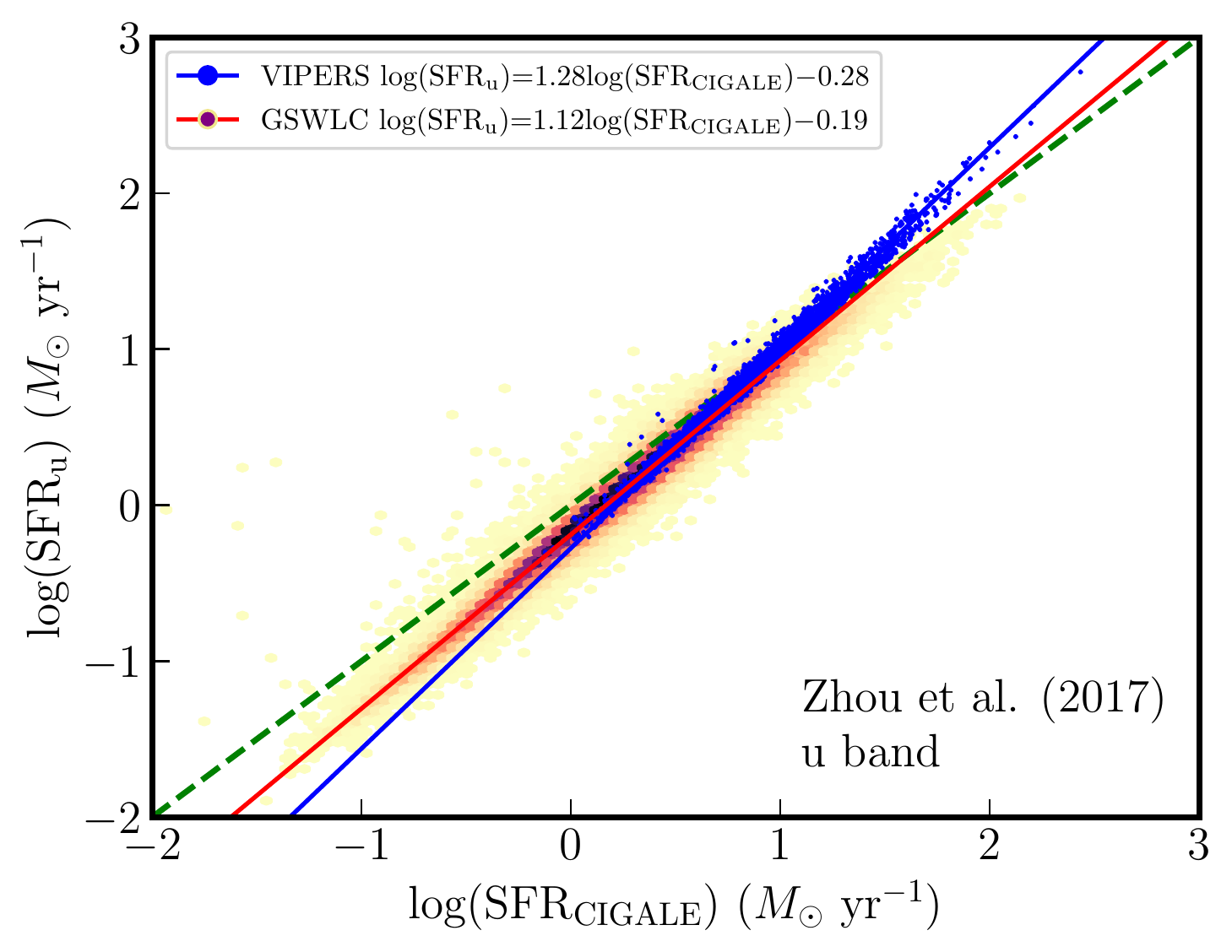}\includegraphics[angle=0,width=0.43\textwidth]{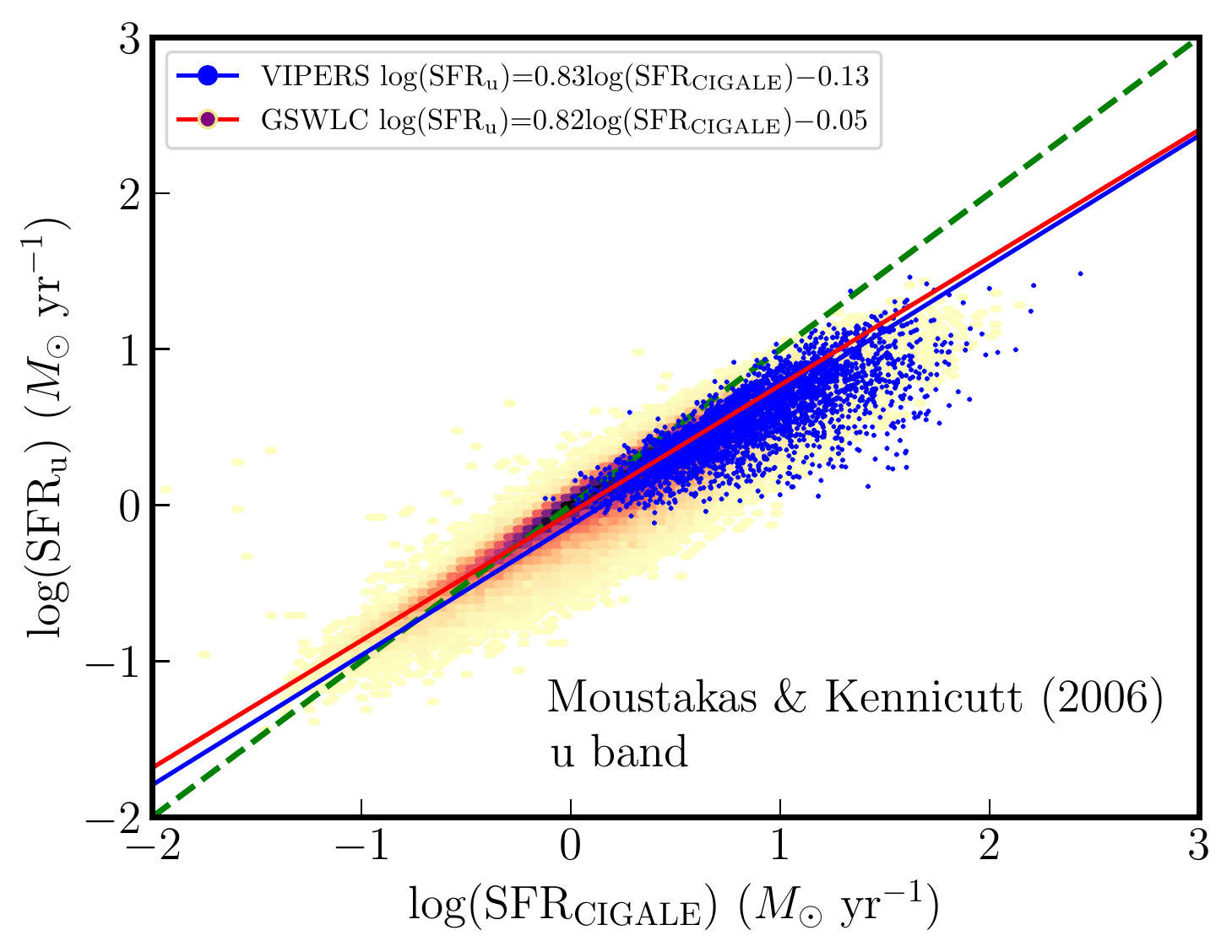}
\includegraphics[angle=0,width=0.43\textwidth]{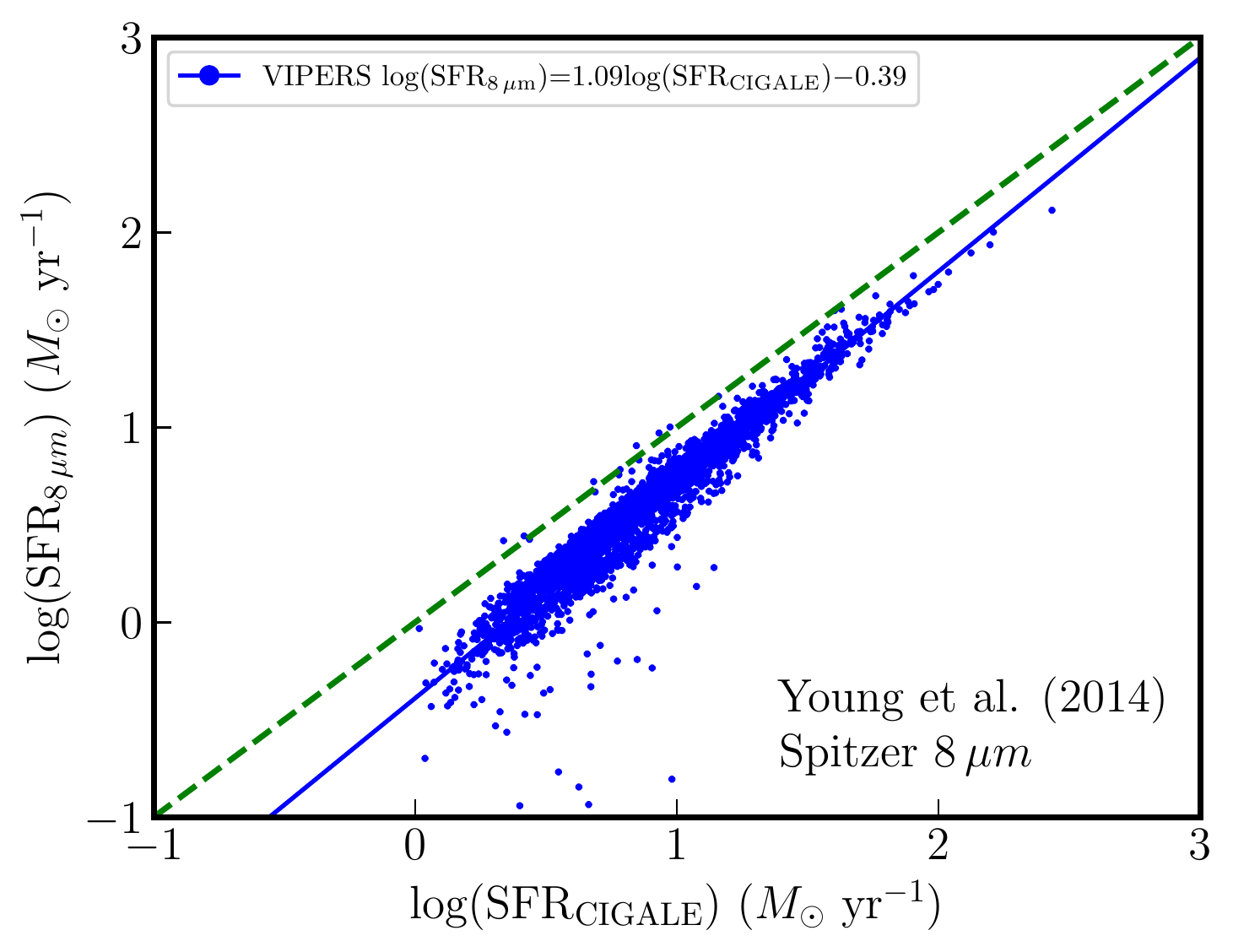}\includegraphics[angle=0,width=0.43\textwidth]{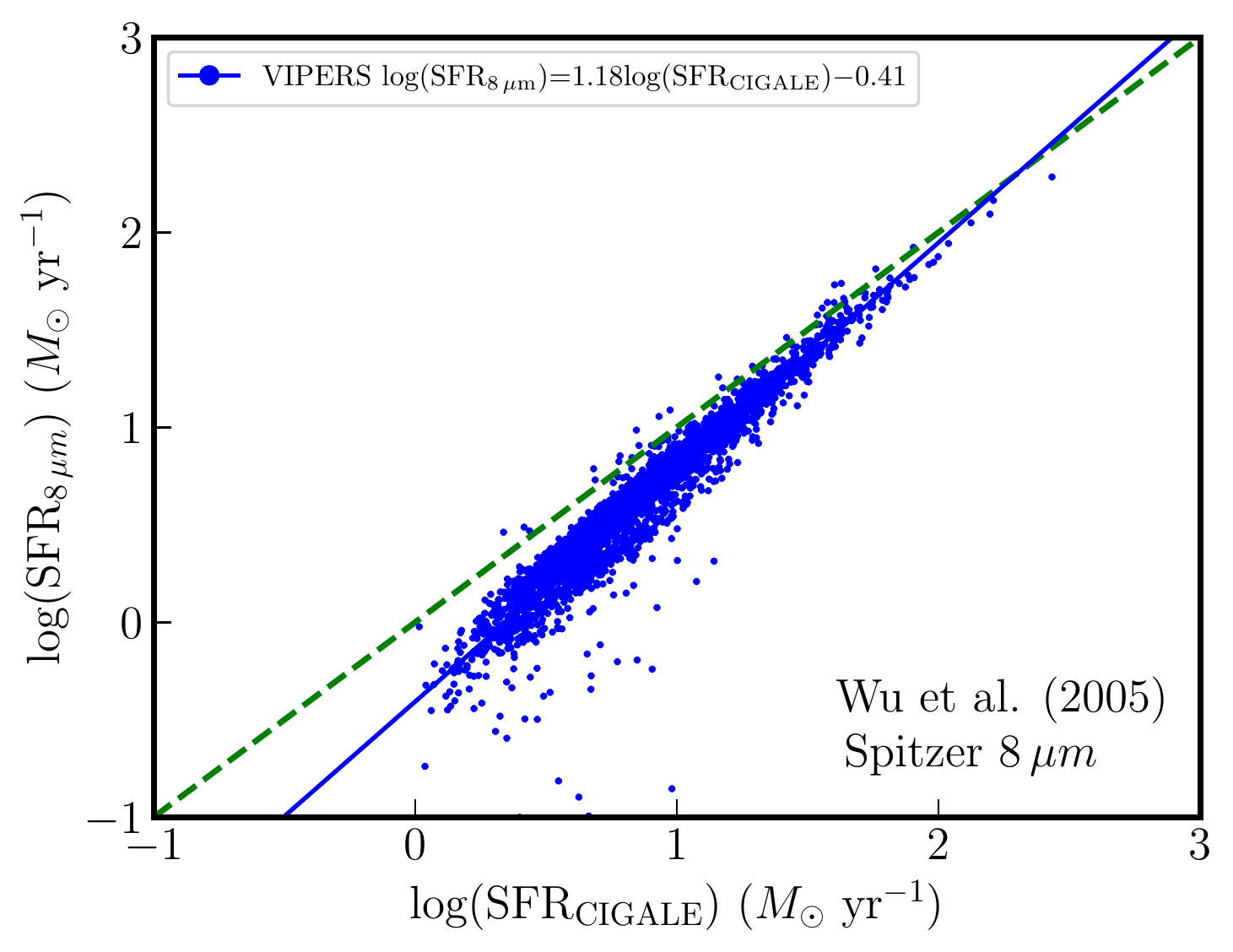}
\caption{SFRs derived using different calibrations (written in the bottom right in each plot) for VIPERS (blue points) and GSWLC (density pot). The dashed blue, red, and green lines are the fit for VIPERS, GSWLC, and the one-to-one relation, respectively.} 
\label{Fig_NUV_Ros02}
\end{figure*} 

\begin{figure*}[h!]
\centering
\includegraphics[angle=0,width=0.43\textwidth]{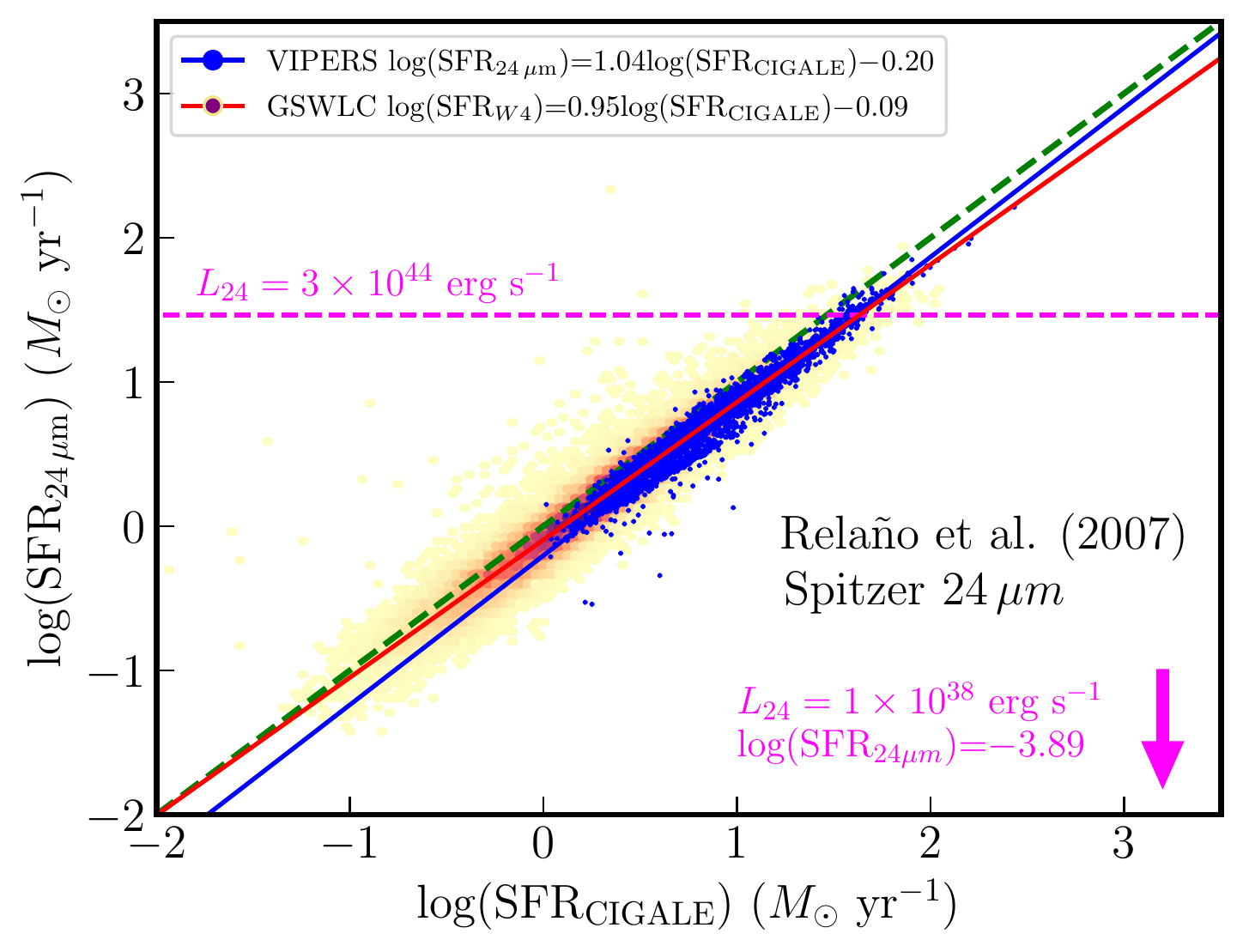}\includegraphics[angle=0,width=0.43\textwidth]{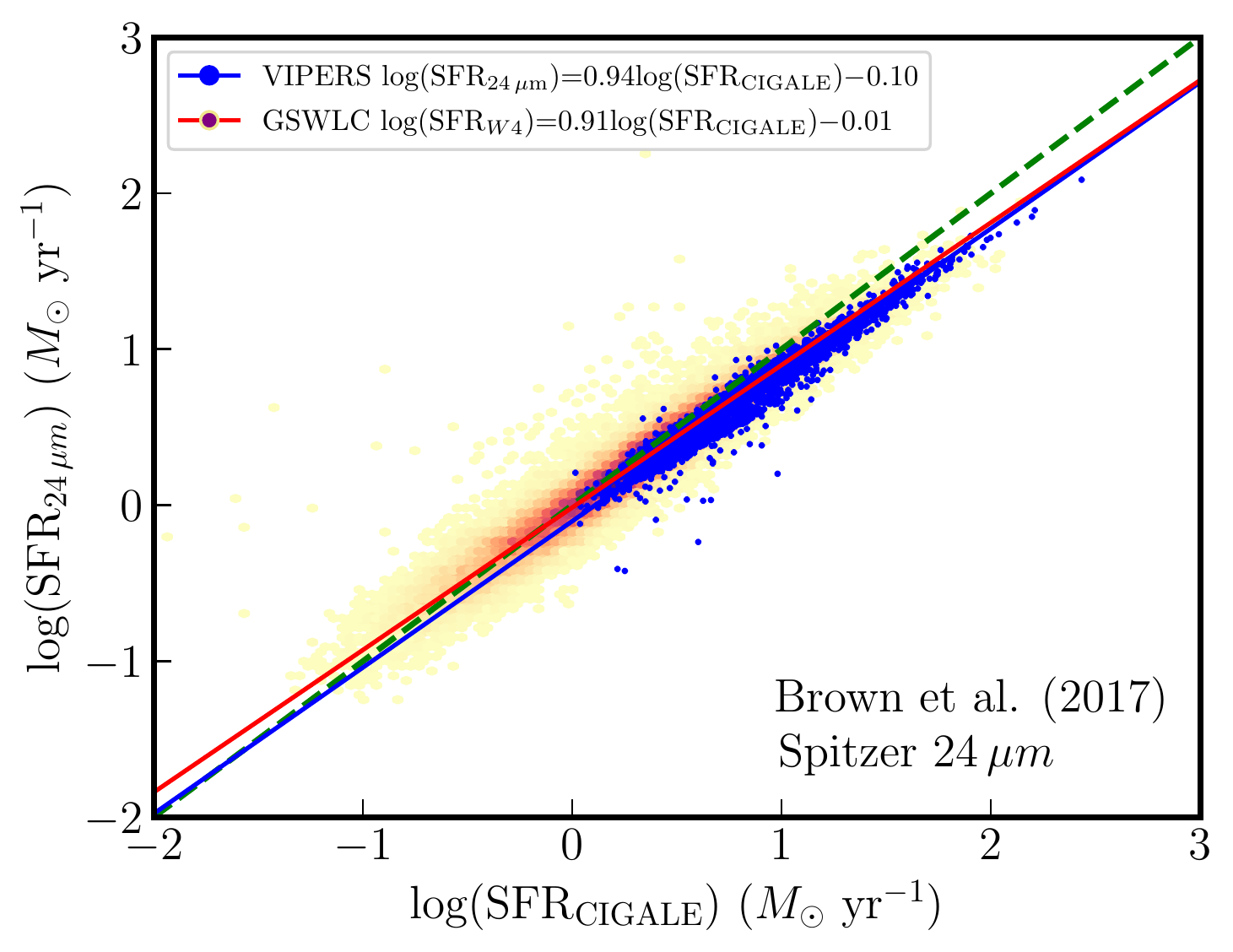}

\includegraphics[angle=0,width=0.43\textwidth]{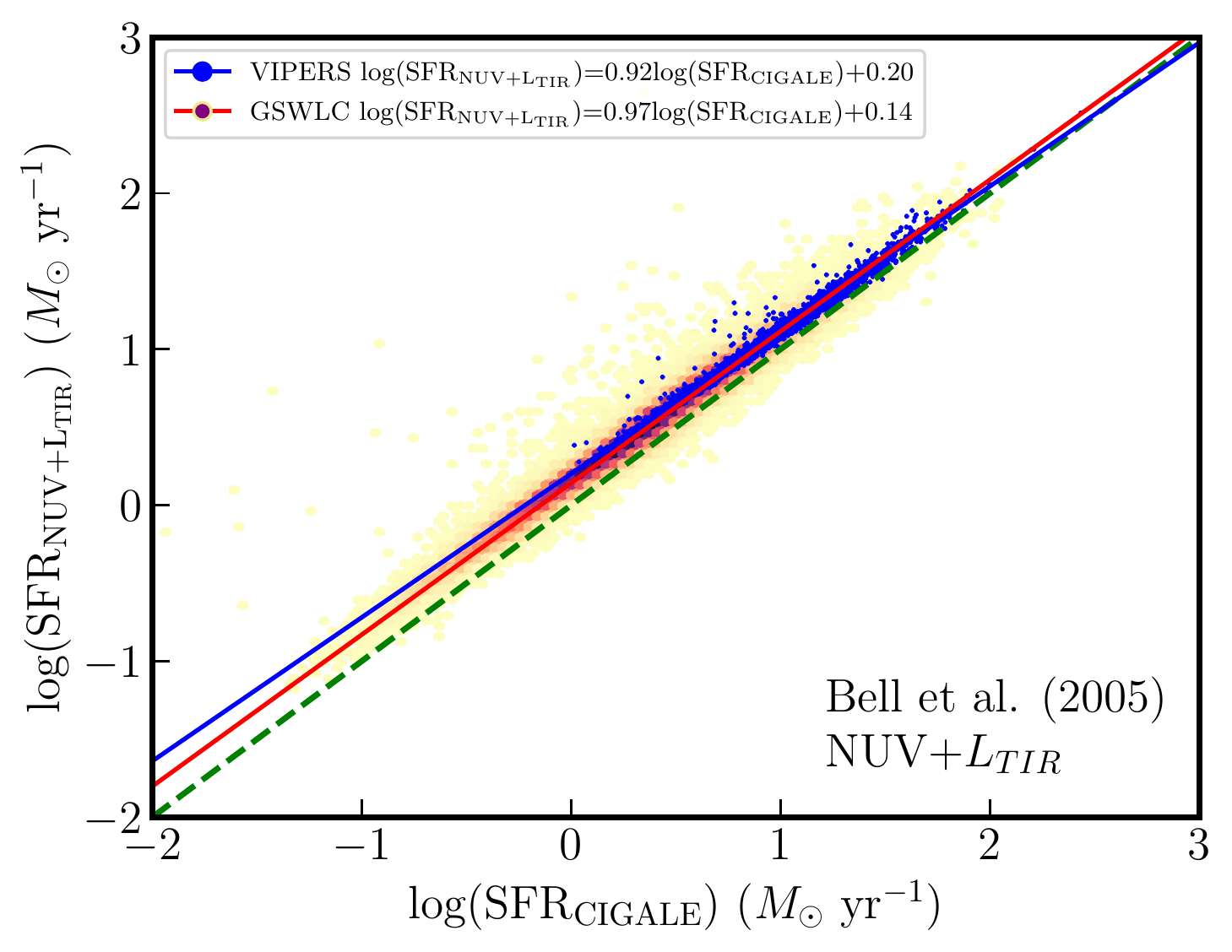}\includegraphics[angle=0,width=0.43\textwidth]{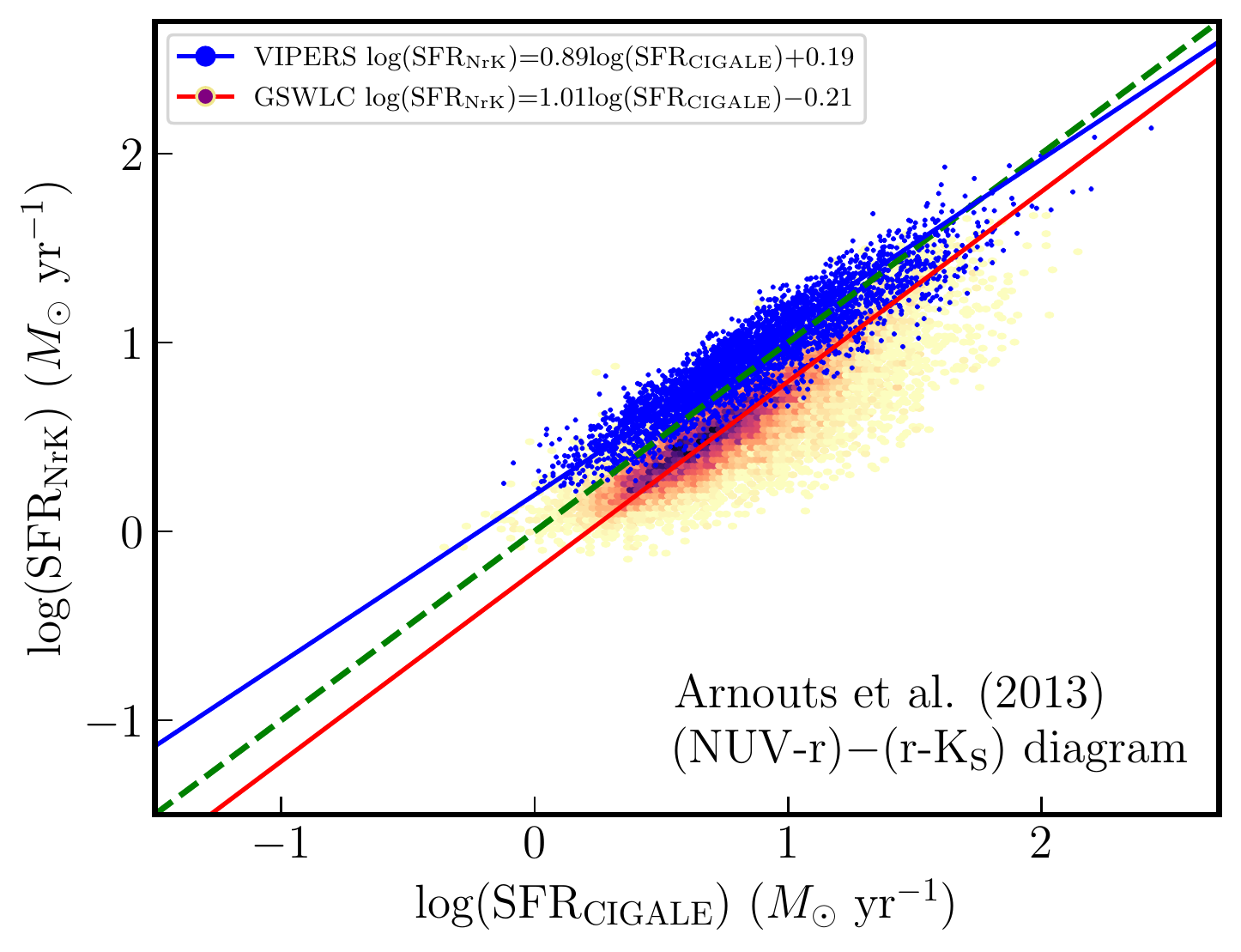}

\centering
\includegraphics[angle=0,width=0.43\textwidth]{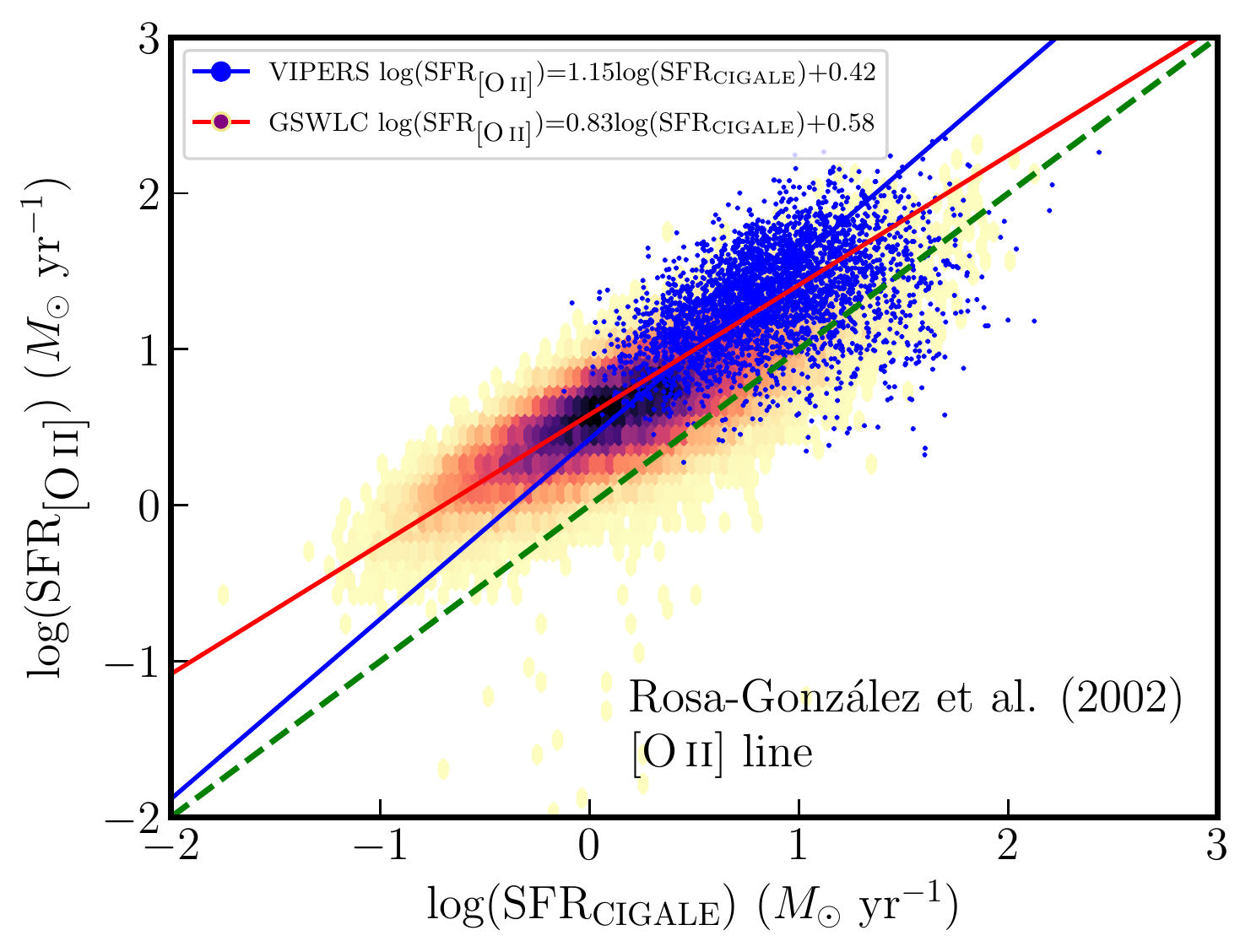}\includegraphics[angle=0,width=0.43\textwidth]{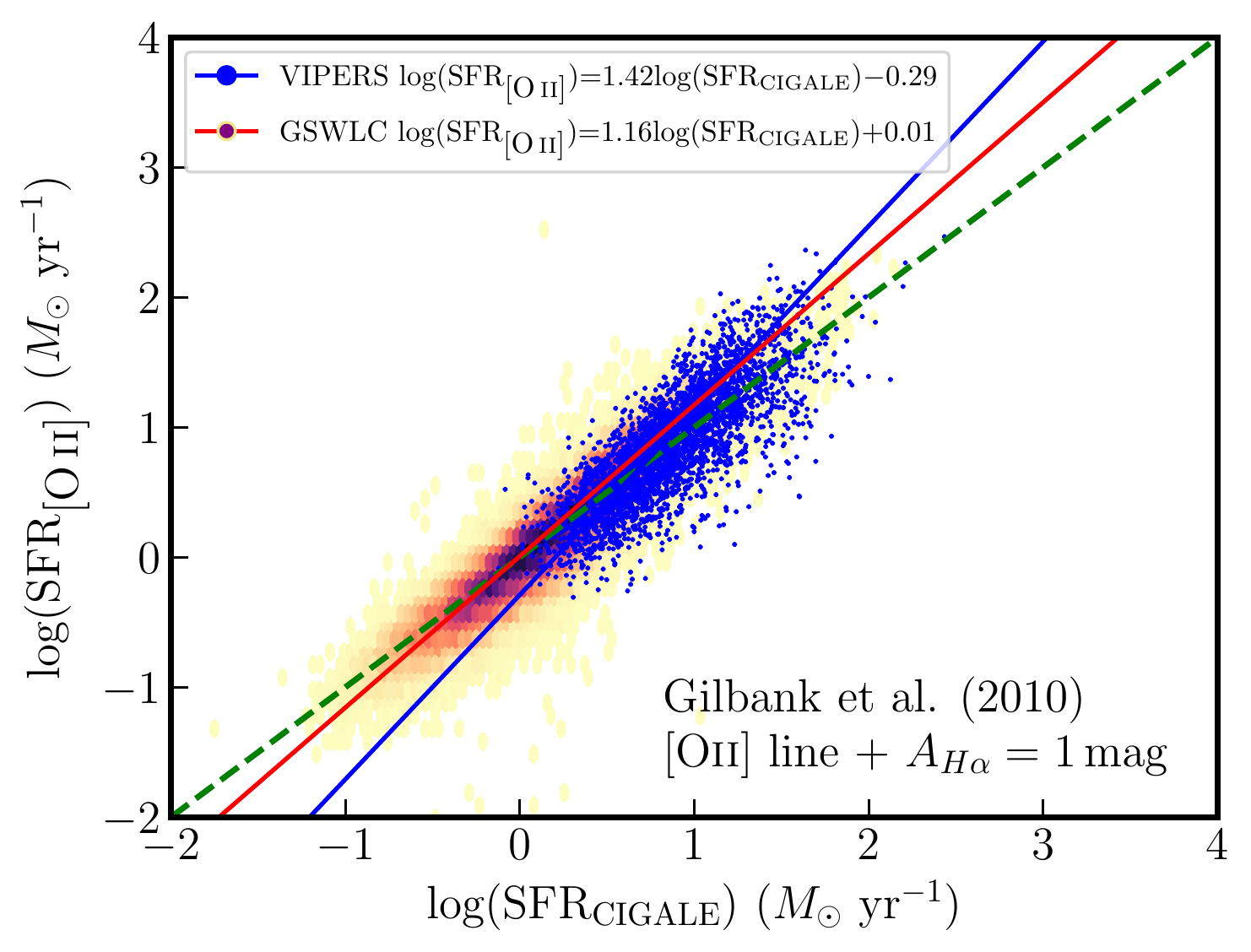}
\caption{Same as for Fig.~\ref{Fig_NUV_Ros02}. The pink arrow indicates that the lowest $L_{24}$ used to calibrate the \citet{rel07} relation is out of the range of the image. For GSWLC, the WISE-4 luminosity is converted to a Spitzer 24-$\mu$m luminosity.} 
\label{Fig_Cal_Brown}
\end{figure*}

\clearpage

\section{WISE-3, WISE-4, and \textit{Spitzer} 24-$\mu$m as SFR calibrators at $0<z<0.9$}\label{Subsect:wise4}
\begin{figure}[h!]
\includegraphics[width=0.5\linewidth]{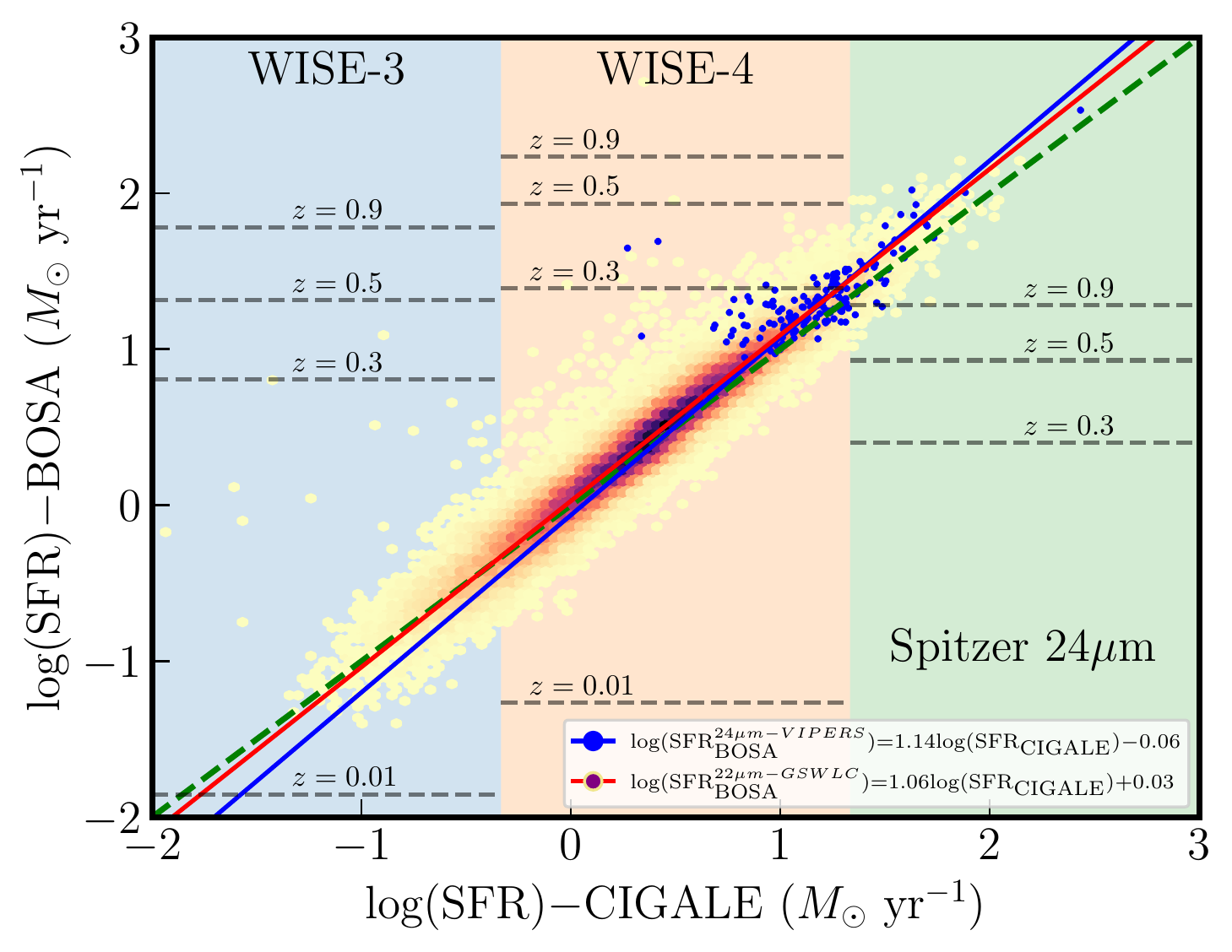}
\caption{Comparison of SFRs from the observed fluxes of \textit{Spitzer} (blue, VIPERS) and WISE-4 (red, GSWLC) using the BOSA templates with SFRs derived from CIGALE. The limits, computed for the redshift limits of VIPERS and GSWLC, above which WISE-3 (blue area), WISE-4 (orange area), and \textit{Spitzer} 24~$\mu$m (green area) can be used to reliably estimate the SFR are shown (dashed black lines). In units of log(SFR [$M_{\odot}$~yr$^{-1}$]), the limits, for increasing redshift, are: -1.9, 0.8, 1.31, and 1.8 (WISE-3), -1.3, 1.4, 1.9, and 2.2 (WISE-4), and -2.2, 0.4, 0.9, and 1.3 (\textit{Spitzer} 24~$\mu$m).}
\label{Fig:SFR_WISE_Spitzer}
\end{figure}

The SFR calibrations based on the monochromatic luminosity around 20~$\mu$m give a good estimation of the SFR because the luminosity at this wavelength traces reasonably well $L_{TIR}$, which is itself a good SFR tracer. Since most of the WISE-4 measurements are upper limits at $z>0.5$, we did not calibrate the WISE-4 band as previously done in the local Universe (e.g., \citealt{clu14,clu17}.\\ Using SED templates, it is possible to obtain reliable measurements of $L_{TIR}$ and the SFR based on the observed flux in the WISE-3, WISE-4, and the \textit{Spitzer} 24-$\mu$m band. Figure~\ref{Fig:SFR_WISE_Spitzer} shows the SFR estimated from the observed flux from WISE-4 (from unWISE) for GSWLC and \textit{Spitzer} 24~$\mu$m for VIPERS, using the BOSA templates (\citealt{boq21}, valid up to $z=4$). Flux uncertainties are propagated through the templates to estimate the uncertainties on the SFR. For GSWLC, the agreement with CIGALE SFR is good, with a mean and scatter of 0.05 and 0.13~dex. Using \textit{Spitzer} 24-$\mu$m measurements, the mean and scatter are equal to 0.02 and 0.20~dex for VIPERS galaxies, but only galaxies for which $L_{TIR}>3.5\times 10^{10}$~$L_{\odot}$ (estimated with CIGALE) are detected at 24~$\mu$m.\\ 
To estimate at which $L_{TIR}$, WISE-3, WISE-4 and \textit{Spitzer} 24-$\mu$m measurements can be reliably used as SFR tracers, we used the BOSA templates to derive the $3\sigma$ detection limit at the redshift limits of GSWLC and VIPERS. Horizontal dashed lines in Figure~\ref{Fig:SFR_WISE_Spitzer} show the $3\sigma$ limits at $z=0.01,~0.3,~0.5,$ and $0.9$ in terms of log(SFR), above which each of the bands can be used as an SFR tracer. The derivation of SFR from \textit{Spitzer} 24~$\mu$m at $z=0.01-0.3$ can be performed for most of the galaxies ($3.9\times 10^{7}-2.6\times 10^{10}$~$L_{\odot}$), although it is reliable only for luminous galaxies ($9.7\times 10^{10}-2.4\times 10^{11}$~$L_{\odot}$) at $z=0.5-0.9$. On the contrary, the sensitivity of the WISE-4 band is too low, and only very luminous galaxies ($1.2\times 10^{12}-2.6\times 10^{12}$~$L_{\odot}$) can be observed at $z=0.5-0.9,$ while this band remains a good SFR tracer for GSWLC ($4.1\times 10^{8}-3.1\times 10^{11}$~$L_{\odot}$). The sensitivity of WISE-3 is in between \textit{Spitzer} 24~$\mu$m and WISE-4 but this band remains unuseful for most of the VIPERS galaxies ($2.6\times 10^{11}-8.3\times 10^{11}$~$L_{\odot}$), while the SFR of most of the GSWLC galaxies can be derived using this band ($9.3\times 10^{7}-7.2\times 10^{10}$~$L_{\odot}$). Good detections (i.e., 3$\sigma$) of all galaxies in our VIPERS sample ($L>1.9\times 10^{9}$~$L_{\odot}$) would require a $3\sigma$ sensitivity at 12 and 24~$\mu$m around 0.58 and 1.7~$\mu$Jy, respectively.

\newpage
\renewcommand{\arraystretch}{1.3}
\section{Parameters of the fit performed in this work}\label{Appendix:Fit_parameters}
{
\small
\begin{longtable}{l|l|r|r|r|r|r|r|r|c}
\caption{Parameters and properties of each fit presented in this work.}\\
Reference & Catalog $-$ Band & \multicolumn{1}{|c|}{N} & \multicolumn{1}{|c|}{m} & \multicolumn{1}{|c|}{b}& Pearson & Mean &
Scatter & CCC & CCC$_{GV}$ \\
(1)&(2)&(3)&(4)&(5)&(6)&(7)&(8)&(9)&(10)\\
\hline
\hline
\endfirsthead
\caption[]{(continued)}\\
\hline\hline
Reference & Catalog $-$ Band & \multicolumn{1}{|c|}{N} & \multicolumn{1}{|c|}{m} & \multicolumn{1}{|c|}{b}& Pearson & Mean &
Scatter & CCC & CCC$_{GV}$ \\
(1)&(2)&(3)&(4)&(5)&(6)&(7)&(8)&(9)&(10)\\
\hline
\hline
\endhead
\multicolumn{10}{c}{\textbf{FUV}} \\
\hline
\multirow{4}{*}{\citet{bro17}\tablefootmark{$2$}}&V $-$ (Calzetti)& 3~457 & 1.21 & $-$0.28 & 0.99 & 0.09 & 0.10 & 0.94 & \\ 
&G $-$ (Calzetti)& 91~533 & 0.95 & $-$0.06 & 0.83 & 0.02 & 0.30 & 0.83 & \multirow{-2}{*}{0.91}\\ 
&\cellcolor{blue!25}V $-$ (Hao)& \cellcolor{blue!25}3~457 & \cellcolor{blue!25}1.05 & \cellcolor{blue!25}$-$0.04 & \cellcolor{blue!25}0.99 & \cellcolor{blue!25}0.01 & \cellcolor{blue!25}0.05 & \cellcolor{blue!25}0.99 & \cellcolor{magenta!40}\\ 
&\cellcolor{blue!25}G $-$ (Hao)& \cellcolor{blue!25}91~533 & \cellcolor{blue!25}0.95 & \cellcolor{blue!25}0.02 & \cellcolor{blue!25}0.88 & \cellcolor{blue!25}$-$0.02 & \cellcolor{blue!25}0.25 & \cellcolor{blue!25}0.88 & \multirow{-2}{*}{\cellcolor{magenta!40}0.94}\\ 
\hline
\multirow{2}{*}{\citet{dav16}}&V & 3~457 & 0.82 & 0.0 & 0.99 & 0.16 & 0.07 & 0.87 & \multirow{2}{*}{0.88}\\ 
&G & 91~533 & 0.67 & 0.12 & 0.85 & $-$0.07 & 0.25 & 0.82 & \\
\hline
\multirow{2}{*}{\citet{sal07}} &\cellcolor{blue!25}V & \cellcolor{blue!25}3~457 & \cellcolor{blue!25}1.0 & \cellcolor{blue!25}$-$0.01 & \cellcolor{blue!25}1.00 & \cellcolor{blue!25}0.02 & \cellcolor{blue!25}0.03 & \cellcolor{blue!25}0.99 &\cellcolor{magenta!40} \\ 
&G & 91~533  & 0.88 & 0.09 & 0.87 & $-$0.08 & 0.24 & 0.86 & \multirow{-2}{*}{\cellcolor{magenta!40}0.94}\\
\hline
\multicolumn{10}{c}{\textbf{NUV}} \\
\hline
\multirow{2}{*}{\citet{dav16}}&V & 3~457 & 0.80 & $-$0.13 & 0.97 & 0.29 & 0.10 & 0.67 & \multirow{2}{*}{0.79}\\
&G & 91~533 & 0.59 & 0.07 & 0.80 & $-$0.01 & 0.29 & 0.76 &\\
\hline
\multirow{2}{*}{\citet{sal07}}&\cellcolor{blue!25}V & \cellcolor{blue!25}3~457 & \cellcolor{blue!25}1.0 & \cellcolor{blue!25}$-$0.01 & \cellcolor{blue!25}1.00 & \cellcolor{blue!25}0.02 & \cellcolor{blue!25}0.03 & \cellcolor{blue!25}0.99 & \cellcolor{magenta!40}\\
&\cellcolor{blue!25}G & \cellcolor{blue!25}91~533 & \cellcolor{blue!25}0.89 & \cellcolor{blue!25}0.09 & \cellcolor{blue!25}0.87 & \cellcolor{blue!25}$-$0.08 & \cellcolor{blue!25}0.24 & \cellcolor{blue!25}0.86 &\multirow{-2}{*}{\cellcolor{magenta!40}0.93}\\
\hline
\multirow{2}{*}{\citet{ros02}} &V & 3~457 & 1.09 & 0.08 & 0.47 & $-$0.11 & 0.32 & 0.43 & \multirow{2}{*}{0.78}\\
&G & 91~533 & 0.83 & 0.33 & 0.86 & $-$0.20 & 0.25 & 0.77 &\\
\hline
\multicolumn{10}{c}{\textbf{\textit{u}-band}}\\
\hline
\multirow{2}{*}{\citet{dav16}\tablefootmark{$3$}}&V & 3~457 & 0.86 & $-$0.05 & 0.99 & 0.18 & 0.06 & 0.85 & \multirow{2}{*}{0.81}\\
&G  & 90~965\tablefootmark{$4$} & 0.83 & $-$0.33 & 0.89 & 0.42 & 0.22 & 0.62 &\\
\hline
\multirow{2}{*}{\citet{hop03}}&V & 3~457 & 1.23 & $-$0.26 & 0.99 & 0.07 & 0.08 & 0.96 & \cellcolor{magenta!40}\\
&\cellcolor{blue!25}G & \cellcolor{blue!25}91~533 & \cellcolor{blue!25}1.06 & \cellcolor{blue!25}$-$0.17 & \cellcolor{blue!25}0.99 & \cellcolor{blue!25}0.15 & \cellcolor{blue!25}0.09 & \cellcolor{blue!25}0.94 & \multirow{-2}{*}{\cellcolor{magenta!40}0.97}\\
\hline
\multirow{2}{*}{\citet{zho17}}&\cellcolor{blue!25}V & \cellcolor{blue!25}3~457 & \cellcolor{blue!25}1.28 & \cellcolor{blue!25}$-$0.28 & \cellcolor{blue!25}0.99 & \cellcolor{blue!25}0.04 & \cellcolor{blue!25}0.10 & \cellcolor{blue!25}0.97 & \multirow{2}{*}{0.96}\\ 
&G & 91~533 & 1.12 & $-$0.19 & 0.99 & 0.16 & 0.10 & 0.93 &\\ 
\hline
\multirow{2}{*}{\citet{mou06b}}&V & 3~457 & 0.83 & $-$0.13 & 0.79 & 0.31 & 0.21 & 0.50 & \multirow{2}{*}{0.80}\\
&G & 91~533 & 0.82 & $-$0.05 & 0.93 & 0.14 & 0.18 & 0.87 &\\
\hline
\multirow{2}{*}{\citet{dav16}}&V & 3~457 & 0.96 & $-$0.12 & 0.99 & 0.17 & 0.04 & 0.88 & \\
&\cellcolor{blue!25}G & \cellcolor{blue!25}91~533 & \cellcolor{blue!25}0.83 & \cellcolor{blue!25}$-$0.05 & \cellcolor{blue!25}0.99 & \cellcolor{blue!25}0.09 & \cellcolor{blue!25}0.11 & \cellcolor{blue!25}0.94 & \multirow{-2}{*}{0.95}\\
\hline
\multicolumn{10}{c}{\textbf{8~$\mu$m}}\\
\hline
\citet{bro17}&\cellcolor{blue!25}V & \cellcolor{blue!25}3~385\tablefootmark{$5$} & \cellcolor{blue!25}0.91 & \cellcolor{blue!25}$-$0.03 & \cellcolor{blue!25}0.96 & \cellcolor{blue!25}0.10 & \cellcolor{blue!25}0.10 & \cellcolor{blue!25}0.91 &\\ 
\hline
\citet{per06}&V & 3~385\tablefootmark{$5$} & 1.03 & $-$0.33 & 0.96 & 0.29 & 0.11 & 0.71 &\\
\hline
\citet{you14}&V & 3~385\tablefootmark{$5$} & 1.09 & $-$0.39 & 0.96 & 0.30 & 0.12 & 0.72 &\\
\hline
\citet{wu05}&V & 3~385\tablefootmark{$5$} & 1.18 & $-$0.41 & 0.96 & 0.24 & 0.14 & 0.78 &\\
\hline
\multicolumn{10}{c}{\textbf{24~$\mu$m}}\\
\hline
\multirow{2}{*}{\citet{rie09}}&V & 3~385\tablefootmark{$5$} & 1.28 & $-$0.28 & 0.97 & 0.004 & 0.13 & 0.95 & \multirow{2}{*}{0.96}\\
&G & 68~911\tablefootmark{$6$} & 1.16 & $-$0.13 & 0.97 & 0.07 & 0.16 & 0.94 & \\
\hline
\multirow{2}{*}{\citet{zhu08}} &\cellcolor{blue!25}V & \cellcolor{blue!25}3~385\tablefootmark{$5$} & \cellcolor{blue!25}1.06 & \cellcolor{blue!25}$-$0.11 & \cellcolor{blue!25}0.97 & \cellcolor{blue!25}0.04 & \cellcolor{blue!25}0.08 & \cellcolor{blue!25}0.97 & \cellcolor{magenta!40}\\
&\cellcolor{blue!25}G & \cellcolor{blue!25}68~911\tablefootmark{$6$} & \cellcolor{blue!25}0.98 & \cellcolor{blue!25}0.0 & \cellcolor{blue!25}0.97 & \cellcolor{blue!25}0.01 & \cellcolor{blue!25}0.12 & \cellcolor{blue!25}0.97 &\multirow{-2}{*}{\cellcolor{magenta!40}0.98}\\
\hline
\multirow{2}{*}{\citet{rel07}}&V & 3~385\tablefootmark{$5$}  & 1.04 & $-$0.20 & 0.97 & 0.15 & 0.08 & 0.89 & \multirow{2}{*}{0.94}\\
&G & 68~911\tablefootmark{$6$} & 0.95 & $-$0.09 & 0.97 & 0.11 & 0.12 & 0.94 &\\
\hline
\multirow{2}{*}{\citet{bro17}}&V & 3~385\tablefootmark{$5$} & 0.94 & $-$0.10 & 0.98 & 0.14 & 0.08 & 0.89 & \multirow{2}{*}{0.95}\\ 
&G & 68~911\tablefootmark{$6$} & 0.91 & $-$0.01 & 0.97 & 0.04 & 0.12 & 0.96 &\\ 
\hline
\multicolumn{10}{c}{\textbf{$L_{\rm{\textbf{TIR}}}$}} \\
\hline
\multirow{2}{*}{\citet{ken98a}}&V & 3~385\tablefootmark{$5$} & 1.22 & $-$0.26 & 0.98 & 0.04 & 0.11 & 0.95 & \multirow{2}{*}{0.97}\\
&G & 68~911\tablefootmark{$6$} & 1.08 & 0.03 & 0.96 & 0.04 & 0.13 & 0.96 & \\
\hline
\multicolumn{10}{c}{\textbf{Composite tracers}} \\
\hline
\multirow{5}{*}{\citet{ken09}}&V $-$ [O\,{\sc{ii}}]+8~$\mu$m & 3~385\tablefootmark{$5$} & 0.93 & $-$0.04 & 0.96 & 0.12 & 0.09 & 0.91 &\\
&V $-$ [O\,{\sc{ii}}]+24~$\mu$m & 3~385\tablefootmark{$5$} & 1.0 & 0.04 & 0.98 & $-$0.04 & 0.06 & 0.98 & \\
&G $-$ [O\,{\sc{ii}}]+WISE-4 & 35~922\tablefootmark{$6,7$} & 1.0 & $-$0.01 & 0.97 & 0.02 & 0.10 & 0.97 &\multirow{-2}{*}{0.98}\\
&V $-$ [O\,{\sc{ii}}]+$L_{TIR}$ & 3~385\tablefootmark{$5$}& 0.98 & 0.0 & 0.98 & 0.01 & 0.07 & 0.98 & \multirow{2}{*}{0.96}\\
&G $-$ [O\,{\sc{ii}}]+$L_{TIR}$ & 35~922\tablefootmark{$6,7$} & 1.05 & 0.08 & 0.92 & $-$0.06 & 0.18 & 0.91 &\\
\hline
\multirow{2}{*}{\citet{arn13}}&V $-$ NUVrK$_{\rm{s}}$ & 3~457 & 0.89 & 0.19 & 0.92 & $-$0.07 & 0.14 & 0.89 & \multirow{2}{*}{0.80}\\
&G $-$ NUVrK$_{\rm{s}}$& 19~438\tablefootmark{$8$} & 1.01 & $-$0.21 & 0.87 & 0.21 & 0.15 & 0.68 & \\
\hline
\multirow{2}{*}{\citet{bel05}}&V $-$ NUV+$L_{TIR}$ & 3~385\tablefootmark{$5$} & 0.92 & 0.20 & 0.99 & $-$0.13 & 0.04 & 0.92 & \multirow{2}{*}{0.97}\\ 
&G $-$ NUV+$L_{TIR}$ & 68~911\tablefootmark{$6$} & 0.97 & 0.14 & 0.98 & $-$0.14 & 0.09 & 0.93 & \\
\hline
\multirow{2}{*}{\citet{cla15}}&V $-$ FUV+24~$\mu$m & 3~385\tablefootmark{$5$} & 1.01 & 0.07 & 1.0 & $-$0.08 & 0.03 & 0.97 & \multirow{2}{*}{0.97}\\
&G $-$ FUV+WISE-4 & 68~911\tablefootmark{$6$} & 1.01 & $-$0.11 & 0.98 & 0.12 & 0.10 & 0.95 & \\
\hline
\multirow{2}{*}{\citet{boq14}}&\cellcolor{blue!25}V $-$ FUV+24~$\mu$m & \cellcolor{blue!25}3~385\tablefootmark{$5$}& \cellcolor{blue!25}1.01 & \cellcolor{blue!25}$-$0.01 & \cellcolor{blue!25}1.00 & \cellcolor{blue!25}$-$0.01 & \cellcolor{blue!25}0.03 & \cellcolor{blue!25}0.99 & \cellcolor{magenta!40}\\
&\cellcolor{blue!25}G $-$ FUV+WISE-4 & \cellcolor{blue!25}68~911\tablefootmark{$6$} & \cellcolor{blue!25}1.07 & \cellcolor{blue!25}$-$0.05 & \cellcolor{blue!25}0.98 & \cellcolor{blue!25}0.02 & \cellcolor{blue!25}0.10 & \cellcolor{blue!25}0.98 & \multirow{-2}{*}{\cellcolor{magenta!40}0.99}\\
\hline
\multicolumn{9}{c}{} \\
\hline
\multicolumn{10}{c}{\textbf{Spectral lines}} \\
\hline
\multirow{2}{*}{\citet{ken98a}}&V $-$ H$\beta$ & 3~457 & 1.26 & $-$0.25 & 0.88 & $-$0.01 & 0.19 & 0.87 & \multirow{2}{*}{0.94}\\
&G $-$ H$\beta$ & 91~533 & 1.04 & $-$0.08 & 0.94 & 0.07 & 0.19 & 0.92 &\\
\hline
\multirow{6}{*}{\citet{ken98a}}&V $-$ [O\,{\sc{ii}}] & 3~457 & 1.03 & 0.12 & 0.85 & $-$0.11 & 0.19 & 0.81 & \multirow{2}{*}{0.91}\\
&G $-$ [O\,{\sc{ii}}] & 48~845\tablefootmark{$7$} & 0.84 & 0.01 & 0.86 & 0.08 & 0.23 & 0.84 &\\
&V $-$ [O\,{\sc{ii}}] Reddening & 3~457 & 1.12 & $-$0.11 & 0.90 & 0.01 & 0.17 & 0.89 & \multirow{2}{*}{0.93}\\
&G $-$ [O\,{\sc{ii}}] Reddening & 48~845\tablefootmark{$7$} & 0.92 & $-$0.10 & 0.88 & 0.12 & 0.22 & 0.86 & \\
&\cellcolor{blue!25}V $-$ [O\,{\sc{ii}}] Metal & \cellcolor{blue!25}3~192\tablefootmark{$9,10$} & \cellcolor{blue!25}1.25 & \cellcolor{blue!25}$-$0.27 & \cellcolor{blue!25}0.93 & \cellcolor{blue!25}0.05 & \cellcolor{blue!25}0.17 & \cellcolor{blue!25}0.90 & \cellcolor{magenta!40}\\
&G $-$ [O\,{\sc{ii}}] Metal & 35~403\tablefootmark{$9,10,11$} & 0.94 & $-$0.07 & 0.91 & 0.06 & 0.20 & 0.90 & \multirow{-2}{*}{\cellcolor{magenta!40}0.94}\\
\hline
\multirow{4}{*}{\citet{ken98a}}&V $-$ [O\,{\sc{iii}}] & 3~457 & 1.20 & $-$0.14 & 0.65 & $-$0.04 & 0.31 & 0.64 & \multirow{2}{*}{0.81}\\
&G $-$ [O\,{\sc{iii}}] & 91~533 & 0.88 & 0.07 & 0.78 & $-$0.17 & 0.32 & 0.73 & \\
&V $-$ [O\,{\sc{iii}}] Metal & 3~192\tablefootmark{$9,10$} & 1.25 & $-$0.22 & 0.77 & 0.05 & 0.26 & 0.75 & \cellcolor{magenta!40}\\
&\cellcolor{blue!25}G $-$ [O\,{\sc{iii}}] Metal & \cellcolor{blue!25}35~396\tablefootmark{$9,10,11$} & \cellcolor{blue!25}1.16 & \cellcolor{blue!25}$-$0.05 & \cellcolor{blue!25}0.93 & \cellcolor{blue!25}0.04 & \cellcolor{blue!25}0.20 & \cellcolor{blue!25}0.92 & \multirow{-2}{*}{\cellcolor{magenta!40}0.91}\\
\hline
\multirow{2}{*}{\citet{vil21}}&\cellcolor{blue!25}V $-$ [O\,{\sc{iii}}] & \cellcolor{blue!25}3~382\tablefootmark{$9$} & \cellcolor{blue!25}1.10 & \cellcolor{blue!25}0.04 & \cellcolor{blue!25}0.79 & \cellcolor{blue!25}0.09 & \cellcolor{blue!25}0.24 & \cellcolor{blue!25}0.76 & \multirow{2}{*}{0.89}\\
&G $-$ [O\,{\sc{iii}}] & 35~451\tablefootmark{$9$} & 0.95 & $-$0.13 & 0.87 & $-$0.19 & 0.24 & 0.80\\
\hline
\multirow{2}{*}{\citet{ros02}}&V $-$ [O\,{\sc{ii}}] & 3~457 & 1.15 & 0.42 & 0.44 & $-$0.46 & 0.35 & 0.22 & \multirow{2}{*}{0.57}\\
&G $-$ [O\,{\sc{ii}}] & 48~845\tablefootmark{$7$} & 0.83 & 0.58 & 0.76 & $-$0.44 & 0.30 & 0.48 &\\
\hline
\multirow{2}{*}{\citet{gil10}}&V $-$ [O\,{\sc{ii}}] & 3~457 & 1.42 & $-$0.29 & 0.80 & $-$0.00 & 0.26 & 0.78 & \multirow{2}{*}{0.91}\\
&\cellcolor{blue!25}G $-$ [O\,{\sc{ii}}] & \cellcolor{blue!25}48~845\tablefootmark{$7$} & \cellcolor{blue!25}1.16 & \cellcolor{blue!25}0.01 & \cellcolor{blue!25}0.92 & \cellcolor{blue!25}0.01 & \cellcolor{blue!25}0.22 & \cellcolor{blue!25}0.91 & \\
\hline
\hline
\end{longtable}
\tablefoot{(1) Reference to the literature, (2) Catalog used (V: VIPERS, G: GSWLC) and bands (for composite and spectral line tracers) used to derive the SFR, (3) Number of galaxies used to perform the fit, (4) and (5) coefficients of the linear fit ($y=mx+b$)\tablefootmark{$1$}, (6) Pearson coefficient, (7) Mean, (8) Scatter estimated from calibrations; additional scatter from template bias can be found in Sect.~\ref{Subsect:absoluteluminosity}, (9) Concordance correlation coefficient, and (10) Concordance correlation coefficient for VIPERS+GSWLC. For each band, the preferred calibration for GSWLC and VIPERS is shown in blue, and in magenta for GSWLC+VIPERS.\\
\tablefoottext{$1$}{Uncertainties on the coefficient m and b are $\sim$10$^{-2}$.}
\tablefoottext{$2$}{Using the \citet{cal00} and \citet{hao11} attenuation recipes.}
\tablefoottext{$3$}{Assuming an old-stellar population decontamination.}
\tablefoottext{$4$}{The decontamination cannot be applied to all GSWLC galaxies.}
\tablefoottext{$5$}{Galaxies whose SED fitting was incorrect in the IR were removed.}
\tablefoottext{$6$}{Galaxies with available $L_{W4}$.}
\tablefoottext{$7$}{[O\,{\sc{ii}}] sample (see
Tab.~\ref{tab:GSWLC_flags_selection}).}
\tablefoottext{$8$}{Galaxies at $z<0.1$ and with $NRK_{sSFR}$>1.9 were excluded (see Sect.~\ref{Sect:composite_tracers}).}
\tablefoottext{$9$}{Galaxies for which [O\,{\sc{iii}}]$\lambda$5007/[O\,{\sc{ii}}]$\lambda 3727<2$ (see Sect.~\ref{Subsect:OII}).}
\tablefoottext{$10$}{Galaxies for which log(O/H)+$12<8.4$ were excluded (see Sect.~\ref{Subsect:OII}).}
\tablefoottext{$11$}{[O\,{\sc{ii}}]+[O\,{\sc{iii}}] sample (see Tab.~\ref{tab:GSWLC_flags_selection}).}
}

}


\end{appendix}

\end{document}